%% file: JIOHS2023_senRev.tex
\documentclass[10pt,letterpaper,twocolumn]{article}

\usepackage{authblk}
\usepackage[margin=0.75in]{geometry}

\usepackage{amsfonts,amssymb,amsmath}
\usepackage{mathtools}
\usepackage{threeparttable}
\usepackage[hidelinks]{hyperref}
\usepackage{color}
\usepackage{graphicx}
\usepackage[section]{placeins}

\usepackage{layouts}
\usepackage{todonotes}
\usepackage{regexpatch}
\makeatletter
\xpatchcmd{\@todo}{\setkeys{todonotes}{#1}}{\setkeys{todonotes}{inline,#1}}{}{}
\makeatother

\usepackage{GB_glossaries}
\newacronym[type=symbolslist]{mtg}{\ensuremath{\overline{(\bar{t})}_{gates}}}{average of gate center times}
\newacronym[type=symbolslist]{dtg}{\ensuremath{\Delta (\bar{t})_{gates}}}{differance between gate center times}

\preto\subsection{\FloatBarrier}
\preto\subsubsection{\FloatBarrier}
\preto\paragraph{\FloatBarrier}
\preto\subparagraph{\FloatBarrier}

\usepackage{siunitx}
\DeclareSIUnit\year{yr}

\usepackage[super,sort]{natbib}
\newcommand{\tblCite}[1]{\citeauthor{#1}\space(\citeyear{#1})\cite{#1}}
\newcommand{\mapFig}{\textsuperscript{\textdagger}}
\newcommand{\volFig}{\textsuperscript{\textdaggerdbl}}
\newcommand{\as}[1]{\acrshort{#1}}

\newcommand{\ie}[1]{(\textit{i.e.,} #1)}

\usepackage[numbered, final]{matlab-prettifier}
\lstMakeShortInline"
\title{Spatial Sensitivity to Absorption Changes\\ for Various Near-Infrared Spectroscopy Methods:\\A Compendium Review}
\author{Giles~Blaney,\footnote{\href{mailto:Giles.Blaney@tufts.edu}{Giles.Blaney@tufts.edu}} Angelo~Sassaroli, \& Sergio~Fantini}
\affil{Department of Biomedical Engineering, Tufts University\\
	4 Colby St, Medford, MA 02155, USA}
\date{Compiled \today}

\begin{document}
\renewcommand*{\thefootnote}{\fnsymbol{footnote}}

\maketitle
\begin{abstract}
	This compendium review focuses on the spatial distribution of sensitivity to localized absorption changes in optically diffuse media, particularly for measurements relevant to near-infrared spectroscopy.
	The three temporal domains, continuous-wave, frequency-domain, and time-domain, each obtain different optical data-types whose changes may be related to effective homogeneous changes in the absorption coefficient. 
	Sensitivity is the relationship between a localized perturbation and the recovered effective homogeneous absorption change.
	Therefore, spatial sensitivity maps representing the perturbation location can be generated for the numerous optical data-types in the three temporal domains.
	The review first presents a history of the past 30 years of work investigating this sensitivity in optically diffuse media.
	These works are experimental and theoretical, presenting 1-, 2-, and 3-dimensional sensitivity maps for different near-infrared spectroscopy methods, domains, and data-types. 
	Following this history, we present a compendium of sensitivity maps organized by temporal domain and then data-type.
	This compendium provides a valuable tool to compare the spatial sensitivity of various measurement methods and parameters in one document.
	Methods for one to generate these maps are provided in the appendix, including code.
	This historical review and comprehensive sensitivity map compendium provides a single source researchers may use to visualize, investigate, compare, and generate sensitivity to localized absorption change maps. 
\end{abstract}

\renewcommand*{\thefootnote}{\alph{footnote}}
\setcounter{footnote}{0}

\section{Introduction}
\nocite{Weiss_J.Mod.Opt.89_StatisticsPenetration, Barbour_10thAnnu.Int.Symp.Geosci.RemoteSens.90_Model3D, Schweiger_Math.MethodsMed.Imaging92_ApplicationFinite, Schotland_AO93_PhotonHitting, Sevick_AO94_LocalizationAbsorbers, Mitic_AO94_TimegatedTransillumination, Patterson_AO95_AbsorptionSpectroscopy, Feng_AO95_PhotonMigration, Arridge_AO95_PhotonmeasurementDensity, Okada_Phys.Med.Bio.95_EffectOverlying, Boas_AO97_DetectionCharacterization, Weiss_Opt.Comm.98_ContinuoustimeRandom, Dehghani_AO00_NearinfraredSpectroscopy, Cox_AO01_SpatialSampling, Lyubimov_Phys.Med.Biol.02_ApplicationPhoton, Liebert_AO04_TimeresolvedMultidistance, Bevilacqua_Phys.Rev.E04_SamplingTissue, Torricelli_Phys.Rev.Lett.05_TimeresolvedReflectance, Hayakawa_SIAMJ.Appl.Math.07_CoupledForwardAdjoint, Eames_OE07_EfficientJacobian, Pifferi_Phys.Rev.Lett.08_TimeResolvedDiffuse, Haeussinger_P.ONE11_SimulationNearinfrared, Sawosz_Phys.Med.Bio.12_ExperimentalEstimation, Mazurenka_OE12_NoncontactTimeresolved}
\nocite{Jelzow_BOE14_SeparationSuperficial, Gardner_JBO14_CoupledForwardadjoint, Gunadi_BOE14_SpatialSensitivity, Brigadoi_NPh15_HowShort, Wu_BOE15_FastEfficient, Martelli_S.Rep.16_TherePlenty, Milej_BOE16_SubtractionbasedApproach, Binzoni_BOE17_DepthSensitivity, Niwayama_JBO18_VoxelbasedMeasurement, Yao_BOE18_DirectApproach}
\nocite{Sawosz_BOE19_MethodImprove, Sassaroli_JOSAA19_DualslopeMethod, Fantini_JIOHS19_TransformationalChange, Wabnitz_BOE20_DepthselectiveData, Blaney_JBio20_PhaseDualslopes, Perkins_JBO21_QuantitativeEvaluation, Fan_NPh21_InvestigationEffect, Sassaroli_BOE23_NovelData}

Over the past \SI{30}{\year}, before writing this review, the spatial distribution of energy associated with photons that have entered an optically diffuse medium at one location and exited at another\footnote{Cases where the photons enter and exit the medium at the same location have also been considered.} has been studied.\cite{Weiss_J.Mod.Opt.89_StatisticsPenetration}\textsuperscript{-}\cite{Sassaroli_BOE23_NovelData}
These studies are often intended to assess the \gls{sen} of \gls{NIRS} techniques in biomedical diffuse optics; however, the results are still applicable to any methods that make measurements in a diffuse medium.
\par

\Gls{sen} may be interpreted by the idea of photon visitation probability\footnote{Photon visitation probability is the probability that a photon emitted from the source and detected by the detector visited a spatial location within the diffuse medium.} when \gls{CW} \gls{I} is considered, however the idea of \gls{sen} can be extended beyond this for other data-types.\cite{Ayaz_NPh22_OpticalImaging}
These other data-types may include the \gls{phi} in \gls{FD} or the \gls{var} in \gls{TD}, for example.\cite{Fantini_FNINS20_FrequencyDomainTechniques, Torricelli_NI14_TimeDomain} 
For these cases, there are methods to obtain a \gls{dmua} \ie{$\as{dmua}_{,\as{Y}}$}, from a change in the data-type \ie{$\Delta\as{Y}$}.\cite{Sassaroli_BOE23_NovelData}
Furthermore, \gls{Y} need not be some data-type recovered from a single pair of sources and detectors \ie{\gls{SD}} but instead from a spatially resolved \ie{over \glspl{rho}} measurement of optical data.
These methods have been variously referred to as \gls{SRS},\cite{Bevilacqua_AO99_VivoLocal,Doornbos_Phys.Med.Biol.99_DeterminationVivohuman} \gls{SS},\cite{Sassaroli_JOSAA19_DualslopeMethod} or \gls{DS}.\cite{Sassaroli_JOSAA19_DualslopeMethod}\textsuperscript{;}\footnote{Note that \gls{DS} has further requirements on the arrangement of optodes beyond simply a measurement of optical data across \gls{rho}.\cite{Fantini_JIOHS19_TransformationalChange,Blaney_RSI20_DesignSource}\label{foot:DSreq}}
Regardless of the choice of measurement method, in the case of small \gls{mua} perturbations the measured $\Delta\as{Y}$ is related to the recovered $\as{dmua}_{,\as{Y}}$ through some proportionality factor, often referred to as \glspl{DPF}\cite{Arridge_Phys.Med.Biol.92_TheoreticalBasis} or \glspl{DSF}.\cite{Blaney_JBio20_PhaseDualslopes}
\par

The true \glspl{dmua} can be spatially localized \ie{$\as{dmua}_{,pert}\left(\as{r}\right)$} or homogeneous \ie{$\as{dmua}_{,pert,homo}$}.
In the homogeneous case, the recovered $\as{dmua}_{,\as{Y}}$ is equal to the true perturbation $\as{dmua}_{,pert,homo}$.\footnote{This is ensured by the use of the correct proportionality factor \ie{\gls{DPF} or \gls{DSF}} for the data-type \gls{Y}.}
However, in the case of a spatially localized perturbation, $\as{dmua}_{,\as{Y}}$ is almost never equal to $\as{dmua}_{,pert}\left(\as{r}\right)$.
Instead $\as{dmua}_{,\as{Y}}$ represents the homogeneous change that would be consistent with the measured $\Delta\as{Y}$.
Therefore, we sometimes name $\as{dmua}_{,\as{Y}}$ the recovered effective homogeneous change.
How the recovered $\as{dmua}_{,\as{Y}}$ is related to the true local $\as{dmua}_{,pert}\left(\as{r}\right)$ is described by \gls{sen}.
Therefore, the \gls{sen} for a particular data-type \ie{\as{Y}} is a function of the spatial location of the localized perturbation \ie{$\as{sen}_{\as{Y}}\left(\as{r}\right)$}.
The exact definition of \gls{sen} and why it can be interpreted in this way is further explained in Section~\ref{sec:defS}.
\par

This review aims to gather results from a variety of previous works, both experimental and theoretical, to provide a reference source for readers interested specifically in spatial maps of \gls{sen}.
Importantly, we also remake and present \gls{2D} and \gls{3D} spatial maps \gls{sen} for various \gls{NIRS} methods and data-types.
This is to provide the reader with a single source for \gls{sen} maps pertaining to most \gls{NIRS} methods, including the \gls{MATLAB} code utilized to generate the maps (found in the Appendix) so that one may recreate or modify them.\cite{Blaney_23_CodeSensitivity}
In doing so, we present the majority of these \gls{sen} maps on the same color-scale so that they may all be compared.
\par

To achieve this, we give a brief historical review of the work investigating \gls{sen} and discuss the various methods and data-types studied (Section~\ref{sec:hist}).
Then, in Section~\ref{sec:sen} we formally define \gls{sen} (Section~\ref{sec:defS}) and explain the methods used to generate the \gls{sen} maps (Appendix~\ref{app:calS}) in this compendium.
The maps themselves are presented as both \gls{2D} cross-sections and \gls{3D} iso-surface volumes in Section~\ref{sec:maps}.
\Gls{CW}, \gls{FD}, and \gls{TD} are covered in Sections~\ref{sec:maps:CW},\ref{sec:maps:FD},\&\ref{sec:maps:TD}, respectively.
Each section explores data-types specific to that temporal domain and includes both measurements at a single \gls{rho}, that is \gls{SD}, or across multiple \glspl{rho}, that is \gls{SS} or \gls{DS}.\cite{Sassaroli_JOSAA19_DualslopeMethod}\textsuperscript{;\ref{foot:DSreq}}
\par

\subsection{Nomenclature}
\subsubsection{Measurement Methods}
The \gls{NIRS} methods considered in our review can be classified in various ways.
The first is the temporal domain of the measurement method: \gls{TD}\cite{Torricelli_NI14_TimeDomain}, \gls{FD}\cite{Fantini_FNINS20_FrequencyDomainTechniques}, or \gls{CW}\cite{Scholkmann_Physiol.Meas.14_MeasuringTissue}.
The next is the optode arrangement which can be either: \gls{SD}, a measurement between a single source and detector spaced by a \gls{rho}; \gls{SS}, a measurement of optical data as a function of \gls{rho} by using multiple sources or multiple detectors;\cite{Niwayama_JBO18_VoxelbasedMeasurement, Fantini_Phys.Med.Biol.99_NoninvasiveOptical} and \gls{DS}, which also measures the optical data dependence on \gls{rho} but from a symmetric set of sources and detectors meeting specific geometric requirements.\cite{Blaney_RSI20_DesignSource, Fantini_JIOHS19_TransformationalChange}
\par

Lastly, we can categorize each temporal domain further, considering measurements possible in each.
In \gls{CW} there is only one data type, the \gls{I}.\cite{Scholkmann_Physiol.Meas.14_MeasuringTissue}
For \gls{FD} there is amplitude, referred to as \gls{FD} \gls{I}\footnote{\Gls{FD} \gls{I} reduces to \gls{CW} \gls{I} in the case of \gls{fmod} being zero.} and \gls{phi}.\cite{Fantini_FNINS20_FrequencyDomainTechniques} 
\Gls{TD} becomes more complex as various kinds of data can be extracted from the \gls{TPSF}.\cite{Wabnitz_BOE20_DepthselectiveData, Torricelli_NI14_TimeDomain}
In this review we focus on the \gls{mTOF}\footnote{The \gls{mTOF} from \gls{TD} can be approximated by \gls{phi} in \gls{FD}.\cite{Sassaroli_BOE23_NovelData}} and the \gls{var}.\footnote{The \gls{var} from \gls{TD} can be approximated by the \gls{FD} \gls{I} normalized by the \gls{CW} \gls{I}.\cite{Sassaroli_BOE23_NovelData}}
In addition to these, we may also consider the \gls{TD} \gls{I} from different \gls{t} gated portions of the \gls{TPSF}.\footnote{For a infinitely large \gls{t} gate of the \gls{TPSF}, the \gls{TD} \gls{I} is equivalent to the \gls{CW} \gls{I}.}
All of these data types exhibit different spatial \gls{sen} which we show in Section~\ref{sec:maps}.
\par

\subsubsection{Modeling Methods}
Various methods have been used to obtain \gls{sen} maps or ones similarly interpreted to \gls{sen}.
Two less common methods are \gls{LRW}\cite{Weiss_Opt.Comm.98_ContinuoustimeRandom, Weiss_94_AspectsApplications, Gandjbakhche_AO93_ScalingRelationships} and modeling of photon paths,\cite{Ueda_Opt.Rev.05_DiffuseOptical, Polishchuk_OL97_AverageMostprobable, Perelman_Phys.Rev.Lett.94_PhotonMigration} which are briefly mentioned in Section~\ref{sec:hist}.
Many more commonly used methods rely on \gls{DT}, which may be formulated in various ways.\cite{Jacques_JBO08_TutorialDiffuse, Contini_AO97_PhotonMigration}
\Gls{DT} solutions may be arrived at numerically or analytically.
One possible numerical method is the \gls{FEM} allowing for complex volumes of scattering media.\cite{Dehghani_CNM09_InfraredOptical, Arridge_Med.Phys.93_FiniteElement}
Analytical solutions for \gls{DT} typically must consider more simple media (\textit{e.g.}, infinite or semi-infinite); a majority of the maps presented in Section~\ref{sec:maps} entail this method which is explained in Appendix~\ref{app:DT}.
Finally, photon propagation may be simulated directly with equally common \gls{MC} methods.\cite{Fang_OE09_MonteCarlo, Wang_CMPB95_MCMLMonte}
Using \gls{MC} one may calculate the times relative to when a photon was launched that each photon spends in each voxel of a simple or complex medium allowing direct calculation of \gls{sen}.\footnote{The average time a photon spends in each voxel (proportional to the \gls{lpath}) normalized by the \gls{mTOF} (proportional to the \gls{Lpath}) is the \gls{sen} for \gls{CW} \gls{I}, for other data-types the distribution of photon voxel times and photon arrival times must be considered to yield \gls{sen}.\cite{Sassaroli_BOE23_NovelData}}
Aside from direct calculation, adjoint methods for \gls{MC} have also been developed providing a more computationally efficient estimation of \gls{sen}; some maps in Section~\ref{sec:maps} utilize adjoint \gls{MC} which is explained in Appendix~\ref{app:MC}.\cite{Gardner_JBO14_CoupledForwardadjoint, Hayakawa_SIAMJ.Appl.Math.07_CoupledForwardAdjoint}
\par

\section{History of Previous Work on Sensitivity}\label{sec:hist}
Table~\ref{tab:papers} shows the select works that studied the spatial distribution of something similar to the \gls{sen} in chronological order.
Many of these works may not present \gls{sen} according to the specific definition in Section~\ref{sec:defS} for our maps in Section~\ref{sec:maps}, but all the works present some results that can be interpreted in a similar way.
This chronological list helps show how the work has progressed over the last \SI{30}{\year}.
\par

\begin{table*}
	\begin{threeparttable}
		\caption{Select works which study sensitivity to absorption changes in optically diffuse media \label{tab:papers}}
		\begin{tabular*}{\linewidth}{@{\extracolsep{\fill}}l|@{}r@{}r@{}r@{}}
			Citation & Temporal Domain & Primary Method & Relevant Figs. \\
			\hline
			\tblCite{Weiss_J.Mod.Opt.89_StatisticsPenetration} 							& \as{CW} 	& \as{LRW}			 	& 2 \\
			
			\tblCite{Barbour_10thAnnu.Int.Symp.Geosci.RemoteSens.90_Model3D}			& \as{CW}	& \as{MC}				& 2\mapFig \\
			\tblCite{Schweiger_Math.MethodsMed.Imaging92_ApplicationFinite}				& \as{TD}	& \as{FEM}				& 5 \\
			\tblCite{Schotland_AO93_PhotonHitting} 										& \as{TD} 	& \as{DT} 				& (4 \& 7)\mapFig \\
			\tblCite{Sevick_AO94_LocalizationAbsorbers} 								& \as{FD}	& Exp.\			 		& 5-6, \& 8 \\
			\tblCite{Mitic_AO94_TimegatedTransillumination}								& \as{TD}	& Exp.\ 				& 8-13, \& 15 \\
			\tblCite{Patterson_AO95_AbsorptionSpectroscopy}								& \as{TD}	& Exp.\					& 6 \\
			\tblCite{Feng_AO95_PhotonMigration} 										& \as{CW} 	& \as{DT}			 	& (2-4)\volFig\ \& 5 \\
			\tblCite{Arridge_AO95_PhotonmeasurementDensity} 							& \as{TD}	& \as{DT}				& (5-6)\volFig\ \& (7-10)\mapFig \\
			\tblCite{Okada_Phys.Med.Bio.95_EffectOverlying}								& \as{CW}	& \as{MC}				& (4-7)\mapFig \\
			\tblCite{Boas_AO97_DetectionCharacterization}								& \as{FD}	& \as{DT}				& 3, 5 \\
			\tblCite{Weiss_Opt.Comm.98_ContinuoustimeRandom}							& \as{TD}	& \as{LRW}				& 5 \\
			
			\tblCite{Dehghani_AO00_NearinfraredSpectroscopy}							& \as{CW}	& \as{FEM}				& 4-9, \& (3 \& 10)\mapFig \\
			\tblCite{Cox_AO01_SpatialSampling} 											& \as{CW}	& \as{DT}			 	& 6\mapFig \\
			\tblCite{Lyubimov_Phys.Med.Biol.02_ApplicationPhoton}						& \as{CW}	& PP					& 2\mapFig \\
			\tblCite{Liebert_AO04_TimeresolvedMultidistance} 							& \as{TD} 	& \as{MC} 				& 2-5 \\
			\tblCite{Bevilacqua_Phys.Rev.E04_SamplingTissue} 							& \as{FD}  	& \as{MC}				& 2-3, 8-9, \& (1 \& 7)\mapFig \\
			\tblCite{Torricelli_Phys.Rev.Lett.05_TimeresolvedReflectance} 				& \as{TD} 	& \as{DT}				& 1\mapFig\ \& 2 \\
			\tblCite{Hayakawa_SIAMJ.Appl.Math.07_CoupledForwardAdjoint}					& \as{CW}	& \as{MC}				& 5.1\mapFig \\
			\tblCite{Eames_OE07_EfficientJacobian}										& \as{CW}	& \as{FEM}				& 1\mapFig \\
			\tblCite{Pifferi_Phys.Rev.Lett.08_TimeResolvedDiffuse}						& \as{TD}	& Exp.\					& 2 \\
			
			\tblCite{Haeussinger_P.ONE11_SimulationNearinfrared} 						& \as{CW} 	& \as{MC}				& 1\volFig\ \& 9\mapFig \\
			\tblCite{Sawosz_Phys.Med.Bio.12_ExperimentalEstimation} 					& \as{TD} 	& Exp.\ 				& (2-6)\mapFig \\
			\tblCite{Mazurenka_OE12_NoncontactTimeresolved}								& \as{TD}	& Exp.\					& 3-4 \\
			\tblCite{Jelzow_BOE14_SeparationSuperficial} 								& \as{TD} 	& \as{DT}				& 4 \\
			\tblCite{Gardner_JBO14_CoupledForwardadjoint}								& \as{CW}	& \as{MC}				& (4-8)\mapFig \\
			\tblCite{Gunadi_BOE14_SpatialSensitivity}									& \as{TD}	& Exp.\					& 4, 6, \& (3 \& 5)\mapFig \\
			\tblCite{Brigadoi_NPh15_HowShort} 											& \as{CW} 	& \as{MC}				& 4\mapFig \\
			\tblCite{Wu_BOE15_FastEfficient}											& \as{CW}	& \as{FEM}				& 1\mapFig \\
			\tblCite{Martelli_S.Rep.16_TherePlenty} 									& \as{TD} 	& \as{DT}				& 2-5 \\
			\tblCite{Milej_BOE16_SubtractionbasedApproach} 								& \as{TD} 	& \as{MC}				& 2-3 \\
			\tblCite{Binzoni_BOE17_DepthSensitivity} 									& \as{FD} 	& \as{DT} 				& 1-6 \\
			\tblCite{Niwayama_JBO18_VoxelbasedMeasurement} 								& \as{CW} 	& \as{MC}				& 2\mapFig \\
			\tblCite{Yao_BOE18_DirectApproach}											& \as{TD}	& \as{MC}				& (2-3)\mapFig \\
			\tblCite{Sawosz_BOE19_MethodImprove} 										& \as{TD} 	& \as{DT}				& (2-5 \& 7-9)\mapFig \\
			\tblCite{Sassaroli_JOSAA19_DualslopeMethod} 								& \as{FD} 	& \as{DT}				& 2, 6, 8, 10-11, \& (5, 7, 9, 12, \& 16-17)\mapFig \\
			\tblCite{Fantini_JIOHS19_TransformationalChange} 							& \as{FD} 	& \as{DT}				& 1-13\volFig \\
			
			\tblCite{Wabnitz_BOE20_DepthselectiveData} 									& \as{TD} 	& \as{DT}				& 5, 9-10, \& (2-4 \& 8)\mapFig \\
			\tblCite{Blaney_JBio20_PhaseDualslopes} 									& \as{FD} 	& \as{DT}				& 3-4\mapFig \\
			\tblCite{Perkins_JBO21_QuantitativeEvaluation}								& \as{FD}	& \as{FEM}				& 3\mapFig \\
			\tblCite{Fan_NPh21_InvestigationEffect} 									& \as{FD} 	& \as{FEM}				& 2\mapFig \\
			\tblCite{Sassaroli_BOE23_NovelData} 										& \as{FD} 	& \as{DT}				& 3-8 \& 10-14 \\
		\end{tabular*}
		\begin{tablenotes}
			\item Acronyms: \Acrfull{CW}, \acrfull{LRW}, \acrfull{MC}, \acrfull{TD}, analytical \acrfull{DT}, \acrfull{FD}, Experimental~(Exp.), \acrfull{FEM}, Photon Path (PP)
			\item\mapFig Presents \acrfull{2D} maps
			\item\volFig Presents \acrfull{3D} volumes
		\end{tablenotes}
	\end{threeparttable}
\end{table*}

\subsection{1980's}
The first work we chose to review is \tblCite{Weiss_J.Mod.Opt.89_StatisticsPenetration} which used the \gls{LRW} method to present the probability density function for photon visitation as a function of depth for photons measured by a \gls{SD} in \gls{CW}.
The \gls{LRW} method of solving photon migration in tissue has its own distinct solutions, however it has been shown\cite{Weiss_94_AspectsApplications,Zauderer_06_PartialDifferential} that one can carry out asymptotic limits to the random walk equations \ie{for lattice spacing approaching zero and rate at which steps are made to approach infinity} and obtain either the diffusion equation \ie{for uncorrelated random walks} or the telegrapher’s equation \ie{for correlated random walks}.
Therefore, it is expected, at least in those situations where the diffusion conditions are fulfilled, that a solution obtained with the uncorrelated random walk will yield substantially the same solution as analytical \gls{DT}. 
We also remind that both diffusion equation and telegrapher’s equation \ie{the P\textsubscript{1} approximation\cite{Heizler_NSE10_AsymptoticTelegrapher}} are derived from the more general radiative transfer equation \ie{under some approximations for the radiance}, which is usually solved by \gls{MC} methods.
\par

\subsection{1990's}
Moving into the 1990's we find many works which are beginning to use different methods and create \gls{2D} or \gls{3D} volumes of \gls{sen} like quantities.
First, a \gls{2D} map of something similar to \gls{sen} in \tblCite{Barbour_10thAnnu.Int.Symp.Geosci.RemoteSens.90_Model3D} which used \gls{MC} simulations.
Then \tblCite{Schweiger_Math.MethodsMed.Imaging92_ApplicationFinite} showed the change in \gls{TD} data types from a localized perturbation in a cylindrical medium; this is an early publication using \gls{FEM} with a comparison to \gls{MC}.
\tblCite{Schotland_AO93_PhotonHitting} used eigenfunction analytical expansions of \gls{DT} to arrive at \gls{2D} maps of photon hitting densities for a homogeneous box, and numerical solutions to result in similar maps but for a heterogeneous medium.
Next, \tblCite{Sevick_AO94_LocalizationAbsorbers} and \tblCite{Mitic_AO94_TimegatedTransillumination} show \gls{FD} and \gls{TD} results from experiments on phantoms, respectively.
Finally, to wrap-up a concentration of experimental works, \tblCite{Patterson_AO95_AbsorptionSpectroscopy} conducted similar experiments in \gls{TD} but instead focusing on penetration depth.
\par

\tblCite{Feng_AO95_PhotonMigration} in their seminal work coined the term \emph{banana} for the shape of the \gls{sen} region for a \gls{SD} arrangement in \gls{CW}; this paper may be considered the first which really allowed good visualization of these banana \gls{sen} regions and include \gls{3D} volumes.
This work was closely followed by \tblCite{Arridge_AO95_PhotonmeasurementDensity} which showed a plethora of analytical solutions to \gls{DT} with a focus on tomographic reconstruction.
Next, \tblCite{Okada_Phys.Med.Bio.95_EffectOverlying} explored the \gls{sen} regions on a geometry meant to match a head by incorporating the consideration of two heterogeneous layers.
\tblCite{Boas_AO97_DetectionCharacterization} considered some practical considerations by analyzing these ideas in terms of \gls{SNR} which is important to determine feasibility.
Finally, wrapping up the 1990's is \tblCite{Weiss_Opt.Comm.98_ContinuoustimeRandom} with one of the later \gls{LRW} papers, this time applied to \gls{TD}.
\par

\subsection{2000's}
In the 2000's we first have \tblCite{Dehghani_AO00_NearinfraredSpectroscopy} who investigated how the optical properties in a more realistic head model affect \gls{sen} regions using a \gls{FEM} solution to \gls{DT}.
Next, \tblCite{Cox_AO01_SpatialSampling} carried out a more detailed investigation of the \gls{DT} solutions with a comparison to \gls{MC}.
\tblCite{Lyubimov_Phys.Med.Biol.02_ApplicationPhoton} studied photon visitation distributions in a less common way by considering the average photon paths within a diffuse medium.
\tblCite{Liebert_AO04_TimeresolvedMultidistance} focused on \gls{TD} and a multiple \gls{rho} approach \ie{\gls{SS}} in an attempt to separate deep versus superficial perturbations.
In the same year, \tblCite{Bevilacqua_Phys.Rev.E04_SamplingTissue} utilized \gls{MC} to investigate scaling relations on variables that affect the penetration depth associated with \gls{FD} data including conditions of short \gls{rho} or high \gls{mua} for which \gls{DT} may not be applicable.
\par

In the second half of the 2000's, \tblCite{Torricelli_Phys.Rev.Lett.05_TimeresolvedReflectance} proposed a \gls{TD} method which utilized a \gls{rho} of \SI{0}{\milli\meter} \ie{null distance} showing its corresponding \gls{sen} region.
To tackle the computation time taken to calculate \gls{sen} maps with \gls{MC}, \tblCite{Hayakawa_SIAMJ.Appl.Math.07_CoupledForwardAdjoint} developed an adjoint method.
As the field developed, more emphasis was put on generating \gls{sen} for image reconstruction; for example \tblCite{Eames_OE07_EfficientJacobian} investigated an efficient reduction \ie{removing imaging regions which do not significantly effect the result} of the \gls{sen} \ie{Jacobian}.
Wrapping up the 2000's we highlight \tblCite{Pifferi_Phys.Rev.Lett.08_TimeResolvedDiffuse} who continued work to make \gls{TD} feasible at small \gls{rho} and showed an experimentally generated \gls{sen} map.
\par

\subsection{2010's}
Continuing to the 2010's we find \tblCite{Haeussinger_P.ONE11_SimulationNearinfrared} who studied how the \gls{sen} region changes on a subject specific basis by using realistic models.
\tblCite{Sawosz_Phys.Med.Bio.12_ExperimentalEstimation} were able to utilize a time-gated camera for \gls{TD} detection to experimentally create maps of photon visitation.
Continuing the vein of interesting instrumentation and experimental work, \tblCite{Mazurenka_OE12_NoncontactTimeresolved} investigated the depth probed by a proposed non-contact \gls{TD} method.
In the same year, \tblCite{Jelzow_BOE14_SeparationSuperficial} presented a method to achieve depth selectivity in a \gls{TD} measurement by leveraging the different \glspl{sen} of different \gls{TD} moments.
Then, \tblCite{Gardner_JBO14_CoupledForwardadjoint} extended the earlier work on adjoint \gls{MC} and generated \gls{sen} for \gls{3D} tissue models.
Also, in the same year \tblCite{Gunadi_BOE14_SpatialSensitivity} explored the spatial \gls{sen} of three different \gls{NIRS} instruments experimentally.
\par

Moving to the second half of the 2010's we review \tblCite{Brigadoi_NPh15_HowShort} who investigated the \gls{sen} of short \glspl{rho} for subtraction methods.
As the field continued to move to image reconstruction methods, \tblCite{Wu_BOE15_FastEfficient} proposed fast and efficient image reconstruction on the human brain by leveraging \gls{FEM}.
\tblCite{Martelli_S.Rep.16_TherePlenty} conducted a more theoretical investigation of photon penetration by deriving expressions for the statistics of photon depth.
In the same year, a technology development paper was published by \tblCite{Milej_BOE16_SubtractionbasedApproach}, who proposed a subtraction-based \gls{TD} approach and investigated its \gls{sen} depth.
The following year \tblCite{Binzoni_BOE17_DepthSensitivity} published another theoretical paper that reported analytical expressions for \gls{sen} with totally absorbing defects.
\tblCite{Niwayama_JBO18_VoxelbasedMeasurement} investigated the \gls{SS} approach and the type of \gls{sen} distributions it achieves.
Meanwhile, in the same year, a different \gls{MC} method was presented by \tblCite{Yao_BOE18_DirectApproach} for investigating perturbations using an approach to re-simulate specific photon paths.
Finally, wrapping up the 2010's was the proposal of the \gls{DS} method for \gls{TD} by \tblCite{Sawosz_BOE19_MethodImprove} and for \gls{FD} by \tblCite{Sassaroli_JOSAA19_DualslopeMethod} with \tblCite{Fantini_JIOHS19_TransformationalChange} showing many \gls{sen} maps for different implementations of the method.
\par

\subsection{2020's}
The most recent decade, the 2020's, continued the new trend of using various data types with \tblCite{Wabnitz_BOE20_DepthselectiveData} investigating the \gls{sen} of various methods in \gls{TD}.
Then, \tblCite{Blaney_JBio20_PhaseDualslopes} continued the work on \gls{DS} and presented the calculation of its \gls{sen} and \gls{SNR}.
The next year, \tblCite{Perkins_JBO21_QuantitativeEvaluation} expanded work on imaging by leveraging the \gls{sen} of different \gls{FD} data-types.
\tblCite{Fan_NPh21_InvestigationEffect} also conducted investigations of the \gls{sen} of \gls{FD} data, this time with emphasis on the impact of \gls{fmod}.
Finally, the most recent paper we review is by \tblCite{Sassaroli_BOE23_NovelData}, who showed various new \gls{FD} data types that achieve features, such as \gls{sen}, similar to those associated with higher order \gls{TD} moments.
\par

\section{Sensitivity Compendium}\label{sec:sen}
\subsection{Definition of Sensitivity}\label{sec:defS}
Consider an optically diffuse medium with some \gls{mua}.
This medium is measured by some optical measurement technique which yields \gls{Y}.
If the \gls{mua} of the medium changes, a change in \gls{Y} will be measured.
The \gls{dmua} may be global \ie{homogeneous throughout the medium} or local at \gls{r} with a perturbation volume ($V_{pert}$); we denote these two cases $\as{dmua}_{,pert,homo}$ and $\as{dmua}_{,pert}(\as{r})$, respectively.
Furthermore, relationships between the measured change in \gls{Y} and \gls{dmua} are described by the global Jacobian $\partial \as{Y} / \partial \as{mua}_{,pert,homo}$ and the local Jacobian $\partial \as{Y} / \partial \as{mua}_{,pert}(\as{r})$, respectively.
\Gls{sen} is defined as the ratio of this local and global Jacobian:
\begin{equation}\label{equ:senDef}
	\as{sen}_{\as{Y}}\left(\as{r}\right)=
		\frac{\partial \as{Y} / \partial \as{mua}_{,pert}(\as{r})}
		{\partial \as{Y} / \partial \as{mua}_{,pert,homo}}
\end{equation}
This definition shows that $\as{sen}_{\as{Y}}\left(\as{r}\right)$ quantifies how a local \gls{mua} perturbation \ie{$\as{dmua}_{,pert}(\as{r})$} affects the measurement \gls{Y} compared to how a global \gls{mua} perturbation \ie{$\as{dmua}_{,pert,homo}$} affects \gls{Y}.
\par

If we remember that $\Delta\as{Y}$ can be converted to a recovered $\as{dmua}_{,\as{Y}}$ using proportionality constants,\cite{Arridge_Phys.Med.Biol.92_TheoreticalBasis, Blaney_JBio20_PhaseDualslopes}\textsuperscript{;}\footnote{These proportionality constants are the \gls{DPF} or \gls{DSF} which must be either assumed or calculated using the known absolute \gls{mua} and \gls{musp} of the medium.} we can interpret Eq.~(\ref{equ:senDef}) differently.
$\as{dmua}_{,\as{Y}}$ represents the effective homogeneous \gls{dmua} which would cause the measured change in \gls{Y}.
Considering considering this, $\as{sen}_{\as{Y}}\left(\as{r}\right)$ can be written as follows:
\begin{equation}\label{equ:interSen}
	\as{sen}_{\as{Y}}\left(\as{r}\right)=\frac{\as{dmua}_{,\as{Y}}}{\as{dmua}_{,pert}\left(\as{r}\right)}
\end{equation}
\noindent where, we can now interpret $\as{sen}_{\as{Y}}\left(\as{r}\right)$ as the ratio between the recovered $\as{dmua}_{,\as{Y}}$ and the true perturbation $\as{dmua}_{,pert}\left(\as{r}\right)$. Appendix~\ref{app:calS} describes how we generate \gls{senMat} containing values of \gls{sen} based on Eq.~(\ref{equ:senDef}) for the maps in Section~\ref{sec:maps}.
\par

\subsection{Sensitivity Maps}\label{sec:maps}
In the following subsections we re-create \gls{sen} maps for various different \gls{NIRS} data types and methods.
The methods to generate a majority of the maps \ie{unless otherwise specified} can be found in Appendices~\ref{app:calS},\ref{app:DT},\&\ref{app:MC} and our previous publications.\cite{Sassaroli_BOE23_NovelData, Blaney_23_DualratioApproach, Blaney_JBio20_PhaseDualslopes, Sassaroli_JOSAA19_DualslopeMethod}
All maps were created assuming a homogeneous semi-infinite medium with \gls{mua} of \SI{0.011}{\per\milli\meter} ("optProp.mua" in Listing~\ref{lst:makeS}), a \gls{musp} of \SI{1.1}{\per\milli\meter} ("optProp.musp" in Listing~\ref{lst:makeS}), an external \gls{n} of \num{1} ("optProp.nout" in Listing~\ref{lst:makeS}), and an internal \gls{n} of \num{1.333} ("optProp.nin" in Listing~\ref{lst:makeS}).
Also, most maps consider a perturbation measuring $\SI{10}{\milli\meter}\times\SI{10}{\milli\meter}\times\SI{2}{\milli\meter}$ ("'pert'" in Listing~\ref{lst:makeS}) scanned by \SI{0.1}{\milli\meter} or \SI{0.5}{\milli\meter} ("'dr'" in Listing~\ref{lst:makeS}) \ie{with overlap}.
A large perturbation is used to show \gls{sen} values which can be more easily interpreted since this size is more plausible.
Smaller perturbation sizes create less smooth maps and qualitatively similar maps.
\par

Further we repeat that, almost all maps consider the same color-scale so that they can be compared throughout the compendium.
Additionally, unless otherwise specified \gls{sen} vales were generated using perturbation \gls{DT} (see "'simTyp'" of "'DT'" in Listing~\ref{lst:makeS} \& Appendix~\ref{app:DT}).
Notice that we have chosen to saturate the color-maps so that values where $\as{sen}>\num{0.07}$ are white and $\as{sen}<\num{-0.03}$ below are black.
Contour lines on the maps correspond to the values of the ticks on the color-scale; and the maximum and minimum \gls{sen} values in the map \ie{if they are below \num{-0.03} or above \num{0.07}} are written on the left and right sides of the scale, respectively.
\par

\subsubsection{\Acrlong{CW}}\label{sec:maps:CW}
In \gls{CW} changes in the \gls{I} relative to a baseline \gls{I} can be converted to \gls{dmua}.
This method is based on knowledge of the \gls{Lpath} of the medium, which is often represented as the \gls{DPF} method.\cite{Ayaz_NPh22_OpticalImaging, Sassaroli_Phys.Med.Biol.04_CommentModified}
We refer to methods of a measurement of \gls{dmua} using a single source and a single detector as \gls{SD}.\cite{Blaney_JBio20_PhaseDualslopes, Sassaroli_JOSAA19_DualslopeMethod}
Therefore, the most basic type of \gls{NIRS} measurement is named \gls{SD} \gls{CW} \gls{I} in this review.
\par

If multiple \glspl{rho} are used to acquire \gls{I} as a function of \gls{rho} in \gls{CW}, temporal changes in the slope of the \gls{lnr2I} versus \gls{rho} can also be converted to \gls{dmua}.\cite{Fantini_Phys.Med.Biol.99_NoninvasiveOptical}
This method, which we refer to as \gls{SS},\footnote{\Gls{SS} is sometimes also referred to as \gls{SRS}.} depends on knowledge of the derivative of the slope versus \gls{dmua} which is sometimes referred to as the \gls{DSF}.\cite{Sassaroli_JOSAA19_DualslopeMethod, Blaney_JBio20_PhaseDualslopes}
\par

Finally, one may also recover \gls{dmua} using the \gls{DS} method.\cite{Sassaroli_JOSAA19_DualslopeMethod, Blaney_JBio20_PhaseDualslopes}
\Gls{DS} uses the same measurement of slope as \gls{SS} but it employs a special symmetric arrangement of two sources and two detectors.\cite{Blaney_RSI20_DesignSource}
As such, calculation of \gls{dmua} for \gls{DS} is also dependent on knowledge of the \gls{DSF}.
\par

\paragraph{\Acrlong{SD} \Acrlong{I}}
Figure~\ref{fig:CW_SD_I_3rd_rho0_MC_vox} shows our first and most basic \gls{sen} map. 
This map was generated with adjoint \gls{MC}\cite{Blaney_23_DualratioApproach, Gardner_JBO14_CoupledForwardadjoint, Fang_BOE10_MeshbasedMonte} due to the small \gls{rho} considered of \SI{0}{\milli\meter} (see "'simTyp'" of "'MC'" in Listing~\ref{lst:makeS} \& Appendix~\ref{app:MC}).
Figure~\ref{fig:CW_SD_I_3rd_rho0_MC_vox} considers a perturbation of $\SI{0.1}{\milli\meter}\times\SI{0.1}{\milli\meter}\times\SI{0.1}{\milli\meter}$, thus the numerical values for \gls{sen} are not comparable to those in the other figures which consider a larger perturbation.
\par

Next we show Fig.~\ref{fig:CW_SD_I_3rd_rho0_MC} which was also generated with adjoint \gls{MC},\cite{Blaney_23_DualratioApproach, Fang_BOE10_MeshbasedMonte, Gardner_JBO14_CoupledForwardadjoint} but this time with a $\SI{10}{\milli\meter}\times\SI{10}{\milli\meter}\times\SI{2}{\milli\meter}$ perturbation. 
Therefore, map's color-scale is comparable to most other figures in this review. 
\par

The classic banana shape\cite{Feng_AO95_PhotonMigration} emerges when we investigate the \gls{sen} for a \gls{SD} measurement with a non-zero \gls{rho}.
Figure~\ref{fig:CW_SD_I_3rd} shows this case.
Next, Fig.~\ref{fig:CW_SD_I_rho} shows \gls{CW} \gls{SD} \gls{I} for \num{10} different values of \gls{rho}.
Now that \gls{rho} is non-zero for Fig.~\ref{fig:CW_SD_I_rho} the map is calculated using a analytical \gls{DT} model (see "'simTyp'" of "'DT'" in Listing~\ref{lst:makeS} \& Appendix~\ref{app:DT}).\cite{Blaney_23_DualratioApproach, Blaney_JBio20_PhaseDualslopes, Sassaroli_JOSAA19_DualslopeMethod}
\par

\input{CW_SD_I_3rd_rho0_MC_vox.tex}
\input{CW_SD_I_3rd_rho0_MC.tex}
\input{CW_SD_I_3rd.tex}
\input{CW_SD_I_rho.tex}

\paragraph{Spatially Resolved \Acrlong{I}}
Now we move to \gls{sen} maps for \gls{SS} and \gls{DS} methods. 
Figures~\ref{fig:CW_SS_I_3rd}\&\ref{fig:CW_DS_I_3rd} show \gls{3D} volumes of these maps while Figs.~\ref{fig:CW_SS_I_mrho},\ref{fig:CW_DS_I_mrho},\ref{fig:CW_SS_I_drho},\&\ref{fig:CW_DS_I_drho} show maps for various \glspl{rho} making up the \gls{SS} or \gls{DS}.
The \glspl{rho} in a \gls{SS} or \gls{DS} set can be described using \gls{mrho} and \gls{drho} as follows:
\begin{equation}\label{equ:mrhoSS}
	\as{mrho}_{\text{\as{SS}}}=\frac{\as{rho}_1+\as{rho}_2}{2}
\end{equation}
\begin{equation}\label{equ:mrhoDS}
	\as{mrho}_{\text{\as{DS}}}=\frac{\as{rho}_1+\as{rho}_2+\as{rho}_3+\as{rho}_4}{4}
\end{equation}
\begin{equation}\label{equ:drhoSS}
	\as{drho}_{\text{\as{SS}}}=\as{rho}_2-\as{rho}_1
\end{equation}
\begin{equation}\label{equ:drhoDS}
	\as{drho}_{\text{\as{DS}}}=\as{rho}_2-\as{rho}_1=\as{rho}_3-\as{rho}_4
\end{equation}
\noindent where $\as{rho}_1<\as{rho}_2$ and $\as{rho}_3>\as{rho}_4$.
For symmetric \gls{DS} sets \ie{all the ones in this article} $\as{rho}_1=\as{rho}_4$ and $\as{rho}_2=\as{rho}_3$.
Figures~\ref{fig:CW_SS_I_mrho}\&\ref{fig:CW_DS_I_mrho} explore various \glspl{mrho}; similarly Figs.~\ref{fig:CW_SS_I_drho}\&\ref{fig:CW_DS_I_drho} explore various \glspl{drho}.
\par

\input{CW_SS_I_3rd.tex}
\input{CW_DS_I_3rd.tex}
\input{CW_SS_I_mrho.tex}
\input{CW_DS_I_mrho.tex}
\input{CW_SS_I_drho.tex}
\input{CW_DS_I_drho.tex}

\subsubsection{\Acrlong{FD}}\label{sec:maps:FD}
Similar to \gls{CW}, \gls{FD} also measures an \gls{I} data-type.
In \gls{FD} the \gls{I} is the amplitude of the measured photon density waves.
At low \gls{fmod} (around \SI{100}{\mega\hertz} or below), \gls{FD} \gls{I} well estimates \gls{CW} \gls{I};\cite{Sassaroli_BOE23_NovelData} as such, we do not show the same maps for \gls{FD} \gls{I} over various \gls{rho}, \gls{mrho}, and \gls{drho} that were already shown for \gls{CW} \gls{I} in Sec.~\ref{sec:maps:CW}. 
However, \gls{FD} does utilize the additional parameter of \gls{fmod}.
For this reason \gls{FD} \gls{I} is explored over \gls{fmod} in this section.
\par

The additional data-type that \gls{FD} gives access to is \gls{phi}. 
\Gls{phi} can be used to recover \gls{dmua} in all the \gls{SD}, \gls{SS}, and \gls{DS} measurement methods using a generalized \gls{Lpath} (Eq.~(\ref{equ:genL})), generalized \gls{DPF} \ie{for \gls{SD}}, or generalized \gls{DSF} \ie{for \gls{SS} or \gls{DS}}.\cite{Blaney_JBio20_PhaseDualslopes}
As with all other maps the methods to generate them can be found in Appendix~\ref{app:calS} and Listing~\ref{lst:makeS}.
\par

\paragraph{\Acrlong{SD}}
\subparagraph{\Acrlong{I}}
Figure~\ref{fig:FD_SD_I_3rd} shows the \gls{3D} \gls{sen} volume for \gls{FD} \gls{SD} \gls{I}.
This is followed by Fig.~\ref{fig:FD_SD_I_fmod} which shows the \gls{sen} map over various \gls{fmod}.
Notice that at high \gls{fmod} a negative \gls{sen} for \gls{FD} \gls{SD} \gls{I} is observed deep in the medium.\cite{Binzoni_BOE17_DepthSensitivity}
\par

\input{FD_SD_I_3rd.tex}
\input{FD_SD_I_fmod.tex}

\subparagraph{\Acrlong{phi}}
Figure~\ref{fig:FD_SD_P_3rd} shows the \gls{3D} \gls{sen} volume for \gls{FD} \gls{SD} \gls{phi}; and Figs.~\ref{fig:FD_SD_P_rho}\&\ref{fig:FD_SD_P_fmod} explore varying \gls{rho} and \gls{fmod}, respectively.
\par

\input{FD_SD_P_3rd.tex}
\input{FD_SD_P_rho.tex}
\input{FD_SD_P_fmod.tex}

\paragraph{Spatially Resolved}
\subparagraph{\Acrlong{I}}
Now we move to \gls{SS} and \gls{DS} type measurements for \gls{FD} \gls{I}.
First, we show the example \gls{3D} volumes for \gls{SS} and \gls{DS} in Figs.~\ref{fig:FD_SS_I_3rd}\&\ref{fig:FD_DS_I_3rd}.
Then the case of \gls{SS} or \gls{DS} and their dependence on \gls{fmod} is displayed in Figs.~\ref{fig:FD_SS_I_fmod}\&\ref{fig:FD_DS_I_fmod}, respectively.
\par

\input{FD_SS_I_3rd.tex}
\input{FD_DS_I_3rd.tex}
\input{FD_SS_I_fmod.tex}
\input{FD_DS_I_fmod.tex}

\subparagraph{\Acrlong{phi}}
To finish our section on \gls{FD} we show the \gls{SS} and \gls{DS} \gls{sen} maps for \gls{FD} \gls{phi}.
First come the \gls{3D} volumes in Figs.~\ref{fig:FD_SS_P_3rd}\&\ref{fig:FD_DS_P_3rd}.
Then for varying \gls{mrho} (Eq.~(\ref{equ:mrhoSS})\&(\ref{equ:mrhoDS})) in Figs.~\ref{fig:FD_SS_P_mrho}\&\ref{fig:FD_DS_P_mrho}; similarly for various \gls{drho} (Eq.~(\ref{equ:drhoSS})\&(\ref{equ:drhoDS})) in Figs.~\ref{fig:FD_SS_P_drho}\&\ref{fig:FD_DS_P_drho}.
Finally, exploring different values of \gls{fmod} for \gls{SS} and \gls{DS} in Figs.~\ref{fig:FD_SS_P_fmod}\&\ref{fig:FD_DS_P_fmod}, respectively.
\par

\input{FD_SS_P_3rd.tex}
\input{FD_DS_P_3rd.tex}
\input{FD_SS_P_mrho.tex}
\input{FD_DS_P_mrho.tex}
\input{FD_SS_P_drho.tex}
\input{FD_DS_P_drho.tex}
\input{FD_SS_P_fmod.tex}
\input{FD_DS_P_fmod.tex}

\subsubsection{\Acrlong{TD}}\label{sec:maps:TD}
\Gls{TD} offers access to the most data-types of any \gls{NIRS} temporal domain.\cite{Torricelli_NI14_TimeDomain}
These \gls{TD} data-types can be split into two categories; those of the \gls{TPSF} moments or time-gated portions of the \gls{TPSF} \ie{gated \gls{TD} \gls{I}}.\cite{Wabnitz_BOE20_DepthselectiveData}
All of these data-types, whether \gls{SD} or spatially resolved \ie{\gls{SS} or \gls{DS}}, can be used to measure \gls{dmua} by the same aforementioned methods \ie{generalized \gls{Lpath}} which utilize the derivative of the data-type with respect to \gls{mua} (Eq.~(\ref{equ:genL})).\cite{Sassaroli_BOE23_NovelData}
\par

The 0\textsuperscript{th} \gls{TD} moment is the \gls{TD} \gls{I} and if the whole \gls{TPSF} is considered, this moment is the same as \gls{CW} \gls{I}.
Next is the 1\textsuperscript{st} moment, the \gls{mTOF}; this moment is well approximated by the \gls{FD} \gls{phi}.\footnote{\Gls{mTOF} is approximated by the \gls{FD} \gls{phi} via the \gls{omega} using the following relation: $\as{phi}\approx\as{omega}\as{mTOF}$.}\textsuperscript{;}\cite{Sassaroli_BOE23_NovelData}
Therefore, we will not show the non-gated \gls{TD} \gls{I} \ie{the integral of the whole \gls{TPSF}} as it would be redundant with \gls{CW} \gls{I} and also not show \gls{TD} \gls{mTOF} over \gls{rho}, \gls{mrho}, and \gls{drho} since it would be redundant with the \gls{FD} \gls{phi} \gls{sen} maps.
\par

The data-type related to the 2\textsuperscript{nd} moment is the \gls{TD} \gls{var}, which will be reviewed in this section.
We would like to note that the \gls{FD} modulation depth \ie{the \gls{FD} \gls{I} which is an amplitude, divided by the \gls{CW} \gls{I}} well approximates the \gls{TD} \gls{var}.\cite{Sassaroli_BOE23_NovelData}
Similar to other data-types, \gls{var} can be considered for \gls{SD}, \gls{SS}, and \gls{DS}.\cite{Sawosz_BOE19_MethodImprove}
\par

The other category of \gls{TD} data is the gated \gls{TD} \gls{I} \ie{the gated \gls{TPSF}}.
This measurement selects photons which have particular sets of path-lengths.
Additionally, a data-type may be constructed which is a difference of the gated \gls{TD} \glspl{I} from different gates.\cite{Wabnitz_BOE20_DepthselectiveData}
For example, subtracting early gated photon data from late gated, selects for \gls{sen} of deep regions where photons have longer paths.
As with all else, we also review these \gls{sen} maps for these data-types applied to \gls{SD}, \gls{SS}, and \gls{DS}.
\par

\paragraph{\Acrlong{SD}}
\subparagraph{Gated \Acrlong{I}}
Now, we begin the gated \gls{TD} \gls{I} section with the \gls{3D} \gls{sen} volumes for null \gls{rho} \ie{$\as{rho}=\SI{0}{\milli\meter}$} in Figs.~\ref{fig:TD_SD_GI_3rd_rho0_MC_vox}\&\ref{fig:TD_SD_GI_3rd_rho0_MC}. 
This is contrasted to the \gls{CW} examples of null \gls{rho} in Figs.~\ref{fig:CW_SD_I_3rd_rho0_MC_vox}\&\ref{fig:CW_SD_I_3rd_rho0_MC} since late \gls{TD} gated \gls{I} can yield relatively deep \gls{sen} unlike \gls{CW}.\cite{Torricelli_Phys.Rev.Lett.05_TimeresolvedReflectance}
We follow the maps of null \gls{rho} with Fig.~\ref{fig:TD_SD_GI_3rd} which presents the \gls{3D} volume for a non-zero \gls{rho}.
Finally, capping out the maps which show \gls{TD} \gls{SD} gated \gls{I} at various \glspl{rho} is Fig.~\ref{fig:TD_SD_GI_rho} which explores various non-zero \glspl{rho} in its subplots.
\par

Next, we look at the \gls{sen} map dependence on the gate \gls{t}.
To do this we define the gate \gls{mt} and the gate \gls{dt} as follows:
\begin{equation}\label{equ:mt}
	\as{mt}=\frac{\as{t}_{start}+\as{t}_{end}}{2}
\end{equation}
\begin{equation}\label{equ:dt}
	\as{dt}=\as{t}_{end}-\as{t}_{start}
\end{equation}
\noindent where, $\as{t}_{start}$ and $\as{t}_{end}$ are the start and end of the gate, respectively.
Considering these variables, we show the \gls{TD} \gls{SD} gated \gls{I} \gls{sen} maps as a over various \gls{mt} in Figs.~\ref{fig:TD_SD_GI_mt_rho0_MC}\&\ref{fig:TD_SD_GI_mt} and over various \gls{dt} in Figs.~\ref{fig:TD_SD_GI_dt_rho0_MC}\&\ref{fig:TD_SD_GI_dt}.
Multiple maps are considered for each since Figs.~\ref{fig:TD_SD_GI_mt_rho0_MC}\&\ref{fig:TD_SD_GI_dt_rho0_MC} and Figs.~\ref{fig:TD_SD_GI_mt}\&\ref{fig:TD_SD_GI_dt} display null \gls{rho} and non-zero \gls{rho}, respectively.
\par

Whenever considering \gls{TD} \gls{SD} gated \gls{I} with a null \gls{rho} (Figs.~\ref{fig:TD_SD_GI_3rd_rho0_MC_vox},\ref{fig:TD_SD_GI_3rd_rho0_MC},\ref{fig:TD_SD_GI_mt_rho0_MC},\&\ref{fig:TD_SD_GI_dt_rho0_MC}), \gls{MC} was used to generate the \gls{sen} map (see "'simTyp'" of "'MC'" in Listing~\ref{lst:makeS} \& Appendix~\ref{app:MC}).
All other maps in this section utilized \gls{DT} (see "'simTyp'" of "'DT'" in Listing~\ref{lst:makeS} \& Appendix~\ref{app:DT}).
\par

\input{TD_SD_GI_3rd_rho0_MC_vox.tex}
\input{TD_SD_GI_3rd_rho0_MC.tex}
\input{TD_SD_GI_3rd.tex}
\input{TD_SD_GI_rho.tex}
\input{TD_SD_GI_mt_rho0_MC.tex}
\input{TD_SD_GI_mt.tex}
\input{TD_SD_GI_dt_rho0_MC.tex}
\input{TD_SD_GI_dt.tex}

\subparagraph{\Acrlong{mTOF}}
As mentioned in the beginning of this section, we do not consider \gls{TD} \gls{mTOF} as a function of the parameter \gls{rho} since it would be redundant with \gls{FD} \gls{phi}.
However, to display an example of \gls{TD} \gls{SD} \gls{mTOF} we provide Fig.~\ref{fig:TD_SD_T_3rd} which is the \gls{3D} volume of the \gls{sen}.
\par

\input{TD_SD_T_3rd.tex}

\subparagraph{Differences in Gated \Acrlong{I}}
Differences in the \gls{I} measured by different \gls{TD} \gls{t} gates may also be considered.\cite{Wabnitz_BOE20_DepthselectiveData}
Here we first present this data-type's \gls{3D} \gls{sen} volume for the null \gls{rho} \ie{$\as{rho}=\SI{0}{\milli\meter}$} in Figs.~\ref{fig:TD_SD_DGI_3rd_rho0_MC_vox}\&\ref{fig:TD_SD_DGI_3rd_rho0_MC}; note these figures present different perturbation sizes.
Following the null \gls{rho} case, we show the non-zero \gls{rho} case's \gls{3D} volume in Fig.~\ref{fig:TD_SD_DGI_3rd}; then the \gls{2D} map over various \glspl{rho} in Fig.~\ref{fig:TD_SD_DGI_rho}.
\par

To begin investigating the dependence on late and early gate \glspl{t}, we define \gls{mtg} and \gls{dtg} as follows:
\begin{equation}\label{equ:mtg}
	\as{mtg}=\frac{\as{mt}_{early}+\as{mt}}{2}
\end{equation}
\begin{equation}
	\as{dtg}=\as{mt}-\as{mt}_{early}
\end{equation}\label{equ:dtg}
\noindent where, $\as{mt}_{early}$ and $\as{mt}$ are the mean times of the early and late time gates, respectively (Eq.~(\ref{equ:mt})).
Note that we only consider the same \gls{dt} for both the early and late \gls{t} gates.
Now considering these parameters, we explore the \gls{TD} \gls{SD} difference in gated \gls{I} \gls{sen} maps over \gls{mtg} in Figs.~\ref{fig:TD_SD_DGI_mtg_rho0_MC}\&\ref{fig:TD_SD_DGI_mtg} and over \gls{dtg} in Figs.~\ref{fig:TD_SD_DGI_dtg_rho0_MC}\&\ref{fig:TD_SD_DGI_dtg}.
In this case Figs.~\ref{fig:TD_SD_DGI_mtg_rho0_MC}\&\ref{fig:TD_SD_DGI_dtg_rho0_MC} and Figs.~\ref{fig:TD_SD_DGI_mtg}\&\ref{fig:TD_SD_DGI_dtg} display a null \gls{rho} and a non-zero \gls{rho}, respectively.
\par

Whenever considering \gls{TD} \gls{SD} difference in gated \gls{I} with a null \gls{rho} (Figs.~\ref{fig:TD_SD_DGI_3rd_rho0_MC_vox},\ref{fig:TD_SD_DGI_3rd_rho0_MC},\ref{fig:TD_SD_DGI_mtg_rho0_MC},\&\ref{fig:TD_SD_DGI_dtg_rho0_MC}), \gls{MC} was used to generate the \gls{sen} map (see "'simTyp'" of "'MC'" in Listing~\ref{lst:makeS} \& Appendix~\ref{app:MC}).
All other maps in this section utilized \gls{DT} (see "'simTyp'" of "'DT'" in Listing~\ref{lst:makeS} \& Appendix~\ref{app:DT}).
\par

\input{TD_SD_DGI_3rd_rho0_MC_vox.tex}
\input{TD_SD_DGI_3rd_rho0_MC.tex}
\input{TD_SD_DGI_3rd.tex}
\input{TD_SD_DGI_rho.tex}
\input{TD_SD_DGI_mtg_rho0_MC.tex}
\input{TD_SD_DGI_mtg.tex}
\input{TD_SD_DGI_dtg_rho0_MC.tex}
\input{TD_SD_DGI_dtg.tex}

\subparagraph{\Acrlong{var}}
An additional data-type that \gls{TD} gives access to is \gls{var}.
We show the \gls{TD} \gls{SD} \gls{var} \gls{3D} \gls{sen} volume in Fig.~\ref{fig:TD_SD_V_3rd}.
Then we finish the \gls{TD} \gls{SD} section with the \gls{TD} \gls{SD} \gls{var} investigated over various \glspl{rho} in Fig.~\ref{fig:TD_SD_V_rho}.
\par

\input{TD_SD_V_3rd.tex}
\input{TD_SD_V_rho.tex}

\paragraph{Spatially Resolved}
Now, we review the \gls{sen} maps for \gls{SS} and \gls{DS} in \gls{TD}.
We do this for the data-types of \gls{t} gated \gls{I}, \gls{mTOF}, and \gls{var} in the following subsections.
\par

\subparagraph{Time Gated \Acrlong{I}}
We show plots of the \gls{3D} \gls{sen} for \gls{SS} and \gls{DS} \gls{TD} gated \gls{I} in Figs.~\ref{fig:TD_SS_GI_3rd}\&\ref{fig:TD_DS_GI_3rd}, respectively.
Then, \gls{SS} and \gls{DS} \gls{TD} gated \gls{I} over various \glspl{mrho} (Eq.~(\ref{equ:mrhoSS})\&(\ref{equ:mrhoDS})) in Figs.~\ref{fig:TD_SS_GI_mrho}\&\ref{fig:TD_DS_GI_mrho}, respectively.
This is followed by plots over various \glspl{drho} (Eq.~(\ref{equ:drhoSS})\&(\ref{equ:drhoDS})) Figs.~\ref{fig:TD_SS_GI_drho}\&\ref{fig:TD_DS_GI_drho}, respectively.
\par

Following the maps which investigate spatial parameters are the ones which vary over the temporal parameters \gls{mt} (Eq.~(\ref{equ:mt})) and \gls{dt} (Eq.~(\ref{equ:dt})).
Figures~\ref{fig:TD_SS_GI_mt}\&\ref{fig:TD_DS_GI_mt} which show the \gls{sen} maps over \gls{mt} for \gls{SS} and \gls{DS}, respectively.
Finally, we have Figs.~\ref{fig:TD_SS_GI_dt}\&\ref{fig:TD_DS_GI_dt} which vary over \gls{dt} in their various subplots for \gls{SS} and \gls{DS}, respectively.

\input{TD_SS_GI_3rd.tex}
\input{TD_DS_GI_3rd.tex}
\input{TD_SS_GI_mrho.tex}
\input{TD_DS_GI_mrho.tex}
\input{TD_SS_GI_drho.tex}
\input{TD_DS_GI_drho.tex}
\input{TD_SS_GI_mt.tex}
\input{TD_DS_GI_mt.tex}
\input{TD_SS_GI_dt.tex}
\input{TD_DS_GI_dt.tex}

\subparagraph{\Acrlong{mTOF}}
As mentioned before in this section, plotting \gls{TD} \gls{mTOF} over \gls{mrho} and \gls{drho} would be redundant with the plots presented for \gls{FD} \gls{phi}.
However, we still present the \gls{SS} and \gls{DS} \gls{TD} \gls{mTOF} \gls{3D} \gls{sen} volumes in Figs.~\ref{fig:TD_SS_T_3rd}\&\ref{fig:TD_DS_T_3rd}, respectively.
\par

\input{TD_SS_T_3rd.tex}
\input{TD_DS_T_3rd.tex}

\subparagraph{\Acrlong{var}}
The final set of maps we review in this article are the ones for \gls{SS} and \gls{DS} \gls{TD} \gls{var}.
The first of this set is the \gls{3D} volumes in Figs.~\ref{fig:TD_SS_V_3rd}\&\ref{fig:TD_DS_V_3rd}, respectively.
We then show the \gls{2D} \gls{sen} maps varying the spatial variables \gls{mrho} (Equations~\ref{equ:mrhoSS}\&\ref{equ:mrhoDS}) and \gls{drho} (Equations~\ref{equ:drhoSS}\&\ref{equ:drhoDS}).
Figures~\ref{fig:TD_SS_V_mrho}\&\ref{fig:TD_DS_V_mrho} explore various \gls{mrho} for \gls{SS} and \gls{DS}, respectively.
Finally, Figs.~\ref{fig:TD_SS_V_drho}\&\ref{fig:TD_DS_V_drho} explore various \gls{drho} for \gls{SS} and \gls{DS}, respectively.

\input{TD_SS_V_3rd.tex}
\input{TD_DS_V_3rd.tex}
\input{TD_SS_V_mrho.tex}
\input{TD_DS_V_mrho.tex}
\input{TD_SS_V_drho.tex}
\input{TD_DS_V_drho.tex}

\section{Conclusion}
This compendium of \gls{sen} seeks to provide the reader with a single document containing spatial maps for a large variety of \gls{NIRS} methods. 
Importantly, all maps present \gls{sen} for the same background optical properties and for the majority, the same perturbation size. 
Additionally, almost all maps are presented on the same color-scale so that they may be compared. 
If a reader wishes to recreate or generate these maps for themselves, they can find the \gls{MATLAB} code available in the Appendix (primarily in Listing~\ref{lst:makeS}) and at Reference~\citenum{Blaney_23_CodeSensitivity}.
\par

In addition to the collection of \gls{sen} maps, we also created a timeline of relevant works that presented \gls{sen} in the history of \gls{NIRS} research.
This set of references serves as a representation of the previous work to generate \gls{sen} maps, which this review recreates.
We hope that this document serves as a go-to reference to visualize \gls{sen} maps for a large variety of \gls{NIRS} data-types.
\par

\subsection*{Code Availability}
Supporting code can be found in the Appendix and accessed at the following link:\cite{Blaney_23_CodeSensitivity} \\\textcolor{blue}{
\href{https://github.com/DOIT-Lab/DOIT-Public/tree/master/SensitivityCompendium}
{github.com/DOIT-Lab/DOIT-Public/\\tree/master/SensitivityCompendium}}

\subsection*{Conflicts of Interest}
The authors declare no conflicts of interest.

\subsection*{Acknowledgments}
The authors acknowledge funding from the \gls{NIH} grant R01-EB029414.
G.B. would like to acknowledge funding from the \gls{NIH} \gls{IRACDA} program grant K12-GM133314. 
The content is solely the authors’ responsibility and does not necessarily represent the official views of the awarding institutions.

\section{Appendices}\label{app}
\onecolumn
\subsection{Calculation of Sensitivity}\label{app:calS}
Listing~\ref{lst:makeS} shows the main \gls{MATLAB} function used to generate \gls{sen}.
This code allows one to input the data-type along with the requisite parameters and returns a \gls{3D} \gls{senMat} ("S" in code).
The method is based on the use of generalized \gls{lpath} and \gls{Lpath} ("l" or "ll" and "L" in code, respectively). 
\gls{Lpath} and \gls{lpath} are described in detail in Appendix~\ref{app:genL}.
Furthermore, for difference data-types like \gls{SS}, \gls{DS}, or differences of gated \gls{TD} \gls{t} the \gls{sen} the method in Listing~\ref{lst:makeS} also utilizes with \gls{Lpath} and \gls{lpath} as is described in Appendix~\ref{app:genL_diffs}.
\par

Depending on the data-type a different function is needed to calculate \gls{Lpath} and \gls{lpath}.
The handling of choosing the correct \gls{Lpath} or \gls{lpath} functions can be found in the case structure starting the following line in Listing~\ref{lst:makeS}:
\begin{lstlisting}[firstnumber=203, firstline=203, lastline=204, frame=none]
    %% Switch over measurement type and calculate generalized path-lengths
    switch typ{1}
\end{lstlisting}
\noindent Functions for "L" or "l" and also \gls{Y} ("Y" in the code) in Listing~\ref{lst:makeS} can be found in Appendices~\ref{app:DT}\&\ref{app:MC}.
\par

\begin{lstlisting}[
	caption=Main \as{MATLAB} code for generation of \acrfull{sen}. \label{lst:makeS}
	]
function [S, params, Svox] = makeS(typStr, rs, rd, optProp, NVA)
% [S, params, Svox] = makeS(typStr, rs, rd, optProp, NVA)
% 
% Giles Blaney Ph.D. Summer 2023
% 
% Inputs:
%   typStr   - String setting the type of data considered in the format:
%               AA_BB_C where, [e.g., CW_SD_I or FD_DS_P]
%               AA is the temporal domain either: 
%                   CW  - Continuous Wave
%                   FD  - Frequency-Domain 
%                   TD  - Time-Domain 
%               BB is the arrangement type either:
%                   SD  - Single-Distance 
%                   SS  - Single-Slope 
%                   DS  - Dual-Slope 
%               CC is the data-type either:
%                   I   - Intensity for CW or FD (Amplitude for FD)
%                   GI  - Gated Intensity for TD
%                   DGI - Difference in Gated Intensity for TD 
%                   P   - Phase for FD
%                   T   - Mean time-of-flight for TD
%                   V   - Variance for TD
%   rs       - Pencil beam source corrdinates [x, y, z]. (mm)
%   rd       - Detector corrdinates in [x, y, z]. (mm)
%   optProp  - (OPTIONAL) Struct of optical properties with the following
%                fields:
%                    nin  - (default=1.333) Index of refraction inside. (-)
%                    nout - (default=1) Index of refraction outside. (-)
%                    musp - (default=1.1 1/mm) Reduced scattering. (1/mm)
%                    g    - (default=0.9) Anisotropy for MC. (-)
%                    mua  - (default=0.011 1/mm) Absorption. (1/mm)
% 
% Name Value Arguments:
%   'pert'   - (default=[1, 1, 1] mm) Size of perturbation in [x, y, z], 
%               must be a multiple of voxel size (i.e., dr). (mm)
%   'dr'     - (default=1 mm) Voxel side length. (mm)
%   'xl'     - (default: 10 mm past the extent of the optodes) Limits in the
%               x direction in [min, max]. (mm)
%   'yl'     - (default: 10 mm past the extent of the optodes) Limits in the
%               y direction in [min, max]. (mm)
%   'zl'     - (default=[0, 25] mm) Limits in the y direction in [min, max].
%               (mm)
%   'fmod'   - (default=100e6 Hz) Modulation frequency for FD. (Hz)
%   'ndt'    - (default=10e3) Number of time bins for TD or MC.
%   'tend'   - (default=10e3 ps) Max time bin edge for TD or MC. (ps) 
%   'tg'     - (default=[1000, 2000] ps) Gate start and end time for TD GI.
%               (ps)
%   'tgE     - (default=[0, 1000] ps) Early gate start and end time for 
%               TD DGI. (ps)
%   'simTyp' - (default='DT'): String to switch between 'DT' 
%               (Diffusion Theory) and 'MC' (Monte Carlo w/ MCX) 
%               simulation type.
%   'detNA'  - (default=0.5) Detector NA for MC.
%   'np'     - (default=1e8): Number of photons for MC.
%   'usePar' - (default=true): Use parfoor loops.
%   'FFTconv'- (default=true): Use ifft(fft*fft) as conv.
% 
% Outputs:
%   S        - Sensitivity (unit-less)
%   params   - Struct with the following fields
%               x        - X-axis vector. (mm)
%               y        - Y-axis vector. (mm)
%               z        - Z-axis vector. (mm)
%               xl_valid - Valid x-axis limits with perturbation size. (mm)
%               yl_valid - Valid y-axis limits with perturbation size. (mm)
%               zl_valid - Valid z-axis limits with perturbation size. (mm)
%   Svox     - Voxelized sensitivity not considering the perturbation size.
%               (unit-less)

    arguments
        typStr (1,:) string;
        rs (:,3) double; %mm [x, y, z]
        rd (:,3) double; %mm [x, y, z]
        
        optProp struct = [];
        
        NVA.pert (1,3) double = [1, 1, 1]; %mm
        NVA.dr (1,1) double = 1; %mm
        NVA.xl (1,2) double = [NaN, NaN]; %mm
        NVA.yl (1,2) double = [NaN, NaN]; %mm
        NVA.zl (1,2) double = [0, 25]; %mm
        
        NVA.fmod (1,1) double = 100e6; %Hz
        
        NVA.ndt (1,1) double = 10e3;
        NVA.tend (1,1) double = 10e3; %ps

        NVA.tg (2,1) double = [1000; 2000]; %ps
        NVA.tgE (2,1) double = [500; 1000]; %ps

        NVA.simTyp (1,:) string = 'DT';
        NVA.detNA (1,1) double = 0.5; 

        NVA.np (1,1) double = 1e8;

        NVA.usePar (1,1) logical = true;
        NVA.FFTconv (1,1) logical = true;
    end
    
    typ=split(upper(typStr), '_');
    
    %% Check Inputs
    % Default optical properties if needed
    if isempty(optProp)
        clear optProp;

        optProp.nin=1.333;
        optProp.nout=1;
        optProp.musp=1.1; %1/mm
        optProp.mua=0.011; %1/mm
        
        warning('Default optical properties used');
    end
    
    % Check rs and rd based on arrangement type
    % Then make iso source and rep for num meas pairs
    if strcmp(NVA.simTyp, 'MC')
        z0=0;
    else
        z0=1/optProp.musp;
    end
    switch typ{2}
        case 'SD'
            if size(rs, 1)~=1 || size(rd, 1)~=1
                error('Incorrect number of optodes for %s', typ{2});
            end
            rSrcs=rs+[0, 0, z0];
            rDets=rd;
        case 'SS'
            if (size(rs, 1)==1 && size(rd, 1)==2)
                rSrcs=[rs; rs]+[0, 0, z0];
                rDets=rd;
            elseif (size(rs, 1)==2 && size(rd, 1)==1)
                rSrcs=rs+[0, 0, z0];
                rDets=[rd; rd];
            else
                error('Incorrect number of optodes for %s', typ{2});
            end
            
        case 'DS'
            if size(rs, 1)~=2 || size(rd, 1)~=2
                error('Incorrect number of optodes for %s', typ{2});
            end
            rSrcs=[rs([1, 1], :); rs([2, 2], :)]+[0, 0, z0];
            rDets=[rd; flipud(rd)];
        otherwise
            error('Unknown arrangement type %s', typ{2});
    end
    
    % Check pert is a multiple of dr
    if any(mod(NVA.pert, NVA.dr)~=0)
        error('pert size must be a multiple of dr')
    end

    % Default xl and yl if needed
    if any(isnan(NVA.xl))
        NVA.xl=[min([rs(:, 1); rd(:, 1)])-10, ...
            max([rs(:, 1); rd(:, 1)])+10];
    end
    if any(isnan(NVA.yl))
        NVA.yl=[min([rs(:, 2); rd(:, 2)])-10, ...
            max([rs(:, 1); rd(:, 2)])+10];
    end
    
    % Check if MC can be run
    if strcmp(NVA.simTyp, 'MC')
        if strcmp(typ{1}, 'FD')
            error('MC not supported for FD at this time');
        end

        if ~(strcmp(typ{3}, 'I') || strcmp(typ{3}, 'GI') || ...
                strcmp(typ{3}, 'DGI'))
            error('MC only supported for I and GI at this time');
        end
    end
    
    %% Check if run MC and make coordinates
    if strcmp(NVA.simTyp, 'MC')
        for i=1:size(rSrcs, 1)
            [adjoint(i), MCXparams(i), ~]=myMCXLAB_adjoint( ...
                rSrcs(i, :), rDets(i, :), optProp, ...
                'np', NVA.np, 'dr', NVA.dr, ...
                'xl', NVA.xl, 'yl', NVA.yl, 'zl', NVA.zl, ...
                'ndt', NVA.ndt, 'tend', NVA.tend, ...
                'detNA', NVA.detNA);
        end
        XX=MCXparams(1).XX;
        params.x=MCXparams(1).x;
        params.y=MCXparams(1).y;
        params.z=MCXparams(1).z;
    else        
        [YY, XX, ZZ]=meshgrid( ...
            NVA.yl(1):NVA.dr:NVA.yl(2), ...
            NVA.xl(1):NVA.dr:NVA.xl(2), ...
            NVA.zl(1):NVA.dr:NVA.zl(2));
        params.x=squeeze(XX(:, 1, 1));
        params.y=squeeze(YY(1, :, 1));
        params.z=squeeze(ZZ(1, 1, :));
        r=[XX(:), YY(:), ZZ(:)];
    end

    %% Switch over measurement type and calculate generalized path-lengths
    switch typ{1}
        case 'CW'
            switch typ{3}
                case 'I'
                    L=NaN(size(rSrcs, 1), 1);
                    Y=NaN(size(rSrcs, 1), 1);
                    ll=NaN([size(XX), size(rSrcs, 1)]);
                    for i=1:size(rSrcs, 1)
                        Y(i)=1;
                        switch NVA.simTyp
                            case 'DT'
                                L(i)=continuousTotPathLen(...
                                    rSrcs(i, :), rDets(i, :), optProp);
                                
                                l=continuousPartPathLen( ...
                                    rSrcs(i, :), r, rDets(i, :), ...
                                    NVA.dr^3, optProp);
                                l(isnan(l))=0;
                                ll(:, :, :, i)=reshape(l, size(XX));
                            case 'MC'
                                [ll(:, :, :, i), L(i)]=...
                                    continuousPathLen_MCadjoint(...
                                    adjoint(i), MCXparams(i).t, NVA.dr^3);
                            otherwise
                                error('Unknown simTyp %s', NVA.simTyp);
                        end
                    end
                otherwise
                    error('Unkown data-type %s for %s', typ{3}, typ{1});
            end
        
        case 'FD'
            switch typ{3}
                case 'I'
                    L=NaN(size(rSrcs, 1), 1);
                    Y=NaN(size(rSrcs, 1), 1);
                    ll=NaN([size(XX), size(rSrcs, 1)]);
                    for i=1:size(rSrcs, 1)
                        L(i)=real(complexTotPathLen(...
                            rSrcs(i, :), rDets(i, :), ...
                            2*pi*NVA.fmod, optProp));
                        Y(i)=1;
                        
                        l=real(complexPartPathLen( ...
                            rSrcs(i, :), r, rDets(i, :), ...
                            NVA.dr^3, 2*pi*NVA.fmod, optProp));
                        l(isnan(l))=0;
                        
                        ll(:, :, :, i)=reshape(l, size(XX));
                    end
                case 'P'                   
                    L=NaN(size(rSrcs, 1), 1);
                    Y=NaN(size(rSrcs, 1), 1);
                    ll=NaN([size(XX), size(rSrcs, 1)]);
                    for i=1:size(rSrcs, 1)
                        L(i)=imag(complexTotPathLen(...
                            rSrcs(i, :), rDets(i, :), ...
                            2*pi*NVA.fmod, optProp));
                        Y(i)=1;

                        l=imag(complexPartPathLen( ...
                            rSrcs(i, :), r, rDets(i, :), ...
                            NVA.dr^3, 2*pi*NVA.fmod, optProp));
                        l(isnan(l))=0;
                        
                        ll(:, :, :, i)=reshape(l, size(XX));
                    end
                otherwise
                    error('Unkown data-type %s for %s', typ{3}, typ{1});
            end
        
        case 'TD'
            switch typ{3}
                case 'GI'                    
                    L=NaN(size(rSrcs, 1), 1);
                    Y=NaN(size(rSrcs, 1), 1);
                    ll=NaN([size(XX), size(rSrcs, 1)]);
                    for i=1:size(rSrcs, 1)
                        Y(i)=1;
                        switch NVA.simTyp
                            case 'DT'
                                L(i)=temporalGateTotPathLen( ...
                                    rSrcs(i, :), rDets(i, :), ...
                                    NVA.tg, optProp, ...
                                    'conv_t', NVA.tend, ...
                                    'conv_dt', NVA.tend/NVA.ndt);
        
                                l=temporalGatePartPathLen( ...
                                    rSrcs(i, :), r, rDets(i, :), ...
                                    NVA.dr^3, NVA.tg, optProp, ...
                                    'conv_t', NVA.tend, ...
                                    'conv_dt', NVA.tend/NVA.ndt, ...
                                    'usePar', NVA.usePar, ...
                                    'FFTconv', NVA.FFTconv);
                                l(isnan(l))=0;
                                ll(:, :, :, i)=reshape(l, size(XX));
                            case 'MC'
                                [ll(:, :, :, i), L(i)]=...
                                    temporalGatePathLen_MCadjoint(...
                                    adjoint(i), MCXparams(i).t, NVA.dr^3,...
                                    NVA.tg);
                            otherwise
                                error('Unknown simTyp %s', NVA.simTyp);
                        end
                    end
                case 'DGI'
                    if size(rSrcs, 1)~=1
                        error('DGI only supported for SD at this time');
                    end
                    typ{2}=[typ{2}, '_DIFF'];
                    tg_all=[NVA.tgE, NVA.tg];

                    L=NaN(size(tg_all, 2), 1);
                    Y=NaN(size(tg_all, 2), 1);
                    ll=NaN([size(XX), size(tg_all, 2)]);
                    for i=1:size(tg_all, 2)
                        Y(i)=1;
                        switch NVA.simTyp
                            case 'DT'
                                L(i)=temporalGateTotPathLen(...
                                    rSrcs, rDets, ...
                                    tg_all(:, i), optProp, ...
                                    'conv_t', NVA.tend, ...
                                    'conv_dt', NVA.tend/NVA.ndt);
        
                                l=temporalGatePartPathLen( ...
                                    rSrcs, r, rDets, ...
                                    NVA.dr^3, tg_all(:, i), optProp, ...
                                    'conv_t', NVA.tend, ...
                                    'conv_dt', NVA.tend/NVA.ndt, ...
                                    'usePar', NVA.usePar, ...
                                    'FFTconv', NVA.FFTconv);
                                l(isnan(l))=0;
                                ll(:, :, :, i)=reshape(l, size(XX));
                            case 'MC'
                                [ll(:, :, :, i), L(i)]=...
                                    temporalGatePathLen_MCadjoint(...
                                    adjoint, MCXparams.t, NVA.dr^3,...
                                    tg_all(:, i));
                            otherwise
                                error('Unknown simTyp %s', NVA.simTyp);
                        end
                    end
                case 'T'                    
                    L=NaN(size(rSrcs, 1), 1);
                    Y=NaN(size(rSrcs, 1), 1);
                    ll=NaN([size(XX), size(rSrcs, 1)]);
                    for i=1:size(rSrcs, 1)
                        L(i)=temporalKthMomTotPathLen(...
                            rSrcs(i, :), rDets(i, :), 1, optProp);
                        Y(i)=temporalKthMoment(...
                            rSrcs(i, :), rDets(i, :), 1, optProp);
                        
                        l=temporalKthMomPartPathLen( ...
                            rSrcs(i, :), r, rDets(i, :), ...
                            NVA.dr^3, 1, optProp, ...
                            'conv_t', NVA.tend, ...
                            'conv_dt', NVA.tend/NVA.ndt, ...
                            'usePar', NVA.usePar, ...
                            'FFTconv', NVA.FFTconv);
                        l(isnan(l))=0;
                        
                        ll(:, :, :, i)=reshape(l, size(XX));
                    end
                case 'V'                    
                    L=NaN(size(rSrcs, 1), 1);
                    Y=NaN(size(rSrcs, 1), 1);
                    ll=NaN([size(XX), size(rSrcs, 1)]);
                    for i=1:size(rSrcs, 1)
                        L(i)=temporalVarTotPathLen(...
                            rSrcs(i, :), rDets(i, :), optProp);
                        Y(i)=temporalVar(...
                            rSrcs(i, :), rDets(i, :), optProp);
                        
                        l=temporalVarPartPathLen( ...
                            rSrcs(i, :), r, rDets(i, :), ...
                            NVA.dr^3, optProp, ...
                            'conv_t', NVA.tend, ...
                            'conv_dt', NVA.tend/NVA.ndt, ...
                            'usePar', NVA.usePar, ...
                            'FFTconv', NVA.FFTconv);
                        l(isnan(l))=0;
                        
                        ll(:, :, :, i)=reshape(l, size(XX));
                    end
                otherwise
                    error('Unkown data-type %s for %s', typ{3}, typ{1});
            end
            
        otherwise
            error('Unknown temporal domain %s', typ{1});
    end
    
    %% Switch over arrangement type and calculate S
    switch typ{2}
        case 'SD'
            Svox=ll/L;
        case {'SS', 'SD_DIFF'}
            Svox=(Y(2)*ll(:, :, :, 2)-Y(1)*ll(:, :, :, 1))/...
                (Y(2)*L(2)-Y(1)*L(1));
        case 'DS'
            Svox=((Y(2)*ll(:, :, :, 2)-Y(1)*ll(:, :, :, 1))+...
                (Y(4)*ll(:, :, :, 4)-Y(3)*ll(:, :, :, 3)))/...
                ((Y(2)*L(2)-Y(1)*L(1))+...
                (Y(4)*L(4)-Y(3)*L(3)));
        otherwise
            error('Unknown arrangement type %s', typ{2});
    end

    %% Apply pert size
    H=ones(NVA.pert/NVA.dr);
    S=convn(Svox, H, 'same');
    
    if all(NVA.pert==1)
        params.xl_valid=params.x([1, end]);
        params.yl_valid=params.y([1, end]);
        params.zl_valid=params.z([1, end]);
    else
        params.xl_valid=[1, -1]*NVA.pert(1)/2+params.x([1, end]);
        params.yl_valid=[1, -1]*NVA.pert(2)/2+params.y([1, end]);
        params.zl_valid=[1, -1]*NVA.pert(3)/2+params.z([1, end]);
    end

end
\end{lstlisting}
	{\small Available at:\cite{Blaney_23_CodeSensitivity}\\
		\href{https://github.com/DOIT-Lab/DOIT-Public/tree/master/SensitivityCompendium/deps/makeS.m}
		{github.com/DOIT-Lab/DOIT-Public/tree/master/SensitivityCompendium/deps/makeS.m}}

\subsubsection{Generalized Path-length}\label{app:genL}
It may be helpful to express the Jacobians with respect to \gls{mua} as normalized which we name generalize optical path-lengths for \gls{Y}.
These path-lengths all have units of length but are only real path-lengths in the cases of \gls{CW} \gls{I} and \gls{TD} \gls{I}.
In the global case the generalized \gls{Lpath} is defined as follows:
\begin{equation}\label{equ:genL}
	\as{Lpath}_{\as{Y}}=
	\begin{dcases} 
		-\frac{\partial \as{Y} / \partial \as{mua}_{,pert,homo}}{{\as{Y}}} & \text{Dimensioned \as{Y}}\\
		-\partial \as{Y} / \partial \as{mua}_{,pert,homo} & \text{Dimensionless \as{Y}}
	\end{dcases}
\end{equation}
\noindent For the local case, the generalized \gls{lpath} is defined as:
\begin{equation}
	\as{lpath}_{\as{Y}}\left(\as{r}\right)=
	\begin{dcases} 
		-\frac{\partial \as{Y} / \partial \as{dmua}_{,pert}\left(\as{r}\right)}{{\as{Y}}} & \text{Dimensioned \as{Y}}\\
		-\partial \as{Y} / \partial \as{dmua}_{,pert}\left(\as{r}\right) & \text{Dimensionless \as{Y}}
	\end{dcases}
\end{equation}
\noindent These piecewise functions depend on whether \gls{Y} does or does not have units. For example, the \gls{lnI} \ie{\gls{I} data-types} and \gls{phi} are dimensionless while \gls{mTOF} and \gls{var} are dimensioned.
For \gls{FD}, \gls{Y} can be the \gls{RCom} which makes the path-lengths the \gls{lpathCom} and \gls{LpathCom} where the path-length for \gls{phi} is the imaginary part and the path-length for \gls{I} is the real part of complex path-length.\cite{Sassaroli_JOSAA19_DualslopeMethod, Blaney_JBio20_PhaseDualslopes}
\par

Now using these definitions of generalized path-lengths we can rewrite Eq.~(\ref{equ:senDef}) as:
\begin{equation}
	\as{sen}_{\as{Y}}\left(\as{r}\right)=\frac{
		\as{lpath}_{\as{Y}}\left(\as{r}\right)}
	{\as{Lpath}_{\as{Y}}}
\end{equation}
\noindent This is seen in Listing~\ref{lst:makeS} in the following lines:
\begin{lstlisting}[firstnumber=399, firstline=399, lastline=400, frame=none]
        case 'SD'
            Svox=ll/L;
\end{lstlisting}
\noindent Also note that we can use this concept of \gls{Lpath} to write the recovered $\as{dmua}_{,\as{Y}}$ as:
\begin{equation}
	\as{dmua}_{,\as{Y}}=
	\begin{dcases}
		\frac{-\Delta \as{Y}}{\as{Y}_0 \as{Lpath}} & \text{Dimensioned \as{Y}}\\
		\frac{-\Delta \as{Y}}{\as{Lpath}} & \text{Dimensionless \as{Y}}
	\end{dcases}
\end{equation}
\noindent which is further helpful for interpreting \gls{sen} in terms of Eq.~(\ref{equ:interSen}).

\paragraph{Sensitivity of Differences}\label{app:genL_diffs}
Various data-types utilize the differences of measurements such as \gls{SS} and \gls{DS} which are based on spatial differences or differences in \gls{TD} gated \gls{I} which is temporal differences.
In either of these cases Eq.~(\ref{equ:senDef}) can be used to find \gls{sen} for the data-type.
However, we may also write the \gls{sen} for difference data-types in terms of the generalized path-lengths of the data for which the difference is being taken.
For example, in \gls{SS} a difference is taken between a long \gls{rho} and short \gls{rho} and the \gls{sen} of \gls{SS} can be expressed in terms of the general path-lengths of the \gls{SD} long \gls{rho} and \gls{SD} short \gls{rho}.
\par

This \gls{sen} of differences between $\as{Y}_1$ and $\as{Y}_2$ is expressed in this way as follows:
\begin{equation}
	\as{sen}_{\as{Y}_2-\as{Y}_1}=
	\begin{dcases}
		\frac{
			\as{Y}_2\as{lpath}_{\as{Y}_2}-\as{Y}_1\as{lpath}_{\as{Y}_1}}
		{\as{Y}_2\as{Lpath}_{\as{Y}_2}-\as{Y}_1\as{Lpath}_{\as{Y}_1}} & \text{Dimensioned \as{Y}}\\
		\frac{
			\as{lpath}_{\as{Y}_2}-\as{lpath}_{\as{Y}_1}}
		{\as{Lpath}_{\as{Y}_2}-\as{Lpath}_{\as{Y}_1}} & \text{Dimensionless \as{Y}}
	\end{dcases}
\end{equation}
\noindent which, can be found in the following lines in Listing~\ref{lst:makeS}:
\begin{lstlisting}[firstnumber=401, firstline=401, lastline=403, frame=none]
        case {'SS', 'SD_DIFF'}
            Svox=(Y(2)*ll(:, :, :, 2)-Y(1)*ll(:, :, :, 1))/...
                (Y(2)*L(2)-Y(1)*L(1));
\end{lstlisting}
\noindent Furthermore, \gls{DS} is a special case since it is a average of differences.\cite{Sassaroli_JOSAA19_DualslopeMethod,Blaney_JBio20_PhaseDualslopes}
Consider a \gls{DS} with the four measurements $\as{Y}_1$, $\as{Y}_2$, $\as{Y}_3$, and $\as{Y}_4$, where even subscripts are short \gls{rho} and odd long \gls{rho}.
Then, \gls{sen} for \gls{DS} can be written as:
\begin{equation}
	\as{sen}_{(\as{Y}_4-\as{Y}_3)+(\as{Y}_2-\as{Y}_1)}=
	\begin{dcases}
		\frac{
			\left(\as{Y}_4\as{lpath}_{\as{Y}_4}-\as{Y}_3\as{lpath}_{\as{Y}_3}\right)+\left(\as{Y}_2\as{lpath}_{\as{Y}_2}-\as{Y}_1\as{lpath}_{\as{Y}_1}\right)}
		{\left(\as{Y}_4\as{Lpath}_{\as{Y}_4}-\as{Y}_3\as{Lpath}_{\as{Y}_3}\right)+\left(\as{Y}_2\as{Lpath}_{\as{Y}_2}-\as{Y}_1\as{Lpath}_{\as{Y}_1}\right)} & \text{Dimensioned \as{Y}}\\
		\frac{
			\left(\as{lpath}_{\as{Y}_4}-\as{lpath}_{\as{Y}_3}\right)+\left(\as{lpath}_{\as{Y}_2}-\as{lpath}_{\as{Y}_1}\right)}
		{\left(\as{Lpath}_{\as{Y}_4}-\as{Lpath}_{\as{Y}_3}\right)+\left(\as{Lpath}_{\as{Y}_2}-\as{Lpath}_{\as{Y}_1}\right)}& \text{Dimensionless \as{Y}}
	\end{dcases}
\end{equation}
which, can also be found in Listing~\ref{lst:makeS} at the lines:
\begin{lstlisting}[firstnumber=404, firstline=404, lastline=408, frame=none]
        case 'DS'
            Svox=((Y(2)*ll(:, :, :, 2)-Y(1)*ll(:, :, :, 1))+...
                (Y(4)*ll(:, :, :, 4)-Y(3)*ll(:, :, :, 3)))/...
                ((Y(2)*L(2)-Y(1)*L(1))+...
                (Y(4)*L(4)-Y(3)*L(3)));
\end{lstlisting}

\subsubsection{Applying the Perturbation Size}
Listing~\ref{lst:makeS} first calculates the \gls{sen} for each voxel ("Svox" in the code).
The size of these voxels is controlled by "'dr'" name value argument which specifies the side length of cubic voxels.
However, one may choose a perturbation size larger than the voxel size to get \gls{S} values from a realistic perturbation size which can be better interpreted.
This \gls{sen} considering the perturbation size ("S" in the code) is calculated by convolving "Svox" with a matrix of ones whose size is the perturbation size ("'pert'" in the code).
The convolution is implemented in Listing~\ref{lst:makeS} in the following lines:
\begin{lstlisting}[firstnumber=413, firstline=413, lastline=415, frame=none]
    %% Apply pert size
    H=ones(NVA.pert/NVA.dr);
    S=convn(Svox, H, 'same');
\end{lstlisting}
\noindent Therefore, one may say that the "S" matrix generated by the code represents a perturbation scanned every voxel with overlap.
It should be noted that this method generates \glspl{senMat} where the values near the edges assume no \gls{sen} outside the matrix \ie{the convolution uses zero-padding}.
For this reason \gls{S} values in \gls{senMat} ("S" in the code) are not valid within one half a perturbation size from the edge.
This convolution zero-padding is considered in the "xl_valid", "yl_valid", and "zl_valid" fields of the "params" struct.
\par

\subsection{Diffusion Theory}\label{app:DT}
\subsubsection{Perturbation Theory}\label{app:pertTheo}
\paragraph{Global Jacobian Functions}
\subparagraph{Continuous Wave} \phantom{x}
\begin{lstlisting}[
	caption=\as{MATLAB} code for claculation of \acrfull{CW} \acrfull{Lpath}. \label{lst:continuousTotPathLen}
	]
function [L, R] = continuousTotPathLen(rs, rd, optProp)
% Giles Blaney Ph.D. Spring 2023
% [L, R] = continuousTotPathLen(rs, rd, optProp)
% Inputs:
%   rs      - Source corrdinates. (mm)
%   rd      - Detector corrdinates. (mm)
%   optProp - (OPTIONAL) Struct of optical properties with the following
%             fields:
%                nin  - Index of refraction inside. (-)
%                nout - Index of refraction outside. (-)
%                musp - Reduced scattering. (1/mm)
%                mua  - Absorption. (1/mm)
% Outputs:
%   L       - Total pathlength. (mm)
%   R       - Reflectance. (1/mm^2)

    arguments
        rs (:,3) double;
        rd (:,3) double;

        optProp struct = [];
    end

    [L, R]=complexTotPathLen(rs, rd, 0, optProp);
end
\end{lstlisting}
	{\small Available at:\cite{Blaney_23_CodeSensitivity}\\
		\href{https://github.com/DOIT-Lab/DOIT-Public/tree/master/SensitivityCompendium/deps/continuousTotPathLen.m}
		{github.com/DOIT-Lab/DOIT-Public/tree/master/SensitivityCompendium/deps/continuousTotPathLen.m}}

\subparagraph{Frequency-Domain} \phantom{x}
\begin{lstlisting}[
	caption=\as{MATLAB} code for claculation of \acrfull{FD} \acrfull{LpathCom}. \label{lst:complexTotPathLen}
	]
function [L, R] = complexTotPathLen(rs, rd, omega, optProp)
% Giles Blaney Spring 2019
% Updated Giles Blaney Ph.D. Spring 2023
% [L, R] = complexTotPathLen(rs, rd, omega, optProp)
% Inputs:
%   rs      - Source corrdinates. (mm)
%   rd      - Detector corrdinates. (mm)
%   omega   - (OPTIONAL, default=2*pi*1.40625e8 rad/sec) Angular modulation
%             frequecy. (rad/sec)
%   optProp - (OPTIONAL) Struct of optical properties with the following
%             fields:
%                nin  - (default=1.4) Index of refraction inside. (-)
%                nout - (default=1) Index of refraction outside. (-)
%                musp - (default=1.2 1/mm) Reduced scattering. (1/mm)
%                mua  - (default=0.01 1/mm) Absorption. (1/mm)
% Outputs:
%   L       - Complex total pathlength. (mm)
%   R       - Complex reflectance. (1/mm^2)

    arguments
        rs (:,3) double;
        rd (:,3) double;
        
        omega (1,1) double = [];
        optProp struct = [];
    end
    
    if isempty(omega)
        fmod=1.40625e8; %Hz
        omega=2*pi*fmod; %rad/sec
    end
        
    if isempty(optProp)
        clear optProp;
        
        optProp.nin=1.4;
        optProp.nout=1;
        optProp.musp=1.2; %1/mm
        optProp.mua=0.01; %1/mm
        
        warning(['Default optical properties used, this may be inconsistent'...
            ' with the musp used for source depth']);
    end

    if size(rs, 1)>1 && size(rd, 1)>1
        error('Can not use multiple sources and multiple detectors');
    end
    
    x0=rs(:, 1); %mm
    y0=rs(:, 2); %mm
    z0=rs(:, 3); %mm

    c=2.99792458e11; %mm/sec
    v=c/optProp.nin;

    A=n2A(optProp.nin, optProp.nout);
    D=1/(3*optProp.musp); %mm
    zb=-2*A*D; %mm

    mueff=sqrt(optProp.mua/D-1i*omega/(v*D)); %1/mm

    rsp=[x0, y0, -z0+2*zb]; %mm

    r1=vecnorm(rd-rs, 2, 2);
    r2=vecnorm(rd-rsp, 2, 2);
    
    R=complexReflectance(rs, rd, omega, optProp);
    
    L=((z0./r1).*exp(-mueff.*r1)+((z0-2.*zb)./r2).*exp(-mueff.*r2))./...
        (8*pi*D*R);
end
\end{lstlisting}
	{\small Available at:\cite{Blaney_23_CodeSensitivity}\\
		\href{https://github.com/DOIT-Lab/DOIT-Public/tree/master/SensitivityCompendium/deps/complexTotPathLen.m}
		{github.com/DOIT-Lab/DOIT-Public/tree/master/SensitivityCompendium/deps/complexTotPathLen.m}}

\subparagraph{Time-Domain} \phantom{x}
\begin{lstlisting}[
	caption=\as{MATLAB} code for claculation of gated \acrfull{TD} \acrfull{Lpath}. \label{lst:temporalGateTotPathLen}
	]
function [L] = temporalGateTotPathLen(rs, rd, tg, optProp, NVA)
% Giles Blaney Ph.D. Spring 2023
% [L] = temporalGateTotPathLen(rs, rd, tg, optProp)
% Inputs:
%   rs      - Source corrdinates. (mm)
%   rd      - Detector corrdinates. (mm)
%   tg      - Gate start and end time. (ps)
%   optProp - (OPTIONAL) Struct of optical properties with the following
%             fields:
%                nin  - (default=1.4) Index of refraction inside. (-)
%                nout - (default=1) Index of refraction outside. (-)
%                musp - (default=1.2 1/mm) Reduced scattering. (1/mm)
%                mua  - (default=0.01 1/mm) Absorption. (1/mm)
%   Name Value Arguments:
%           - 'conv_t' (default=10e3 ps): Time window for convolution. (ps)
%           - 'conv_dt' (default=1 ps): Time step for convolution. (ps)
%           - 'simTyp' (default='DT'): String to switch between 'DT' and
%               'MC' simulation type
% Outputs:
%   L       - Total pathlength of gated t. (mm)
    
    arguments
        rs (:,3) double; %mm
        rd (:,3) double; %mm
        tg (2,:) double; %ps

        optProp struct = [];

        NVA.conv_t (1,1) double = 10e3; %ps
        NVA.conv_dt (1,1) = 1; %ps

        NVA.simTyp (1,:) string = 'DT';
    end
        
    if isempty(optProp)
        clear optProp;
        
        optProp.nin=1.4;
        optProp.nout=1;
        optProp.musp=1.2; %1/mm
        optProp.mua=0.01; %1/mm
        
        warning(['Default optical properties used, this may be inconsistent'...
            ' with the musp used for source depth']);
    end

    if size(rs, 1)>1 && size(rd, 1)>1
        error('Can not use multiple sources and multiple detectors');
    end

    c=0.299792458; %mm/ps
    v=c/optProp.nin;

    t=-NVA.conv_t:NVA.conv_dt:NVA.conv_t; %ps
    
    Rsd_t=temporalReflectance(rs, rd, t, optProp, 'simTyp', NVA.simTyp);

    L=NaN(size(Rsd_t, 1), size(tg, 2));
    for j=1:size(tg, 2)
        [~, i1]=min(abs(t-tg(1, j)));
        [~, i2]=min(abs(t-tg(2, j)));
        
        num=trapz(t(i1:i2), v*t(i1:i2).*Rsd_t(:, i1:i2), 2);
        Rsd_g=trapz(t(i1:i2), Rsd_t(:, i1:i2), 2);

        L(:, j)=num./Rsd_g;
    end
end
\end{lstlisting}
	{\small Available at:\cite{Blaney_23_CodeSensitivity}\\
		\href{https://github.com/DOIT-Lab/DOIT-Public/tree/master/SensitivityCompendium/deps/temporalGateTotPathLen.m}
		{github.com/DOIT-Lab/DOIT-Public/tree/master/SensitivityCompendium/deps/temporalGateTotPathLen.m}}
	
\begin{lstlisting}[
	caption=\as{MATLAB} code for claculation of \acrfull{TD} \acrfull{tk} generalized \acrfull{Lpath}. \label{lst:temporalKthMomTotPathLen}
	]
function [L] = temporalKthMomTotPathLen(rs, rd, k, optProp)
% Giles Blaney Ph.D. Spring 2023
% [L] = temporalKthMomTotPathLen(rs, rd, k, optProp)
% Inputs:
%   rs      - Source corrdinates. (mm)
%   rd      - Detector corrdinates. (mm)
%   k       - (OPTIONAL; default=1)Moment order.
%   optProp - (OPTIONAL) Struct of optical properties with the following
%             fields:
%                nin  - (default=1.4) Index of refraction inside. (-)
%                nout - (default=1) Index of refraction outside. (-)
%                musp - (default=1.2 1/mm) Reduced scattering. (1/mm)
%                mua  - (default=0.01 1/mm) Absorption. (1/mm)
% Outputs:
%   L       - Total pathlength of kth momment of t. (mm)
    
    arguments
        rs (:,3) double; %mm
        rd (:,3) double; %mm

        k (1,1) double = 1; 

        optProp struct = [];
    end
        
    if isempty(optProp)
        clear optProp;
        
        optProp.nin=1.4;
        optProp.nout=1;
        optProp.musp=1.2; %1/mm
        optProp.mua=0.01; %1/mm
        
        warning(['Default optical properties used, this may be inconsistent'...
            ' with the musp used for source depth']);
    end
    
    if size(rs, 1)>1 && size(rd, 1)>1
        error('Can not use multiple sources and multiple detectors');
    end

    nin=optProp.nin;

    c=0.299792458; %mm/ps
    v=c/nin;

    t1Mom=temporalKthMoment(rs, rd, 1, optProp); %ps
    tkMom=temporalKthMoment(rs, rd, k, optProp); %ps^k
    tkp1Mom=temporalKthMoment(rs, rd, k+1, optProp); %ps^(k+1)

    L=-(v*(t1Mom.*tkMom-tkp1Mom))./tkMom;
end
\end{lstlisting}
	{\small Available at:\cite{Blaney_23_CodeSensitivity}\\
		\href{https://github.com/DOIT-Lab/DOIT-Public/tree/master/SensitivityCompendium/deps/temporalKthMomTotPathLen.m}
		{github.com/DOIT-Lab/DOIT-Public/tree/master/SensitivityCompendium/deps/temporalKthMomTotPathLen.m}}
	
\begin{lstlisting}[
	caption=\as{MATLAB} code for claculation of \acrfull{TD} \acrfull{var} generalized \acrfull{Lpath}. \label{lst:temporalVarTotPathLen}
	]
function [L] = temporalVarTotPathLen(rs, rd, optProp)
% Giles Blaney Ph.D. Spring 2023
% [L] = temporalVarTotPathLen(rs, rd, optProp)
% Inputs:
%   rs      - Source corrdinates. (mm)
%   rd      - Detector corrdinates. (mm)
%   optProp - (OPTIONAL) Struct of optical properties with the following
%             fields:
%                nin  - (default=1.4) Index of refraction inside. (-)
%                nout - (default=1) Index of refraction outside. (-)
%                musp - (default=1.2 1/mm) Reduced scattering. (1/mm)
%                mua  - (default=0.01 1/mm) Absorption. (1/mm)
% Outputs:
%   L       - Total pathlength of variance. (mm)
    
    arguments
        rs (:,3) double; %mm
        rd (:,3) double; %mm

        optProp struct = [];
    end
    
    if isempty(optProp)
        clear optProp;
        
        optProp.nin=1.4;
        optProp.nout=1;
        optProp.musp=1.2; %1/mm
        optProp.mua=0.01; %1/mm
        
        warning(['Default optical properties used, this may be inconsistent'...
            ' with the musp used for source depth']);
    end
    
    if size(rs, 1)>1 && size(rd, 1)>1
        error('Can not use multiple sources and multiple detectors');
    end
    
    t1Mom=temporalKthMoment(rs, rd, 1, optProp);
    t2Mom=temporalKthMoment(rs, rd, 2, optProp);
    V=t2Mom-t1Mom.^2;

    Lt1=temporalKthMomTotPathLen(rs, rd, 1, optProp);
    Lt2=temporalKthMomTotPathLen(rs, rd, 2, optProp);
    
    L=(t2Mom.*Lt2-2*t1Mom.^2.*Lt1)./V;
end
\end{lstlisting}
	{\small Available at:\cite{Blaney_23_CodeSensitivity}\\
		\href{https://github.com/DOIT-Lab/DOIT-Public/tree/master/SensitivityCompendium/deps/temporalVarTotPathLen.m}
		{github.com/DOIT-Lab/DOIT-Public/tree/master/SensitivityCompendium/deps/temporalVarTotPathLen.m}}

\paragraph{Local Jacobian Functions}
\subparagraph{Continuous Wave} \phantom{x}
\begin{lstlisting}[
	caption=\as{MATLAB} code for claculation of \acrfull{CW} \acrfull{lpath}. \label{lst:continuousPartPathLen}
	]
function [l] = continuousPartPathLen(rs, r, rd, V, optProp)
% Giles Blaney Ph.D. Spring 2023
% [l] = continuousPartPathLen(rs, r, rd, V, optProp)
% Inputs:
%   rs      - Source corrdinates. (mm)
%   r       - Center corrdinate of volume. (mm)
%   rd      - Detector corrdinates. (mm)
%   V       - Volume. (mm^3)
%   optProp - (OPTIONAL) Struct of optical properties with the following
%             fields:
%                nin  - Index of refraction inside. (-)
%                nout - Index of refraction outside. (-)
%                musp - Reduced scattering. (1/mm)
%                mua  - Absorption. (1/mm)
% Outputs:
%   l       - Partial pathlength. (mm)

    arguments
        rs (:,3) double;
        r (:,3) double;
        rd (:,3) double;
        V (1,1) double;

        optProp struct = [];
    end
    
    l=complexPartPathLen(rs, r, rd, V, 0, optProp);
end
\end{lstlisting}
	{\small Available at:\cite{Blaney_23_CodeSensitivity}\\
		\href{https://github.com/DOIT-Lab/DOIT-Public/tree/master/SensitivityCompendium/deps/continuousPartPathLen.m}
		{github.com/DOIT-Lab/DOIT-Public/tree/master/SensitivityCompendium/deps/continuousPartPathLen.m}}

\subparagraph{Frequency-Domain} \phantom{x}
\begin{lstlisting}[
	caption=\as{MATLAB} code for claculation of \acrfull{FD} \acrfull{lpathCom}. \label{lst:complexPartPathLen}
	]
function [l] = complexPartPathLen(rs, r, rd, V, omega, optProp)
% Giles Blaney Spring 2019
% Updated Giles Blaney Ph.D. Spring 2023
% [l] = complexPartPathLen(rs, r, rd, V, omega, optProp)
% Inputs:
%   rs      - Source corrdinates. (mm)
%   r       - Center corrdinate of volume. (mm)
%   rd      - Detector corrdinates. (mm)
%   V       - Volume. (mm^3)
%   omega   - (OPTIONAL, default=2*pi*1.40625e8 rad/sec) Angular modulation
%             frequecy. (rad/sec)
%   optProp - (OPTIONAL) Struct of optical properties with the following
%             fields:
%                nin  - (default=1.4) Index of refraction inside. (-)
%                nout - (default=1) Index of refraction outside. (-)
%                musp - (default=1.2 1/mm) Reduced scattering. (1/mm)
%                mua  - (default=0.01 1/mm) Absorption. (1/mm)
% Outputs:
%   l       - Complex partial pathlength. (mm)

    arguments
        rs (:,3) double;
        r (:,3) double;
        rd (:,3) double;
        V (1,1) double;
        
        omega (1,1) = [];
        optProp struct = [];
    end
    
    if isempty(omega)
        fmod=1.40625e8; %Hz
        omega=2*pi*fmod; %rad/sec
    end
        
    if isempty(optProp)
        clear optProp;
        
        optProp.nin=1.4;
        optProp.nout=1;
        optProp.musp=1.2; %1/mm
        optProp.mua=0.01; %1/mm
        
        warning(['Default optical properties used, this may be inconsistent'...
            ' with the musp used for source depth']);
    end
    
    PHIrs_r=complexFluence(rs, r, omega, optProp);
    Rr_rd=complexReflectance(r, rd, omega, optProp);
    Rrs_rd=complexReflectance(rs, rd, omega, optProp);
    
    l=(PHIrs_r.*Rr_rd.*V)./Rrs_rd;
end
\end{lstlisting}
	{\small Available at:\cite{Blaney_23_CodeSensitivity}\\
		\href{https://github.com/DOIT-Lab/DOIT-Public/tree/master/SensitivityCompendium/deps/complexPartPathLen.m}
		{github.com/DOIT-Lab/DOIT-Public/tree/master/SensitivityCompendium/deps/complexPartPathLen.m}}

\subparagraph{Time-Domain} \phantom{x}
\begin{lstlisting}[
	caption=\as{MATLAB} code for claculation of gated \acrfull{TD} \acrfull{lpath}. \label{lst:temporalGatePartPathLen}
	]
function [l] = temporalGatePartPathLen(rs, r, rd, Vol, tg, optProp, NVA)
% Giles Blaney Ph.D. Spring 2023
% [l] = temporalGatePartPathLen(rs, rd, Vol, tg, optProp, NVA)
% Inputs:
%   rs      - Source corrdinates. (mm)
%   r       - Center corrdinate of volume. (mm)
%   rd      - Detector corrdinates. (mm)
%   Vol     - Volume. (mm^3)
%   tg      - Gate start and end time. (ps)
%   optProp - (OPTIONAL) Struct of optical properties with the following
%             fields:
%                nin  - (default=1.4) Index of refraction inside. (-)
%                nout - (default=1) Index of refraction outside. (-)
%                musp - (default=1.2 1/mm) Reduced scattering. (1/mm)
%                mua  - (default=0.01 1/mm) Absorption. (1/mm)
%   Name Value Arguments:
%           - 'conv_t' (default=10e3 ps): Time window for convolution. (ps)
%           - 'conv_dt' (default=1 ps): Time step for convolution. (ps)
%           - 'usePar' (default=true): Use parfoor loops.
%           - 'FFTconv' (default=true): Use ifft(fft*fft) as conv.
% Outputs:
%   l       - Partial pathlength of gated t. (mm)
    
    arguments
        rs (1,3) double; %mm
        r (:,3) double; %mm
        rd (1,3) double; %mm
        Vol (1,1) double; %mm^3
        tg (2,:) double; %ps

        optProp struct = [];

        NVA.conv_t (1,1) double = 10e3; %ps
        NVA.conv_dt (1,1) = 1; %ps
        NVA.usePar (1,1) logical = true;
        NVA.FFTconv (1,1) logical = true;
    end
        
    if isempty(optProp)
        clear optProp;
        
        optProp.nin=1.4;
        optProp.nout=1;
        optProp.musp=1.2; %1/mm
        optProp.mua=0.01; %1/mm
        
        warning(['Default optical properties used, this may be inconsistent'...
            ' with the musp used for source depth']);
    end
    
    if NVA.usePar
        pp=gcp;
    end

    t=-NVA.conv_t:NVA.conv_dt:NVA.conv_t; %ps
    posInds=t>0;
    
    Rsd_t=temporalReflectance(rs, rd, t, optProp);
    
    tgInd=NaN(size(tg));
    Rsd_g=NaN(1, size(tg, 2));
    for j=1:size(tg, 2)
        [~, tgInd(1, j)]=min(abs(t-tg(1, j)));
        [~, tgInd(2, j)]=min(abs(t-tg(2, j)));
        i1=tgInd(1, j);
        i2=tgInd(2, j);
        
        Rsd_g(1, j)=trapz(t(i1:i2), Rsd_t(i1:i2), 2);
    end
    
    if NVA.usePar
        FFTconv=NVA.FFTconv;
        dt=NVA.conv_dt;
        l=NaN(size(r, 1), size(tg, 2));
        parfor i=1:size(r, 1)     
            
            PHIsi=zeros(size(t));
            Rid=zeros(size(t));
            PHIsi(posInds)=temporalFluence(rs, r(i, :), t(posInds), ...
                optProp);
            Rid(posInds)=temporalReflectance(r(i, :), rd, t(posInds), ...
                optProp);
            
            num=NaN(1, size(tg, 2));
            for j=1:size(tg, 2)
                i1=tgInd(1, j);
                i2=tgInd(2, j);
                
                if FFTconv
                    convPHIsiRid=ifft(fft(PHIsi).*fft(Rid));
                    convPHIsiRid((sum(posInds)+1):end)=[];
                    
                    [~, j1]=min(abs(t(i1)-t(posInds)));
                    [~, j2]=min(abs(t(i2)-t(posInds)));
                    
                    num(1, j)=sum(convPHIsiRid(j1:j2), 2)*dt;
                else
                    convTmp=conv(PHIsi, Rid, 'same')*dt;
                    num(1, j)=trapz(t(i1:i2), convTmp(i1:i2), 2);
                end
            end
            l(i, :)=(num./Rsd_g)*Vol;
        end
    else
        PHIsi=zeros(size(r, 1), length(t));
        Rid=zeros(size(r, 1), length(t));
        PHIsi(:, posInds)=temporalFluence(rs, r, t(posInds), optProp);
        Rid(:, posInds)=temporalReflectance(r, rd, t(posInds), optProp);
        convPR=NaN(size(r, 1), length(t));
        convPHIsiRid=NaN(size(r, 1), sum(posInds));
        
        if NVA.FFTconv
            for i=1:size(r, 1)
                tmp=ifft(fft(PHIsi(i, :)).*fft(Rid(i, :)));
                convPHIsiRid(i, :)=tmp(1:sum(posInds));
            end
            
            num=NaN(size(r, 1), size(tg, 2));
            for j=1:size(tg, 2)
                i1=tgInd(1, j);
                i2=tgInd(2, j);

                [~, j1]=min(abs(t(i1)-t(posInds)));
                [~, j2]=min(abs(t(i2)-t(posInds)));
    
                num(:, j)=sum(convPHIsiRid(:, j1:j2), 2)*NVA.conv_dt;
            end
        else
            for i=1:size(r, 1)
                convPR(i, :)=conv(PHIsi(i, :), Rid(i, :), 'same')*NVA.conv_dt;
            end
            
            num=NaN(size(r, 1), size(tg, 2));
            for j=1:size(tg, 2)
                i1=tgInd(1, j);
                i2=tgInd(2, j);
    
                num(:, j)=trapz(t(i1:i2), convPR(:, i1:i2), 2);
            end
        end
        l=(num./Rsd_g)*Vol;
    end
end
\end{lstlisting}
	{\small Available at:\cite{Blaney_23_CodeSensitivity}\\
		\href{https://github.com/DOIT-Lab/DOIT-Public/tree/master/SensitivityCompendium/deps/temporalGatePartPathLen.m}
		{github.com/DOIT-Lab/DOIT-Public/tree/master/SensitivityCompendium/deps/temporalGatePartPathLen.m}}
	
\begin{lstlisting}[
	caption=\as{MATLAB} code for claculation of \acrfull{TD} \acrfull{tk} generalized \acrfull{lpath}. \label{lst:temporalKthMomPartPathLen}
	]
function [l] = temporalKthMomPartPathLen(rs, r, rd, Vol, k, optProp, NVA)
% Giles Blaney Ph.D. Spring 2023
% [l] = temporalKthMomPartPathLen(rs, rd, tns, optProp)
% Inputs:
%   rs      - Source corrdinates. (mm)
%   r       - Center corrdinate of volume. (mm)
%   rd      - Detector corrdinates. (mm)
%   Vol     - Volume. (mm^3)
%   k       - (OPTIONAL; default=1) Moment order.
%   optProp - (OPTIONAL) Struct of optical properties with the following
%             fields:
%                nin  - (default=1.4) Index of refraction inside. (-)
%                nout - (default=1) Index of refraction outside. (-)
%                musp - (default=1.2 1/mm) Reduced scattering. (1/mm)
%                mua  - (default=0.01 1/mm) Absorption. (1/mm)
%   Name Value Arguments:
%           - 'conv_t' (default=10e3 ps): Time window for convolution. (ps)
%           - 'conv_dt' (default=1 ps): Time step for convolution. (ps)
%           - 'usePar' (default=true): Use parfoor loops.
%           - 'FFTconv' (default=true): Use ifft(fft*fft) as conv.
% Outputs:
%   l       - Partial pathlength of kth momment of t. (mm)
    
    arguments
        rs (1,3) double; %mm
        r (:,3) double; %mm
        rd (1,3) double; %mm
        Vol (1,1) double; %mm^3

        k (1,1) double = 1; 

        optProp struct = [];

        NVA.conv_t (1,1) double = 10e3; %ps
        NVA.conv_dt (1,1) = 1; %ps
        NVA.usePar (1,1) logical = true;
        NVA.FFTconv (1,1) logical = true;
    end
        
    if isempty(optProp)
        clear optProp;
        
        optProp.nin=1.4;
        optProp.nout=1;
        optProp.musp=1.2; %1/mm
        optProp.mua=0.01; %1/mm
        
        warning(['Default optical properties used, this may be inconsistent'...
            ' with the musp used for source depth']);
    end
    
    if NVA.usePar
        pp=gcp;
    end

    t=-NVA.conv_t:NVA.conv_dt:NVA.conv_t; %ps
    posInds=t>0;
    
    tkMom=temporalKthMoment(rs, rd, k, optProp);
    RC=continuousReflectance(rs, rd, optProp);
    
    if NVA.usePar
        FFTconv=NVA.FFTconv;
        dt=NVA.conv_dt;
        l=NaN(size(r, 1), 1);
        parfor i=1:size(r, 1)
            lC=continuousPartPathLen(rs, r(i, :), rd, Vol, optProp);
            
            PHIsi=zeros(size(t));
            Rid=zeros(size(t));
            PHIsi(posInds)=temporalFluence(rs, r(i, :), t(posInds), ...
                optProp);
            Rid(posInds)=temporalReflectance(r(i, :), rd, t(posInds), ...
                optProp);
            
            if FFTconv
                convPHIsiRid=ifft(fft(PHIsi).*fft(Rid));
                convPHIsiRid((sum(posInds)+1):end)=[];
                
                l(i)=-(lC*tkMom-(Vol/RC)*...
                    sum(t(posInds).^k.*...
                    (convPHIsiRid*dt),...
                    2))*dt/tkMom;
            else
                l(i)=-(lC*tkMom-(Vol/RC)*...
                    trapz(t, t.^k.*...
                    (conv(PHIsi, Rid, 'same')*dt),...
                    2))/tkMom;
            end
        end
    else
        PHIsi=zeros(size(r, 1), length(t));
        Rid=zeros(size(r, 1), length(t));
        lC=continuousPartPathLen(rs, r, rd, Vol, optProp);
        PHIsi(:, posInds)=temporalFluence(rs, r, t(posInds), optProp);
        Rid(:, posInds)=temporalReflectance(r, rd, t(posInds), optProp);
        convPR=NaN(size(r, 1), length(t));
        convPHIsiRid=NaN(size(r, 1), sum(posInds));
        
        if NVA.FFTconv
            for i=1:size(r, 1)
                tmp=ifft(fft(PHIsi(i, :)).*fft(Rid(i, :)));
                convPHIsiRid(i, :)=tmp(1:sum(posInds));
            end
            
            dkMomIdmua=lC*tkMom-(Vol/RC)*...
                sum(...
                t(posInds).^k.*(convPHIsiRid*NVA.conv_dt)*NVA.conv_dt, 2);
        else
            for i=1:size(r, 1)
                convPR(i, :)=conv( ...
                    PHIsi(i, :), Rid(i, :), 'same')*NVA.conv_dt;
            end
            
            dkMomIdmua=lC*tkMom-(Vol/RC)*...
                trapz(t, ...
                t.^k.*convPR, 2);
        end
        
        l=-dkMomIdmua/tkMom;
    end
end
\end{lstlisting}
	{\small Available at:\cite{Blaney_23_CodeSensitivity}\\
		\href{https://github.com/DOIT-Lab/DOIT-Public/tree/master/SensitivityCompendium/deps/temporalKthMomPartPathLen.m}
		{github.com/DOIT-Lab/DOIT-Public/tree/master/SensitivityCompendium/deps/temporalKthMomPartPathLen.m}}
	
\begin{lstlisting}[
	caption=\as{MATLAB} code for claculation of \acrfull{TD} \acrfull{var} generalized \acrfull{lpath}. \label{lst:temporalVarPartPathLen}
	]
function [l] = temporalVarPartPathLen(rs, r, rd, Vol, optProp, NVA)
% Giles Blaney Ph.D. Spring 2023
% [l] = temporalVarPartPathLen(rs, r, rd, Vol, optProp, NVA)
% Inputs:
%   rs      - Source corrdinates. (mm)
%   r       - Center corrdinate of volume. (mm)
%   rd      - Detector corrdinates. (mm)
%   Vol     - Volume. (mm^3)
%   optProp - (OPTIONAL) Struct of optical properties with the following
%             fields:
%                nin  - (default=1.4) Index of refraction inside. (-)
%                nout - (default=1) Index of refraction outside. (-)
%                musp - (default=1.2 1/mm) Reduced scattering. (1/mm)
%                mua  - (default=0.01 1/mm) Absorption. (1/mm)
%   Name Value Arguments:
%           - 'conv_t' (default=10e3 ps): Time window for convolution. (ps)
%           - 'conv_dt' (default=1 ps): Time step for convolution. (ps)
%           - 'usePar' (default=true): Use parfoor loops.
%           - 'FFTconv' (default=true): Use ifft(fft*fft) as conv.
% Outputs:
%   l       - Partial pathlength of variance. (mm)
    
    arguments
        rs (1,3) double; %mm
        r (:,3) double; %mm
        rd (1,3) double; %mm
        Vol (1,1) double; %mm^3

        optProp struct = [];

        NVA.conv_t (1,1) double = 10e3; %ps
        NVA.conv_dt (1,1) = 1; %ps
        NVA.usePar (1,1) logical = true;
        NVA.FFTconv (1,1) logical = true;
    end
        
    if isempty(optProp)
        clear optProp;
        
        optProp.nin=1.4;
        optProp.nout=1;
        optProp.musp=1.2; %1/mm
        optProp.mua=0.01; %1/mm
        
        warning(['Default optical properties used, this may be inconsistent'...
            ' with the musp used for source depth']);
    end
    
    if size(rs, 1)>1 && size(rd, 1)>1
        error('Can not use multiple sources and multiple detectors');
    end

    NVAstruct=struct2pairs(NVA);
    
    t1Mom=temporalKthMoment(rs, rd, 1, optProp);
    t2Mom=temporalKthMoment(rs, rd, 2, optProp);
    V=t2Mom-t1Mom.^2;

    lt1=temporalKthMomPartPathLen(rs, r, rd, Vol, 1, optProp, NVAstruct{:});
    lt2=temporalKthMomPartPathLen(rs, r, rd, Vol, 2, optProp, NVAstruct{:});
    
    l=(t2Mom.*lt2-2*t1Mom.^2.*lt1)./V;

    l(isnan(l))=0;
end
\end{lstlisting}
	{\small Available at:\cite{Blaney_23_CodeSensitivity}\\
		\href{https://github.com/DOIT-Lab/DOIT-Public/tree/master/SensitivityCompendium/deps/temporalVarPartPathLen.m}
		{github.com/DOIT-Lab/DOIT-Public/tree/master/SensitivityCompendium/deps/temporalVarPartPathLen.m}}

\subsubsection{Forward Functions}\label{app:DTfwd}
\paragraph{Continuous Wave} \phantom{x}
\begin{lstlisting}[
	caption=\as{MATLAB} code for claculation of \acrfull{CW} \acrfull{R}. \label{lst:continuousReflectance}
	]
function [R] = continuousReflectance(rs, rd, optProp)
% Giles Blaney Ph.D. Spring 2023
% [R] = continuousFluence(rs, rd, optProp)
% Inputs:
%   rs      - Source corrdinates. (mm)
%   rd      - Detector corrdinates. (mm)
%   optProp - (OPTIONAL) Struct of optical properties with the following
%             fields:
%                nin  - Index of refraction inside. (-)
%                nout - Index of refraction outside. (-)
%                musp - Reduced scattering. (1/mm)
%                mua  - Absorption. (1/mm)
% Outputs:
%   R       - Reflectance. (1/mm^2)

    arguments
        rs (:,3) double;
        rd (:,3) double;

        optProp struct = [];
    end

    R=complexReflectance(rs, rd, 0, optProp);
end
\end{lstlisting}
	{\small Available at:\cite{Blaney_23_CodeSensitivity}\\
		\href{https://github.com/DOIT-Lab/DOIT-Public/tree/master/SensitivityCompendium/deps/continuousReflectance.m}
		{github.com/DOIT-Lab/DOIT-Public/tree/master/SensitivityCompendium/deps/continuousReflectance.m}}

\paragraph{Frequency-Domain} \phantom{x}
\begin{lstlisting}[
	caption=\as{MATLAB} code for claculation of \acrfull{FD} \acrfull{PHIcom}. \label{lst:complexFluence}
	]
function [PHI] = complexFluence(rs, rd, omega, optProp)
% Giles Blaney Spring 2019
% Updated Giles Blaney Ph.D. Spring 2023
% [PHI] = complexFluence(rs, rd, omega, optProp)
% Inputs:
%   rs      - Source corrdinates. (mm)
%   rd      - Detector corrdinates. (mm)
%   omega   - (OPTIONAL, default=2*pi*1.40625e8 rad/sec) Angular modulation
%             frequecy. (rad/sec)
%   optProp - (OPTIONAL) Struct of optical properties with the following
%             fields:
%                nin  - (default=1.4) Index of refraction inside. (-)
%                nout - (default=1) Index of refraction outside. (-)
%                musp - (default=1.2 1/mm) Reduced scattering. (1/mm)
%                mua  - (default=0.01 1/mm) Absorption. (1/mm)
% Outputs:
%   PHI     - Complex fluence. (1/mm^2)
    
    arguments
        rs (:,3) double;
        rd (:,3) double;
        
        omega (1,1) double = [];
        optProp struct = [];
    end
    
    if isempty(omega)
        fmod=1.40625e8; %Hz
        omega=2*pi*fmod; %rad/sec
    end
        
    if isempty(optProp)
        clear optProp;
        
        optProp.nin=1.4;
        optProp.nout=1;
        optProp.musp=1.2; %1/mm
        optProp.mua=0.01; %1/mm
        
        warning(['Default optical properties used, this may be inconsistent'...
            ' with the musp used for source depth']);
    end
    
    if size(rs, 1)>1 && size(rd, 1)>1
        error('Can not use multiple sources and multiple detectors');
    end
    
    x0=rs(:, 1); %mm
    y0=rs(:, 2); %mm
    z0=rs(:, 3); %mm

    c=2.99792458e11; %mm/sec
    v=c/optProp.nin;

    A=n2A(optProp.nin, optProp.nout);
    D=1/(3*optProp.musp); %mm
    zb=-2*A*D; %mm

    mueff=sqrt(optProp.mua/D-1i*omega/(v*D)); %1/mm

    rsp=[x0, y0, -z0+2*zb]; %mm

    r1=vecnorm(rd-rs, 2, 2);
    r2=vecnorm(rd-rsp, 2, 2);
    
    PHI=(exp(-mueff.*r1)./r1-exp(-mueff.*r2)./r2)/(4*pi*D);
end
\end{lstlisting}
	{\small Available at:\cite{Blaney_23_CodeSensitivity}\\
		\href{https://github.com/DOIT-Lab/DOIT-Public/tree/master/SensitivityCompendium/deps/complexFluence.m}
		{github.com/DOIT-Lab/DOIT-Public/tree/master/SensitivityCompendium/deps/complexFluence.m}}

\begin{lstlisting}[
	caption=\as{MATLAB} code for claculation of \acrfull{FD} \acrfull{RCom}. \label{lst:complexReflectance}
	]
function [R] = complexReflectance(rs, rd, omega, optProp)
% Giles Blaney Spring 2019
% Updated Giles Blaney Ph.D. Spring 2023
% [R] = complexReflectance(rs, rd, omega, optProp)
% Inputs:
%   rs      - Source corrdinates. (mm)
%   rd      - Detector corrdinates. (mm)
%   omega   - (OPTIONAL, default=2*pi*1.40625e8 rad/sec) Angular modulation
%             frequecy. (rad/sec)
%   optProp - (OPTIONAL) Struct of optical properties with the following
%             fields:
%                nin  - (default=1.4) Index of refraction inside. (-)
%                nout - (default=1) Index of refraction outside. (-)
%                musp - (default=1.2 1/mm) Reduced scattering. (1/mm)
%                mua  - (default=0.01 1/mm) Absorption. (1/mm)
% Outputs:
%   R       - Complex reflectance. (1/mm^2)

    arguments
        rs (:,3) double;
        rd (:,3) double;
        
        omega (1,1) double = [];
        optProp struct = [];
    end
    
    if isempty(omega)
        fmod=1.40625e8; %Hz
        omega=2*pi*fmod; %rad/sec
    end
        
    if isempty(optProp)
        clear optProp;
        
        optProp.nin=1.4;
        optProp.nout=1;
        optProp.musp=1.2; %1/mm
        optProp.mua=0.01; %1/mm
        
        warning(['Default optical properties used, this may be inconsistent'...
            ' with the musp used for source depth']);
    end
    
    if size(rs, 1)>1 && size(rd, 1)>1
        error('Can not use multiple sources and multiple detectors');
    end
    
    x0=rs(:, 1); %mm
    y0=rs(:, 2); %mm
    z0=rs(:, 3); %mm

    c=2.99792458e11; %mm/sec
    v=c/optProp.nin;

    A=n2A(optProp.nin, optProp.nout);
    D=1/(3*optProp.musp); %mm
    zb=-2*A*D; %mm

    mueff=sqrt(optProp.mua/D-1i*omega/(v*D)); %1/mm

    rsp=[x0, y0, -z0+2*zb]; %mm

    r1=vecnorm(rd-rs, 2, 2);
    r2=vecnorm(rd-rsp, 2, 2);
    
    R=(z0.*(1./r1+mueff).*exp(-mueff.*r1)./(r1.^2)+...
        (z0-2*zb).*(1./r2+mueff).*exp(-mueff.*r2)./(r2.^2))/...
        (4*pi);
end
\end{lstlisting}
	{\small Available at:\cite{Blaney_23_CodeSensitivity}\\
		\href{https://github.com/DOIT-Lab/DOIT-Public/tree/master/SensitivityCompendium/deps/complexReflectance.m}
		{github.com/DOIT-Lab/DOIT-Public/tree/master/SensitivityCompendium/deps/complexReflectance.m}}

\paragraph{Time-Domain} \phantom{x}
\begin{lstlisting}[
	caption=\as{MATLAB} code for claculation of \acrfull{TD} \acrfull{PHI}. \label{lst:temporalFluence}
	]
function [PHI] = temporalFluence(rs, rd, t, optProp)
% Giles Blaney Ph.D. Spring 2023
% [PHI] = temporalFluence(rs, rd, tns, optProp)
% Inputs:
%   rs      - Source corrdinates. (mm)
%   rd      - Detector corrdinates. (mm)
%   t       - Time. (ps)
%   optProp - (OPTIONAL) Struct of optical properties with the following
%             fields:
%                nin  - (default=1.4) Index of refraction inside. (-)
%                nout - (default=1) Index of refraction outside. (-)
%                musp - (default=1.2 1/mm) Reduced scattering. (1/mm)
%                mua  - (default=0.01 1/mm) Absorption. (1/mm)
% Outputs:
%   PHI     - Temporal fluence. (1/(ps mm^2))
    
    arguments
        rs (:,3) double; %mm
        rd (:,3) double; %mm
        t (1,:) double; %ps

        optProp struct = [];
    end
        
    if isempty(optProp)
        clear optProp;
        
        optProp.nin=1.4;
        optProp.nout=1;
        optProp.musp=1.2; %1/mm
        optProp.mua=0.01; %1/mm
        
        warning(['Default optical properties used, this may be inconsistent'...
            ' with the musp used for source depth']);
    end
    
    if size(rs, 1)>1 && size(rd, 1)>1
        error('Can not use multiple sources and multiple detectors');
    end

    nin=optProp.nin;
    nout=optProp.nout;
    musp=optProp.musp; %1/mm
    mua=optProp.mua; %1/mm
    
    x0=rs(:, 1); %mm
    y0=rs(:, 2); %mm
    z0=rs(:, 3); %mm

    c=0.299792458; %mm/ps
    v=c/nin;

    A=n2A(nin, nout);
    D=1/(3*musp); %mm
    zb=-2*A*D; %mm

    rsp=[x0, y0, -z0+2*zb]; %mm

    r1=vecnorm(rd-rs, 2, 2);
    r2=vecnorm(rd-rsp, 2, 2);

    posInds=t>0;
    tPos=t(posInds);
    
    PHI=...
        ((v*exp(-mua*v*tPos))./(4*pi*D*v*tPos).^(3/2)).*...
        (exp(-r1.^2./(4*D*v*tPos))-...
        exp(-r2.^2./(4*D*v*tPos))); %1/(ps mm^2)

    PHI=[zeros(size(PHI, 1), sum(~posInds)), PHI];
end
\end{lstlisting}
	{\small Available at:\cite{Blaney_23_CodeSensitivity}\\
		\href{https://github.com/DOIT-Lab/DOIT-Public/tree/master/SensitivityCompendium/deps/temporalFluence.m}
		{github.com/DOIT-Lab/DOIT-Public/tree/master/SensitivityCompendium/deps/temporalFluence.m}}
	
\lstinputlisting[
	caption=\as{MATLAB} code for claculation of \acrfull{TD} \acrfull{R}. \label{lst:temporalReflectance}
	]{temporalReflectance.m}
	{\small Available at:\cite{Blaney_23_CodeSensitivity}\\
		\href{https://github.com/DOIT-Lab/DOIT-Public/tree/master/SensitivityCompendium/deps/temporalReflectance.m}
		{github.com/DOIT-Lab/DOIT-Public/tree/master/SensitivityCompendium/deps/temporalReflectance.m}}
	
\begin{lstlisting}[
	caption=\as{MATLAB} code for claculation of \acrfull{TD} \acrfull{tk}. \label{lst:temporalKthMoment}
	]
function [tkMom] = temporalKthMoment(rs, rd, k, optProp)
% Giles Blaney Ph.D. Spring 2023
% [R] = temporalReflectance(rs, rd, tns, optProp)
% Inputs:
%   rs      - Source corrdinates. (mm)
%   rd      - Detector corrdinates. (mm)
%   k       - (OPTIONAL; default=1)Moment order.
%   optProp - (OPTIONAL) Struct of optical properties with the following
%             fields:
%                nin  - (default=1.4) Index of refraction inside. (-)
%                nout - (default=1) Index of refraction outside. (-)
%                musp - (default=1.2 1/mm) Reduced scattering. (1/mm)
%                mua  - (default=0.01 1/mm) Absorption. (1/mm)
% Outputs:
%   tkMom     - kth Momment of t; <t^k>. (ps^k)
    
    arguments
        rs (:,3) double; %mm
        rd (:,3) double; %mm

        k (1,1) double = 1; 

        optProp struct = [];
    end
        
    if isempty(optProp)
        clear optProp;
        
        optProp.nin=1.4;
        optProp.nout=1;
        optProp.musp=1.2; %1/mm
        optProp.mua=0.01; %1/mm
        
        warning(['Default optical properties used, this may be inconsistent'...
            ' with the musp used for source depth']);
    end
    
    if size(rs, 1)>1 && size(rd, 1)>1
        error('Can not use multiple sources and multiple detectors');
    end

    nin=optProp.nin;
    nout=optProp.nout;
    musp=optProp.musp; %1/mm
    mua=optProp.mua; %1/mm
    
    x0=rs(:, 1); %mm
    y0=rs(:, 2); %mm
    z0=rs(:, 3); %mm

    c=0.299792458; %mm/ps
    v=c/nin;

    A=n2A(nin, nout);
    D=1/(3*musp); %mm
    zb=-2*A*D; %mm

    mueff=sqrt(mua/D);

    rsp=[x0, y0, -z0+2*zb]; %mm

    r1=vecnorm(rd-rs, 2, 2);
    r2=vecnorm(rd-rsp, 2, 2);
    
    R_C=continuousReflectance(rs, rd, optProp);

    switch k
        case 1
            tkMom=...
                ((z0./r1).*exp(-mueff*r1)+((z0-2*zb)./r2).*exp(-mueff*r2))./...
                (8*pi*v*D*R_C);
        case 2
            tkMom=...
                (z0.*exp(-mueff*r1)+(z0-2*zb).*exp(-mueff*r2))./...
                (16*pi*(v*D)^2*mueff*R_C);
        case 3
            tkMom=...
                (z0.*(r1+1/mueff).*exp(-mueff*r1)+...
                (z0-2*zb).*(r2+1/mueff).*exp(-mueff*r2))./...
                (32*pi*D^2*v^3*mua*R_C);
        case 4
            tkMom=...
                (z0.*(r1.^2+3*r1/mueff+3/mueff^2).*exp(-mueff*r1)+...
                (z0-2*zb).*(r2.^2+3*r2/mueff+3/mueff^2).*exp(-mueff*r2))./...
                (64*pi*D^(5/2)*mua^(3/2)*v^4*R_C);
        otherwise
            error('k>4 not supported');
    end
end
\end{lstlisting}
	{\small Available at:\cite{Blaney_23_CodeSensitivity}\\
		\href{https://github.com/DOIT-Lab/DOIT-Public/tree/master/SensitivityCompendium/deps/temporalKthMoment.m}
		{github.com/DOIT-Lab/DOIT-Public/tree/master/SensitivityCompendium/deps/temporalKthMoment.m}}
	
\lstinputlisting[
	caption=\as{MATLAB} code for claculation of \acrfull{TD} \acrfull{var}. \label{lst:temporalVar}
	]{temporalVar.m}
	{\small Available at:\cite{Blaney_23_CodeSensitivity}\\
		\href{https://github.com/DOIT-Lab/DOIT-Public/tree/master/SensitivityCompendium/deps/temporalVar.m}
		{github.com/DOIT-Lab/DOIT-Public/tree/master/SensitivityCompendium/deps/temporalVar.m}}

\subsection{Monte Carlo}\label{app:MC}
\lstinputlisting[
	caption=\as{MATLAB} code to setup and run \acrfull{MC} Extreme.\cite{Fang_BOE10_MeshbasedMonte} \label{lst:myMCXLAB_adjoint}
	]{myMCXLAB_adjoint.m}
	{\small Available at:\cite{Blaney_23_CodeSensitivity}\\
		\href{https://github.com/DOIT-Lab/DOIT-Public/tree/master/SensitivityCompendium/deps/myMCXLAB_adjoint.m}
		{github.com/DOIT-Lab/DOIT-Public/tree/master/SensitivityCompendium/deps/myMCXLAB\_adjoint.m}}

\lstinputlisting[
	caption=\as{MATLAB} code to calculate \acrfull{CW} \acrfull{Lpath} and \acrfull{lpath} from the output of \acrfull{MC} Extreme (Listing~\ref{lst:myMCXLAB_adjoint}) with the adjoint method.\cite{Fang_BOE10_MeshbasedMonte, Gardner_JBO14_CoupledForwardadjoint} \label{lst:continuousPathLen_MCadjoint}
	]{continuousPathLen_MCadjoint.m}
	{\small Available at:\cite{Blaney_23_CodeSensitivity}\\
		\href{https://github.com/DOIT-Lab/DOIT-Public/tree/master/SensitivityCompendium/deps/continuousPathLen_MCadjoint.m}
		{github.com/DOIT-Lab/DOIT-Public/tree/master/SensitivityCompendium/deps/continuousPathLen\_MCadjoint.m}}

\lstinputlisting[
	caption=\as{MATLAB} code to calculate \acrfull{TD} generalized \acrfull{Lpath} and \acrfull{lpath} from the output of \acrfull{MC} Extreme (Listing~\ref{lst:myMCXLAB_adjoint}) with the adjoint method.\cite{Fang_BOE10_MeshbasedMonte, Gardner_JBO14_CoupledForwardadjoint} \label{lst:temporalGatePathLen_MCadjoint}
	]{temporalGatePathLen_MCadjoint.m}
	{\small Available at:\cite{Blaney_23_CodeSensitivity}\\
		\href{https://github.com/DOIT-Lab/DOIT-Public/tree/master/SensitivityCompendium/deps/temporalGatePathLen_MCadjoint.m}
		{github.com/DOIT-Lab/DOIT-Public/tree/master/SensitivityCompendium/deps/temporalGatePathLen\_MCadjoint.m}}

\subsection{Other Supporting MATLAB Functions}
\begin{lstlisting}[
	caption=\as{MATLAB} code to make the saturated color limits for a \acrfull{sen} map. \label{lst:makeCL}
	]
function [cl, cols]=makeCL(x, NVA)
% [cl, cols, cTicks]=makeCL(x)
% 
% Giles Blaney Ph.D. Summer 2023
% 
% Make saturated colormap and limits for a given array 
    
    arguments
        x double;

        NVA.qants=[0.05, 0.95];
    end
    
    cols=jet(100);
    cols(1, :)=0;
    cols(end, :)=1;
    
    cl=[quantile(x(:), NVA.qants(1)), quantile(x(:), NVA.qants(2))];
end
\end{lstlisting}
	{\small Available at:\cite{Blaney_23_CodeSensitivity}\\
		\href{https://github.com/DOIT-Lab/DOIT-Public/tree/master/SensitivityCompendium/deps/makeCL.m}
		{github.com/DOIT-Lab/DOIT-Public/tree/master/SensitivityCompendium/deps/makeCL.m}}

\lstinputlisting[
	caption=\as{MATLAB} code to slice \acrfull{senMat} output from from Listing~\ref{lst:makeS}. \label{lst:sliceS}
	]{sliceS.m}
	{\small Available at:\cite{Blaney_23_CodeSensitivity}\\
		\href{https://github.com/DOIT-Lab/DOIT-Public/tree/master/SensitivityCompendium/deps/sliceS.m}
		{github.com/DOIT-Lab/DOIT-Public/tree/master/SensitivityCompendium/deps/sliceS.m}}
	
\lstinputlisting[
	caption=\as{MATLAB} code to calculate the index of refraction mismatch parameter ($A$).\cite{Martelli_Appl.Opt.97_PhotonMigration} \label{lst:n2A}
	]{n2A.m}
	{\small Available at:\cite{Blaney_23_CodeSensitivity}\\
		\href{https://github.com/DOIT-Lab/DOIT-Public/tree/master/SensitivityCompendium/deps/n2A.m}
		{github.com/DOIT-Lab/DOIT-Public/tree/master/SensitivityCompendium/deps/n2A.m}}
	
\begin{lstlisting}[
	caption=Input parsing function for the \as{MATLAB} code. \label{lst:struct2pairs}
	]
function C=struct2pairs(S)
%Source: https://www.mathworks.com/matlabcentral/answers/469700-how-to-pass-a-struct-as-name-value-pairs-to-a-function#answer_381511
%Turns a scalar struct S into a cell of string-value pairs C
%
%  C=struct2pairs(S)
%
%If S is a cell already, it will be returned unchanged.
if iscell(S)
 C=S; return
elseif length(S)>1
    error 'Input must be a scalar struct or cell';
end
C=[fieldnames(S).'; struct2cell(S).'];
C=C(:).';
\end{lstlisting}
	{\small Available at:\cite{Blaney_23_CodeSensitivity}\\
		\href{https://github.com/DOIT-Lab/DOIT-Public/tree/master/SensitivityCompendium/deps/struct2pairs.m}
		{github.com/DOIT-Lab/DOIT-Public/tree/master/SensitivityCompendium/deps/struct2pairs.m}}

\twocolumn

\bibliographystyle{unsrtnat}
\bibliography{GilesBlaneyZoteroExport230613}

\end{document}

%% file: CW_SD_I_3rd_rho0_MC_vox.tex
\begin{figure*}
	\begin{center}
		\includegraphics{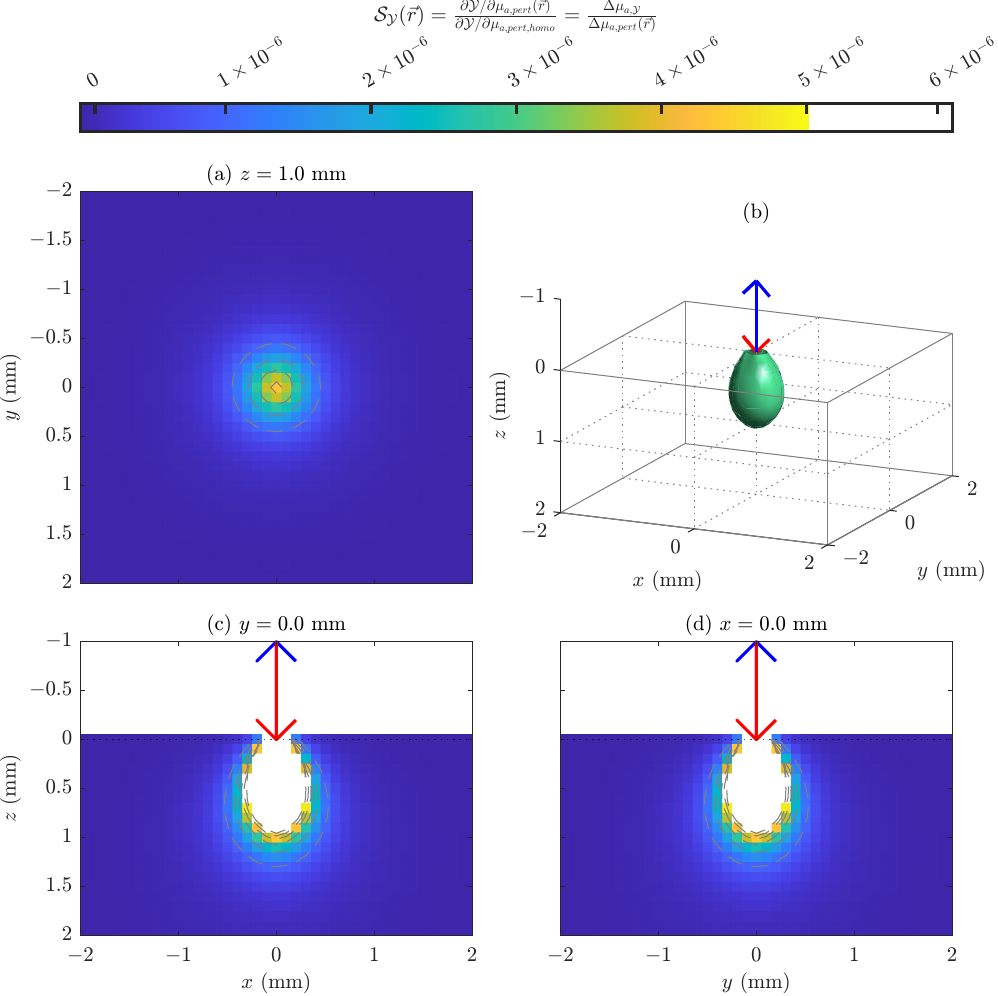}
	\end{center}
	\caption{Third angle projection of the \acrfull{sen} to a $\SI{0.1}{\milli\meter}\times\SI{0.1}{\milli\meter}\times\SI{0.1}{\milli\meter}$ perturbation scanned \SI{0.1}{\milli\meter} measured by \acrfull{CW} \acrfull{SD} \acrfull{I}. (a) $x$-$y$ plane sliced at $z=\SI{1.0}{\milli\meter}$. (b) Iso-surface sliced at $\as{S}=3.000\times 10^{-5}$. (c) $x$-$z$ plane sliced at $y=\SI{0.0}{\milli\meter}$. (d) $y$-$z$ plane sliced at $x=\SI{0.0}{\milli\meter}$. Generated using \acrfull{MC}.\\ 
	\Acrfull{rho}: \SI{0.0}{\milli\meter}\\ 
	\Acrfull{n} inside: \num{1.333};\quad	\Acrfull{n} outside: \num{1.000}\\ 
	\Acrfull{musp}: \SI{1.10}{\per\milli\meter};\quad	\Acrfull{g}: \num{0.9}\\ 
	\Acrfull{mua}: \SI{0.011}{\per\milli\meter}\\ 
	Detector Numerical Aperature (NA): \num{0.5};\quad	Number of photons: \num{1000000000}\\ 
	}\label{fig:CW_SD_I_3rd_rho0_MC_vox}
\end{figure*}

%% file: CW_SD_I_3rd_rho0_MC.tex
\begin{figure*}
	\begin{center}
		\includegraphics{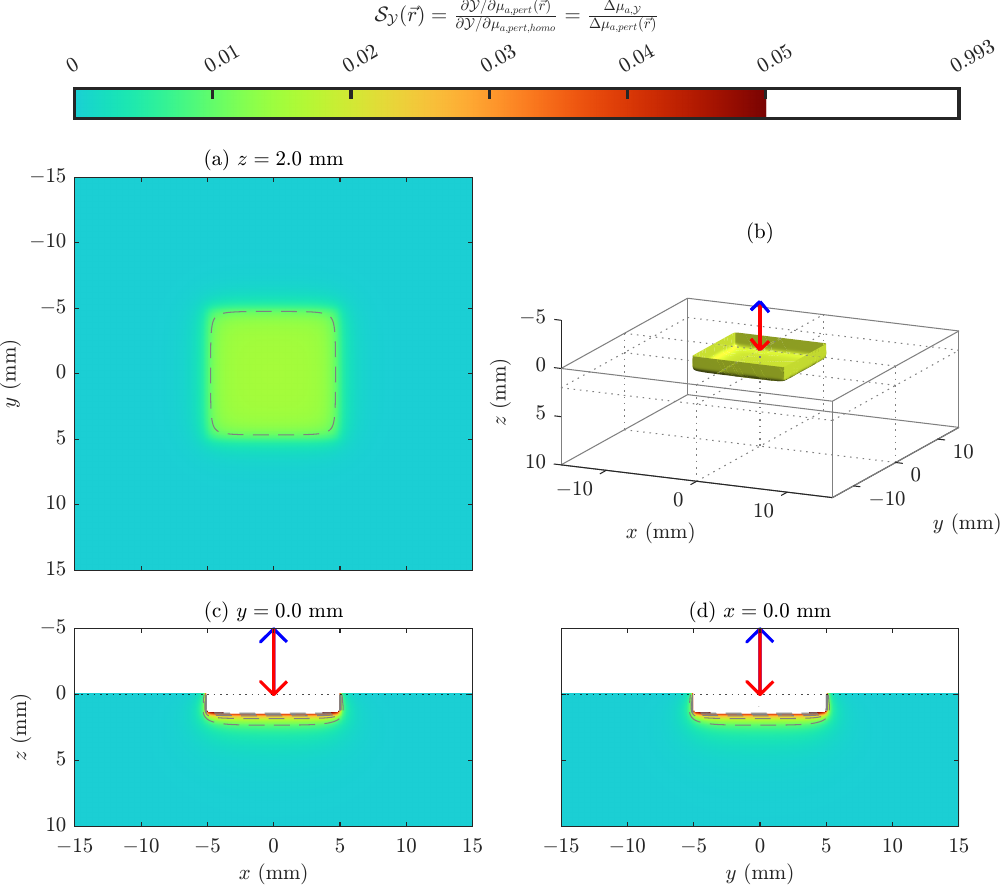}
	\end{center}
	\caption{Third angle projection of the \acrfull{sen} to a $\SI{10.0}{\milli\meter}\times\SI{10.0}{\milli\meter}\times\SI{2.0}{\milli\meter}$ perturbation scanned \SI{0.1}{\milli\meter} measured by \acrfull{CW} \acrfull{SD} \acrfull{I}. (a) $x$-$y$ plane sliced at $z=\SI{2.0}{\milli\meter}$. (b) Iso-surface sliced at $\as{S}=0.020$. (c) $x$-$z$ plane sliced at $y=\SI{0.0}{\milli\meter}$. (d) $y$-$z$ plane sliced at $x=\SI{0.0}{\milli\meter}$. Generated using \acrfull{MC}.\\ 
	\Acrfull{rho}: \SI{0.0}{\milli\meter}\\ 
	\Acrfull{n} inside: \num{1.333};\quad	\Acrfull{n} outside: \num{1.000}\\ 
	\Acrfull{musp}: \SI{1.10}{\per\milli\meter};\quad	\Acrfull{g}: \num{0.9}\\ 
	\Acrfull{mua}: \SI{0.011}{\per\milli\meter}\\ 
	Detector Numerical Aperature (NA): \num{0.5};\quad	Number of photons: \num{1000000000}\\ 
	}\label{fig:CW_SD_I_3rd_rho0_MC}
\end{figure*}

%% file: CW_SD_I_3rd.tex
\begin{figure*}
	\begin{center}
		\includegraphics{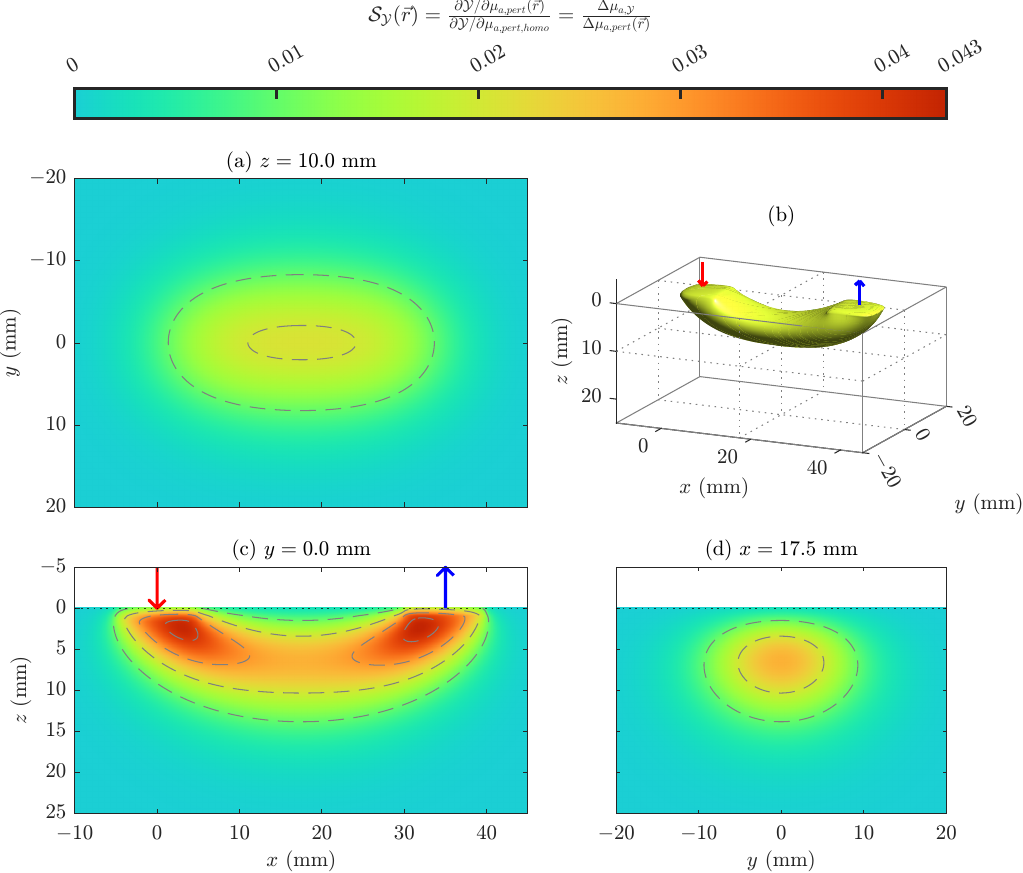}
	\end{center}
	\caption{Third angle projection of the \acrfull{sen} to a $\SI{10.0}{\milli\meter}\times\SI{10.0}{\milli\meter}\times\SI{2.0}{\milli\meter}$ perturbation scanned \SI{0.1}{\milli\meter} measured by \acrfull{CW} \acrfull{SD} \acrfull{I}. (a) $x$-$y$ plane sliced at $z=\SI{10.0}{\milli\meter}$. (b) Iso-surface sliced at $\as{S}=0.020$. (c) $x$-$z$ plane sliced at $y=\SI{0.0}{\milli\meter}$. (d) $y$-$z$ plane sliced at $x=\SI{17.5}{\milli\meter}$. Generated using \acrfull{DT}.\\ 
	\Acrfull{rho}: \SI{35.0}{\milli\meter}\\ 
	\Acrfull{n} inside: \num{1.333}\\ 
	\Acrfull{n} outside: \num{1.000}\\ 
	\Acrfull{musp}: \SI{1.10}{\per\milli\meter}\\ 
	\Acrfull{mua}: \SI{0.011}{\per\milli\meter}\\ 
	}\label{fig:CW_SD_I_3rd}
\end{figure*}

%% file: CW_SD_I_rho.tex
\begin{figure*}
	\begin{center}
		\includegraphics{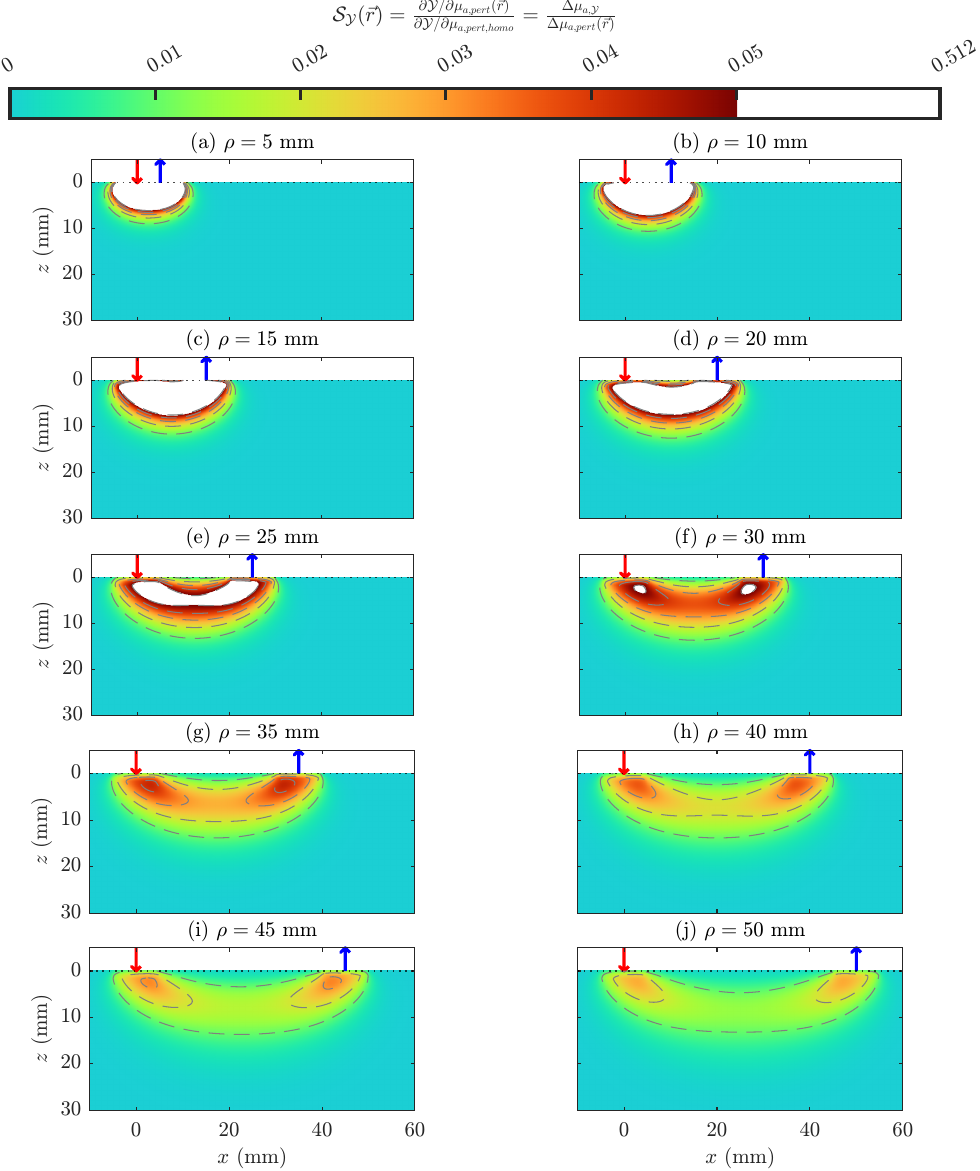}
	\end{center}
	\caption{$x$-$z$ plane of the \acrfull{sen} to a $\SI{10.0}{\milli\meter}\times\SI{10.0}{\milli\meter}\times\SI{2.0}{\milli\meter}$ perturbation scanned \SI{0.1}{\milli\meter} measured by \acrfull{CW} \acrfull{SD} \acrfull{I}. (a)-(j) Different values of \acrfull{rho}. Generated using \acrfull{DT}.\\ 
	\Acrfull{n} inside: \num{1.333};\quad	\Acrfull{n} outside: \num{1.000}\\ 
	\Acrfull{musp}: \SI{1.10}{\per\milli\meter}\quad	\Acrfull{mua}: \SI{0.011}{\per\milli\meter}\\ 
	}\label{fig:CW_SD_I_rho}
\end{figure*}

%% file: CW_SS_I_3rd.tex
\begin{figure*}
	\begin{center}
		\includegraphics{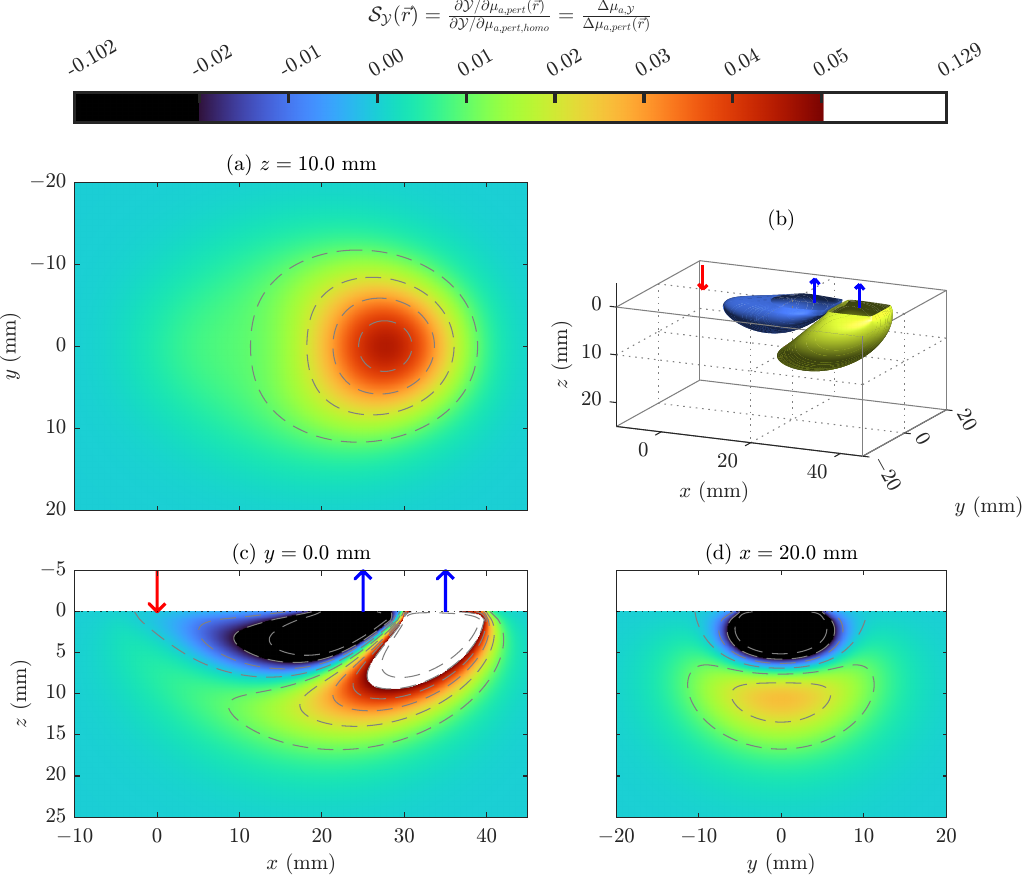}
	\end{center}
	\caption{Third angle projection of the \acrfull{sen} to a $\SI{10.0}{\milli\meter}\times\SI{10.0}{\milli\meter}\times\SI{2.0}{\milli\meter}$ perturbation scanned \SI{0.1}{\milli\meter} measured by \acrfull{CW} \acrfull{SS} \acrfull{I}.(a) $x$-$y$ plane sliced at $z=\SI{10.0}{\milli\meter}$. (b) Iso-surface sliced at $\as{S}=0.020$ and $\as{S}=-0.010$. (c) $x$-$z$ plane sliced at $y=\SI{0.0}{\milli\meter}$. (d) $y$-$z$ plane sliced at $x=\SI{20.0}{\milli\meter}$. Generated using \acrfull{DT}.\\ 
	\Acrfullpl{rho}: [25.0, 35.0]~\si{\milli\meter}\\ 
	\Acrfull{n} inside: \num{1.333}\\ 
	\Acrfull{n} outside: \num{1.000}\\ 
	\Acrfull{musp}: \SI{1.10}{\per\milli\meter}\\ 
	\Acrfull{mua}: \SI{0.011}{\per\milli\meter}\\ 
	}\label{fig:CW_SS_I_3rd}
\end{figure*}

%% file: CW_DS_I_3rd.tex
\begin{figure*}
	\begin{center}
		\includegraphics{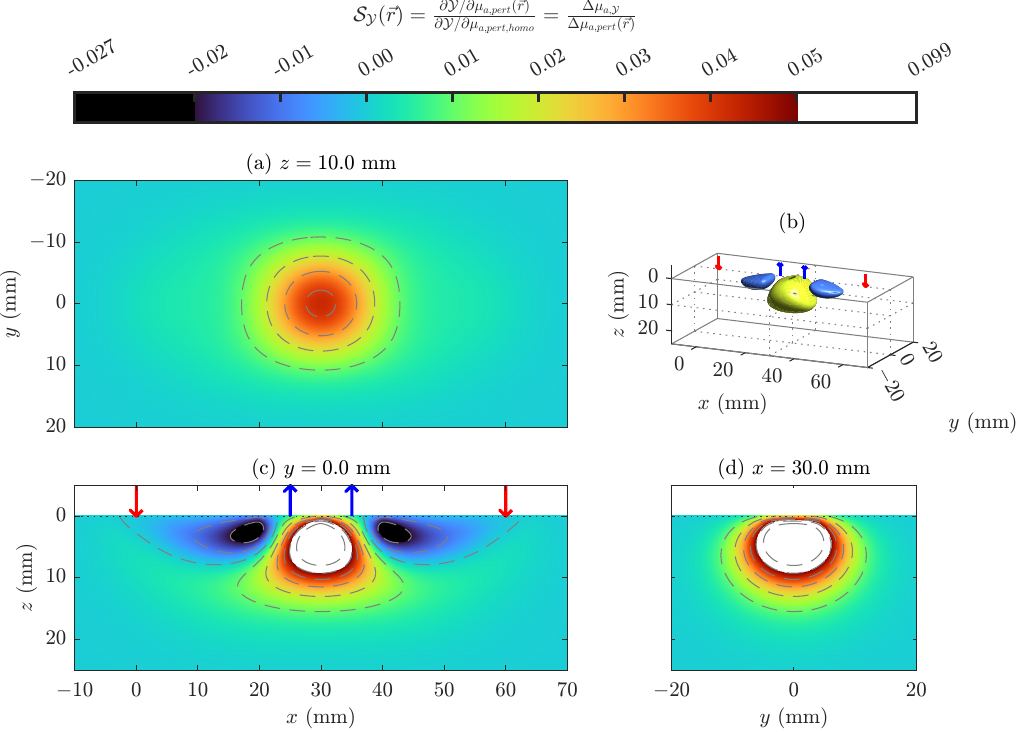}
	\end{center}
	\caption{Third angle projection of the \acrfull{sen} to a $\SI{10.0}{\milli\meter}\times\SI{10.0}{\milli\meter}\times\SI{2.0}{\milli\meter}$ perturbation scanned \SI{0.1}{\milli\meter} measured by \acrfull{CW} \acrfull{DS} \acrfull{I}.(a) $x$-$y$ plane sliced at $z=\SI{10.0}{\milli\meter}$. (b) Iso-surface sliced at $\as{S}=0.020$ and $\as{S}=-0.010$. (c) $x$-$z$ plane sliced at $y=\SI{0.0}{\milli\meter}$. (d) $y$-$z$ plane sliced at $x=\SI{30.0}{\milli\meter}$. Generated using \acrfull{DT}.\\ 
	\Acrfullpl{rho}: [25.0, 35.0, 35.0, 25.0]~\si{\milli\meter}\\ 
	\Acrfull{n} inside: \num{1.333}\\ 
	\Acrfull{n} outside: \num{1.000}\\ 
	\Acrfull{musp}: \SI{1.10}{\per\milli\meter}\\ 
	\Acrfull{mua}: \SI{0.011}{\per\milli\meter}\\ 
	}\label{fig:CW_DS_I_3rd}
\end{figure*}

%% file: CW_SS_I_mrho.tex
\begin{figure*}
	\begin{center}
		\includegraphics{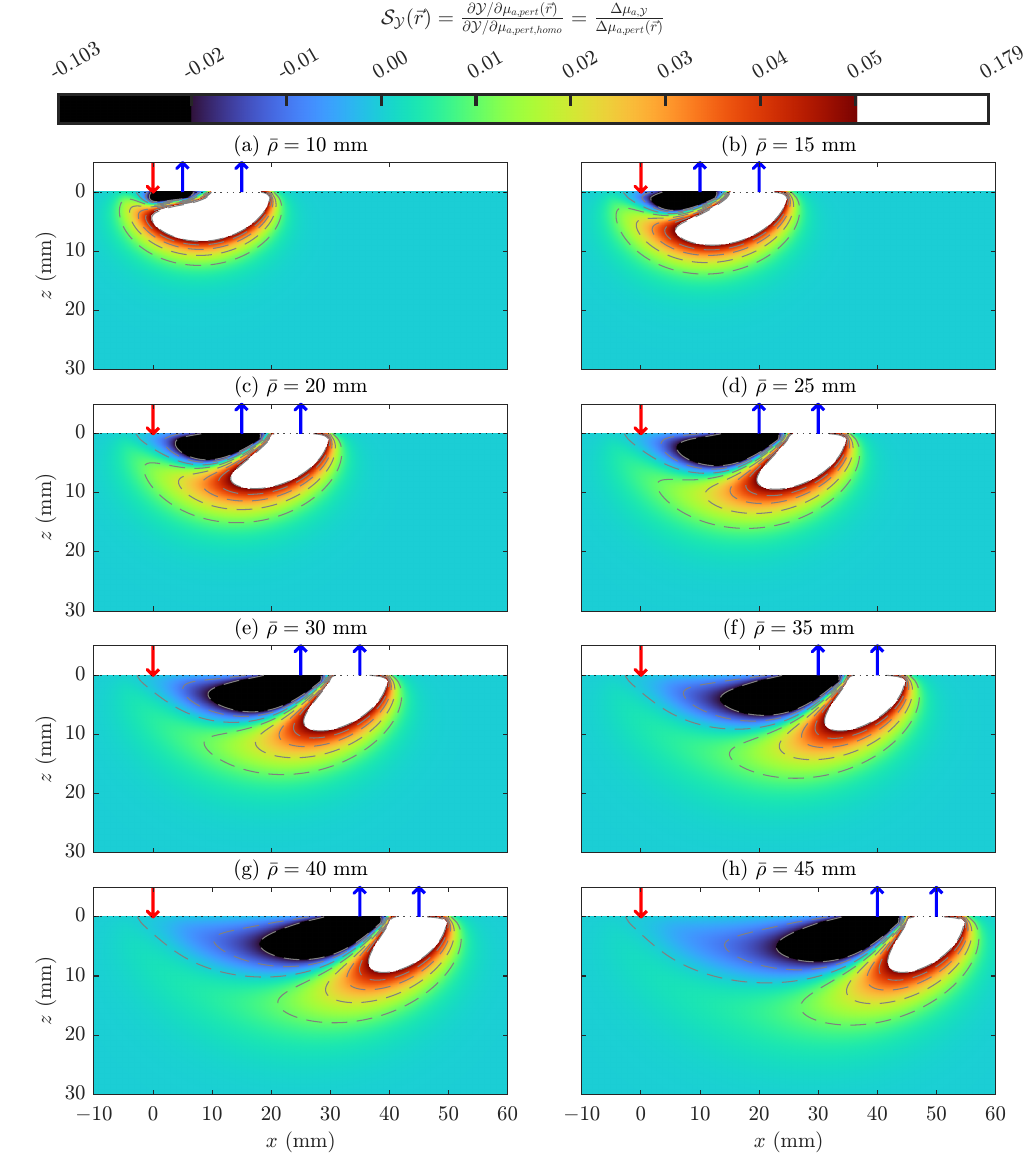}
	\end{center}
	\caption{$x$-$z$ plane of the \acrfull{sen} to a $\SI{10.0}{\milli\meter}\times\SI{10.0}{\milli\meter}\times\SI{2.0}{\milli\meter}$ perturbation scanned \SI{0.1}{\milli\meter} measured by \acrfull{CW} \acrfull{SS} \acrfull{I}. (a)-(h) Different values of \acrfull{mrho}. Generated using \acrfull{DT}.\\ 
	\Acrfull{drho}: \SI{10.0}{\milli\meter}\\ 
	\Acrfull{n} inside: \num{1.333};\quad	\Acrfull{n} outside: \num{1.000}\\ 
	\Acrfull{musp}: \SI{1.10}{\per\milli\meter}\quad	\Acrfull{mua}: \SI{0.011}{\per\milli\meter}\\ 
	}\label{fig:CW_SS_I_mrho}
\end{figure*}

%% file: CW_DS_I_mrho.tex
\begin{figure*}
	\begin{center}
		\includegraphics{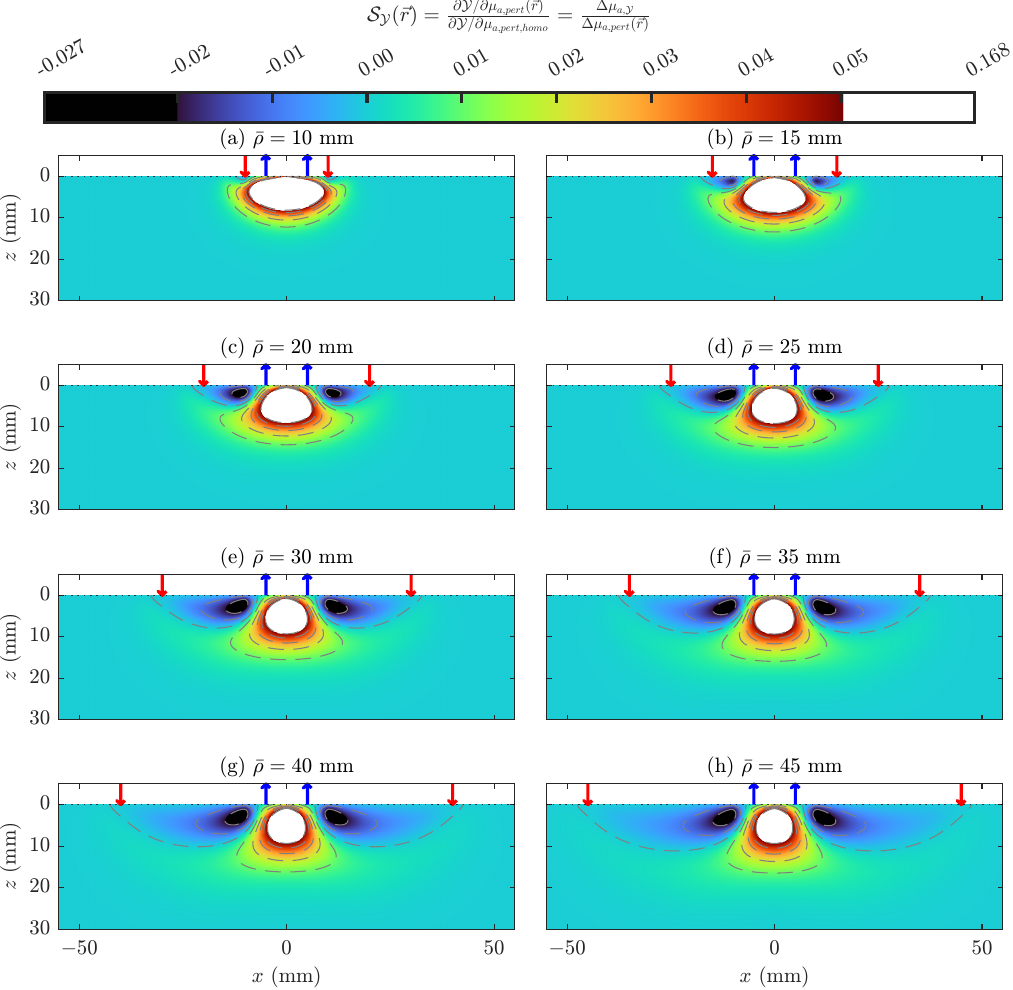}
	\end{center}
	\caption{$x$-$z$ plane of the \acrfull{sen} to a $\SI{10.0}{\milli\meter}\times\SI{10.0}{\milli\meter}\times\SI{2.0}{\milli\meter}$ perturbation scanned \SI{0.1}{\milli\meter} measured by \acrfull{CW} \acrfull{DS} \acrfull{I}. (a)-(h) Different values of \acrfull{mrho}. Generated using \acrfull{DT}.\\ 
	\Acrfull{drho}: \SI{10.0}{\milli\meter}\\ 
	\Acrfull{n} inside: \num{1.333};\quad	\Acrfull{n} outside: \num{1.000}\\ 
	\Acrfull{musp}: \SI{1.10}{\per\milli\meter}\quad	\Acrfull{mua}: \SI{0.011}{\per\milli\meter}\\ 
	}\label{fig:CW_DS_I_mrho}
\end{figure*}

%% file: CW_SS_I_drho.tex
\begin{figure*}
	\begin{center}
		\includegraphics{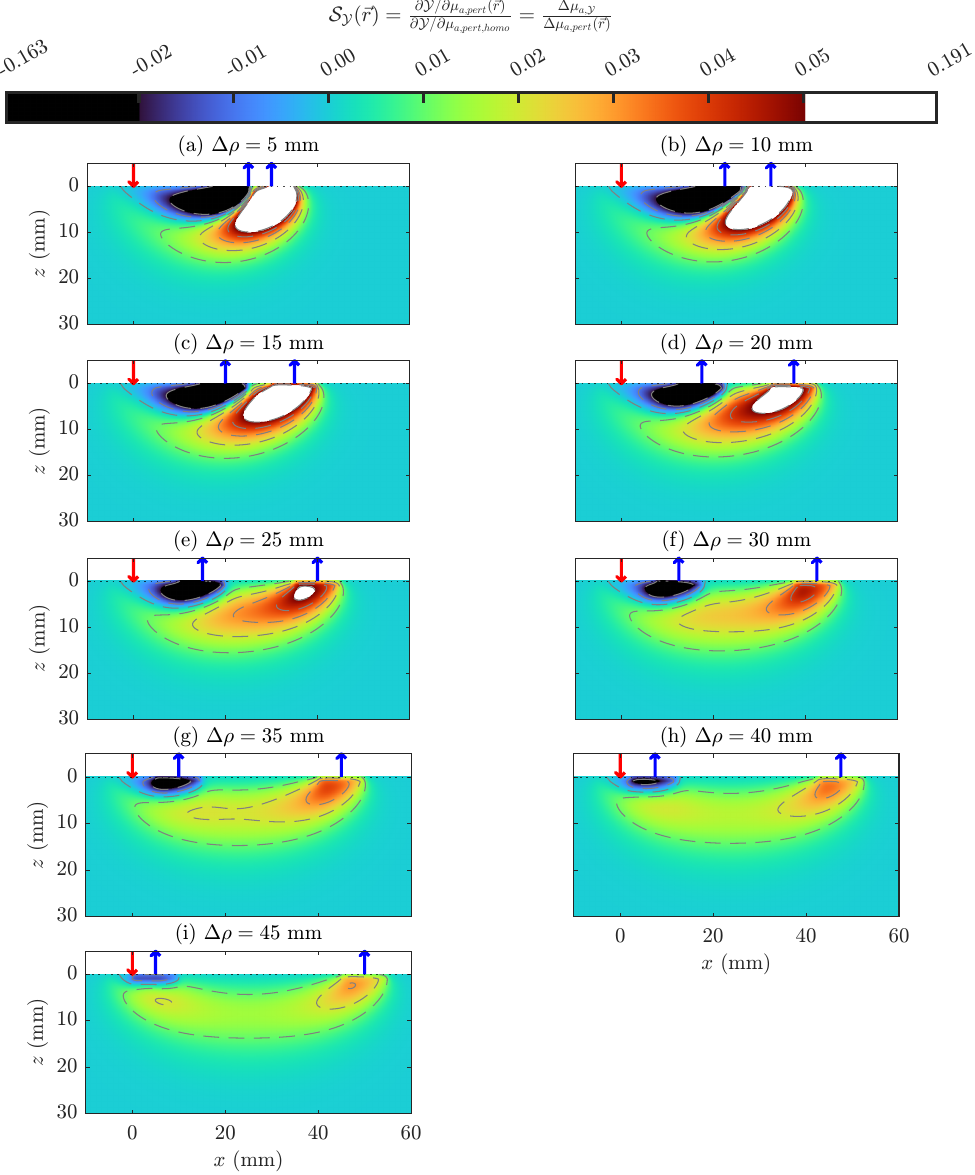}
	\end{center}
	\caption{$x$-$z$ plane of the \acrfull{sen} to a $\SI{10.0}{\milli\meter}\times\SI{10.0}{\milli\meter}\times\SI{2.0}{\milli\meter}$ perturbation scanned \SI{0.1}{\milli\meter} measured by \acrfull{CW} \acrfull{SS} \acrfull{I}. (a)-(i) Different values of \acrfull{drho}. Generated using \acrfull{DT}.\\ 
	\Acrfull{mrho}: \SI{27.5}{\milli\meter}\\ 
	\Acrfull{n} inside: \num{1.333};\quad	\Acrfull{n} outside: \num{1.000}\\ 
	\Acrfull{musp}: \SI{1.10}{\per\milli\meter}\quad	\Acrfull{mua}: \SI{0.011}{\per\milli\meter}\\ 
	}\label{fig:CW_SS_I_drho}
\end{figure*}

%% file: CW_DS_I_drho.tex
\begin{figure*}
	\begin{center}
		\includegraphics{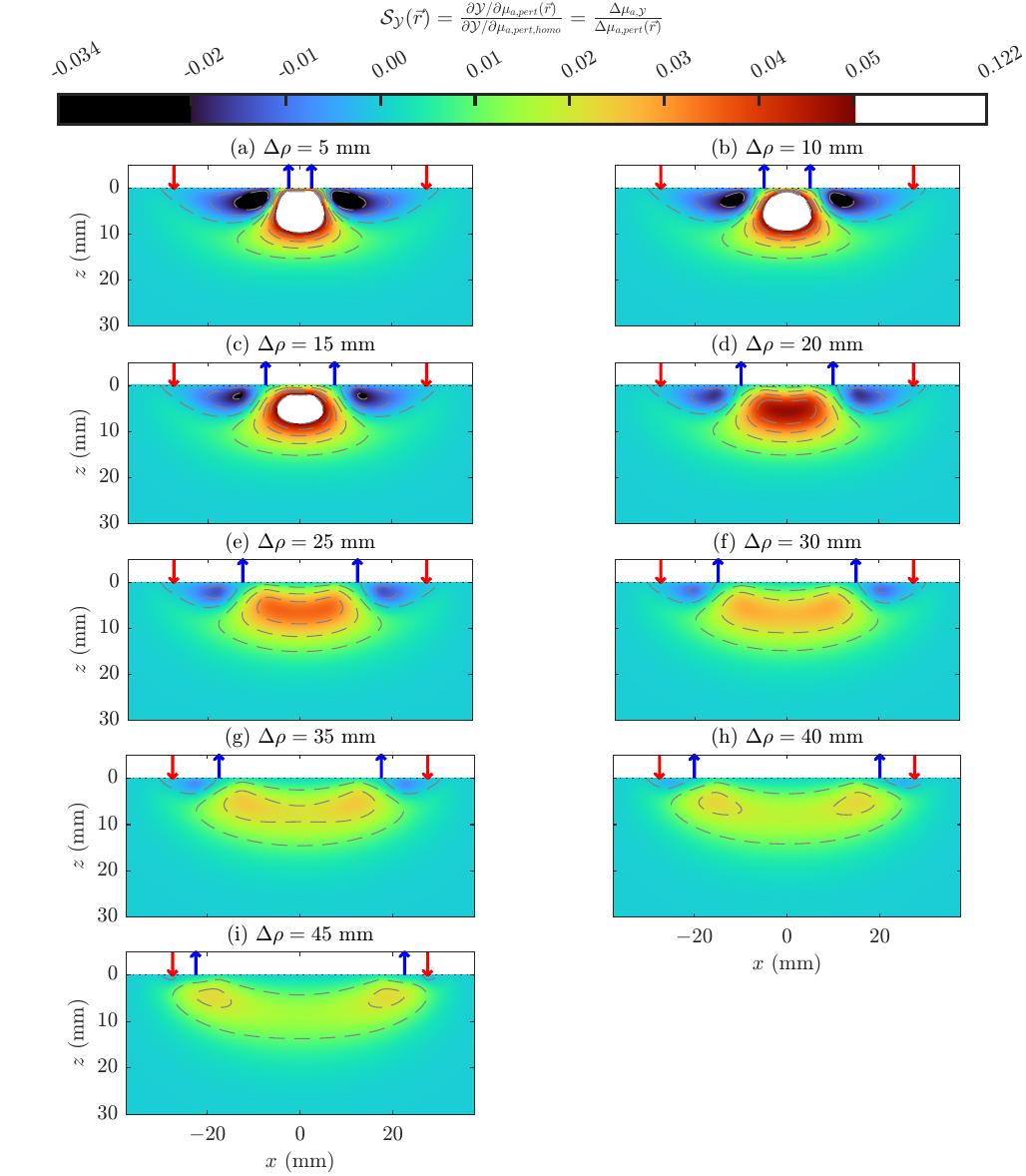}
	\end{center}
	\caption{$x$-$z$ plane of the \acrfull{sen} to a $\SI{10.0}{\milli\meter}\times\SI{10.0}{\milli\meter}\times\SI{2.0}{\milli\meter}$ perturbation scanned \SI{0.1}{\milli\meter} measured by \acrfull{CW} \acrfull{DS} \acrfull{I}. (a)-(i) Different values of \acrfull{drho}. Generated using \acrfull{DT}.\\ 
	\Acrfull{mrho}: \SI{27.5}{\milli\meter}\\ 
	\Acrfull{n} inside: \num{1.333};\quad	\Acrfull{n} outside: \num{1.000}\\ 
	\Acrfull{musp}: \SI{1.10}{\per\milli\meter}\quad	\Acrfull{mua}: \SI{0.011}{\per\milli\meter}\\ 
	}\label{fig:CW_DS_I_drho}
\end{figure*}

%% file: FD_SD_I_3rd.tex
\begin{figure*}
	\begin{center}
		\includegraphics{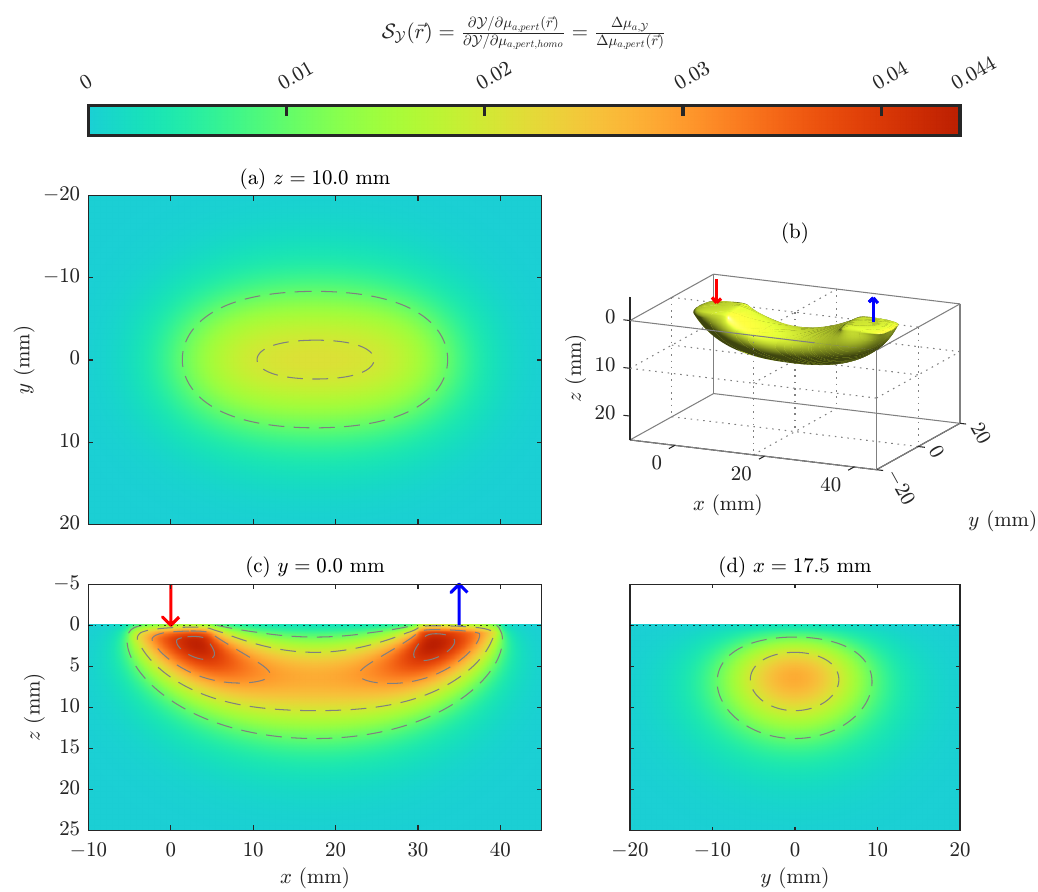}
	\end{center}
	\caption{Third angle projection of the \acrfull{sen} to a $\SI{10.0}{\milli\meter}\times\SI{10.0}{\milli\meter}\times\SI{2.0}{\milli\meter}$ perturbation scanned \SI{0.1}{\milli\meter} measured by \acrfull{FD} \acrfull{SD} \acrfull{I}. (a) $x$-$y$ plane sliced at $z=\SI{10.0}{\milli\meter}$. (b) Iso-surface sliced at $\as{S}=0.020$. (c) $x$-$z$ plane sliced at $y=\SI{0.0}{\milli\meter}$. (d) $y$-$z$ plane sliced at $x=\SI{17.5}{\milli\meter}$. Generated using \acrfull{DT}.\\ 
	\Acrfull{rho}: \SI{35.0}{\milli\meter}\\ 
	\Acrfull{n} inside: \num{1.333}\\ 
	\Acrfull{n} outside: \num{1.000}\\ 
	\Acrfull{musp}: \SI{1.10}{\per\milli\meter}\\ 
	\Acrfull{mua}: \SI{0.011}{\per\milli\meter}\\ 
	\Acrfull{fmod}: \SI{100}{\mega\hertz}\\ 
	}\label{fig:FD_SD_I_3rd}
\end{figure*}

%% file: FD_SD_I_fmod.tex
\begin{figure*}
	\begin{center}
		\includegraphics{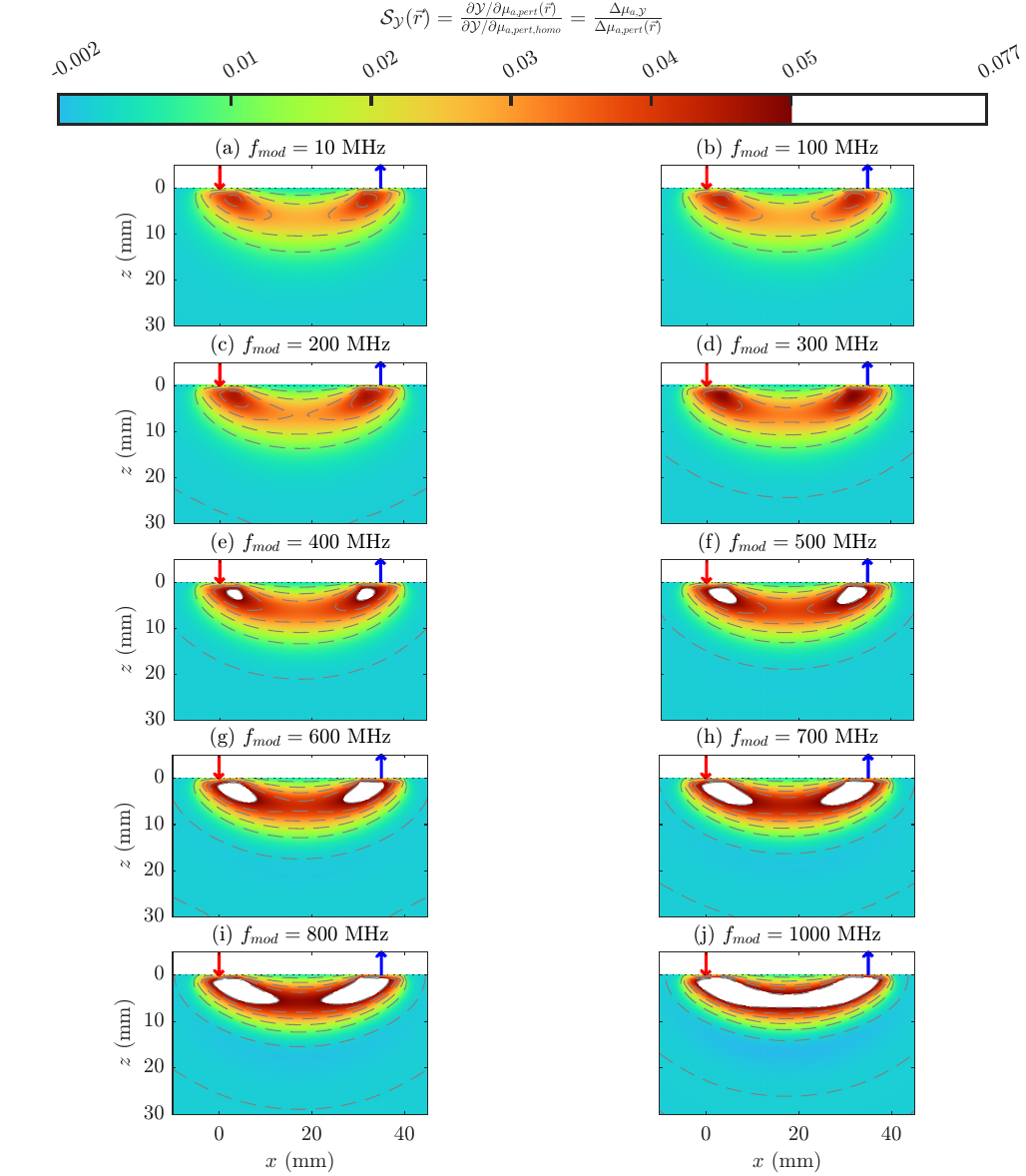}
	\end{center}
	\caption{$x$-$z$ plane of the \acrfull{sen} to a $\SI{10.0}{\milli\meter}\times\SI{10.0}{\milli\meter}\times\SI{2.0}{\milli\meter}$ perturbation scanned \SI{0.1}{\milli\meter} measured by \acrfull{FD} \acrfull{SD} \acrfull{I}. (a)-(j) Different values of \acrfull{fmod}. Generated using \acrfull{DT}.\\ 
	\Acrfull{rho}: \SI{35.0}{\milli\meter}\\ 
	\Acrfull{n} inside: \num{1.333};\quad	\Acrfull{n} outside: \num{1.000}\\ 
	\Acrfull{musp}: \SI{1.10}{\per\milli\meter};\quad	\Acrfull{mua}: \SI{0.011}{\per\milli\meter}\\ 
	}\label{fig:FD_SD_I_fmod}
\end{figure*}

%% file: FD_SD_P_3rd.tex
\begin{figure*}
	\begin{center}
		\includegraphics{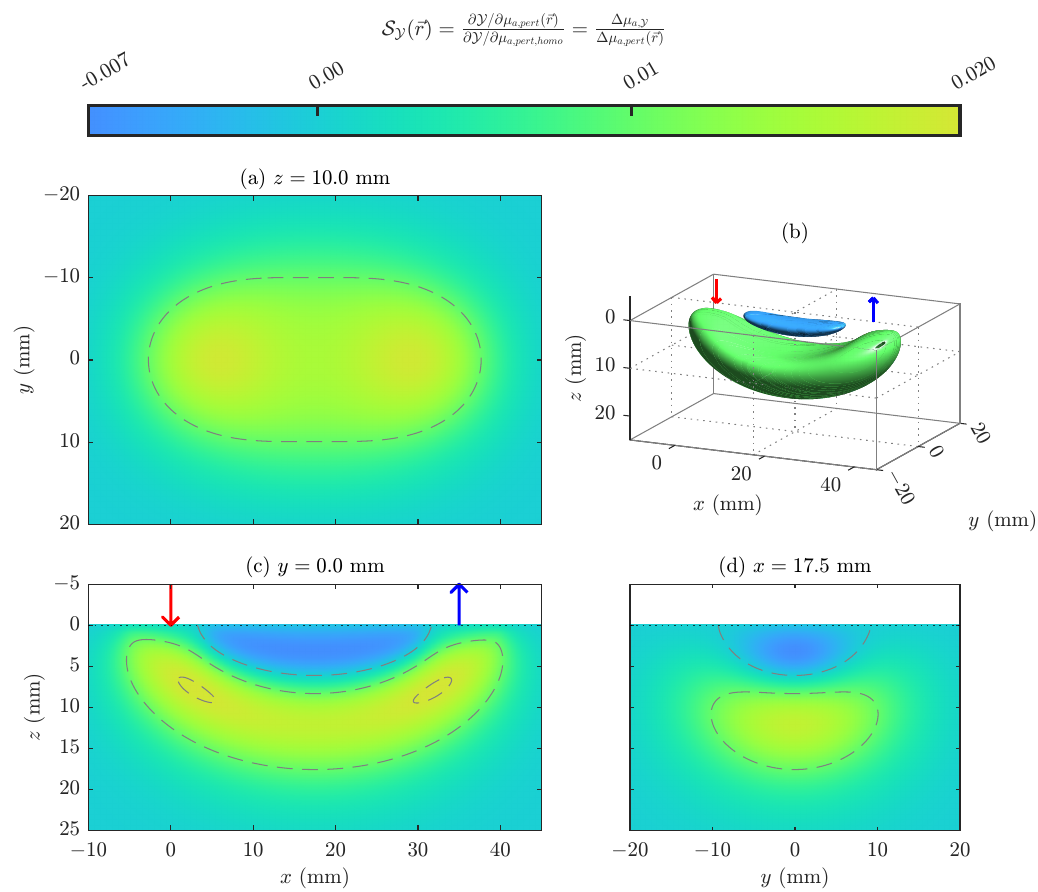}
	\end{center}
	\caption{Third angle projection of the \acrfull{sen} to a $\SI{10.0}{\milli\meter}\times\SI{10.0}{\milli\meter}\times\SI{2.0}{\milli\meter}$ perturbation scanned \SI{0.1}{\milli\meter} measured by \acrfull{FD} \acrfull{SD} \acrfull{phi}. (a) $x$-$y$ plane sliced at $z=\SI{10.0}{\milli\meter}$. (b) Iso-surface sliced at $\as{S}=0.010$ and $\as{S}=-0.005$. (c) $x$-$z$ plane sliced at $y=\SI{0.0}{\milli\meter}$. (d) $y$-$z$ plane sliced at $x=\SI{17.5}{\milli\meter}$. Generated using \acrfull{DT}.\\ 
	\Acrfull{rho}: \SI{35.0}{\milli\meter}\\ 
	\Acrfull{n} inside: \num{1.333};\quad	\Acrfull{n} outside: \num{1.000}\\ 
	\Acrfull{musp}: \SI{1.10}{\per\milli\meter};\quad	\Acrfull{mua}: \SI{0.011}{\per\milli\meter}\\ 
	\Acrfull{fmod}: \SI{100}{\mega\hertz}\\ 
	}\label{fig:FD_SD_P_3rd}
\end{figure*}

%% file: FD_SD_P_rho.tex
\begin{figure*}
	\begin{center}
		\includegraphics{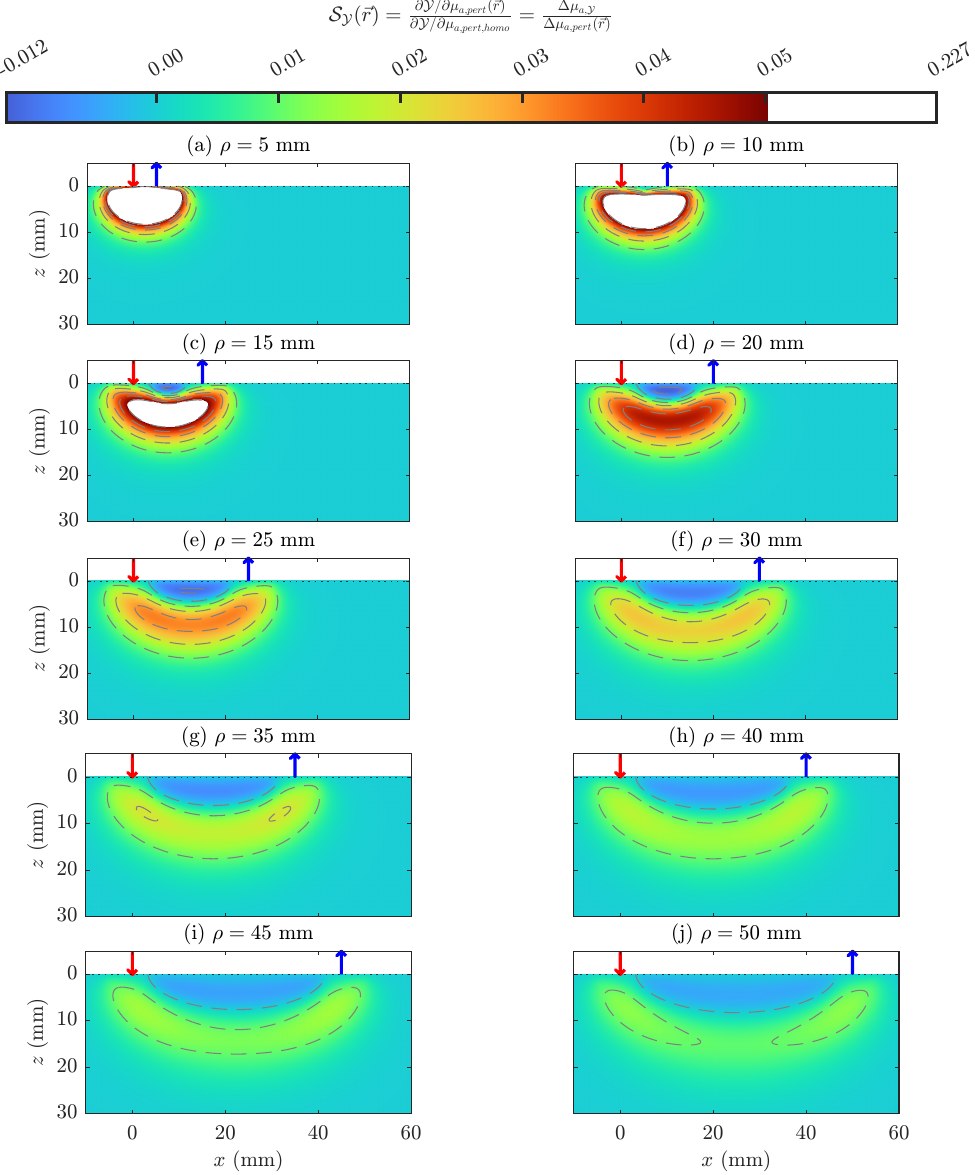}
	\end{center}
	\caption{$x$-$z$ plane of the \acrfull{sen} to a $\SI{10.0}{\milli\meter}\times\SI{10.0}{\milli\meter}\times\SI{2.0}{\milli\meter}$ perturbation scanned \SI{0.1}{\milli\meter} measured by \acrfull{FD} \acrfull{SD} \acrfull{phi}. (a)-(j) Different values of \acrfull{rho}. Generated using \acrfull{DT}.\\ 
	\Acrfull{n} inside: \num{1.333};\quad	\Acrfull{n} outside: \num{1.000}\\ 
	\Acrfull{musp}: \SI{1.10}{\per\milli\meter};\quad	\Acrfull{mua}: \SI{0.011}{\per\milli\meter}\\ 
	\Acrfull{fmod}: \SI{100}{\mega\hertz}\\ 
	}\label{fig:FD_SD_P_rho}
\end{figure*}

%% file: FD_SD_P_fmod.tex
\begin{figure*}
	\begin{center}
		\includegraphics{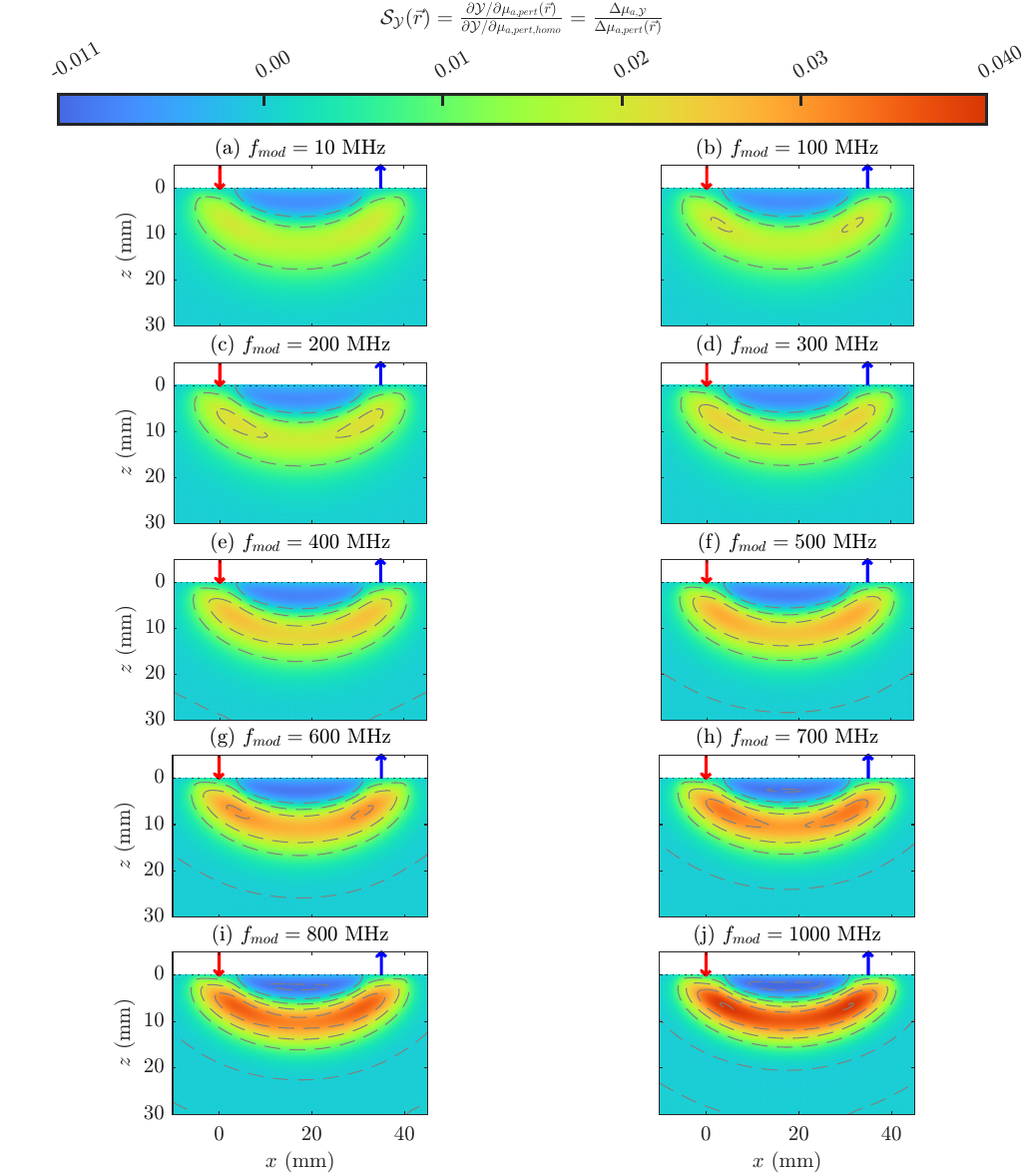}
	\end{center}
	\caption{$x$-$z$ plane of the \acrfull{sen} to a $\SI{10.0}{\milli\meter}\times\SI{10.0}{\milli\meter}\times\SI{2.0}{\milli\meter}$ perturbation scanned \SI{0.1}{\milli\meter} measured by \acrfull{FD} \acrfull{SD} \acrfull{phi}. (a)-(j) Different values of \acrfull{fmod}. Generated using \acrfull{DT}.\\ 
	\Acrfull{rho}: \SI{35.0}{\milli\meter}\\ 
	\Acrfull{n} inside: \num{1.333};\quad	\Acrfull{n} outside: \num{1.000}\\ 
	\Acrfull{musp}: \SI{1.10}{\per\milli\meter};\quad	\Acrfull{mua}: \SI{0.011}{\per\milli\meter}\\ 
	}\label{fig:FD_SD_P_fmod}
\end{figure*}

%% file: FD_SS_I_3rd.tex
\begin{figure*}
	\begin{center}
		\includegraphics{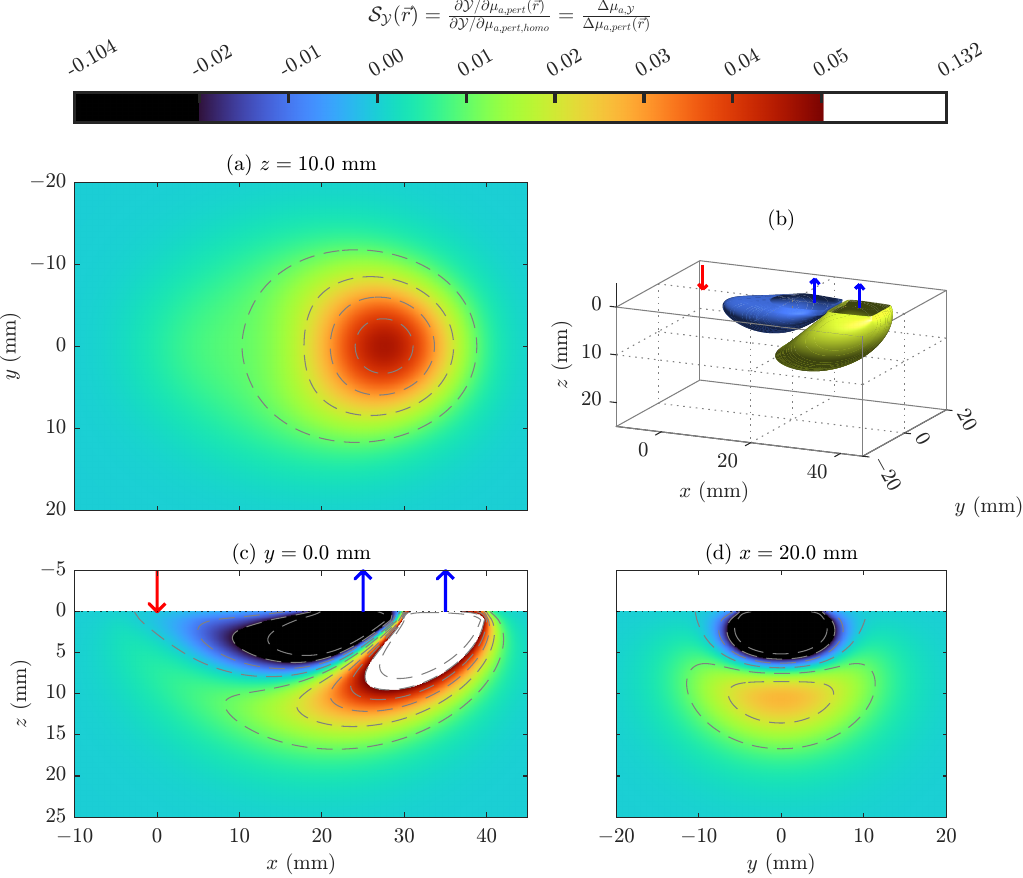}
	\end{center}
	\caption{Third angle projection of the \acrfull{sen} to a $\SI{10.0}{\milli\meter}\times\SI{10.0}{\milli\meter}\times\SI{2.0}{\milli\meter}$ perturbation scanned \SI{0.1}{\milli\meter} measured by \acrfull{FD} \acrfull{SS} \acrfull{I}. (a) $x$-$y$ plane sliced at $z=\SI{10.0}{\milli\meter}$. (b) Iso-surface sliced at $\as{S}=0.020$ and $\as{S}=-0.010$. (c) $x$-$z$ plane sliced at $y=\SI{0.0}{\milli\meter}$. (d) $y$-$z$ plane sliced at $x=\SI{20.0}{\milli\meter}$. Generated using \acrfull{DT}.\\ 
	\Acrfullpl{rho}: [25.0, 35.0]~\si{\milli\meter}\\ 
	\Acrfull{n} inside: \num{1.333}\\ 
	\Acrfull{n} outside: \num{1.000}\\ 
	\Acrfull{musp}: \SI{1.10}{\per\milli\meter}\\ 
	\Acrfull{mua}: \SI{0.011}{\per\milli\meter}\\ 
	\Acrfull{fmod}: \SI{100}{\mega\hertz}\\ 
	}\label{fig:FD_SS_I_3rd}
\end{figure*}

%% file: FD_DS_I_3rd.tex
\begin{figure*}
	\begin{center}
		\includegraphics{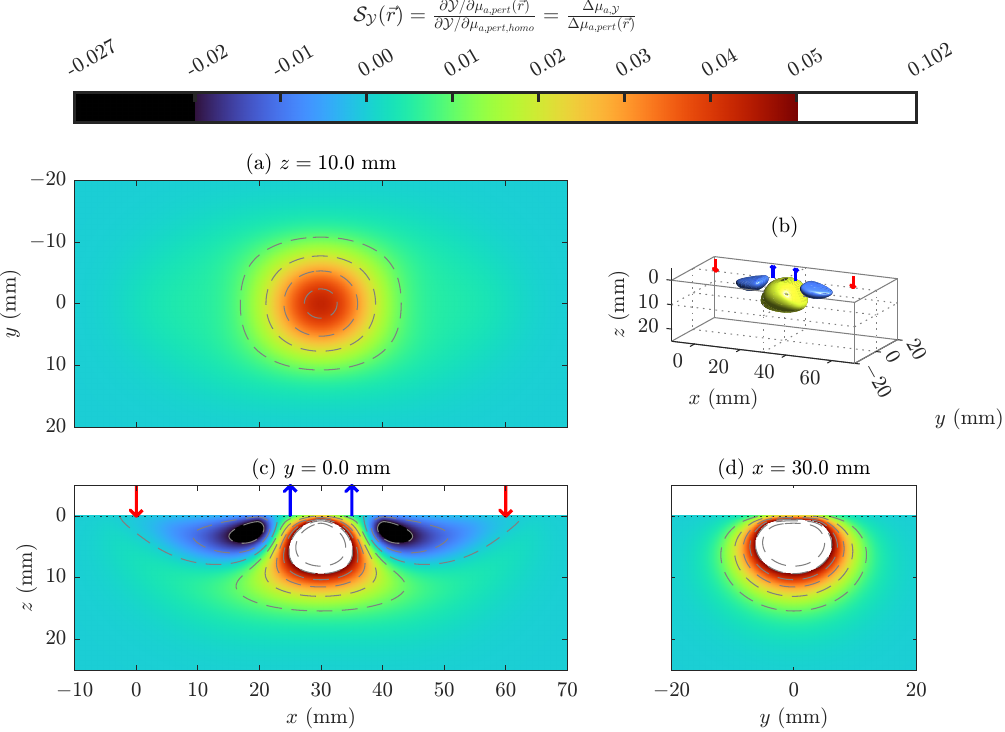}
	\end{center}
	\caption{Third angle projection of the \acrfull{sen} to a $\SI{10.0}{\milli\meter}\times\SI{10.0}{\milli\meter}\times\SI{2.0}{\milli\meter}$ perturbation scanned \SI{0.1}{\milli\meter} measured by \acrfull{FD} \acrfull{DS} \acrfull{I}. (a) $x$-$y$ plane sliced at $z=\SI{10.0}{\milli\meter}$. (b) Iso-surface sliced at $\as{S}=0.020$ and $\as{S}=-0.010$. (c) $x$-$z$ plane sliced at $y=\SI{0.0}{\milli\meter}$. (d) $y$-$z$ plane sliced at $x=\SI{30.0}{\milli\meter}$. Generated using \acrfull{DT}.\\ 
	\Acrfullpl{rho}: [25.0, 35.0, 35.0, 25.0]~\si{\milli\meter}\\ 
	\Acrfull{n} inside: \num{1.333}\\ 
	\Acrfull{n} outside: \num{1.000}\\ 
	\Acrfull{musp}: \SI{1.10}{\per\milli\meter}\\ 
	\Acrfull{mua}: \SI{0.011}{\per\milli\meter}\\ 
	\Acrfull{fmod}: \SI{100}{\mega\hertz}\\ 
	}\label{fig:FD_DS_I_3rd}
\end{figure*}

%% file: FD_SS_I_fmod.tex
\begin{figure*}
	\begin{center}
		\includegraphics{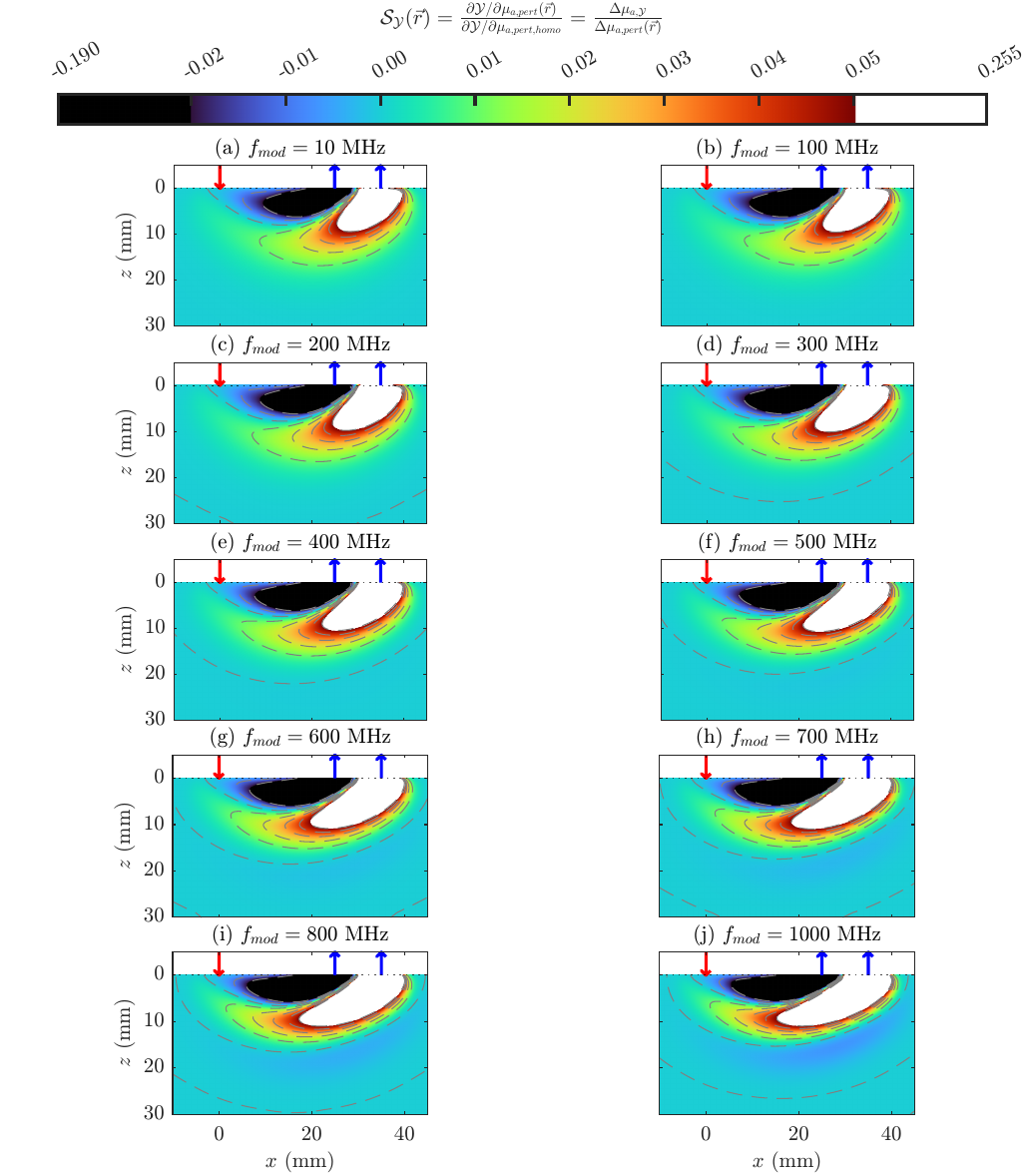}
	\end{center}
	\caption{$x$-$z$ plane of the \acrfull{sen} to a $\SI{10.0}{\milli\meter}\times\SI{10.0}{\milli\meter}\times\SI{2.0}{\milli\meter}$ perturbation scanned \SI{0.1}{\milli\meter} measured by \acrfull{FD} \acrfull{SS} \acrfull{I}. (a)-(j) Different values of \acrfull{fmod}. Generated using \acrfull{DT}.\\ 
	\Acrfullpl{rho}: [25.0, 35.0]~\si{\milli\meter}\\ 
	\Acrfull{n} inside: \num{1.333};\quad	\Acrfull{n} outside: \num{1.000}\\ 
	\Acrfull{musp}: \SI{1.10}{\per\milli\meter};\quad	\Acrfull{mua}: \SI{0.011}{\per\milli\meter}\\ 
	}\label{fig:FD_SS_I_fmod}
\end{figure*}

%% file: FD_DS_I_fmod.tex
\begin{figure*}
	\begin{center}
		\includegraphics{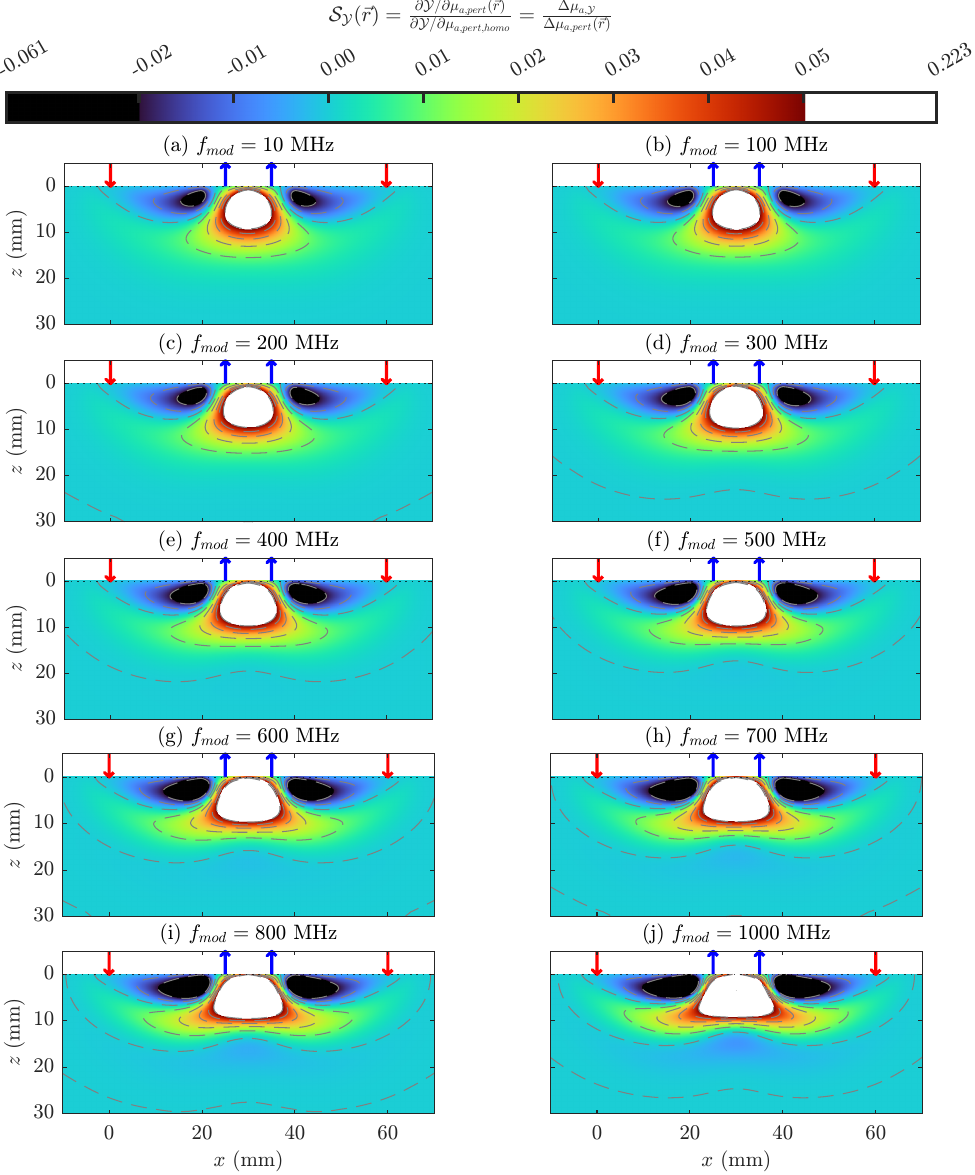}
	\end{center}
	\caption{$x$-$z$ plane of the \acrfull{sen} to a $\SI{10.0}{\milli\meter}\times\SI{10.0}{\milli\meter}\times\SI{2.0}{\milli\meter}$ perturbation scanned \SI{0.1}{\milli\meter} measured by \acrfull{FD} \acrfull{DS} \acrfull{I}. (a)-(j) Different values of \acrfull{fmod}. Generated using \acrfull{DT}.\\ 
	\Acrfullpl{rho}: [25.0, 35.0, 35.0, 25.0]~\si{\milli\meter}\\ 
	\Acrfull{n} inside: \num{1.333};\quad	\Acrfull{n} outside: \num{1.000}\\ 
	\Acrfull{musp}: \SI{1.10}{\per\milli\meter};\quad	\Acrfull{mua}: \SI{0.011}{\per\milli\meter}\\ 
	}\label{fig:FD_DS_I_fmod}
\end{figure*}

%% file: FD_SS_P_3rd.tex
\begin{figure*}
	\begin{center}
		\includegraphics{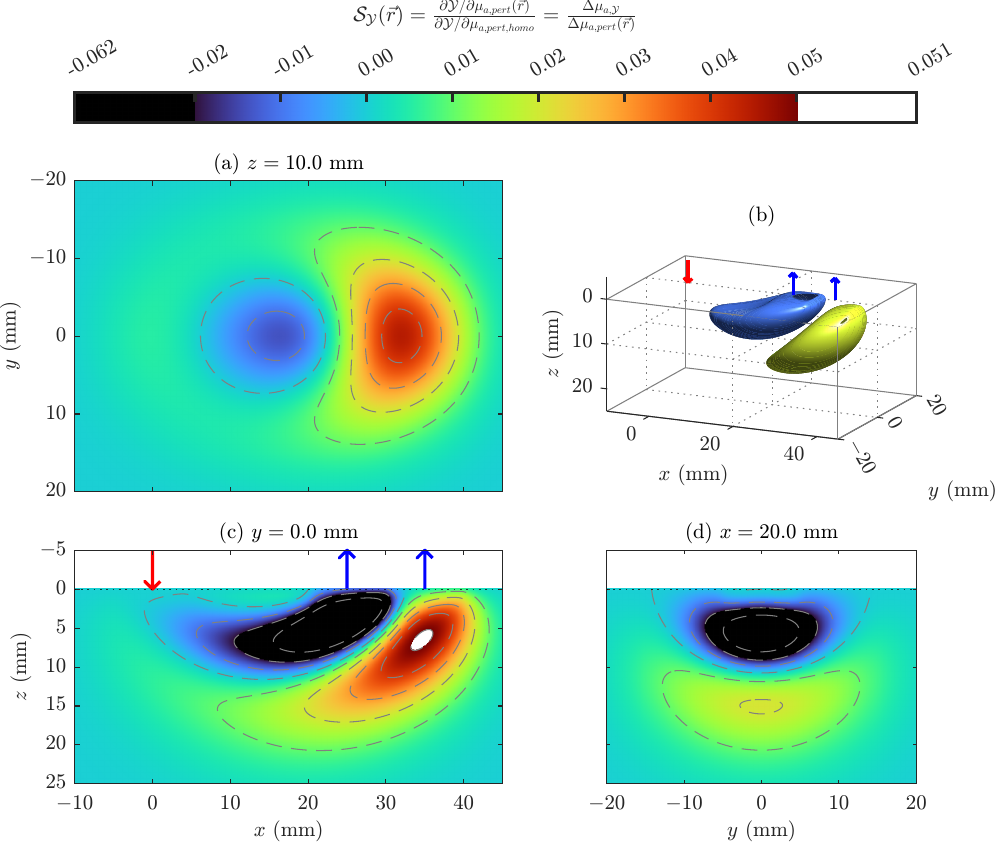}
	\end{center}
	\caption{Third angle projection of the \acrfull{sen} to a $\SI{10.0}{\milli\meter}\times\SI{10.0}{\milli\meter}\times\SI{2.0}{\milli\meter}$ perturbation scanned \SI{0.1}{\milli\meter} measured by \acrfull{FD} \acrfull{SS} \acrfull{phi}. (a) $x$-$y$ plane sliced at $z=\SI{10.0}{\milli\meter}$. (b) Iso-surface sliced at $\as{S}=0.020$ and $\as{S}=-0.010$. (c) $x$-$z$ plane sliced at $y=\SI{0.0}{\milli\meter}$. (d) $y$-$z$ plane sliced at $x=\SI{20.0}{\milli\meter}$. Generated using \acrfull{DT}.\\ 
	\Acrfullpl{rho}: [25.0, 35.0]~\si{\milli\meter}\\ 
	\Acrfull{n} inside: \num{1.333}\\ 
	\Acrfull{n} outside: \num{1.000}\\ 
	\Acrfull{musp}: \SI{1.10}{\per\milli\meter}\\ 
	\Acrfull{mua}: \SI{0.011}{\per\milli\meter}\\ 
	\Acrfull{fmod}: \SI{100}{\mega\hertz}\\ 
	}\label{fig:FD_SS_P_3rd}
\end{figure*}

%% file: FD_DS_P_3rd.tex
\begin{figure*}
	\begin{center}
		\includegraphics{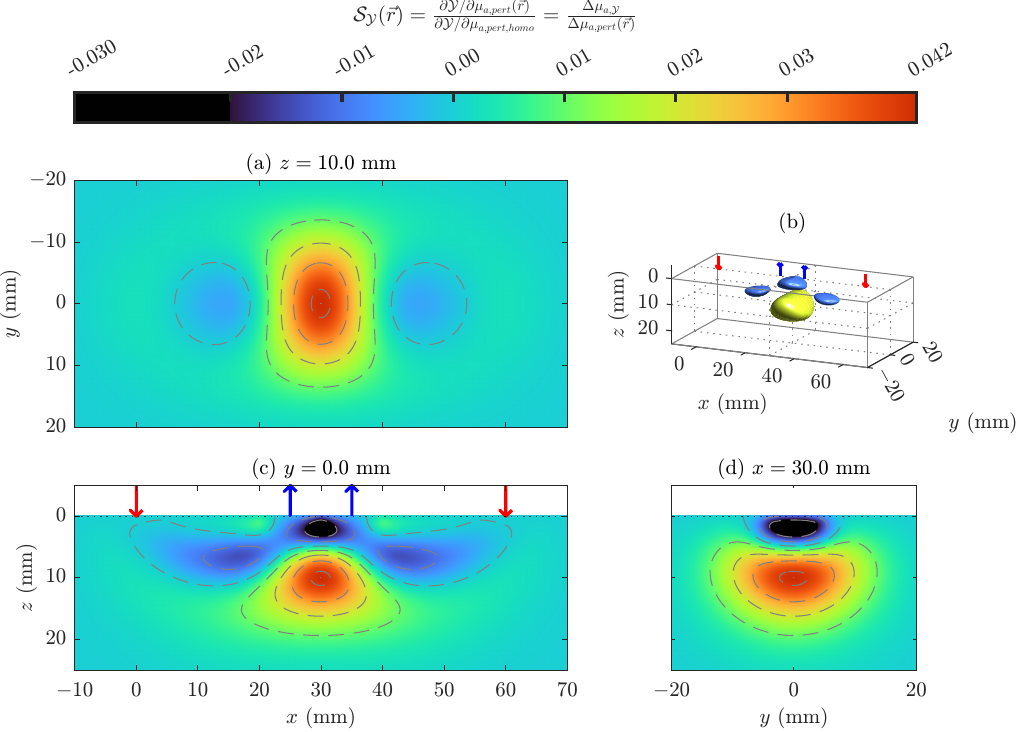}
	\end{center}
	\caption{Third angle projection of the \acrfull{sen} to a $\SI{10.0}{\milli\meter}\times\SI{10.0}{\milli\meter}\times\SI{2.0}{\milli\meter}$ perturbation scanned \SI{0.1}{\milli\meter} measured by \acrfull{FD} \acrfull{DS} \acrfull{phi}. (a) $x$-$y$ plane sliced at $z=\SI{10.0}{\milli\meter}$. (b) Iso-surface sliced at $\as{S}=0.020$ and $\as{S}=-0.010$. (c) $x$-$z$ plane sliced at $y=\SI{0.0}{\milli\meter}$. (d) $y$-$z$ plane sliced at $x=\SI{30.0}{\milli\meter}$. Generated using \acrfull{DT}.\\ 
	\Acrfullpl{rho}: [25.0, 35.0, 35.0, 25.0]~\si{\milli\meter}\\ 
	\Acrfull{n} inside: \num{1.333};\quad	\Acrfull{n} outside: \num{1.000}\\ 
	\Acrfull{musp}: \SI{1.10}{\per\milli\meter};\quad	\Acrfull{mua}: \SI{0.011}{\per\milli\meter}\\ 
	\Acrfull{fmod}: \SI{100}{\mega\hertz}\\ 
	}\label{fig:FD_DS_P_3rd}
\end{figure*}

%% file: FD_SS_P_mrho.tex
\begin{figure*}
	\begin{center}
		\includegraphics{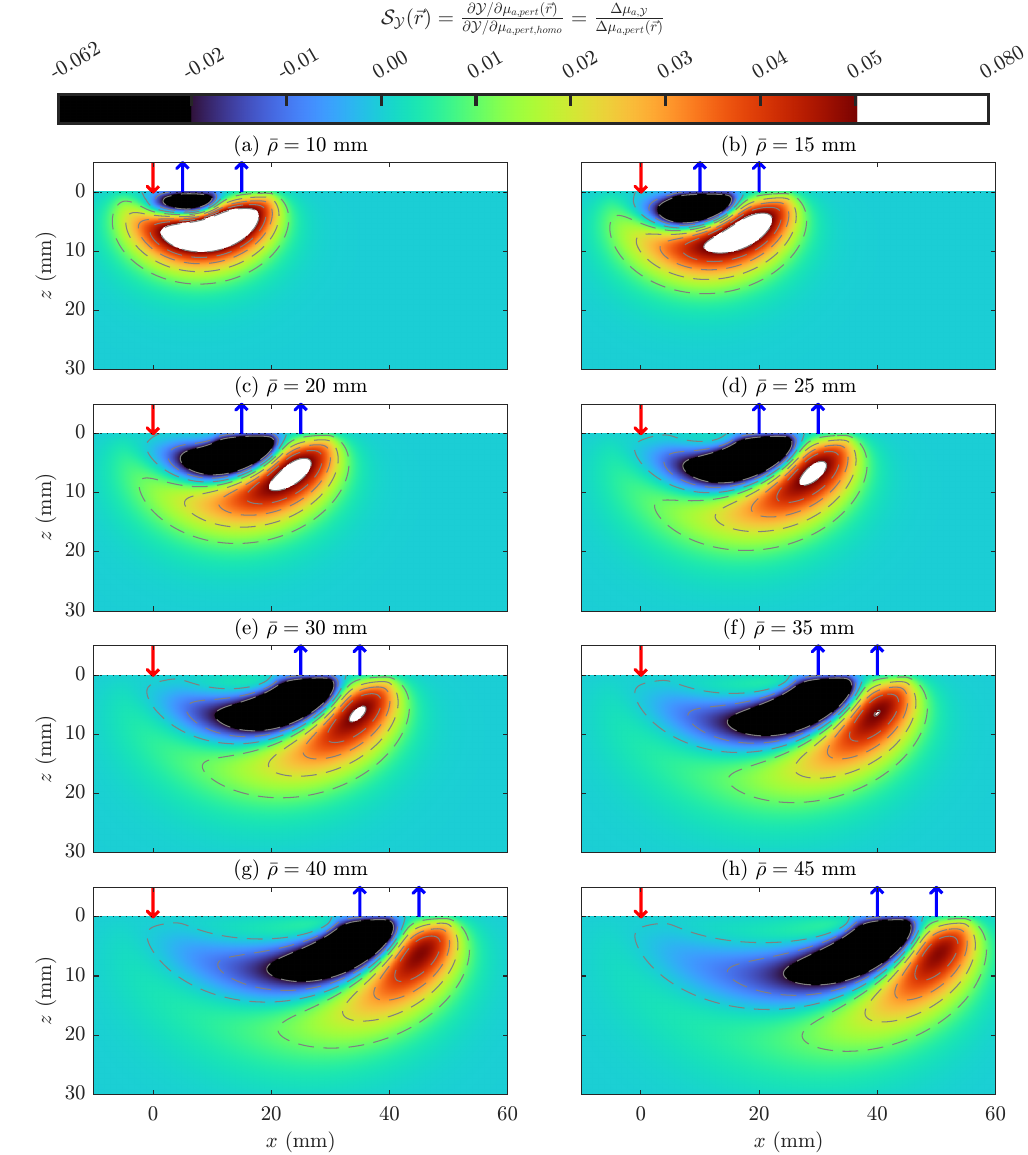}
	\end{center}
	\caption{$x$-$z$ plane of the \acrfull{sen} to a $\SI{10.0}{\milli\meter}\times\SI{10.0}{\milli\meter}\times\SI{2.0}{\milli\meter}$ perturbation scanned \SI{0.1}{\milli\meter} measured by \acrfull{FD} \acrfull{SS} \acrfull{phi}. (a)-(h) Different values of \acrfull{mrho}. Generated using \acrfull{DT}.\\ 
	\Acrfull{drho}: \SI{10.0}{\milli\meter}\\ 
	\Acrfull{n} inside: \num{1.333};\quad	\Acrfull{n} outside: \num{1.000}\\ 
	\Acrfull{musp}: \SI{1.10}{\per\milli\meter};\quad	\Acrfull{mua}: \SI{0.011}{\per\milli\meter}\\ 
	\Acrfull{fmod}: \SI{100}{\mega\hertz}\\ 
	}\label{fig:FD_SS_P_mrho}
\end{figure*}

%% file: FD_DS_P_mrho.tex
\begin{figure*}
	\begin{center}
		\includegraphics{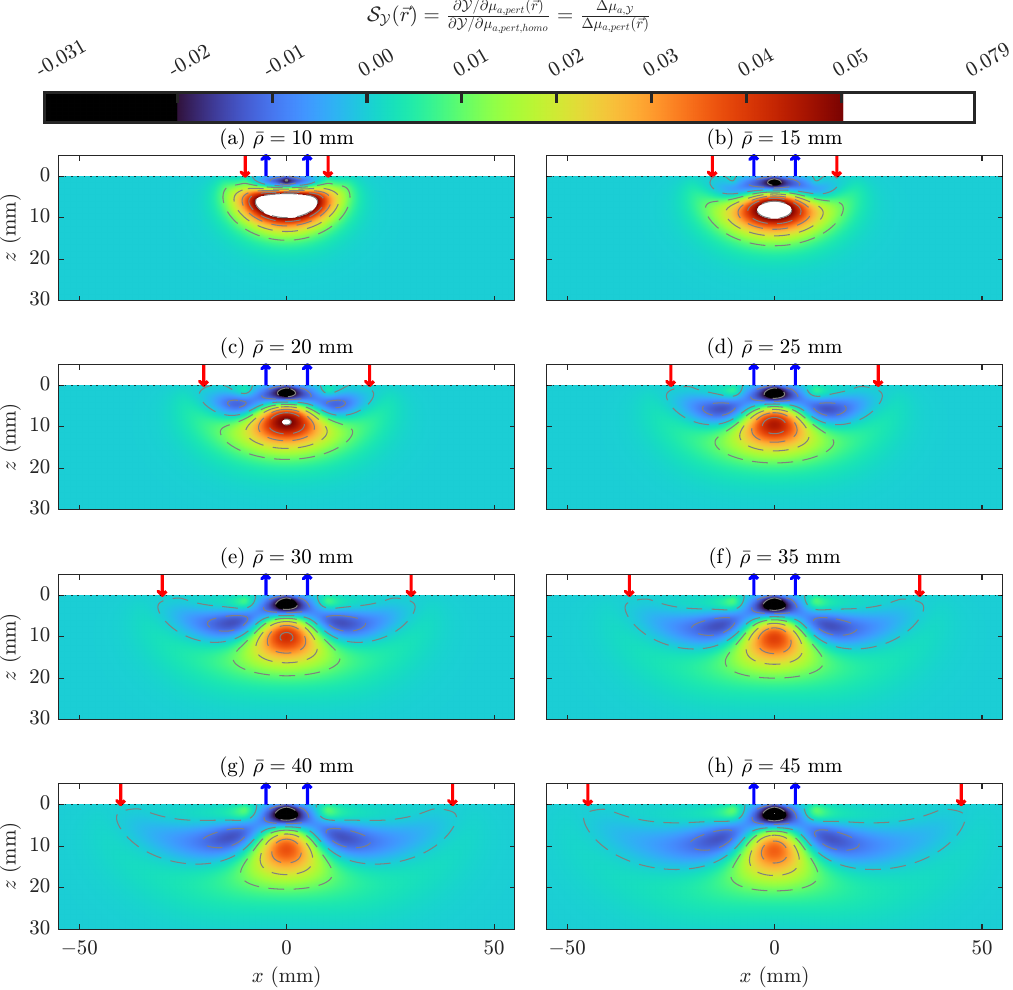}
	\end{center}
	\caption{$x$-$z$ plane of the \acrfull{sen} to a $\SI{10.0}{\milli\meter}\times\SI{10.0}{\milli\meter}\times\SI{2.0}{\milli\meter}$ perturbation scanned \SI{0.1}{\milli\meter} measured by \acrfull{FD} \acrfull{DS} \acrfull{phi}. (a)-(h) Different values of \acrfull{mrho}. Generated using \acrfull{DT}.\\ 
	\Acrfull{drho}: \SI{10.0}{\milli\meter}\\ 
	\Acrfull{n} inside: \num{1.333};\quad	\Acrfull{n} outside: \num{1.000}\\ 
	\Acrfull{musp}: \SI{1.10}{\per\milli\meter};\quad	\Acrfull{mua}: \SI{0.011}{\per\milli\meter}\\ 
	\Acrfull{fmod}: \SI{100}{\mega\hertz}\\ 
	}\label{fig:FD_DS_P_mrho}
\end{figure*}

%% file: FD_SS_P_drho.tex
\begin{figure*}
	\begin{center}
		\includegraphics{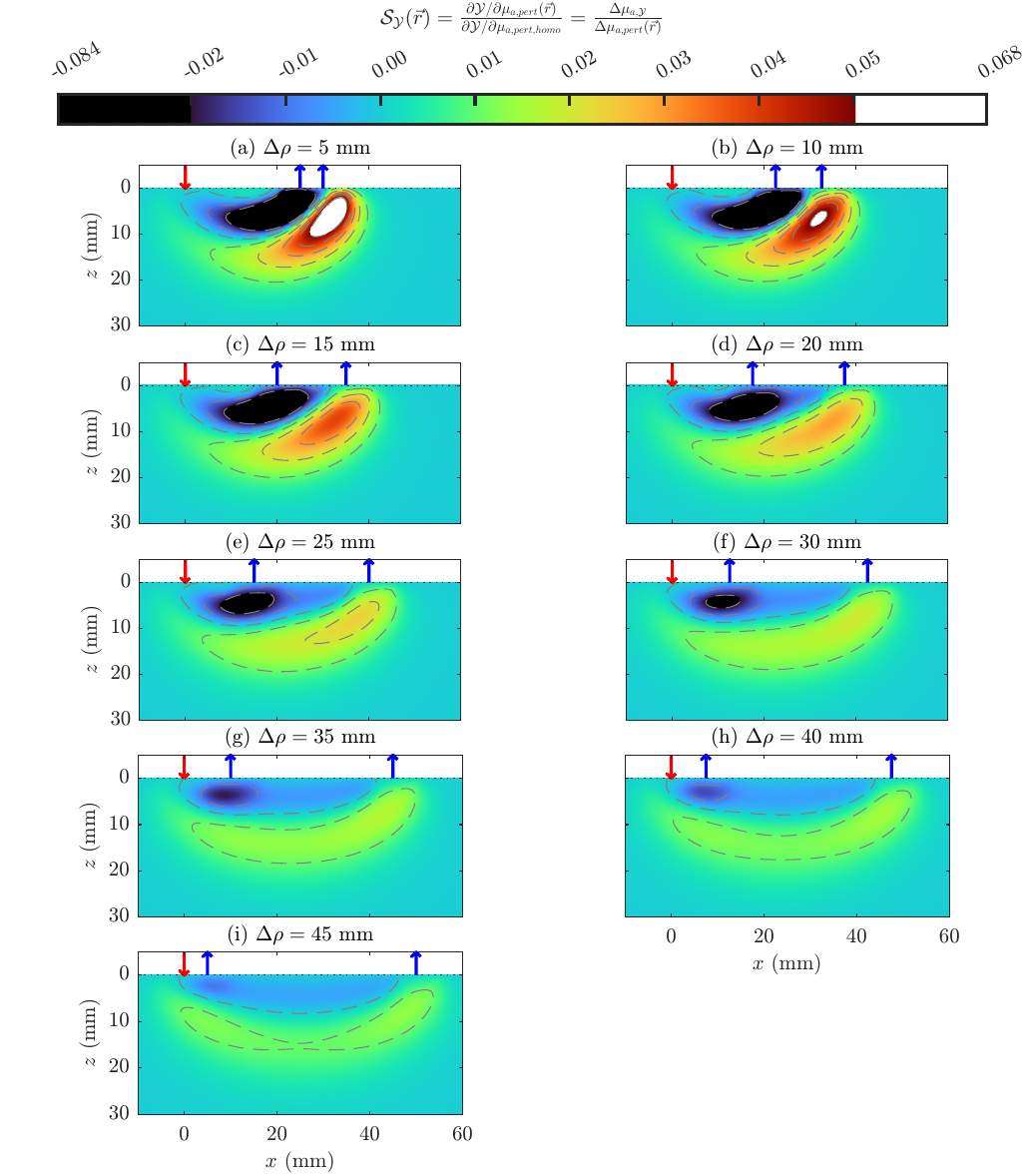}
	\end{center}
	\caption{$x$-$z$ plane of the \acrfull{sen} to a $\SI{10.0}{\milli\meter}\times\SI{10.0}{\milli\meter}\times\SI{2.0}{\milli\meter}$ perturbation scanned \SI{0.1}{\milli\meter} measured by \acrfull{FD} \acrfull{SS} \acrfull{phi}. (a)-(i) Different values of \acrfull{drho}. Generated using \acrfull{DT}.\\ 
	\Acrfull{mrho}: \SI{27.5}{\milli\meter}\\ 
	\Acrfull{n} inside: \num{1.333};\quad	\Acrfull{n} outside: \num{1.000}\\ 
	\Acrfull{musp}: \SI{1.10}{\per\milli\meter};\quad	\Acrfull{mua}: \SI{0.011}{\per\milli\meter}\\ 
	\Acrfull{fmod}: \SI{100}{\mega\hertz}\\ 
	}\label{fig:FD_SS_P_drho}
\end{figure*}

%% file: FD_DS_P_drho.tex
\begin{figure*}
	\begin{center}
		\includegraphics{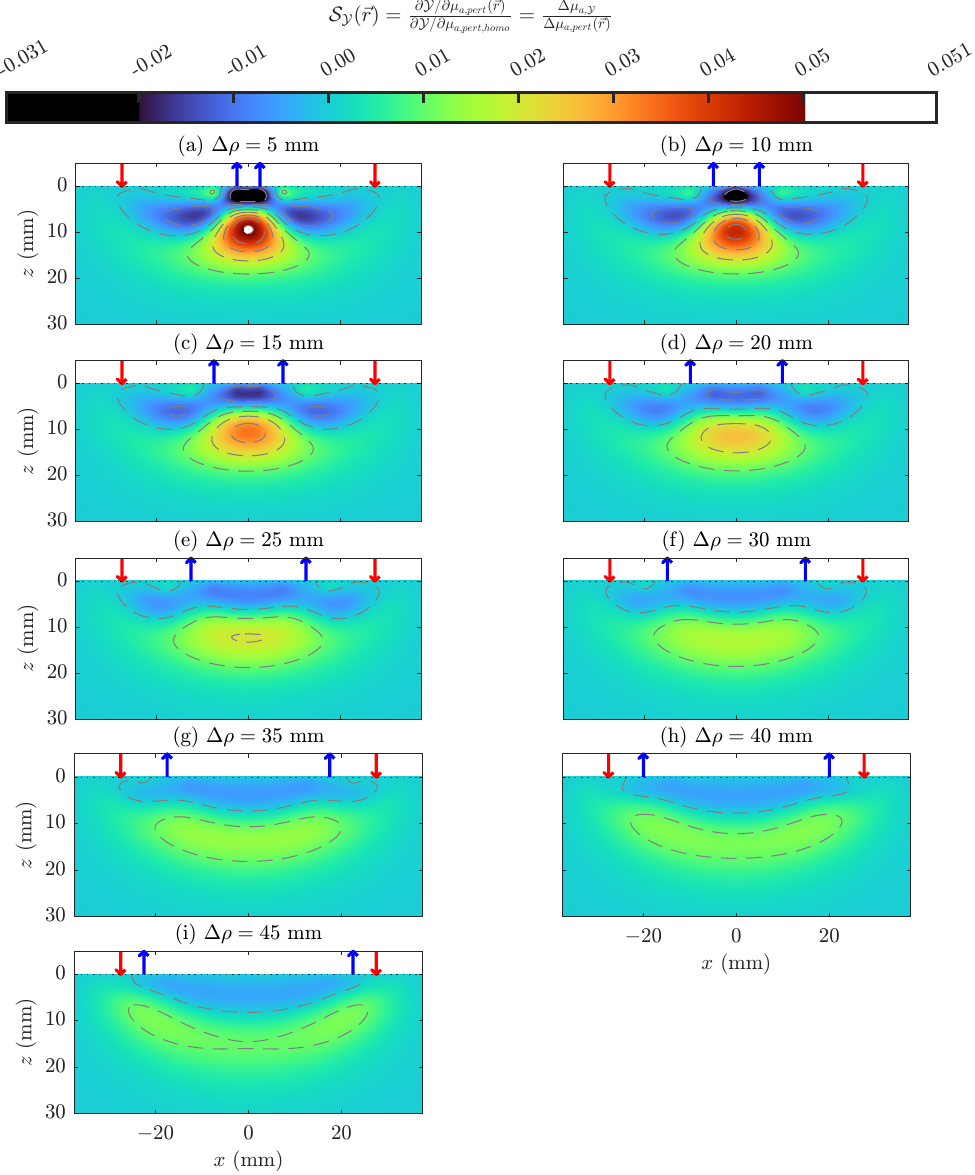}
	\end{center}
	\caption{$x$-$z$ plane of the \acrfull{sen} to a $\SI{10.0}{\milli\meter}\times\SI{10.0}{\milli\meter}\times\SI{2.0}{\milli\meter}$ perturbation scanned \SI{0.1}{\milli\meter} measured by \acrfull{FD} \acrfull{DS} \acrfull{phi}. (a)-(i) Different values of \acrfull{drho}. Generated using \acrfull{DT}.\\ 
	\Acrfull{mrho}: \SI{27.5}{\milli\meter}\\ 
	\Acrfull{n} inside: \num{1.333};\quad	\Acrfull{n} outside: \num{1.000}\\ 
	\Acrfull{musp}: \SI{1.10}{\per\milli\meter};\quad	\Acrfull{mua}: \SI{0.011}{\per\milli\meter}\\ 
	\Acrfull{fmod}: \SI{100}{\mega\hertz}\\ 
	}\label{fig:FD_DS_P_drho}
\end{figure*}

%% file: FD_SS_P_fmod.tex
\begin{figure*}
	\begin{center}
		\includegraphics{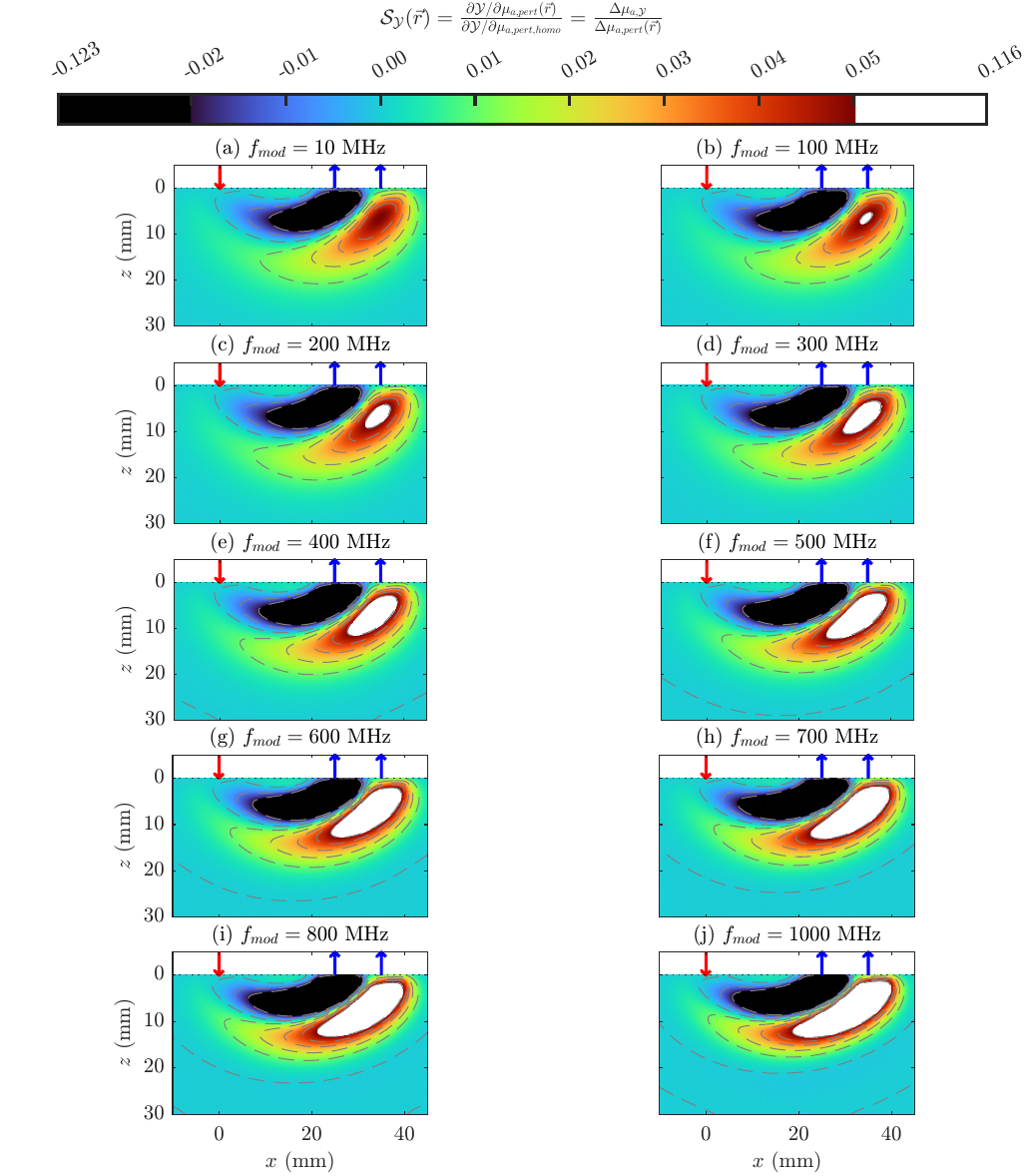}
	\end{center}
	\caption{$x$-$z$ plane of the \acrfull{sen} to a $\SI{10.0}{\milli\meter}\times\SI{10.0}{\milli\meter}\times\SI{2.0}{\milli\meter}$ perturbation scanned \SI{0.1}{\milli\meter} measured by \acrfull{FD} \acrfull{SS} \acrfull{phi}. (a)-(j) Different values of \acrfull{fmod}. Generated using \acrfull{DT}.\\ 
	\Acrfullpl{rho}: [25.0, 35.0]~\si{\milli\meter}\\ 
	\Acrfull{n} inside: \num{1.333};\quad	\Acrfull{n} outside: \num{1.000}\\ 
	\Acrfull{musp}: \SI{1.10}{\per\milli\meter};\quad	\Acrfull{mua}: \SI{0.011}{\per\milli\meter}\\ 
	}\label{fig:FD_SS_P_fmod}
\end{figure*}

%% file: FD_DS_P_fmod.tex
\begin{figure*}
	\begin{center}
		\includegraphics{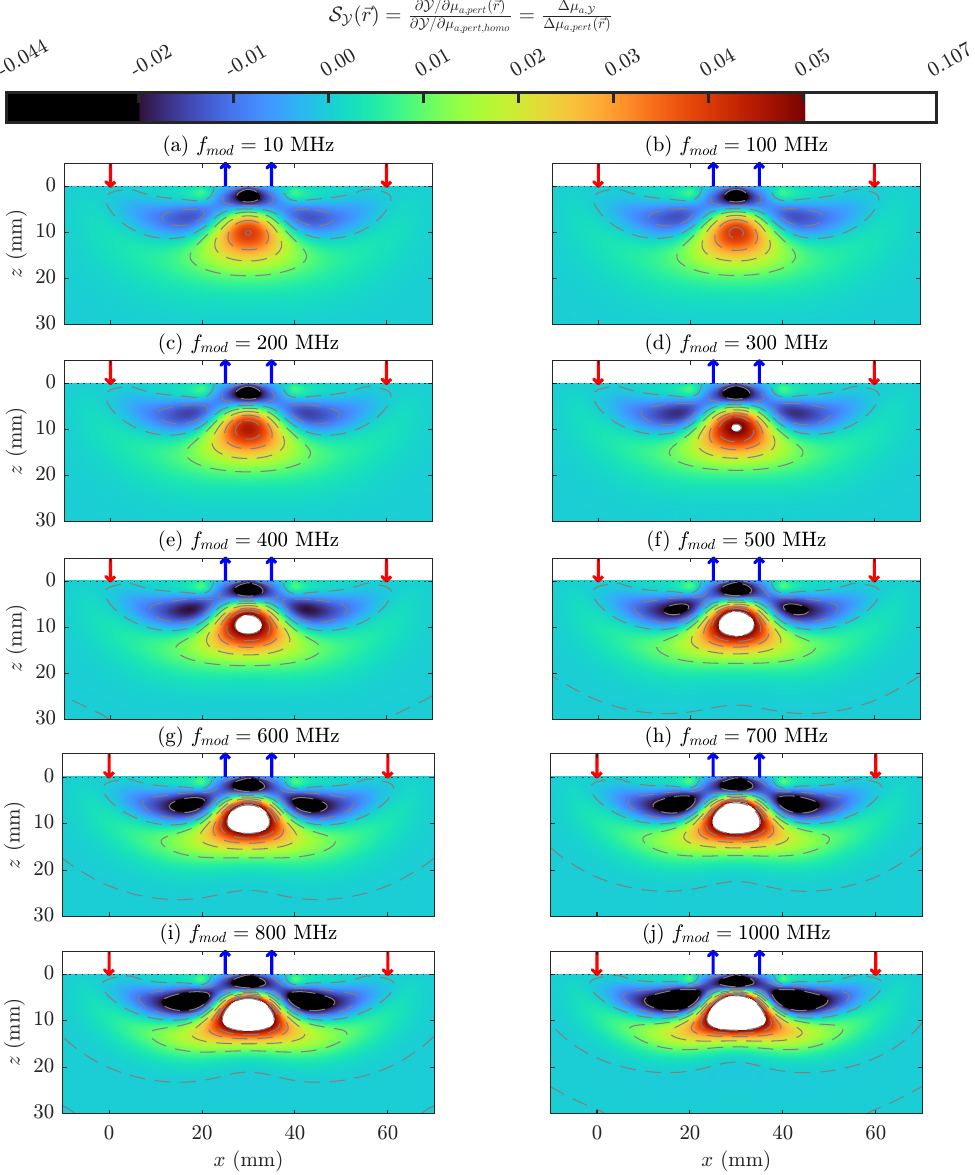}
	\end{center}
	\caption{$x$-$z$ plane of the \acrfull{sen} to a $\SI{10.0}{\milli\meter}\times\SI{10.0}{\milli\meter}\times\SI{2.0}{\milli\meter}$ perturbation scanned \SI{0.1}{\milli\meter} measured by \acrfull{FD} \acrfull{DS} \acrfull{phi}. (a)-(j) Different values of \acrfull{fmod}. Generated using \acrfull{DT}.\\ 
	\Acrfullpl{rho}: [25.0, 35.0, 35.0, 25.0]~\si{\milli\meter}\\ 
	\Acrfull{n} inside: \num{1.333};\quad	\Acrfull{n} outside: \num{1.000}\\ 
	\Acrfull{musp}: \SI{1.10}{\per\milli\meter};\quad	\Acrfull{mua}: \SI{0.011}{\per\milli\meter}\\ 
	}\label{fig:FD_DS_P_fmod}
\end{figure*}

%% file: TD_SD_GI_3rd_rho0_MC_vox.tex
\begin{figure*}
	\begin{center}
		\includegraphics{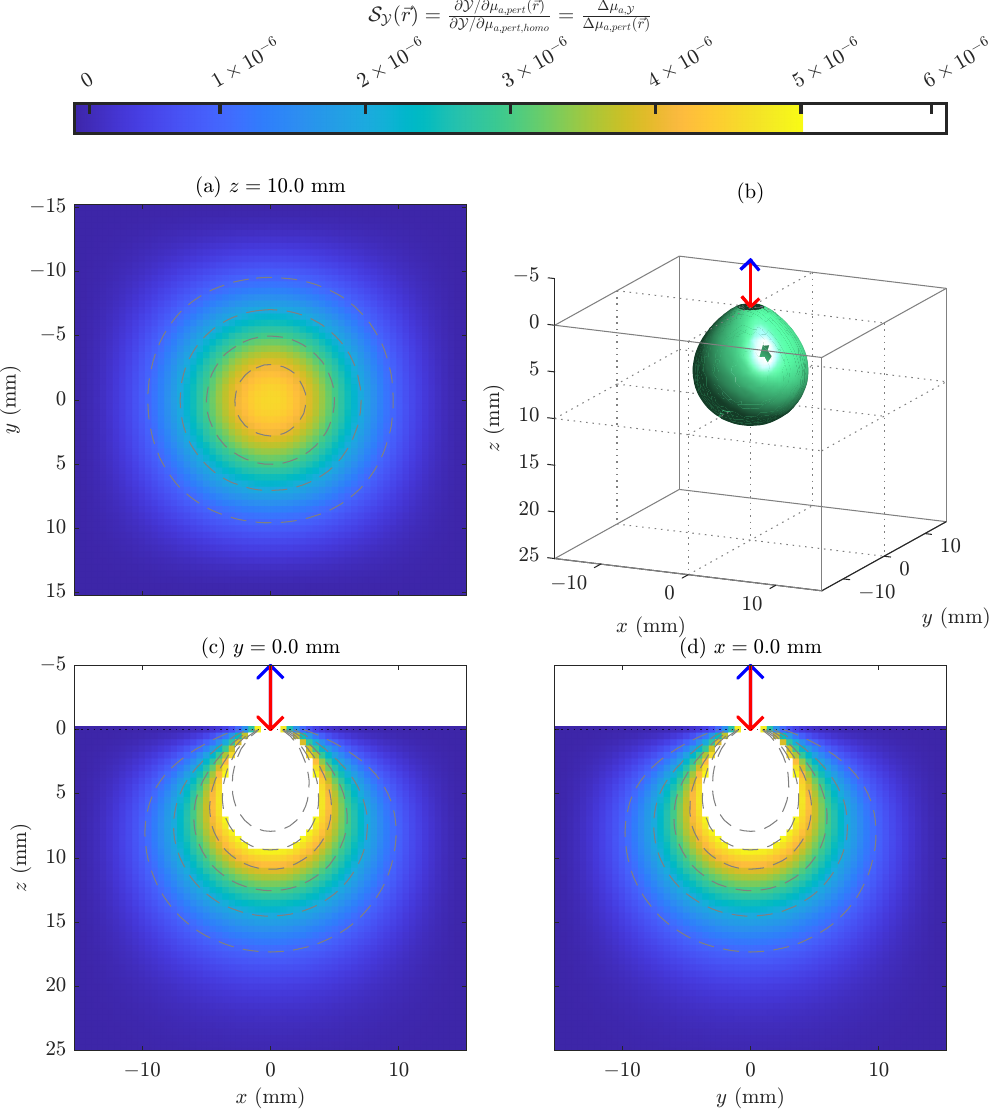}
	\end{center}
	\caption{Third angle projection of the \acrfull{sen} to a $\SI{0.5}{\milli\meter}\times\SI{0.5}{\milli\meter}\times\SI{0.5}{\milli\meter}$ perturbation scanned \SI{0.5}{\milli\meter} measured by \acrfull{TD} \acrfull{SD} gated \acrfull{I}. (a) $x$-$y$ plane sliced at $z=\SI{10.0}{\milli\meter}$. (b) Iso-surface sliced at $\as{S}=3.000\times 10^{-5}$. (c) $x$-$z$ plane sliced at $y=\SI{0.0}{\milli\meter}$. (d) $y$-$z$ plane sliced at $x=\SI{0.0}{\milli\meter}$. Generated using \acrfull{MC}.\\ 
	\Acrfull{rho}: \SI{0.0}{\milli\meter}\\ 
	\Acrfull{n} inside: \num{1.333};\quad	\Acrfull{n} outside: \num{1.000}\\ 
	\Acrfull{musp}: \SI{1.10}{\per\milli\meter};\quad	\Acrfull{g}: \num{0.9}\\ 
	\Acrfull{mua}: \SI{0.011}{\per\milli\meter}\\ 
	\Acrfull{t} gate: [1500, 2000]~\si{\pico\second}\\ 
	Detector Numerical Aperature (NA): \num{0.5};\quad	Number of photons: \num{1000000000}\\ 
	}\label{fig:TD_SD_GI_3rd_rho0_MC_vox}
\end{figure*}

%% file: TD_SD_GI_3rd_rho0_MC.tex
\begin{figure*}
	\begin{center}
		\includegraphics{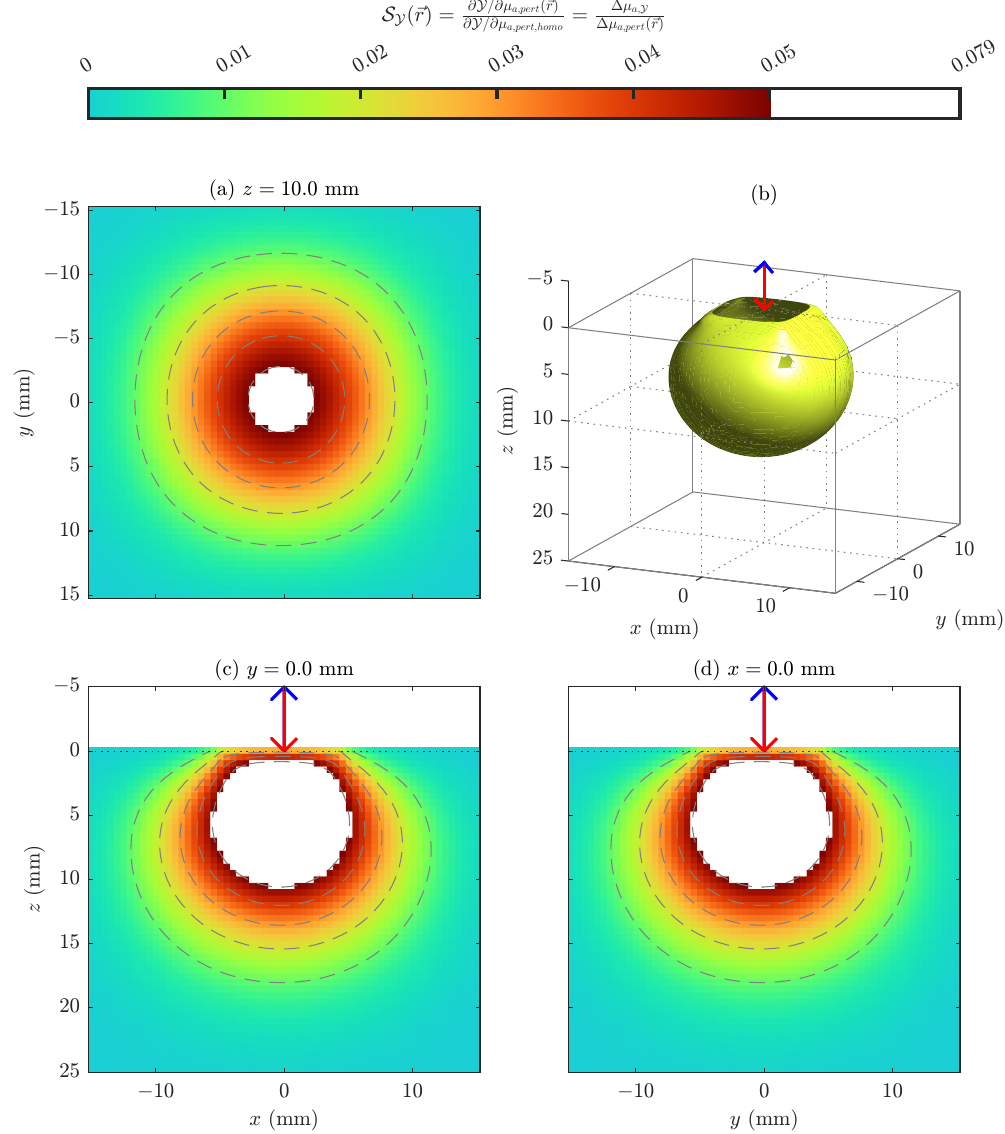}
	\end{center}
	\caption{Third angle projection of the \acrfull{sen} to a $\SI{10.0}{\milli\meter}\times\SI{10.0}{\milli\meter}\times\SI{2.0}{\milli\meter}$ perturbation scanned \SI{0.5}{\milli\meter} measured by \acrfull{TD} \acrfull{SD} gated \acrfull{I}. (a) $x$-$y$ plane sliced at $z=\SI{10.0}{\milli\meter}$. (b) Iso-surface sliced at $\as{S}=0.020$. (c) $x$-$z$ plane sliced at $y=\SI{0.0}{\milli\meter}$. (d) $y$-$z$ plane sliced at $x=\SI{0.0}{\milli\meter}$. Generated using \acrfull{MC}.\\ 
	\Acrfull{rho}: \SI{0.0}{\milli\meter}\\ 
	\Acrfull{n} inside: \num{1.333};\quad	\Acrfull{n} outside: \num{1.000}\\ 
	\Acrfull{musp}: \SI{1.10}{\per\milli\meter};\quad	\Acrfull{g}: \num{0.9}\\ 
	\Acrfull{mua}: \SI{0.011}{\per\milli\meter}\\ 
	\Acrfull{t} gate: [1500, 2000]~\si{\pico\second}\\ 
	Detector Numerical Aperature (NA): \num{0.5};\quad	Number of photons: \num{1000000000}\\ 
	}\label{fig:TD_SD_GI_3rd_rho0_MC}
\end{figure*}

%% file: TD_SD_GI_3rd.tex
\begin{figure*}
	\begin{center}
		\includegraphics{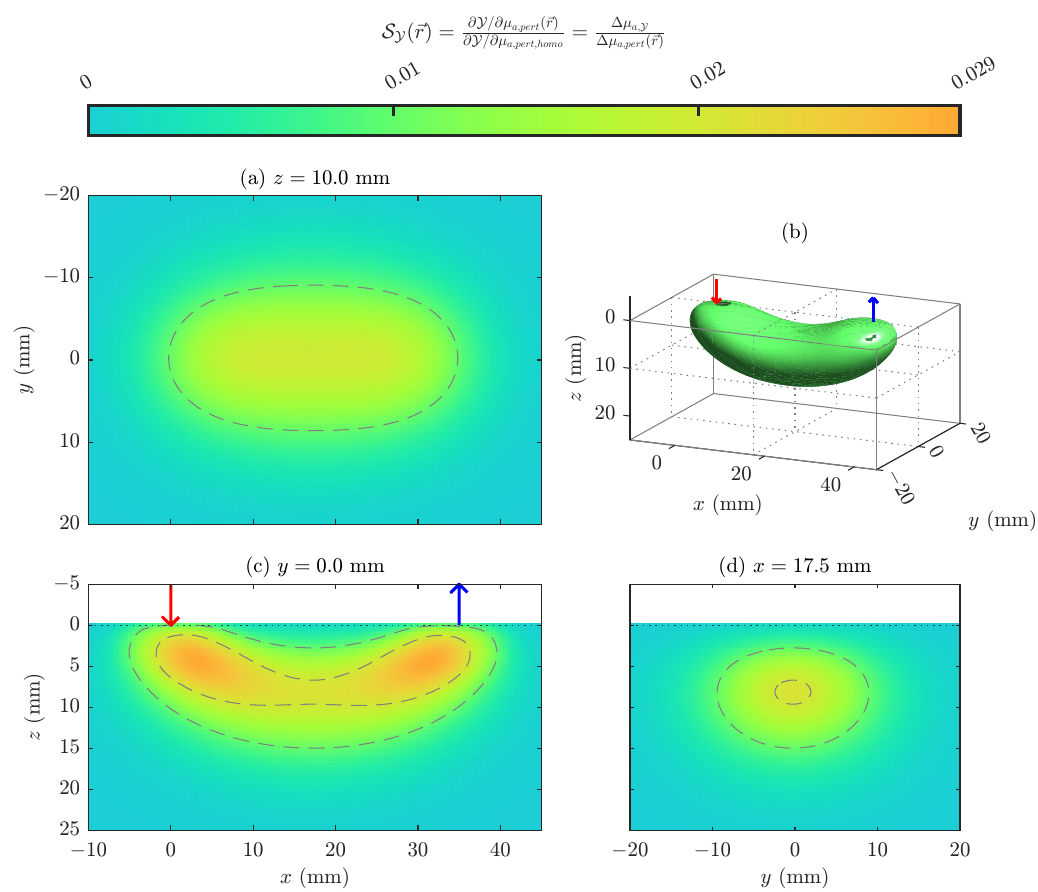}
	\end{center}
	\caption{Third angle projection of the \acrfull{sen} to a $\SI{10.0}{\milli\meter}\times\SI{10.0}{\milli\meter}\times\SI{2.0}{\milli\meter}$ perturbation scanned \SI{0.5}{\milli\meter} measured by \acrfull{TD} \acrfull{SD} gated \acrfull{I}. (a) $x$-$y$ plane sliced at $z=\SI{10.0}{\milli\meter}$. (b) Iso-surface sliced at $\as{S}=0.010$ and $\as{S}=-0.005$. (c) $x$-$z$ plane sliced at $y=\SI{0.0}{\milli\meter}$. (d) $y$-$z$ plane sliced at $x=\SI{17.5}{\milli\meter}$. Generated using \acrfull{DT}.\\ 
	\Acrfull{rho}: \SI{35.0}{\milli\meter}\\ 
	\Acrfull{n} inside: \num{1.333}\\ 
	\Acrfull{n} outside: \num{1.000}\\ 
	\Acrfull{musp}: \SI{1.10}{\per\milli\meter}\\ 
	\Acrfull{mua}: \SI{0.011}{\per\milli\meter}\\ 
	\Acrfull{t} gate: [1500, 2000]~\si{\pico\second}\\ 
	}\label{fig:TD_SD_GI_3rd}
\end{figure*}

%% file: TD_SD_GI_rho.tex
\begin{figure*}
	\begin{center}
		\includegraphics{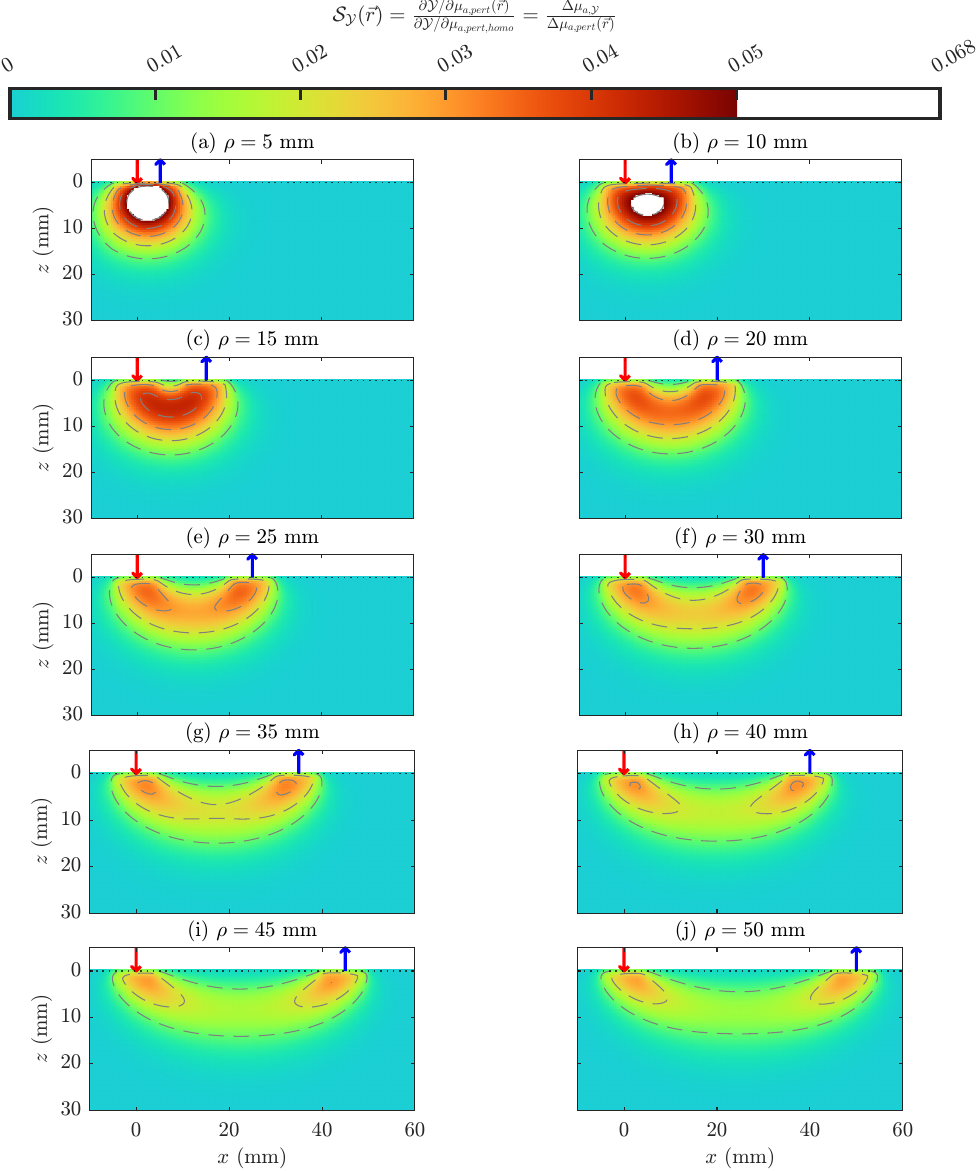}
	\end{center}
	\caption{$x$-$z$ plane of the \acrfull{sen} to a $\SI{10.0}{\milli\meter}\times\SI{10.0}{\milli\meter}\times\SI{2.0}{\milli\meter}$ perturbation scanned \SI{0.5}{\milli\meter} measured by \acrfull{TD} \acrfull{SD} gated \acrfull{I}. (a)-(j) Different values of \acrfull{rho}. Generated using \acrfull{DT}.\\ 
	\Acrfull{n} inside: \num{1.333};\quad	\Acrfull{n} outside: \num{1.000}\\ 
	\Acrfull{musp}: \SI{1.10}{\per\milli\meter};\quad	\Acrfull{mua}: \SI{0.011}{\per\milli\meter}\\ 
	\Acrfull{t} gate: [1500, 2000]~\si{\pico\second}\\ 
	}\label{fig:TD_SD_GI_rho}
\end{figure*}

%% file: TD_SD_GI_mt_rho0_MC.tex
\begin{figure*}
	\begin{center}
		\includegraphics{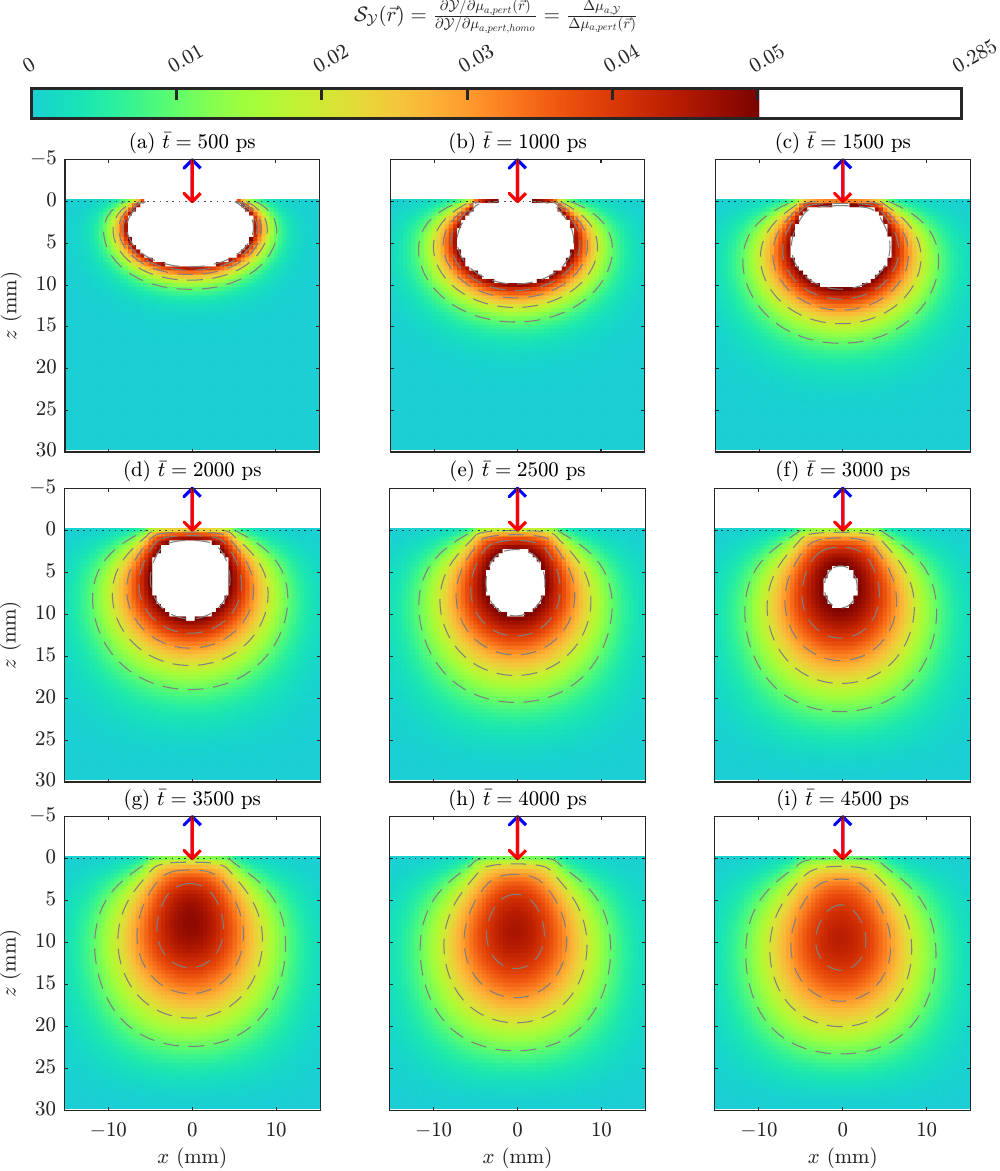}
	\end{center}
	\caption{$x$-$z$ plane of the \acrfull{sen} to a $\SI{10.0}{\milli\meter}\times\SI{10.0}{\milli\meter}\times\SI{2.0}{\milli\meter}$ perturbation scanned \SI{0.5}{\milli\meter} measured by \acrfull{TD} \acrfull{SD} gated \acrfull{I}. (a)-(i) Different values of \acrfull{mt}. Generated using \acrfull{MC}.\\ 
	\Acrfull{rho}: \SI{0.0}{\milli\meter}\\ 
	\Acrfull{n} inside: \num{1.333};\quad	\Acrfull{n} outside: \num{1.000}\\ 
	\Acrfull{musp}: \SI{1.10}{\per\milli\meter};\quad	\Acrfull{g}: \num{0.9}\\ 
	\Acrfull{mua}: \SI{0.011}{\per\milli\meter}\\ 
	\Acrfull{dt} gate: 500~\si{\pico\second}\\ 
	Detector Numerical Aperature (NA): \num{0.5};\quad	Number of photons: \num{1000000000}\\ 
	}\label{fig:TD_SD_GI_mt_rho0_MC}
\end{figure*}

%% file: TD_SD_GI_mt.tex
\begin{figure*}
	\begin{center}
		\includegraphics{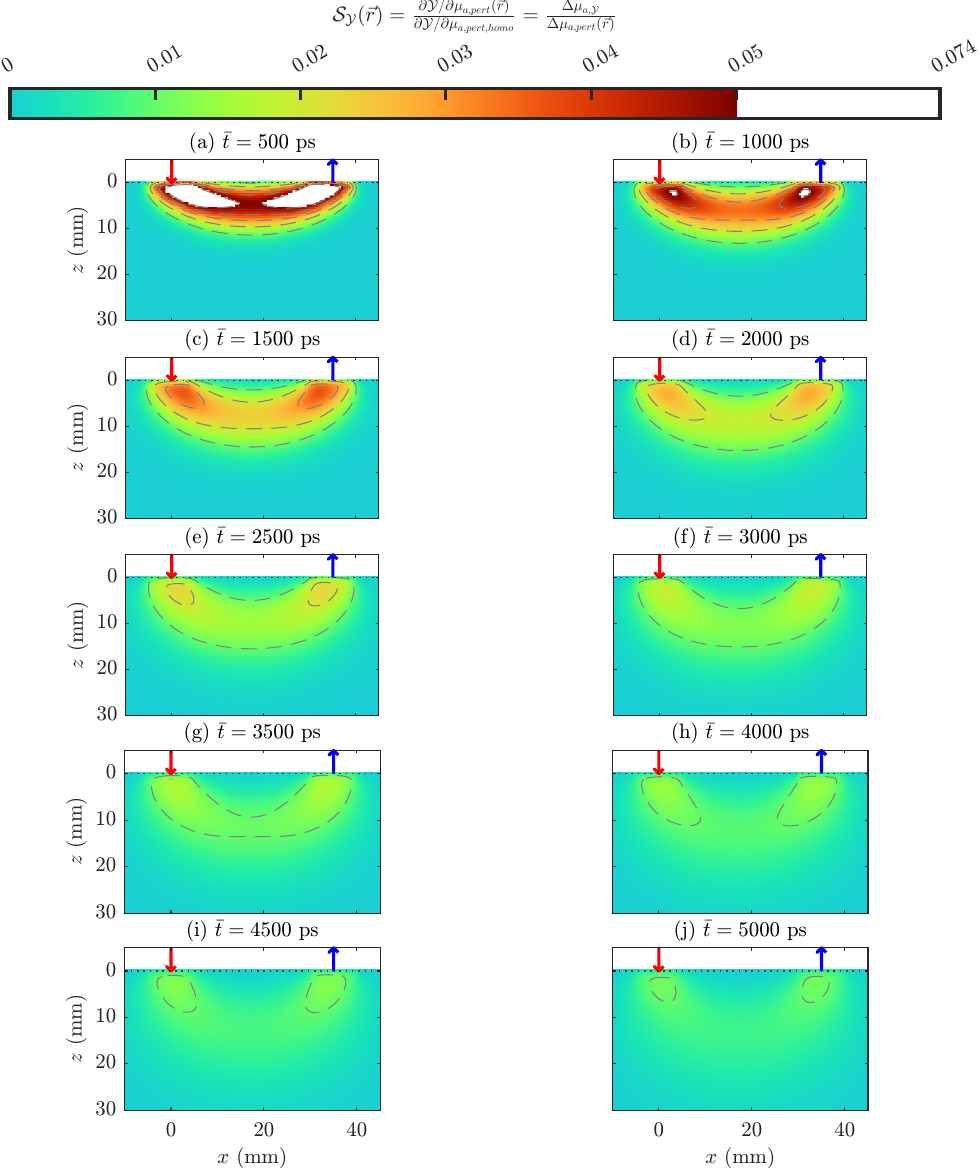}
	\end{center}
	\caption{$x$-$z$ plane of the \acrfull{sen} to a $\SI{10.0}{\milli\meter}\times\SI{10.0}{\milli\meter}\times\SI{2.0}{\milli\meter}$ perturbation scanned \SI{0.5}{\milli\meter} measured by \acrfull{TD} \acrfull{SD} gated \acrfull{I}. (a)-(j) Different values of \acrfull{mt}. Generated using \acrfull{DT}.\\ 
	\Acrfull{rho}: \SI{35.0}{\milli\meter}\\ 
	\Acrfull{n} inside: \num{1.333};\quad	\Acrfull{n} outside: \num{1.000}\\ 
	\Acrfull{musp}: \SI{1.10}{\per\milli\meter};\quad	\Acrfull{mua}: \SI{0.011}{\per\milli\meter}\\ 
	\Acrfull{dt} gate: 500~\si{\pico\second}\\ 
	}\label{fig:TD_SD_GI_mt}
\end{figure*}

%% file: TD_SD_GI_dt_rho0_MC.tex
\begin{figure*}
	\begin{center}
		\includegraphics{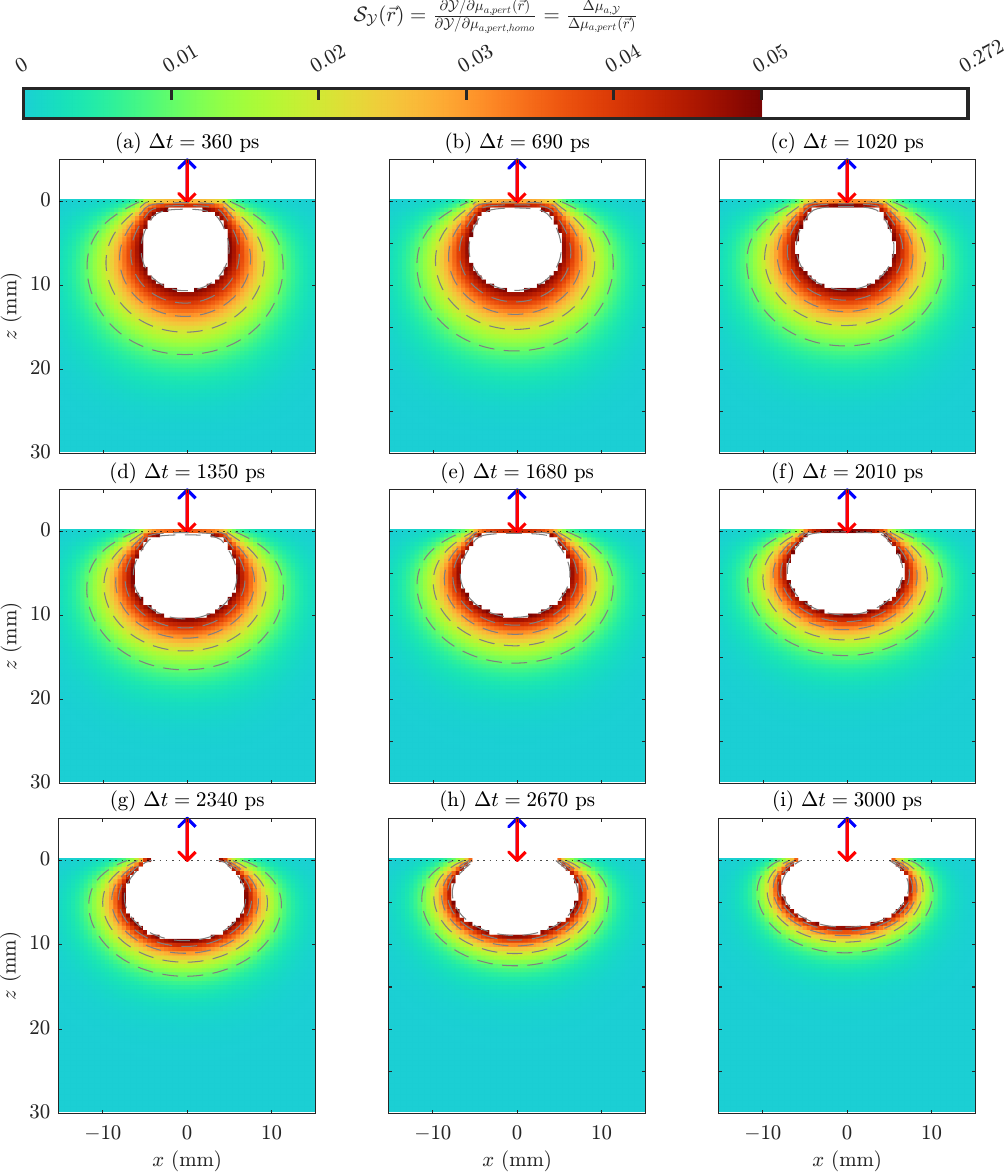}
	\end{center}
	\caption{$x$-$z$ plane of the \acrfull{sen} to a $\SI{10.0}{\milli\meter}\times\SI{10.0}{\milli\meter}\times\SI{2.0}{\milli\meter}$ perturbation scanned \SI{0.5}{\milli\meter} measured by \acrfull{TD} \acrfull{SD} gated \acrfull{I}. (a)-(i) Different values of \acrfull{dt}. Generated using \acrfull{MC}.\\ 
	\Acrfull{rho}: \SI{0.0}{\milli\meter}\\ 
	\Acrfull{n} inside: \num{1.333};\quad	\Acrfull{n} outside: \num{1.000}\\ 
	\Acrfull{musp}: \SI{1.10}{\per\milli\meter};\quad	\Acrfull{g}: \num{0.9}\\ 
	\Acrfull{mua}: \SI{0.011}{\per\milli\meter}\\ 
	\Acrfull{mt} gate: 1750~\si{\pico\second}\\ 
	Detector Numerical Aperature (NA): \num{0.5};\quad	Number of photons: \num{1000000000}\\ 
	}\label{fig:TD_SD_GI_dt_rho0_MC}
\end{figure*}

%% file: TD_SD_GI_dt.tex
\begin{figure*}
	\begin{center}
		\includegraphics{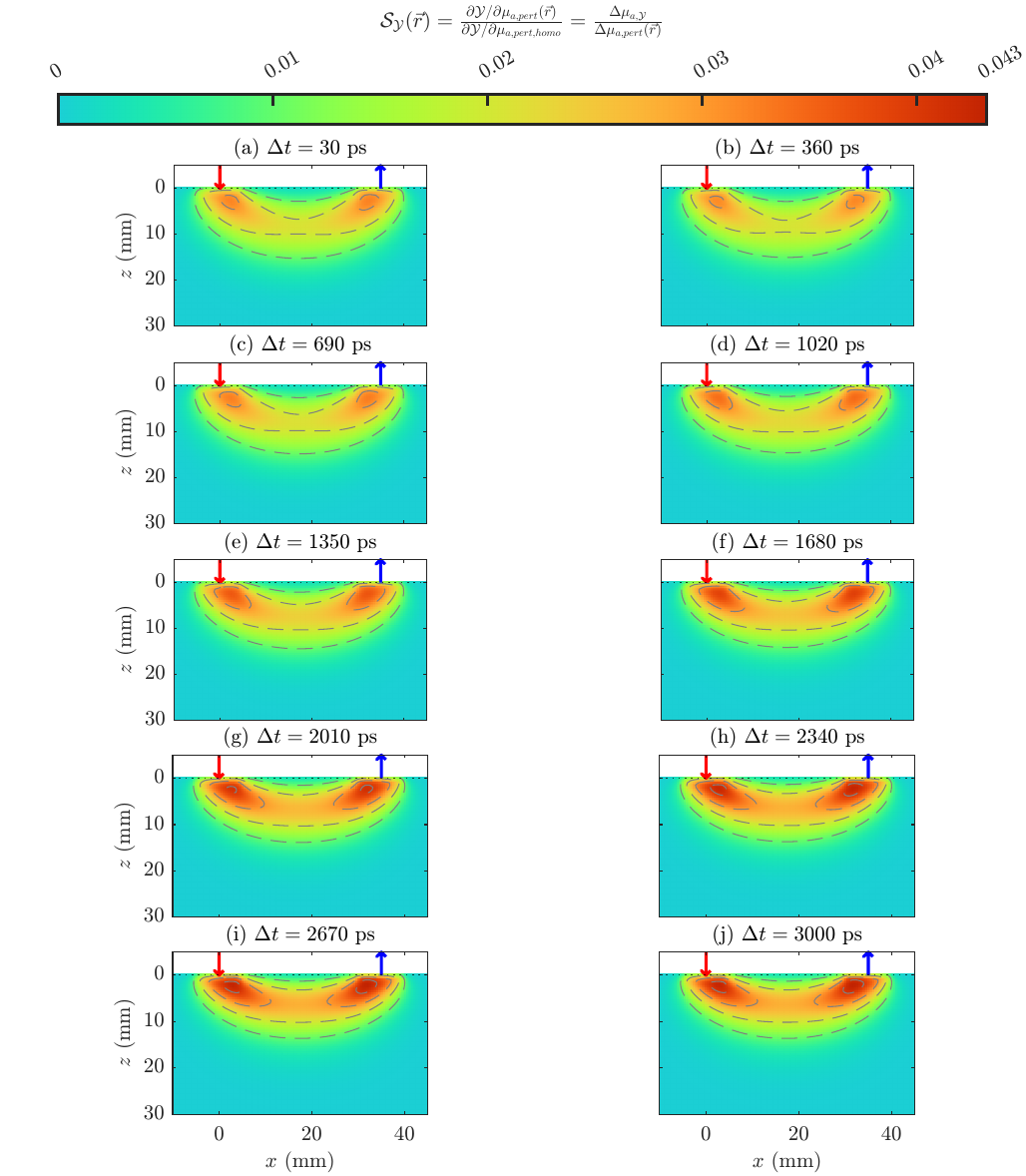}
	\end{center}
	\caption{$x$-$z$ plane of the \acrfull{sen} to a $\SI{10.0}{\milli\meter}\times\SI{10.0}{\milli\meter}\times\SI{2.0}{\milli\meter}$ perturbation scanned \SI{0.5}{\milli\meter} measured by \acrfull{TD} \acrfull{SD} gated \acrfull{I}. (a)-(j) Different values of \acrfull{dt}. Generated using \acrfull{DT}.\\ 
	\Acrfull{rho}: \SI{35.0}{\milli\meter}\\ 
	\Acrfull{n} inside: \num{1.333};\quad	\Acrfull{n} outside: \num{1.000}\\ 
	\Acrfull{musp}: \SI{1.10}{\per\milli\meter};\quad	\Acrfull{mua}: \SI{0.011}{\per\milli\meter}\\ 
	\Acrfull{mt} gate: 1750~\si{\pico\second}\\ 
	}\label{fig:TD_SD_GI_dt}
\end{figure*}

%% file: TD_SD_T_3rd.tex
\begin{figure*}
	\begin{center}
		\includegraphics{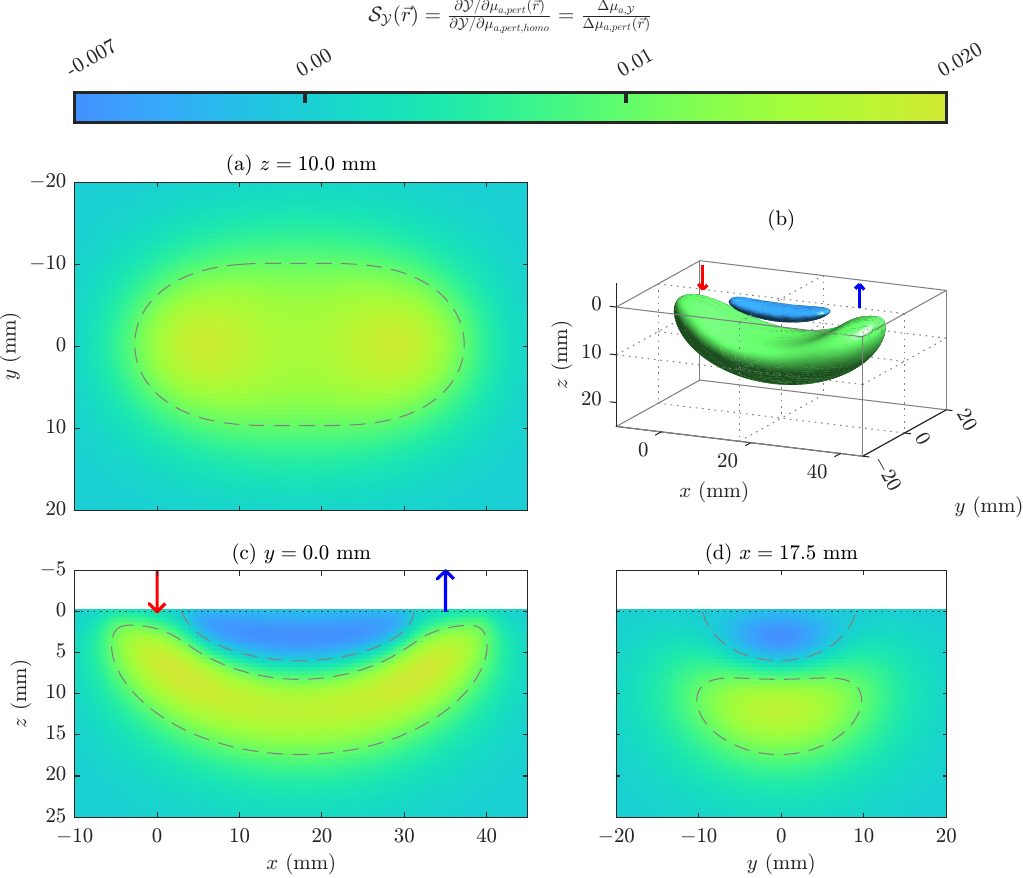}
	\end{center}
	\caption{Third angle projection of the \acrfull{sen} to a $\SI{10.0}{\milli\meter}\times\SI{10.0}{\milli\meter}\times\SI{2.0}{\milli\meter}$ perturbation scanned \SI{0.5}{\milli\meter} measured by \acrfull{TD} \acrfull{SD} \acrfull{mTOF}. (a) $x$-$y$ plane sliced at $z=\SI{10.0}{\milli\meter}$. (b) Iso-surface sliced at $\as{S}=0.010$ and $\as{S}=-0.005$. (c) $x$-$z$ plane sliced at $y=\SI{0.0}{\milli\meter}$. (d) $y$-$z$ plane sliced at $x=\SI{17.5}{\milli\meter}$. Generated using \acrfull{DT}.\\ 
	\Acrfull{rho}: \SI{35.0}{\milli\meter}\\ 
	\Acrfull{n} inside: \num{1.333}\\ 
	\Acrfull{n} outside: \num{1.000}\\ 
	\Acrfull{musp}: \SI{1.10}{\per\milli\meter}\\ 
	\Acrfull{mua}: \SI{0.011}{\per\milli\meter}\\ 
	}\label{fig:TD_SD_T_3rd}
\end{figure*}

%% file: TD_SD_DGI_3rd_rho0_MC_vox.tex
\begin{figure*}
	\begin{center}
		\includegraphics{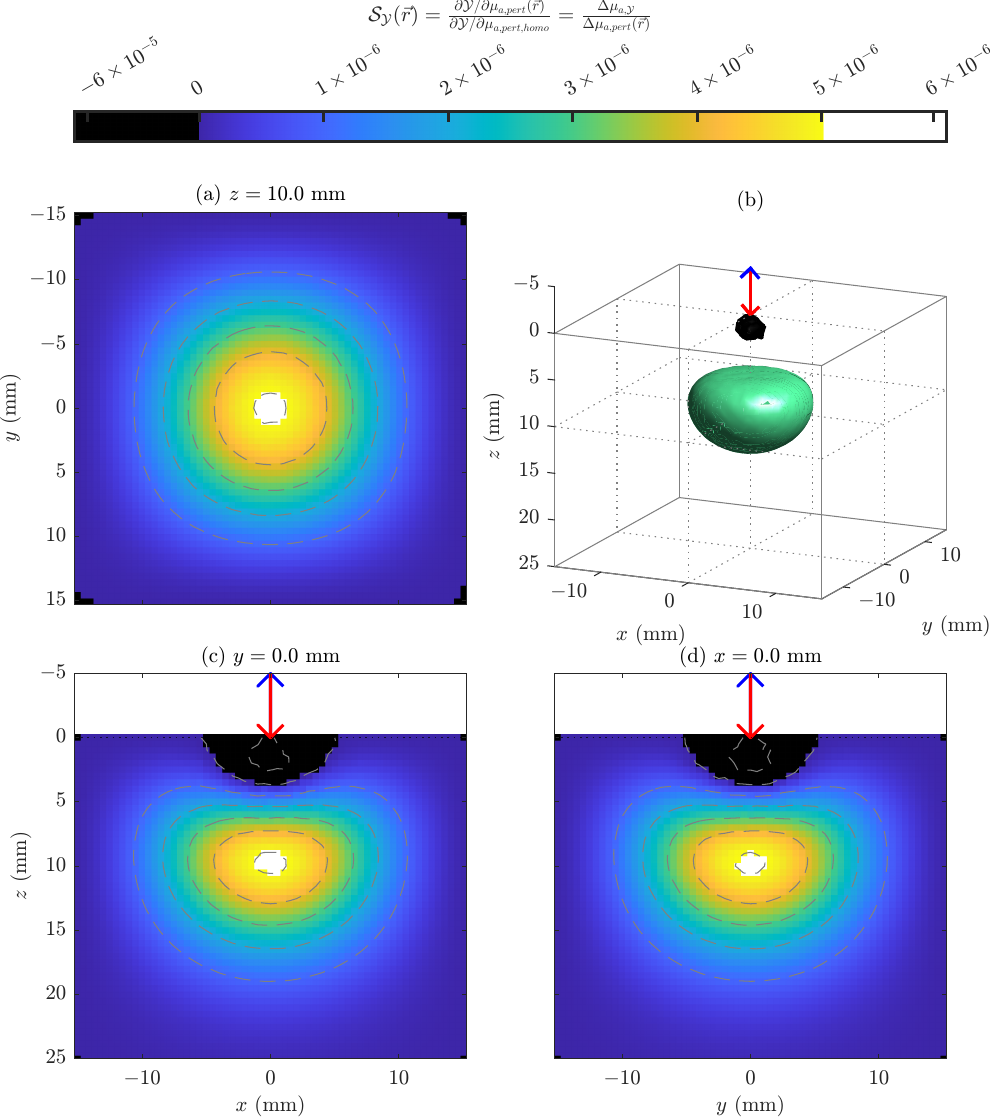}
	\end{center}
	\caption{Third angle projection of the \acrfull{sen} to a $\SI{0.5}{\milli\meter}\times\SI{0.5}{\milli\meter}\times\SI{0.5}{\milli\meter}$ perturbation scanned \SI{0.5}{\milli\meter} measured by \acrfull{TD} \acrfull{SD} difference in gated \acrfull{I}. (a) $x$-$y$ plane sliced at $z=\SI{10.0}{\milli\meter}$. (b) Iso-surface sliced at $\as{S}=3.000\times 10^{-5}$and $\as{S}=-1.000\times 10^{-5}$. (c) $x$-$z$ plane sliced at $y=\SI{0.0}{\milli\meter}$. (d) $y$-$z$ plane sliced at $x=\SI{0.0}{\milli\meter}$. Generated using \acrfull{MC}.\\ 
	\Acrfull{rho}: \SI{0.0}{\milli\meter}\\ 
	\Acrfull{n} inside: \num{1.333};\quad	\Acrfull{n} outside: \num{1.000}\\ 
	\Acrfull{musp}: \SI{1.10}{\per\milli\meter};\quad	\Acrfull{g}: \num{0.9}\\ 
	\Acrfull{mua}: \SI{0.011}{\per\milli\meter}\\ 
	Early \acrfull{t} gate: [500, 1000]~\si{\pico\second};\quad
	\Acrfull{t} gate: [1500, 2000]~\si{\pico\second}\\ 
	Detector Numerical Aperature (NA): \num{0.5};\quad	Number of photons: \num{1000000000}\\ 
	}\label{fig:TD_SD_DGI_3rd_rho0_MC_vox}
\end{figure*}

%% file: TD_SD_DGI_3rd_rho0_MC.tex
\begin{figure*}
	\begin{center}
		\includegraphics{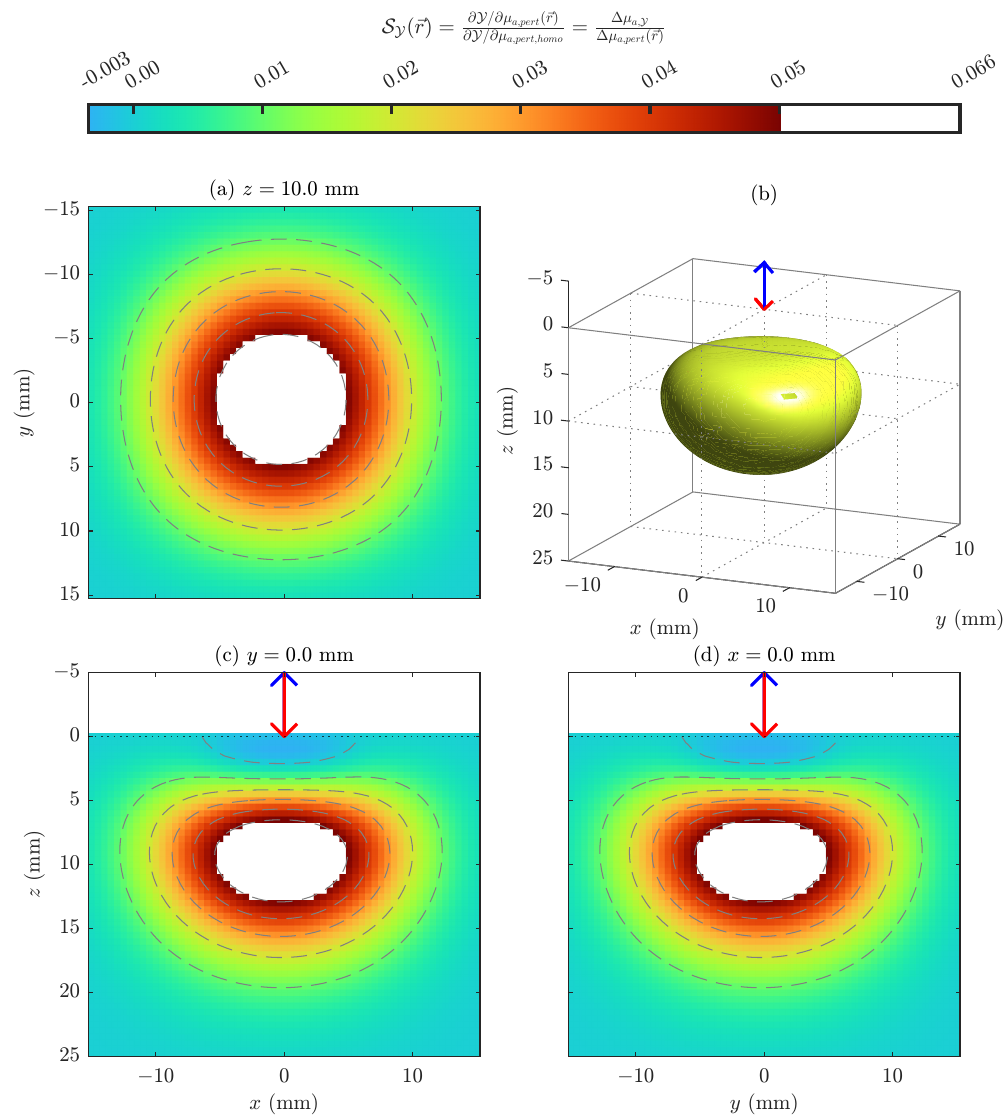}
	\end{center}
	\caption{Third angle projection of the \acrfull{sen} to a $\SI{10.0}{\milli\meter}\times\SI{10.0}{\milli\meter}\times\SI{2.0}{\milli\meter}$ perturbation scanned \SI{0.5}{\milli\meter} measured by \acrfull{TD} \acrfull{SD} difference in gated \acrfull{I}. (a) $x$-$y$ plane sliced at $z=\SI{10.0}{\milli\meter}$. (b) Iso-surface sliced at $\as{S}=0.020$ and $\as{S}=-0.010$. (c) $x$-$z$ plane sliced at $y=\SI{0.0}{\milli\meter}$. (d) $y$-$z$ plane sliced at $x=\SI{0.0}{\milli\meter}$. Generated using \acrfull{MC}.\\ 
	\Acrfull{rho}: \SI{0.0}{\milli\meter}\\ 
	\Acrfull{n} inside: \num{1.333};\quad	\Acrfull{n} outside: \num{1.000}\\ 
	\Acrfull{musp}: \SI{1.10}{\per\milli\meter};\quad	\Acrfull{g}: \num{0.9}\\ 
	\Acrfull{mua}: \SI{0.011}{\per\milli\meter}\\ 
	Early \acrfull{t} gate: [500, 1000]~\si{\pico\second};\quad
	\Acrfull{t} gate: [1500, 2000]~\si{\pico\second}\\ 
	Detector Numerical Aperature (NA): \num{0.5};\quad	Number of photons: \num{1000000000}\\ 
	}\label{fig:TD_SD_DGI_3rd_rho0_MC}
\end{figure*}

%% file: TD_SD_DGI_3rd.tex
\begin{figure*}
	\begin{center}
		\includegraphics{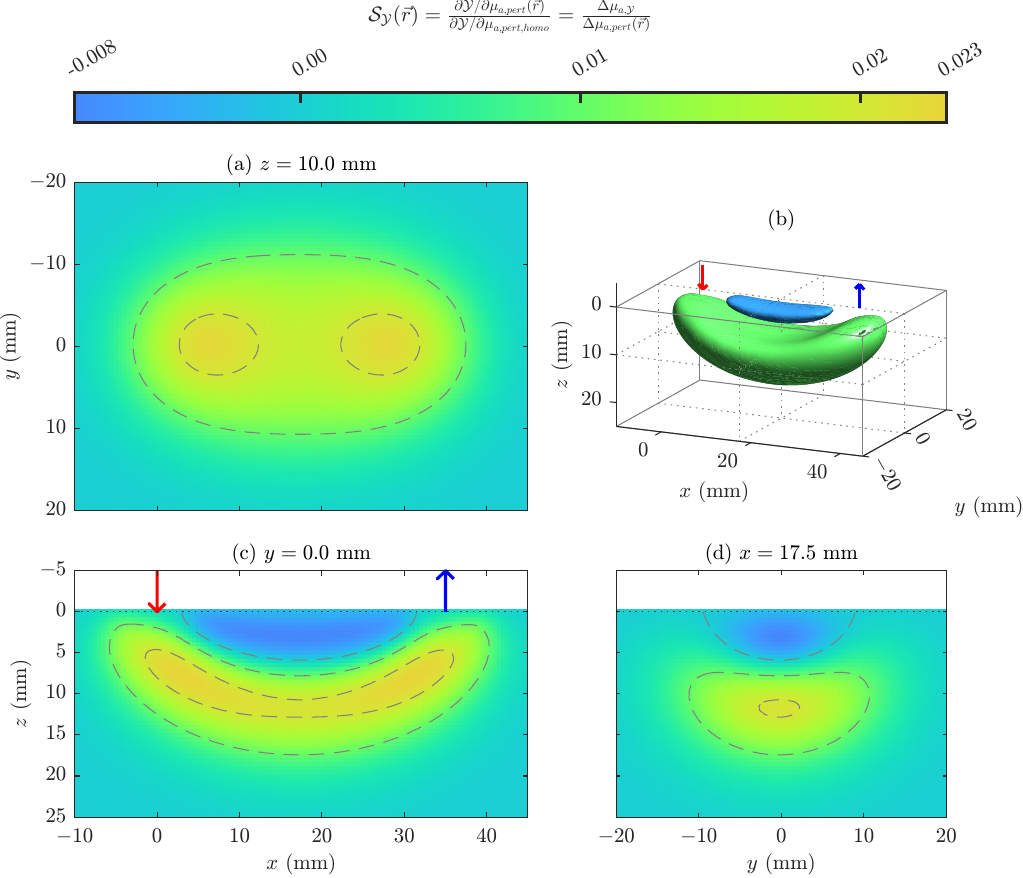}
	\end{center}
	\caption{Third angle projection of the \acrfull{sen} to a $\SI{10.0}{\milli\meter}\times\SI{10.0}{\milli\meter}\times\SI{2.0}{\milli\meter}$ perturbation scanned \SI{0.5}{\milli\meter} measured by \acrfull{TD} \acrfull{SD} difference in gated \acrfull{I}. (a) $x$-$y$ plane sliced at $z=\SI{10.0}{\milli\meter}$. (b) Iso-surface sliced at $\as{S}=0.010$ and $\as{S}=-0.005$. (c) $x$-$z$ plane sliced at $y=\SI{0.0}{\milli\meter}$. (d) $y$-$z$ plane sliced at $x=\SI{17.5}{\milli\meter}$. Generated using \acrfull{DT}.\\ 
	\Acrfull{rho}: \SI{35.0}{\milli\meter}\\ 
	\Acrfull{n} inside: \num{1.333}\\ 
	\Acrfull{n} outside: \num{1.000}\\ 
	\Acrfull{musp}: \SI{1.10}{\per\milli\meter}\\ 
	\Acrfull{mua}: \SI{0.011}{\per\milli\meter}\\ 
	Early \acrfull{t} gate: [500, 1000]~\si{\pico\second}\\ 
	\Acrfull{t} gate: [1500, 2000]~\si{\pico\second}\\ 
	}\label{fig:TD_SD_DGI_3rd}
\end{figure*}

%% file: TD_SD_DGI_rho.tex
\begin{figure*}
	\begin{center}
		\includegraphics{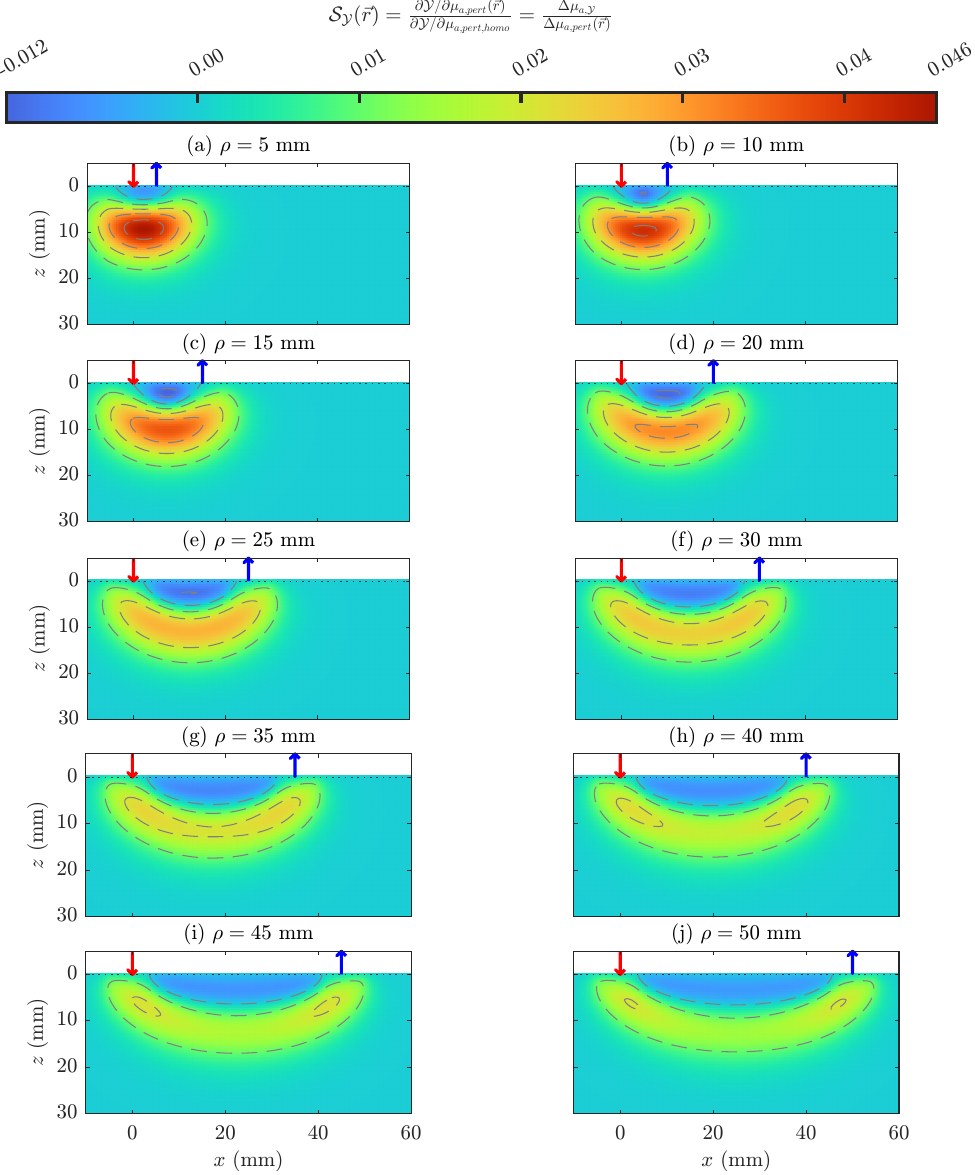}
	\end{center}
	\caption{$x$-$z$ plane of the \acrfull{sen} to a $\SI{10.0}{\milli\meter}\times\SI{10.0}{\milli\meter}\times\SI{2.0}{\milli\meter}$ perturbation scanned \SI{0.5}{\milli\meter} measured by \acrfull{TD} \acrfull{SD} difference in gated \acrfull{I}. (a)-(j) Different values of \acrfull{rho}. Generated using \acrfull{DT}.\\ 
	\Acrfull{n} inside: \num{1.333};\quad	\Acrfull{n} outside: \num{1.000}\\ 
	\Acrfull{musp}: \SI{1.10}{\per\milli\meter};\quad	\Acrfull{mua}: \SI{0.011}{\per\milli\meter}\\ 
	Early \acrfull{t} gate: [500, 1000]~\si{\pico\second}\\ 
	\Acrfull{t} gate: [1500, 2000]~\si{\pico\second}\\ 
	}\label{fig:TD_SD_DGI_rho}
\end{figure*}

%% file: TD_SD_DGI_mtg_rho0_MC.tex
\begin{figure*}
	\begin{center}
		\includegraphics{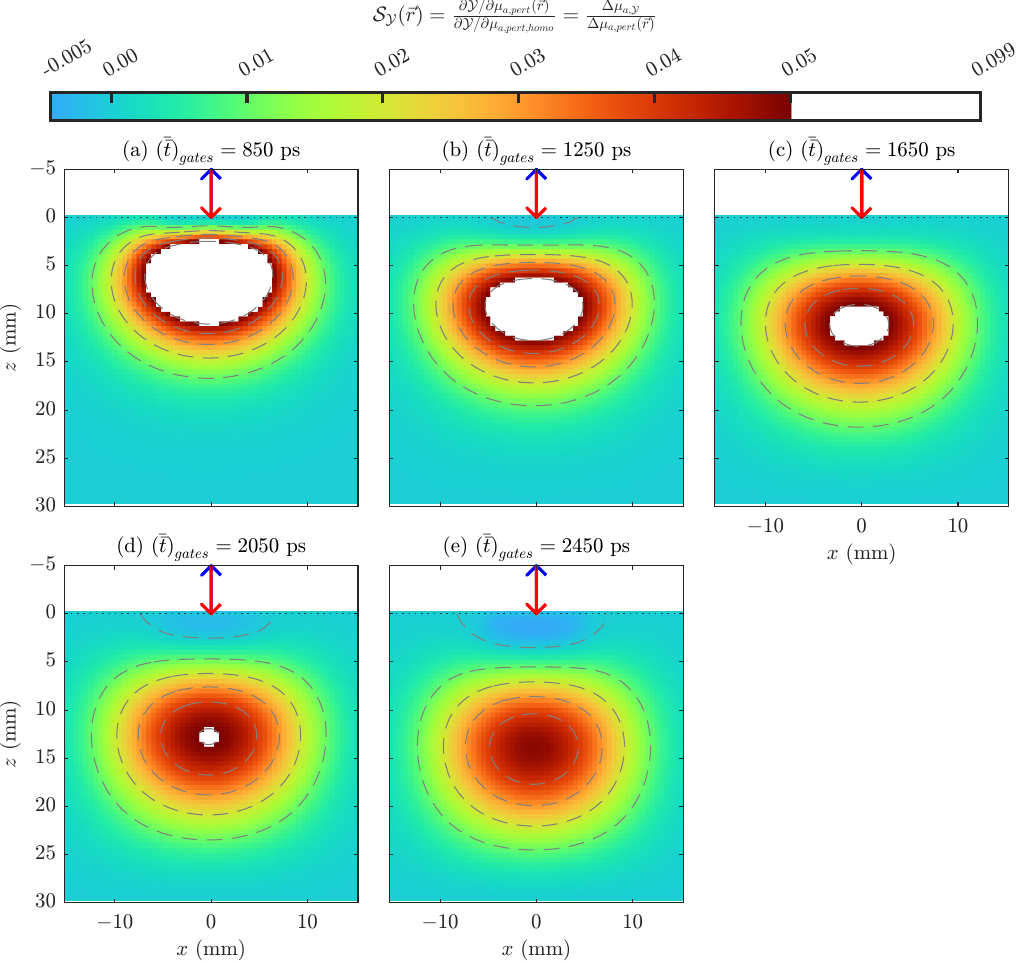}
	\end{center}
	\caption{$x$-$z$ plane of the \acrfull{sen} to a $\SI{10.0}{\milli\meter}\times\SI{10.0}{\milli\meter}\times\SI{2.0}{\milli\meter}$ perturbation scanned \SI{0.5}{\milli\meter} measured by \acrfull{TD} \acrfull{SD} difference in gated \acrfull{I}. (a)-(e) Different values of \acrfull{mtg}. Generated using \acrfull{MC}.\\ 
	\Acrfull{rho}: \SI{0.0}{\milli\meter}\\ 
	\Acrfull{n} inside: \num{1.333};\quad	\Acrfull{n} outside: \num{1.000}\\ 
	\Acrfull{musp}: \SI{1.10}{\per\milli\meter};\quad	\Acrfull{g}: \num{0.9}\\ 
	\Acrfull{mua}: \SI{0.011}{\per\milli\meter}\\ 
	\Acrfull{dt} gates: 500~\si{\pico\second}\quad	\Acrfull{dtg}: 1000~\si{\pico\second}\\ 
	Detector Numerical Aperature (NA): \num{0.5};\quad	Number of photons: \num{1000000000}\\ 
	}\label{fig:TD_SD_DGI_mtg_rho0_MC}
\end{figure*}

%% file: TD_SD_DGI_mtg.tex
\begin{figure*}
	\begin{center}
		\includegraphics{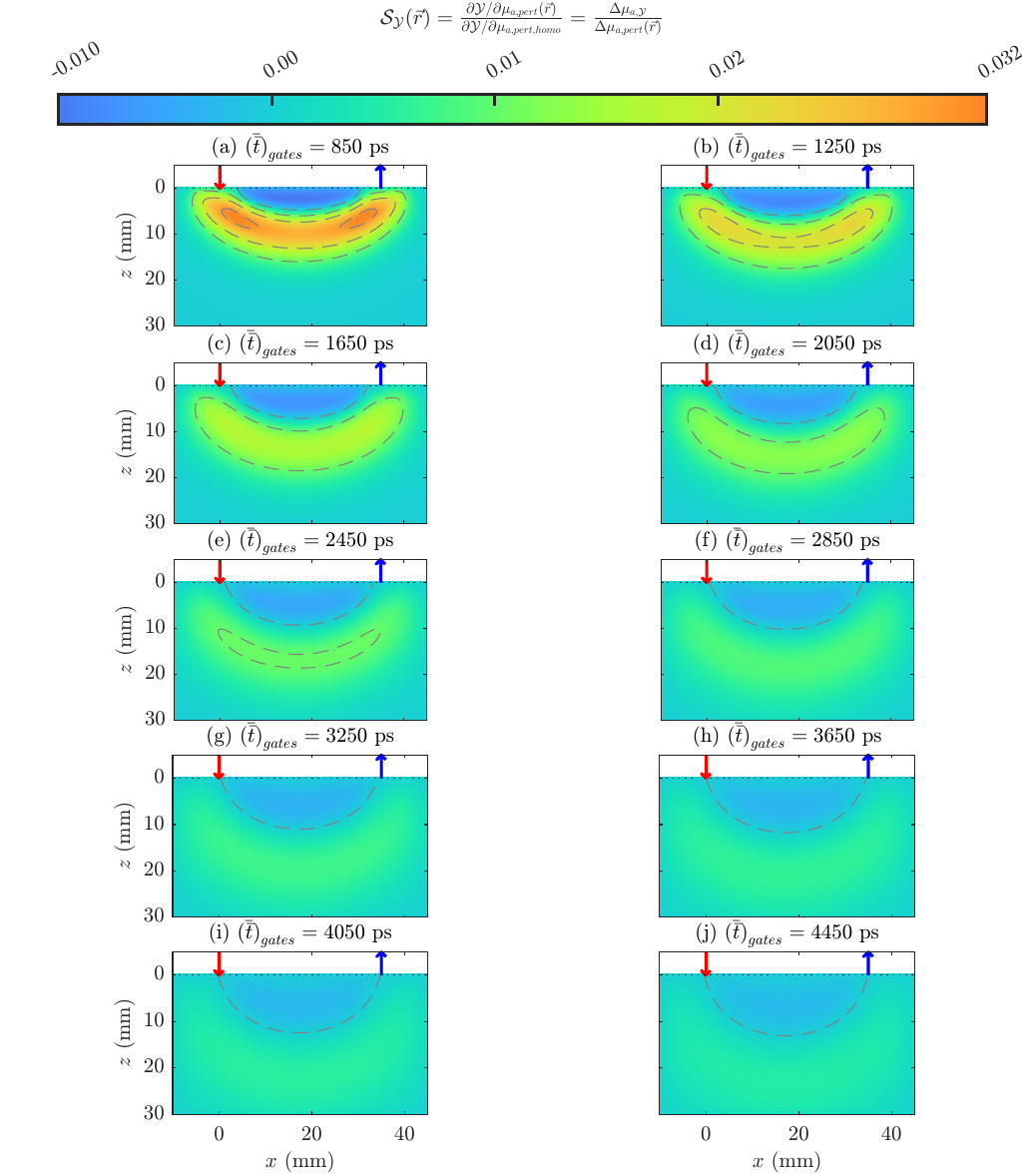}
	\end{center}
	\caption{$x$-$z$ plane of the \acrfull{sen} to a $\SI{10.0}{\milli\meter}\times\SI{10.0}{\milli\meter}\times\SI{2.0}{\milli\meter}$ perturbation scanned \SI{0.5}{\milli\meter} measured by \acrfull{TD} \acrfull{SD} difference in gated \acrfull{I}. (a)-(j) Different values of \acrfull{mtg}. Generated using \acrfull{DT}.\\ 
	\Acrfull{rho}: \SI{35.0}{\milli\meter}\\ 
	\Acrfull{n} inside: \num{1.333};\quad	\Acrfull{n} outside: \num{1.000}\\ 
	\Acrfull{musp}: \SI{1.10}{\per\milli\meter};\quad	\Acrfull{mua}: \SI{0.011}{\per\milli\meter}\\ 
	\Acrfull{dt} gates: 500~\si{\pico\second}\quad	\Acrfull{dtg}: 1000~\si{\pico\second}\\ 
	}\label{fig:TD_SD_DGI_mtg}
\end{figure*}

%% file: TD_SD_DGI_dtg_rho0_MC.tex
\begin{figure*}
	\begin{center}
		\includegraphics{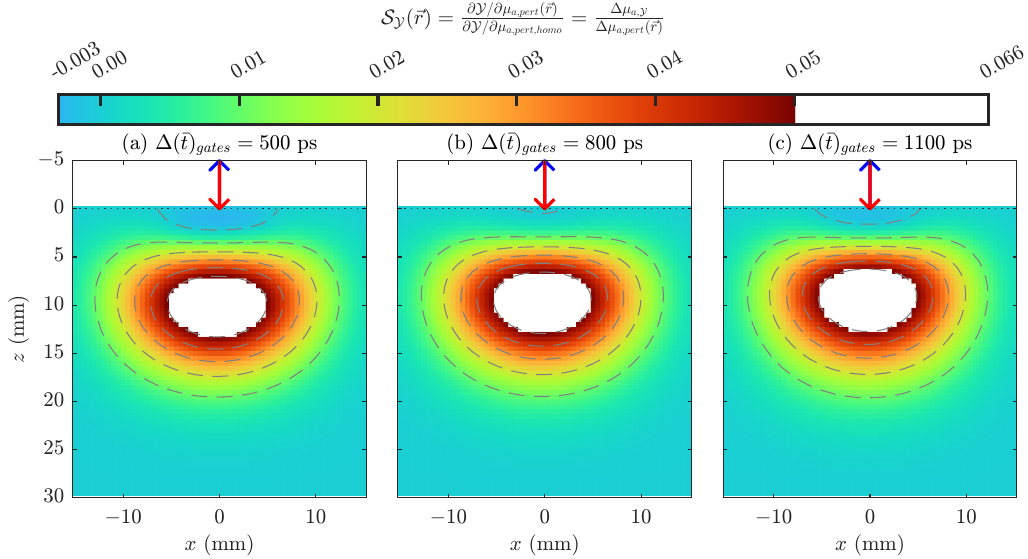}
	\end{center}
	\caption{$x$-$z$ plane of the \acrfull{sen} to a $\SI{10.0}{\milli\meter}\times\SI{10.0}{\milli\meter}\times\SI{2.0}{\milli\meter}$ perturbation scanned \SI{0.5}{\milli\meter} measured by \acrfull{TD} \acrfull{SD} difference in gated \acrfull{I}. (a)-(c) Different values of \acrfull{dtg}. Generated using \acrfull{MC}.\\ 
	\Acrfull{rho}: \SI{0.0}{\milli\meter}\\ 
	\Acrfull{n} inside: \num{1.333};\quad	\Acrfull{n} outside: \num{1.000}\\ 
	\Acrfull{musp}: \SI{1.10}{\per\milli\meter};\quad	\Acrfull{g}: \num{0.9}\\ 
	\Acrfull{mua}: \SI{0.011}{\per\milli\meter}\\ 
	\Acrfull{dt} gates: 500~\si{\pico\second}\quad	\Acrfull{mtg}: 1250~\si{\pico\second}\\ 
	Detector Numerical Aperature (NA): \num{0.5};\quad	Number of photons: \num{1000000000}\\ 
	}\label{fig:TD_SD_DGI_dtg_rho0_MC}
\end{figure*}

%% file: TD_SD_DGI_dtg.tex
\begin{figure*}
	\begin{center}
		\includegraphics{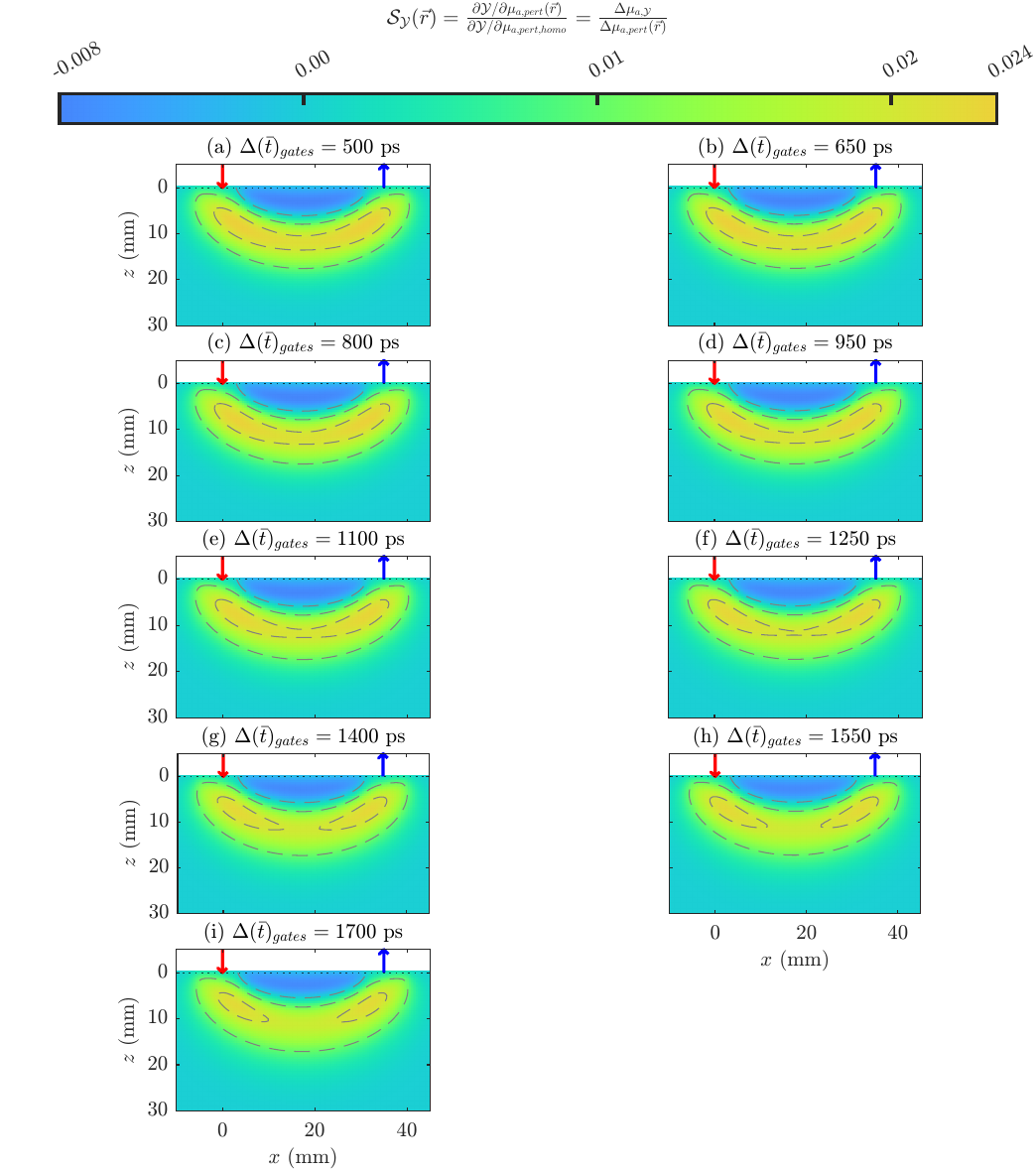}
	\end{center}
	\caption{$x$-$z$ plane of the \acrfull{sen} to a $\SI{10.0}{\milli\meter}\times\SI{10.0}{\milli\meter}\times\SI{2.0}{\milli\meter}$ perturbation scanned \SI{0.5}{\milli\meter} measured by \acrfull{TD} \acrfull{SD} difference in gated \acrfull{I}. (a)-(i) Different values of \acrfull{dtg}. Generated using \acrfull{DT}.\\ 
	\Acrfull{rho}: \SI{35.0}{\milli\meter}\\ 
	\Acrfull{n} inside: \num{1.333};\quad	\Acrfull{n} outside: \num{1.000}\\ 
	\Acrfull{musp}: \SI{1.10}{\per\milli\meter};\quad	\Acrfull{mua}: \SI{0.011}{\per\milli\meter}\\ 
	\Acrfull{dt} gates: 500~\si{\pico\second}\quad	\Acrfull{mtg}: 1250~\si{\pico\second}\\ 
	}\label{fig:TD_SD_DGI_dtg}
\end{figure*}

%% file: TD_SD_V_3rd.tex
\begin{figure*}
	\begin{center}
		\includegraphics{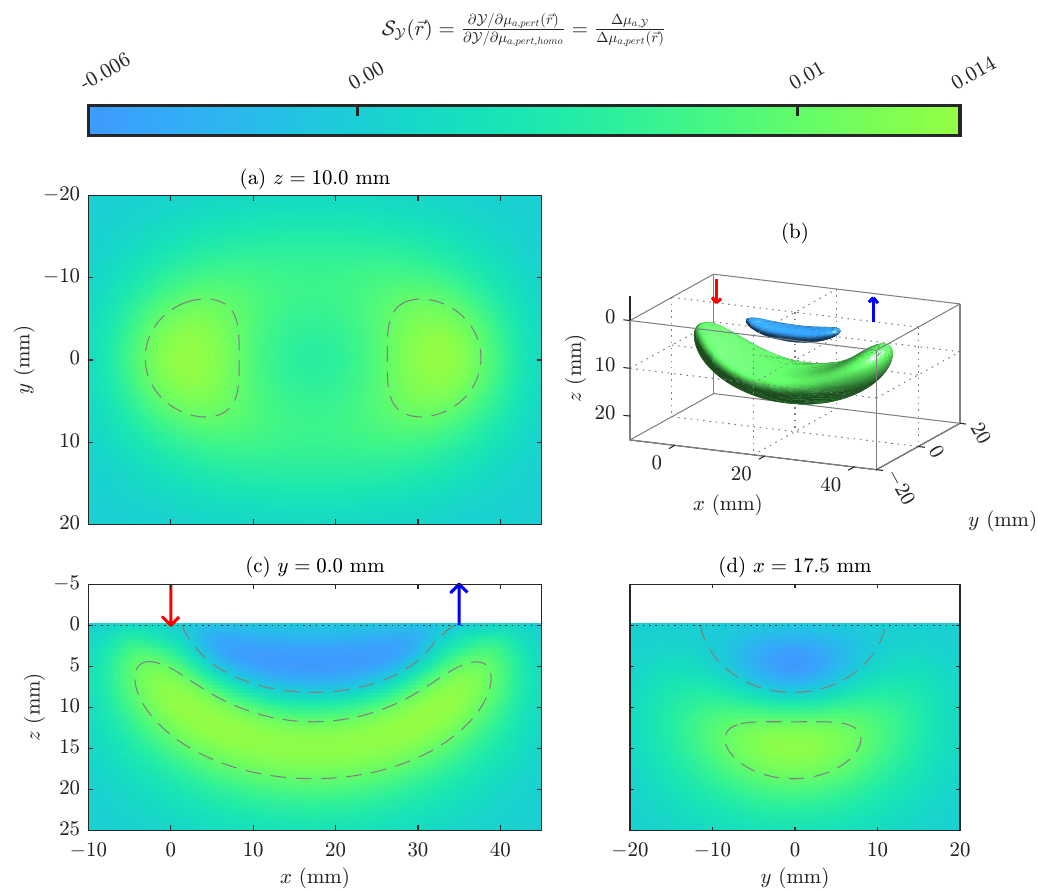}
	\end{center}
	\caption{Third angle projection of the \acrfull{sen} to a $\SI{10.0}{\milli\meter}\times\SI{10.0}{\milli\meter}\times\SI{2.0}{\milli\meter}$ perturbation scanned \SI{0.5}{\milli\meter} measured by \acrfull{TD} \acrfull{SD} \acrfull{var}. (a) $x$-$y$ plane sliced at $z=\SI{10.0}{\milli\meter}$. (b) Iso-surface sliced at $\as{S}=0.010$ and $\as{S}=-0.005$. (c) $x$-$z$ plane sliced at $y=\SI{0.0}{\milli\meter}$. (d) $y$-$z$ plane sliced at $x=\SI{17.5}{\milli\meter}$. Generated using \acrfull{DT}.\\ 
	\Acrfull{rho}: \SI{35.0}{\milli\meter}\\ 
	\Acrfull{n} inside: \num{1.333}\\ 
	\Acrfull{n} outside: \num{1.000}\\ 
	\Acrfull{musp}: \SI{1.10}{\per\milli\meter}\\ 
	\Acrfull{mua}: \SI{0.011}{\per\milli\meter}\\ 
	}\label{fig:TD_SD_V_3rd}
\end{figure*}

%% file: TD_SD_V_rho.tex
\begin{figure*}
	\begin{center}
		\includegraphics{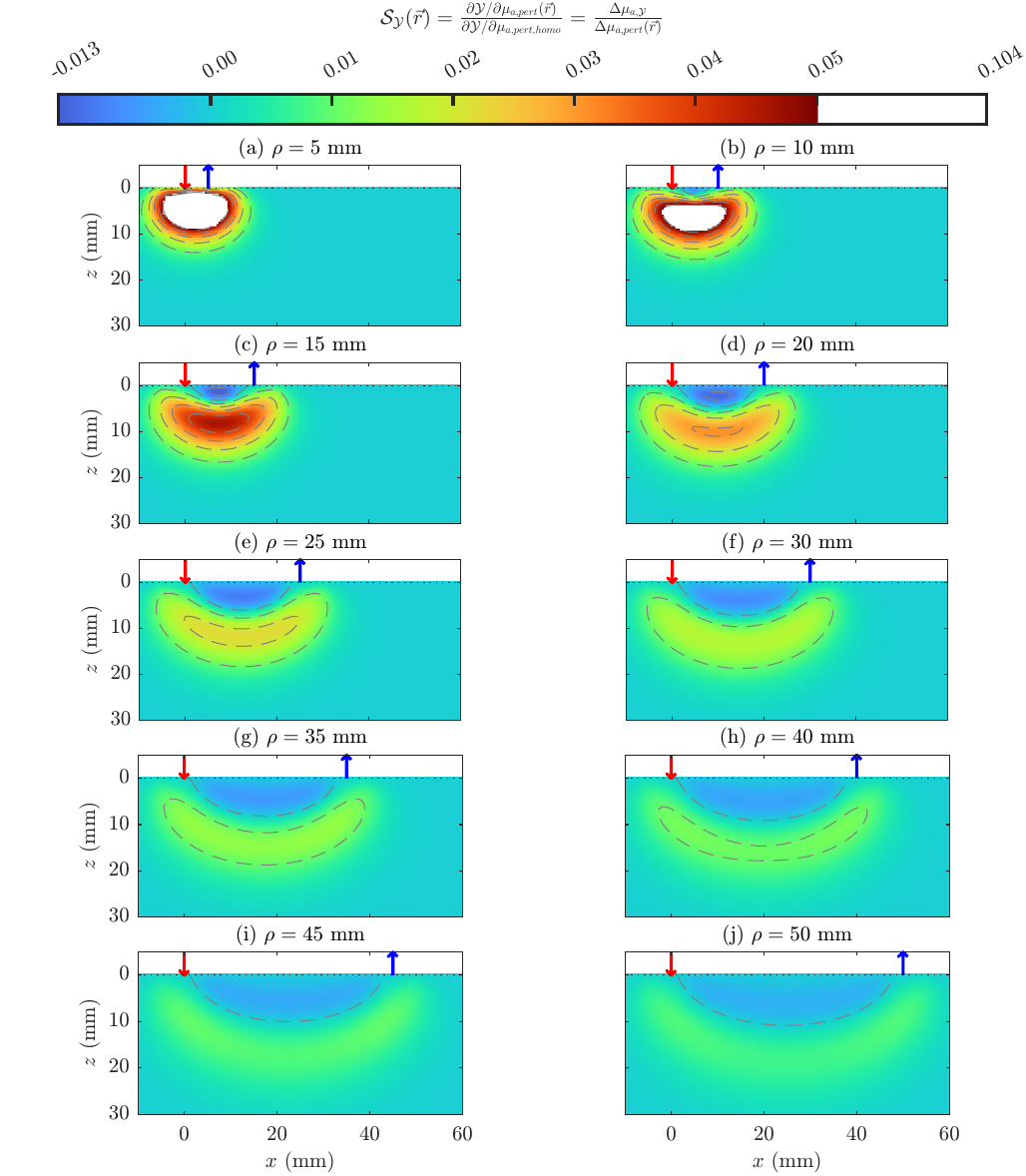}
	\end{center}
	\caption{$x$-$z$ plane of the \acrfull{sen} to a $\SI{10.0}{\milli\meter}\times\SI{10.0}{\milli\meter}\times\SI{2.0}{\milli\meter}$ perturbation scanned \SI{0.5}{\milli\meter} measured by \acrfull{TD} \acrfull{SD} \acrfull{var}. (a)-(j) Different values of \acrfull{rho}. Generated using \acrfull{DT}.\\ 
	\Acrfull{n} inside: \num{1.333};\quad	\Acrfull{n} outside: \num{1.000}\\ 
	\Acrfull{musp}: \SI{1.10}{\per\milli\meter};\quad	\Acrfull{mua}: \SI{0.011}{\per\milli\meter}\\ 
	}\label{fig:TD_SD_V_rho}
\end{figure*}

%% file: TD_SS_GI_3rd.tex
\begin{figure*}
	\begin{center}
		\includegraphics{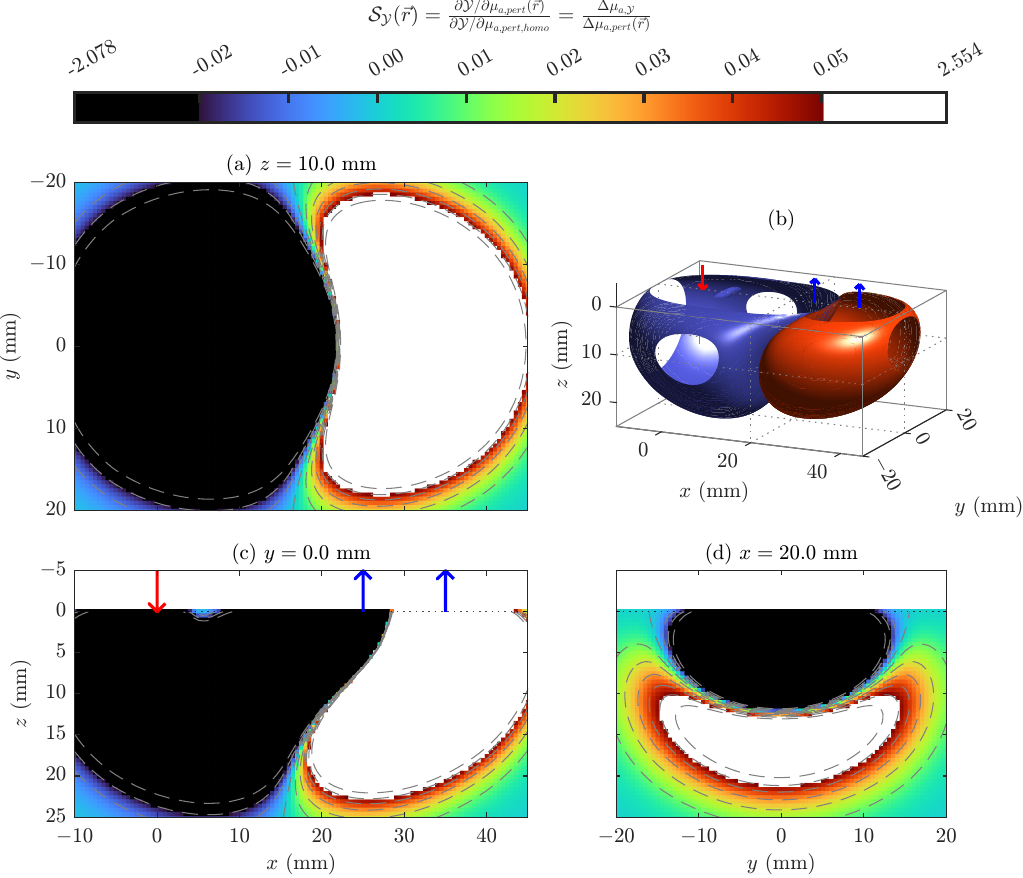}
	\end{center}
	\caption{Third angle projection of the \acrfull{sen} to a $\SI{10.0}{\milli\meter}\times\SI{10.0}{\milli\meter}\times\SI{2.0}{\milli\meter}$ perturbation scanned \SI{0.5}{\milli\meter} measured by \acrfull{TD} \acrfull{SS} gated \acrfull{I}. (a) $x$-$y$ plane sliced at $z=\SI{10.0}{\milli\meter}$. (b) Iso-surface sliced at $\as{S}=0.040$ and $\as{S}=-0.015$. (c) $x$-$z$ plane sliced at $y=\SI{0.0}{\milli\meter}$. (d) $y$-$z$ plane sliced at $x=\SI{20.0}{\milli\meter}$. Generated using \acrfull{DT}.\\ 
	\Acrfullpl{rho}: [25.0, 35.0]~\si{\milli\meter}\\ 
	\Acrfull{n} inside: \num{1.333}\\ 
	\Acrfull{n} outside: \num{1.000}\\ 
	\Acrfull{musp}: \SI{1.10}{\per\milli\meter}\\ 
	\Acrfull{mua}: \SI{0.011}{\per\milli\meter}\\ 
	\Acrfull{t} gate: [1500, 2000]~\si{\pico\second}\\ 
	}\label{fig:TD_SS_GI_3rd}
\end{figure*}

%% file: TD_DS_GI_3rd.tex
\begin{figure*}
	\begin{center}
		\includegraphics{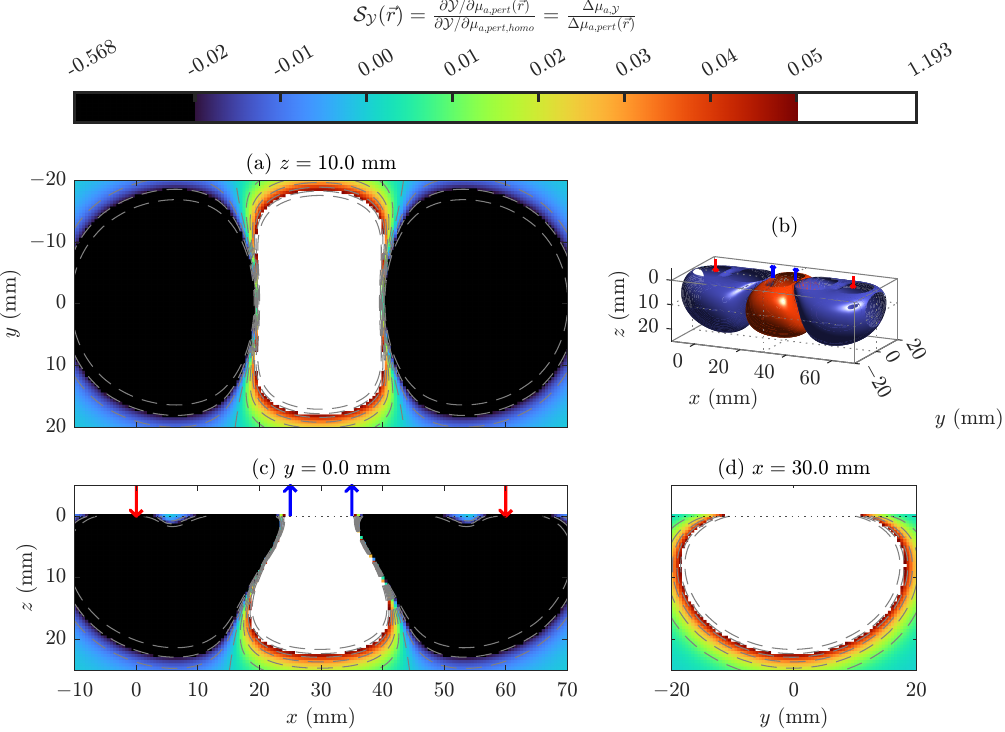}
	\end{center}
	\caption{Third angle projection of the \acrfull{sen} to a $\SI{10.0}{\milli\meter}\times\SI{10.0}{\milli\meter}\times\SI{2.0}{\milli\meter}$ perturbation scanned \SI{0.5}{\milli\meter} measured by \acrfull{TD} \acrfull{DS} gated \acrfull{I}.(a) $x$-$y$ plane sliced at $z=\SI{10.0}{\milli\meter}$. (b) Iso-surface sliced at $\as{S}=0.040$ and $\as{S}=-0.015$. (c) $x$-$z$ plane sliced at $y=\SI{0.0}{\milli\meter}$. (d) $y$-$z$ plane sliced at $x=\SI{30.0}{\milli\meter}$. Generated using \acrfull{DT}.\\ 
	\Acrfullpl{rho}: [25.0, 35.0, 35.0, 25.0]~\si{\milli\meter}\\ 
	\Acrfull{n} inside: \num{1.333}\\ 
	\Acrfull{n} outside: \num{1.000}\\ 
	\Acrfull{musp}: \SI{1.10}{\per\milli\meter}\\ 
	\Acrfull{mua}: \SI{0.011}{\per\milli\meter}\\ 
	\Acrfull{t} gate: [1500, 2000]~\si{\pico\second}\\ 
	}\label{fig:TD_DS_GI_3rd}
\end{figure*}

%% file: TD_SS_GI_mrho.tex
\begin{figure*}
	\begin{center}
		\includegraphics{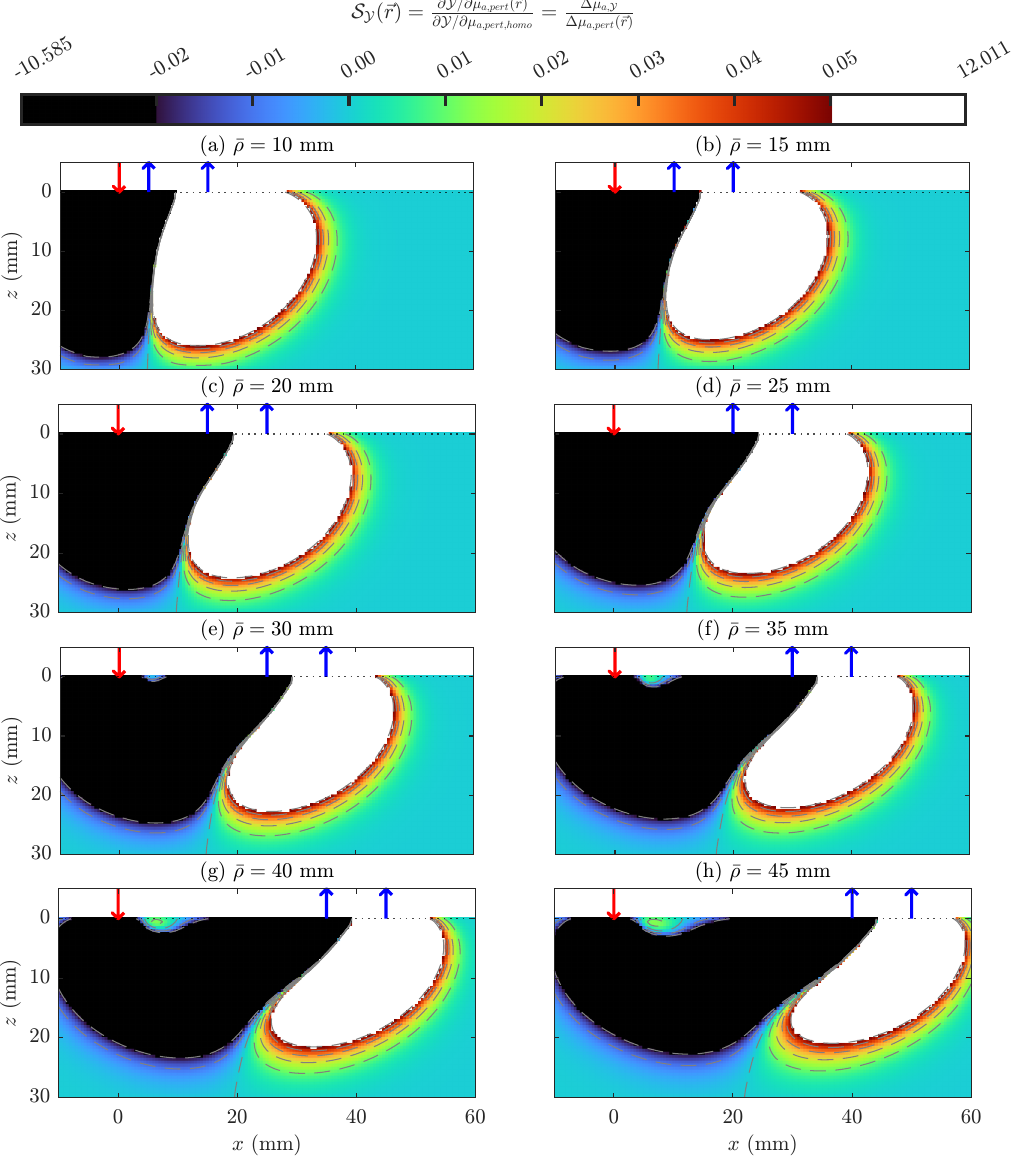}
	\end{center}
	\caption{$x$-$z$ plane of the \acrfull{sen} to a $\SI{10.0}{\milli\meter}\times\SI{10.0}{\milli\meter}\times\SI{2.0}{\milli\meter}$ perturbation scanned \SI{0.5}{\milli\meter} measured by \acrfull{TD} \acrfull{SS} gated \acrfull{I}. (a)-(h) Different values of \acrfull{mrho}. Generated using \acrfull{DT}.\\ 
	\Acrfull{drho}: \SI{10.0}{\milli\meter}\\ 
	\Acrfull{n} inside: \num{1.333};\quad	\Acrfull{n} outside: \num{1.000}\\ 
	\Acrfull{musp}: \SI{1.10}{\per\milli\meter};\quad	\Acrfull{mua}: \SI{0.011}{\per\milli\meter}\\ 
	\Acrfull{t} gate: [1500, 2000]~\si{\pico\second}\\ 
	}\label{fig:TD_SS_GI_mrho}
\end{figure*}

%% file: TD_DS_GI_mrho.tex
\begin{figure*}
	\begin{center}
		\includegraphics{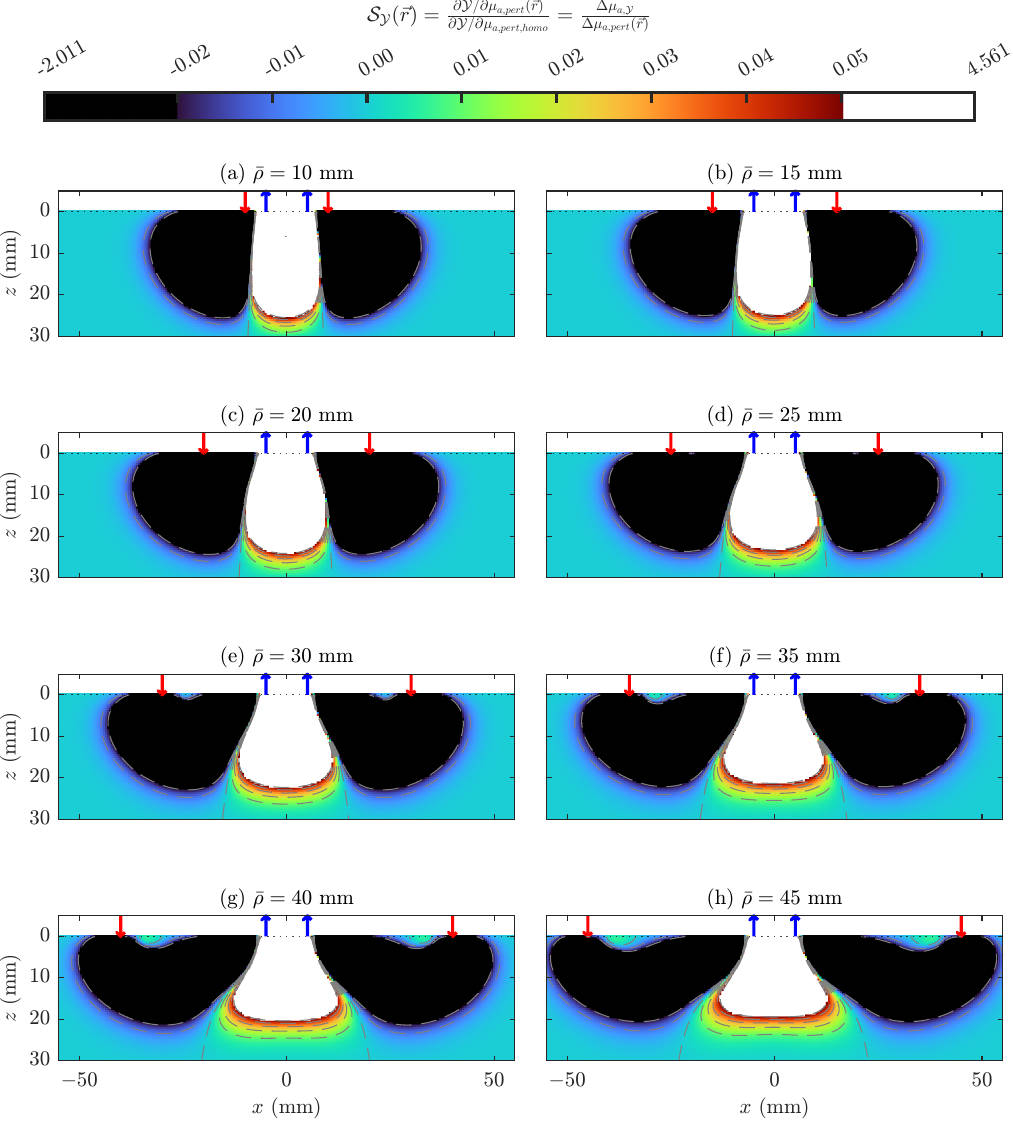}
	\end{center}
	\caption{$x$-$z$ plane of the \acrfull{sen} to a $\SI{10.0}{\milli\meter}\times\SI{10.0}{\milli\meter}\times\SI{2.0}{\milli\meter}$ perturbation scanned \SI{0.5}{\milli\meter} measured by \acrfull{TD} \acrfull{DS} gated \acrfull{I}. (a)-(h) Different values of \acrfull{mrho}. Generated using \acrfull{DT}.\\ 
	\Acrfull{drho}: \SI{10.0}{\milli\meter}\\ 
	\Acrfull{n} inside: \num{1.333};\quad	\Acrfull{n} outside: \num{1.000}\\ 
	\Acrfull{musp}: \SI{1.10}{\per\milli\meter};\quad	\Acrfull{mua}: \SI{0.011}{\per\milli\meter}\\ 
	\Acrfull{t} gate: [1500, 2000]~\si{\pico\second}\\ 
	}\label{fig:TD_DS_GI_mrho}
\end{figure*}

%% file: TD_SS_GI_drho.tex
\begin{figure*}
	\begin{center}
		\includegraphics{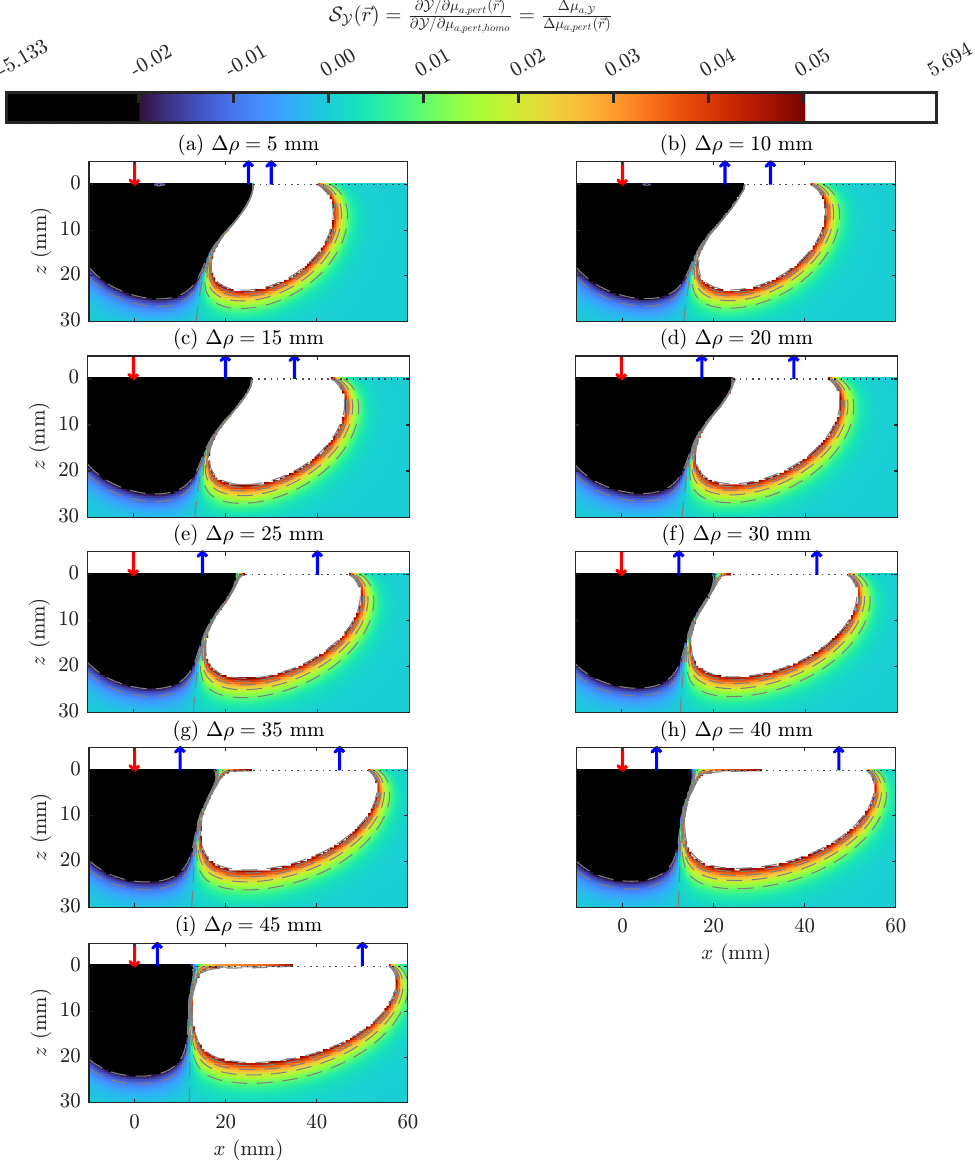}
	\end{center}
	\caption{$x$-$z$ plane of the \acrfull{sen} to a $\SI{10.0}{\milli\meter}\times\SI{10.0}{\milli\meter}\times\SI{2.0}{\milli\meter}$ perturbation scanned \SI{0.5}{\milli\meter} measured by \acrfull{TD} \acrfull{SS} gated \acrfull{I}. (a)-(i) Different values of \acrfull{drho}. Generated using \acrfull{DT}.\\ 
	\Acrfull{mrho}: \SI{27.5}{\milli\meter}\\ 
	\Acrfull{n} inside: \num{1.333};\quad	\Acrfull{n} outside: \num{1.000}\\ 
	\Acrfull{musp}: \SI{1.10}{\per\milli\meter};\quad	\Acrfull{mua}: \SI{0.011}{\per\milli\meter}\\ 
	\Acrfull{t} gate: [1500, 2000]~\si{\pico\second}\\ 
	}\label{fig:TD_SS_GI_drho}
\end{figure*}

%% file: TD_DS_GI_drho.tex
\begin{figure*}
	\begin{center}
		\includegraphics{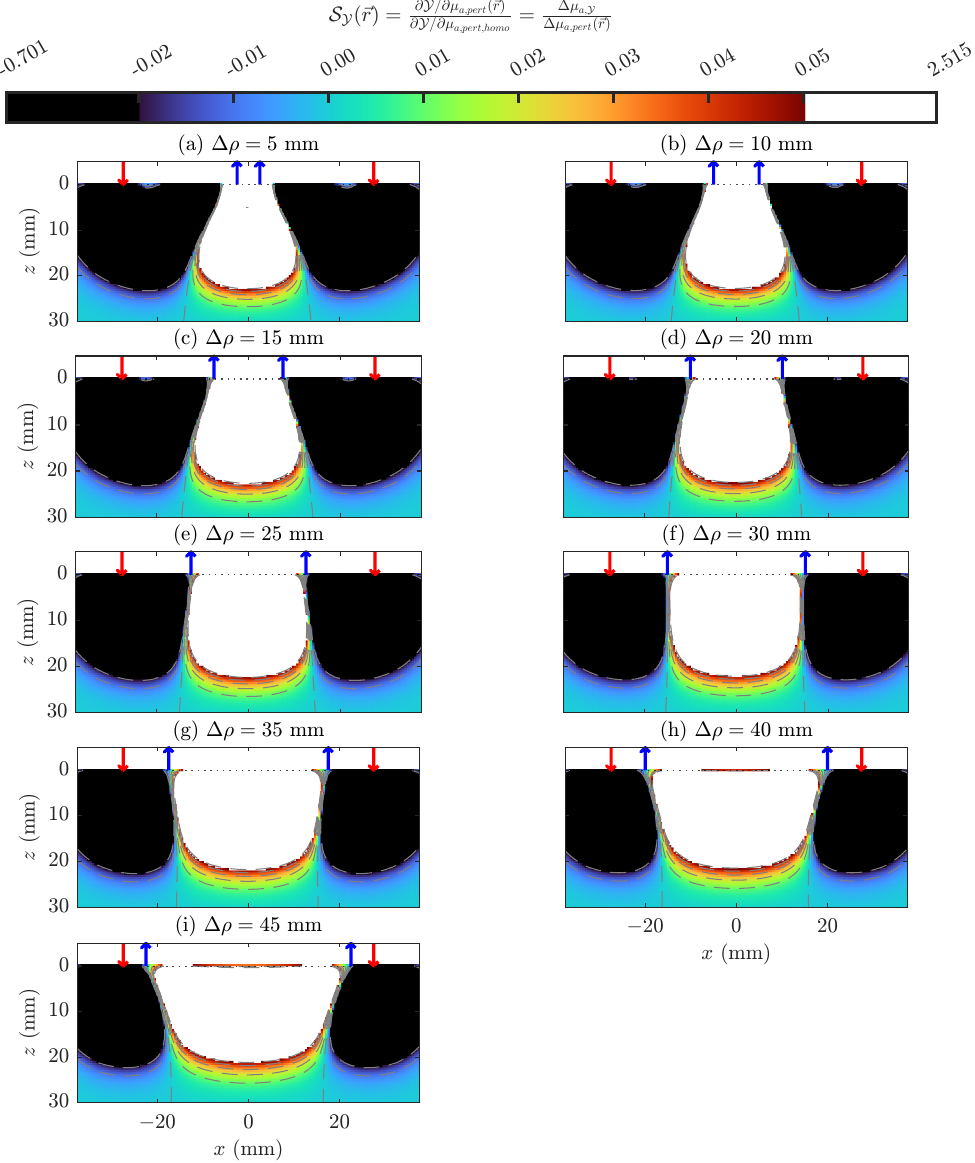}
	\end{center}
	\caption{$x$-$z$ plane of the \acrfull{sen} to a $\SI{10.0}{\milli\meter}\times\SI{10.0}{\milli\meter}\times\SI{2.0}{\milli\meter}$ perturbation scanned \SI{0.5}{\milli\meter} measured by \acrfull{TD} \acrfull{DS} gated \acrfull{I}. (a)-(i) Different values of \acrfull{drho}. Generated using \acrfull{DT}.\\ 
	\Acrfull{mrho}: \SI{27.5}{\milli\meter}\\ 
	\Acrfull{n} inside: \num{1.333};\quad	\Acrfull{n} outside: \num{1.000}\\ 
	\Acrfull{musp}: \SI{1.10}{\per\milli\meter};\quad	\Acrfull{mua}: \SI{0.011}{\per\milli\meter}\\ 
	\Acrfull{t} gate: [1500, 2000]~\si{\pico\second}\\ 
	}\label{fig:TD_DS_GI_drho}
\end{figure*}

%% file: TD_SS_GI_mt.tex
\begin{figure*}
	\begin{center}
		\includegraphics{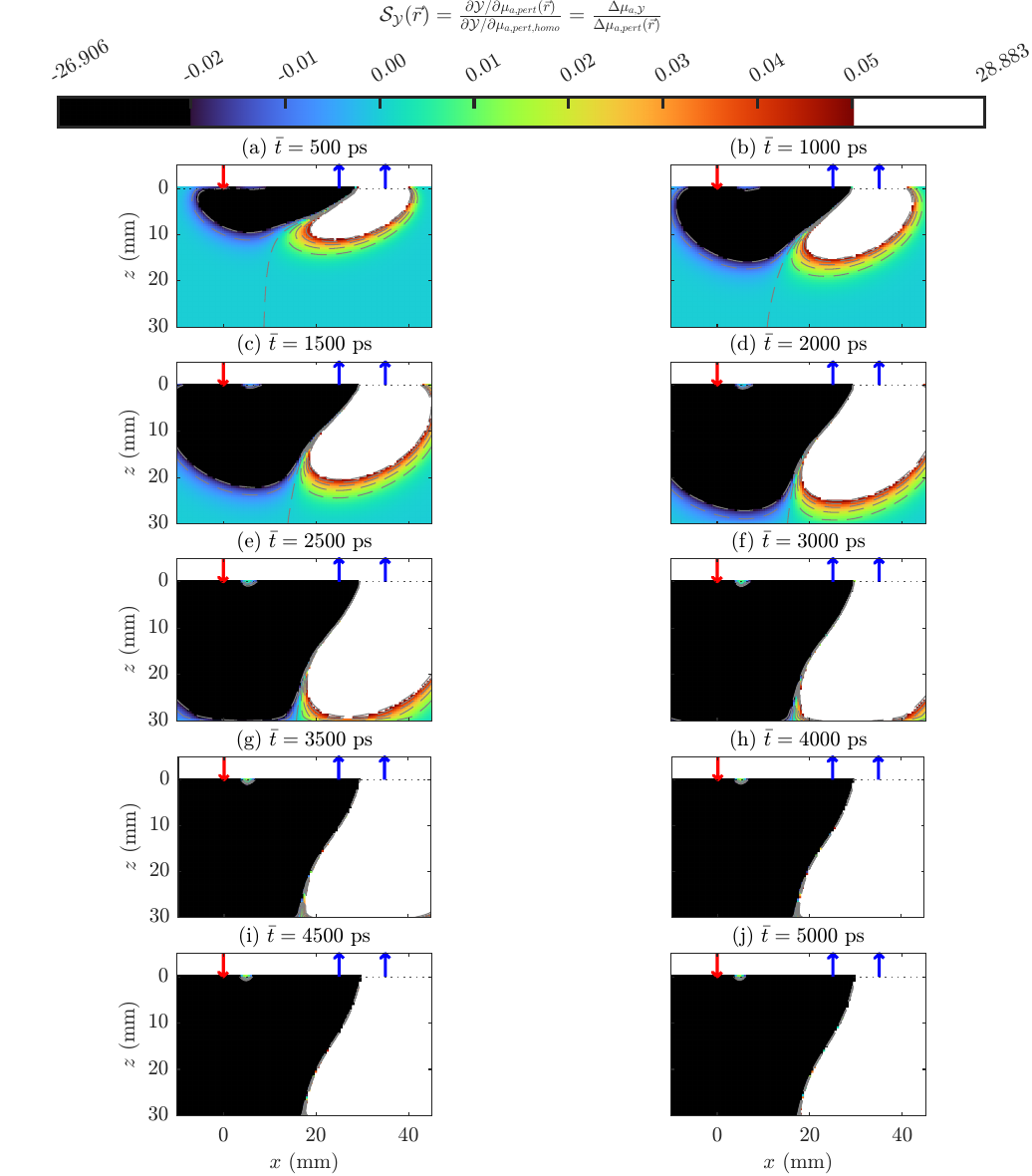}
	\end{center}
	\caption{$x$-$z$ plane of the \acrfull{sen} to a $\SI{10.0}{\milli\meter}\times\SI{10.0}{\milli\meter}\times\SI{2.0}{\milli\meter}$ perturbation scanned \SI{0.5}{\milli\meter} measured by \acrfull{TD} \acrfull{SS} gated \acrfull{I}. (a)-(j) Different values of \acrfull{mt}. Generated using \acrfull{DT}.\\ 
	\Acrfullpl{rho}: [25.0, 35.0]~\si{\milli\meter}\\ 
	\Acrfull{n} inside: \num{1.333};\quad	\Acrfull{n} outside: \num{1.000}\\ 
	\Acrfull{musp}: \SI{1.10}{\per\milli\meter};\quad	\Acrfull{mua}: \SI{0.011}{\per\milli\meter}\\ 
	\Acrfull{dt} gate: 500~\si{\pico\second}\\ 
	}\label{fig:TD_SS_GI_mt}
\end{figure*}

%% file: TD_DS_GI_mt.tex
\begin{figure*}
	\begin{center}
		\includegraphics{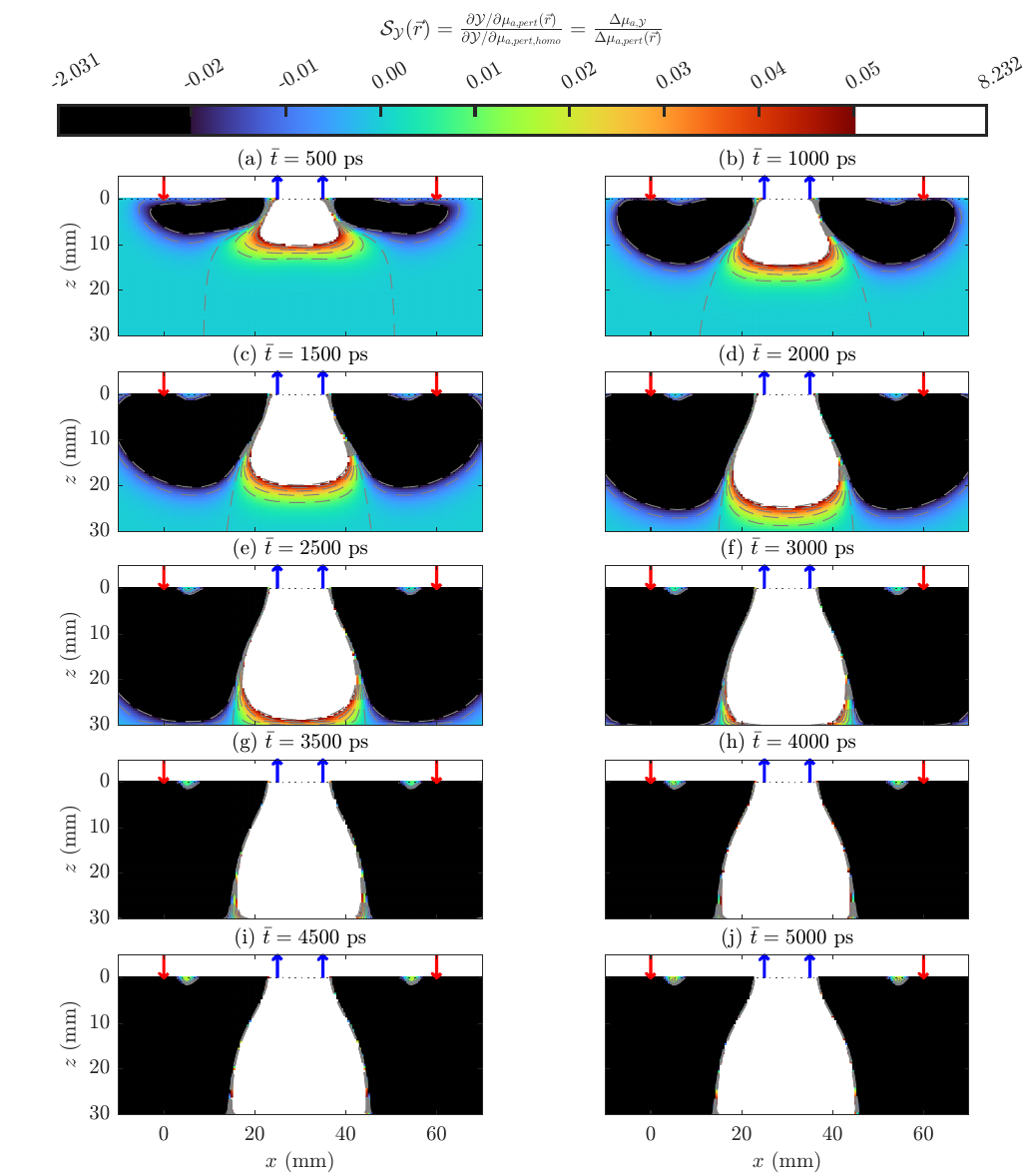}
	\end{center}
	\caption{$x$-$z$ plane of the \acrfull{sen} to a $\SI{10.0}{\milli\meter}\times\SI{10.0}{\milli\meter}\times\SI{2.0}{\milli\meter}$ perturbation scanned \SI{0.5}{\milli\meter} measured by \acrfull{TD} \acrfull{DS} gated \acrfull{I}. (a)-(j) Different values of \acrfull{mt}. Generated using \acrfull{DT}.\\ 
	\Acrfullpl{rho}: [25.0, 35.0, 35.0, 25.0]~\si{\milli\meter}\\ 
	\Acrfull{n} inside: \num{1.333};\quad	\Acrfull{n} outside: \num{1.000}\\ 
	\Acrfull{musp}: \SI{1.10}{\per\milli\meter};\quad	\Acrfull{mua}: \SI{0.011}{\per\milli\meter}\\ 
	\Acrfull{dt} gate: 500~\si{\pico\second}\\ 
	}\label{fig:TD_DS_GI_mt}
\end{figure*}

%% file: TD_SS_GI_dt.tex
\begin{figure*}
	\begin{center}
		\includegraphics{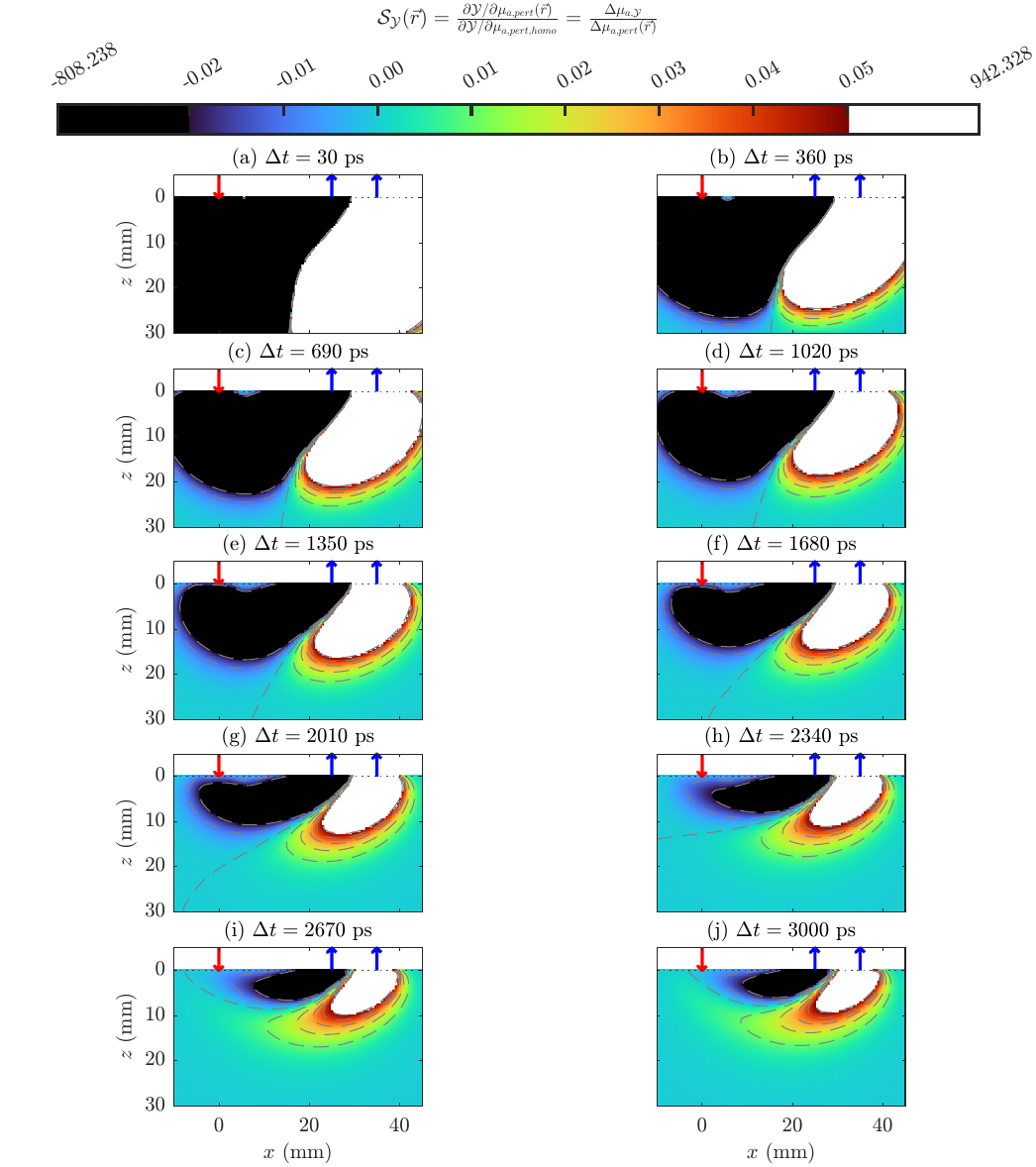}
	\end{center}
	\caption{$x$-$z$ plane of the \acrfull{sen} to a $\SI{10.0}{\milli\meter}\times\SI{10.0}{\milli\meter}\times\SI{2.0}{\milli\meter}$ perturbation scanned \SI{0.5}{\milli\meter} measured by \acrfull{TD} \acrfull{SS} gated \acrfull{I}. (a)-(j) Different values of \acrfull{dt}. Generated using \acrfull{DT}.\\ 
	\Acrfullpl{rho}: [25.0, 35.0]~\si{\milli\meter}\\ 
	\Acrfull{n} inside: \num{1.333};\quad	\Acrfull{n} outside: \num{1.000}\\ 
	\Acrfull{musp}: \SI{1.10}{\per\milli\meter};\quad	\Acrfull{mua}: \SI{0.011}{\per\milli\meter}\\ 
	\Acrfull{mt} gate: 1750~\si{\pico\second}\\ 
	}\label{fig:TD_SS_GI_dt}
\end{figure*}

%% file: TD_DS_GI_dt.tex
\begin{figure*}
	\begin{center}
		\includegraphics{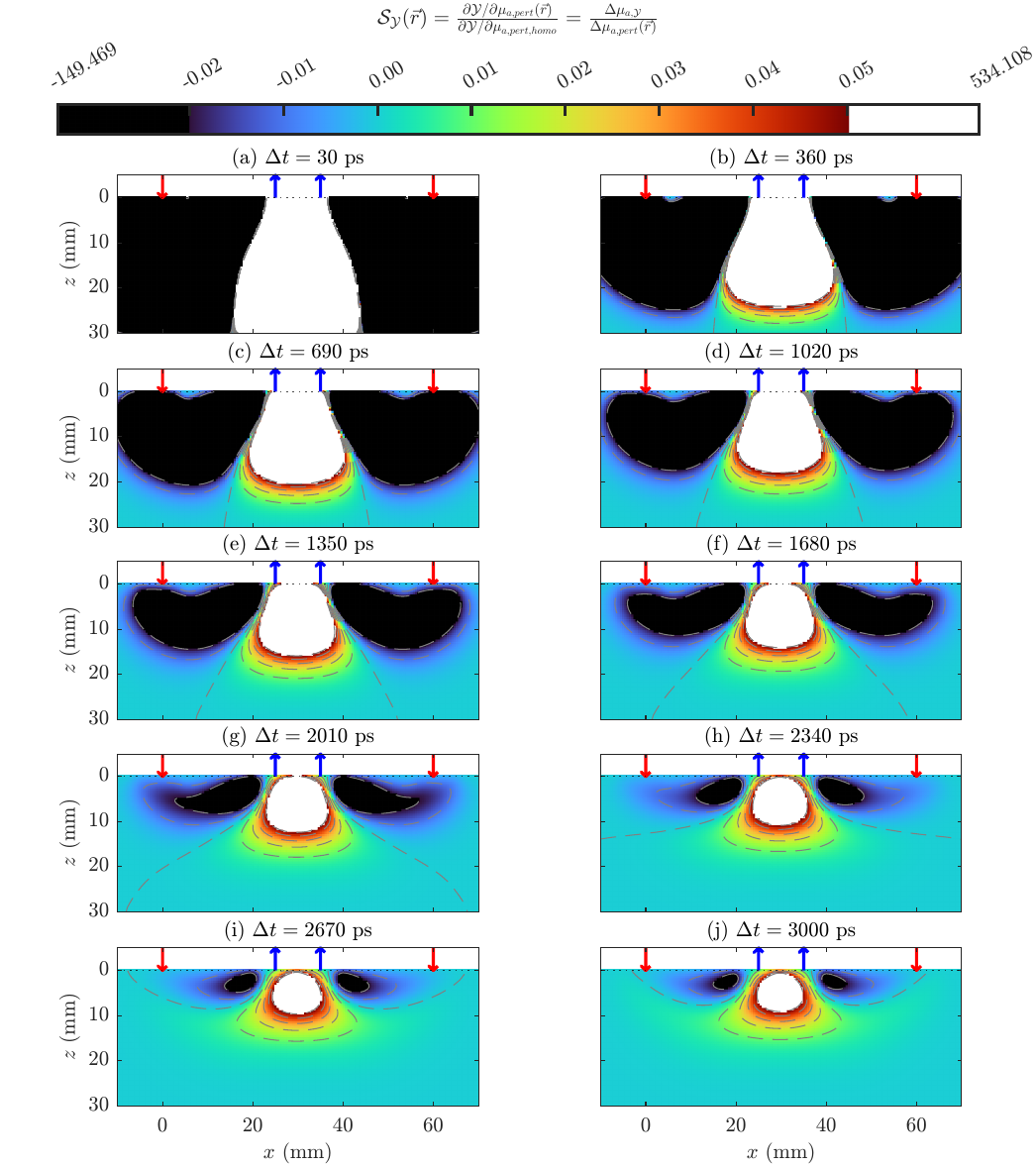}
	\end{center}
	\caption{$x$-$z$ plane of the \acrfull{sen} to a $\SI{10.0}{\milli\meter}\times\SI{10.0}{\milli\meter}\times\SI{2.0}{\milli\meter}$ perturbation scanned \SI{0.5}{\milli\meter} measured by \acrfull{TD} \acrfull{DS} gated \acrfull{I}. (a)-(j) Different values of \acrfull{dt}. Generated using \acrfull{DT}.\\ 
	\Acrfullpl{rho}: [25.0, 35.0, 35.0, 25.0]~\si{\milli\meter}\\ 
	\Acrfull{n} inside: \num{1.333};\quad	\Acrfull{n} outside: \num{1.000}\\ 
	\Acrfull{musp}: \SI{1.10}{\per\milli\meter};\quad	\Acrfull{mua}: \SI{0.011}{\per\milli\meter}\\ 
	\Acrfull{mt} gate: 1750~\si{\pico\second}\\ 
	}\label{fig:TD_DS_GI_dt}
\end{figure*}

%% file: TD_SS_T_3rd.tex
\begin{figure*}
	\begin{center}
		\includegraphics{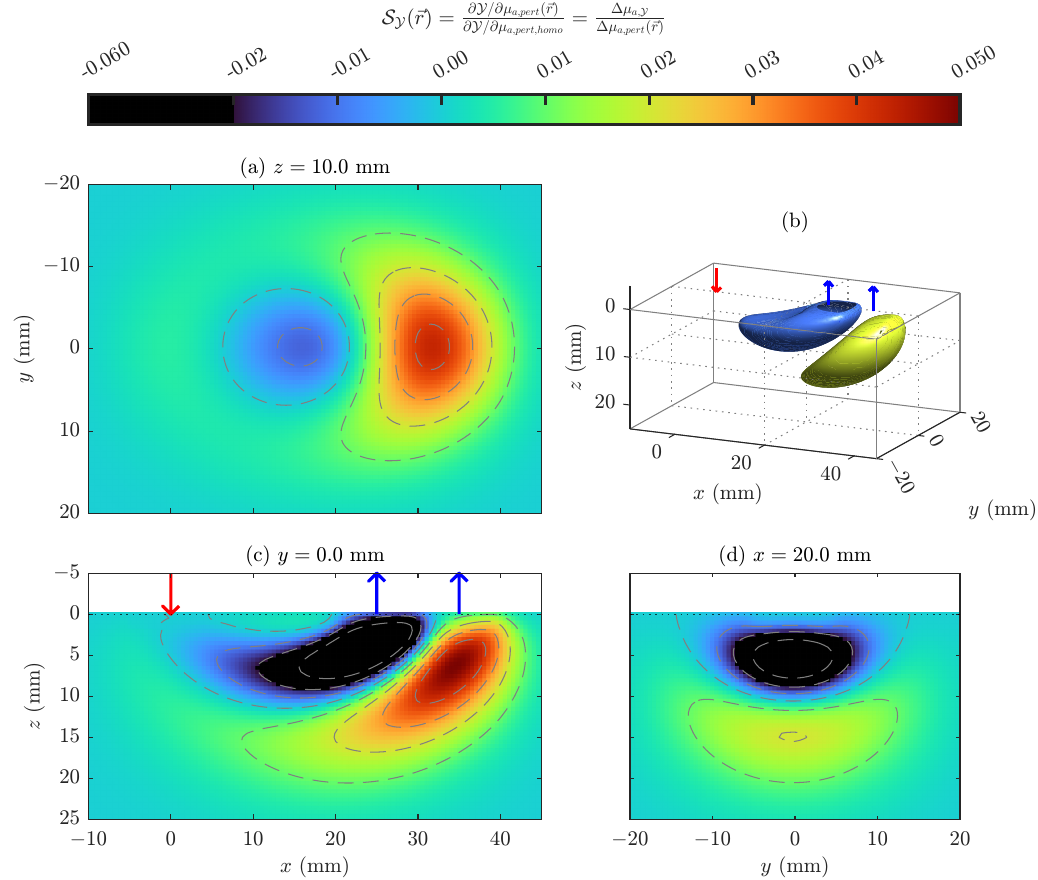}
	\end{center}
	\caption{Third angle projection of the \acrfull{sen} to a $\SI{10.0}{\milli\meter}\times\SI{10.0}{\milli\meter}\times\SI{2.0}{\milli\meter}$ perturbation scanned \SI{0.5}{\milli\meter} measured by \acrfull{TD} \acrfull{SS} \acrfull{mTOF}. (a) $x$-$y$ plane sliced at $z=\SI{10.0}{\milli\meter}$. (b) Iso-surface sliced at $\as{S}=0.020$ and $\as{S}=-0.010$. (c) $x$-$z$ plane sliced at $y=\SI{0.0}{\milli\meter}$. (d) $y$-$z$ plane sliced at $x=\SI{20.0}{\milli\meter}$. Generated using \acrfull{DT}.\\ 
	\Acrfullpl{rho}: [25.0, 35.0]~\si{\milli\meter}\\ 
	\Acrfull{n} inside: \num{1.333}\\ 
	\Acrfull{n} outside: \num{1.000}\\ 
	\Acrfull{musp}: \SI{1.10}{\per\milli\meter}\\ 
	\Acrfull{mua}: \SI{0.011}{\per\milli\meter}\\ 
	}\label{fig:TD_SS_T_3rd}
\end{figure*}

%% file: TD_DS_T_3rd.tex
\begin{figure*}
	\begin{center}
		\includegraphics{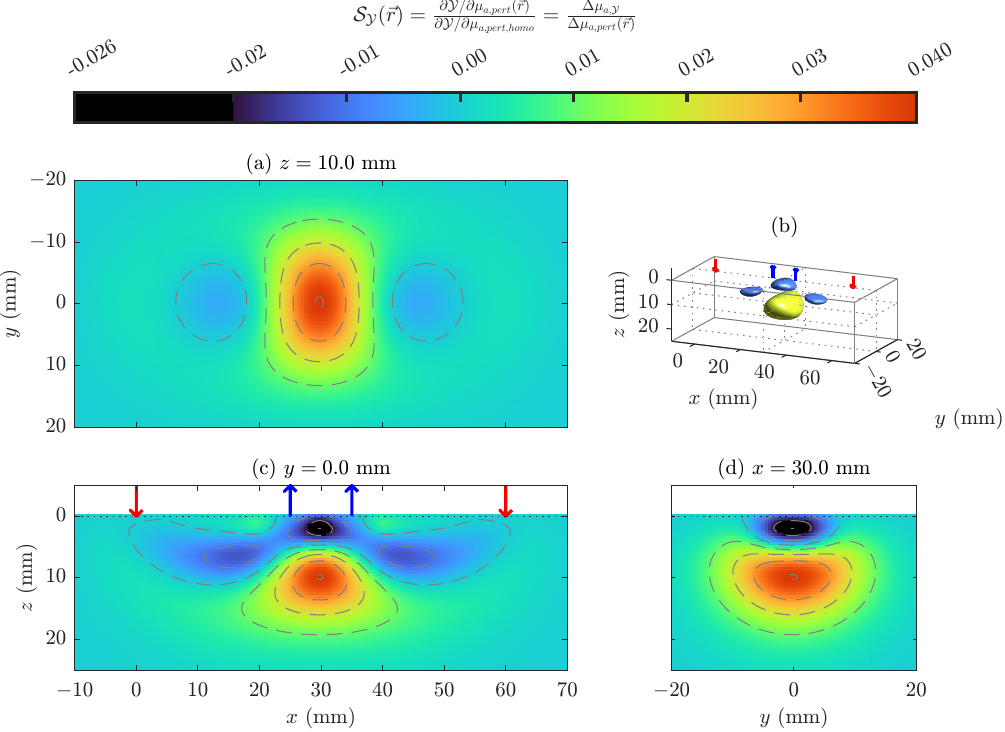}
	\end{center}
	\caption{Third angle projection of the \acrfull{sen} to a $\SI{10.0}{\milli\meter}\times\SI{10.0}{\milli\meter}\times\SI{2.0}{\milli\meter}$ perturbation scanned \SI{0.5}{\milli\meter} measured by \acrfull{TD} \acrfull{DS} \acrfull{mTOF}. (a) $x$-$y$ plane sliced at $z=\SI{10.0}{\milli\meter}$. (b) Iso-surface sliced at $\as{S}=0.020$ and $\as{S}=-0.010$. (c) $x$-$z$ plane sliced at $y=\SI{0.0}{\milli\meter}$. (d) $y$-$z$ plane sliced at $x=\SI{30.0}{\milli\meter}$. Generated using \acrfull{DT}.\\ 
	\Acrfullpl{rho}: [25.0, 35.0, 35.0, 25.0]~\si{\milli\meter}\\ 
	\Acrfull{n} inside: \num{1.333};\quad	\Acrfull{n} outside: \num{1.000}\\ 
	\Acrfull{musp}: \SI{1.10}{\per\milli\meter};\quad	\Acrfull{mua}: \SI{0.011}{\per\milli\meter}\\ 
	}\label{fig:TD_DS_T_3rd}
\end{figure*}

%% file: TD_SS_V_3rd.tex
\begin{figure*}
	\begin{center}
		\includegraphics{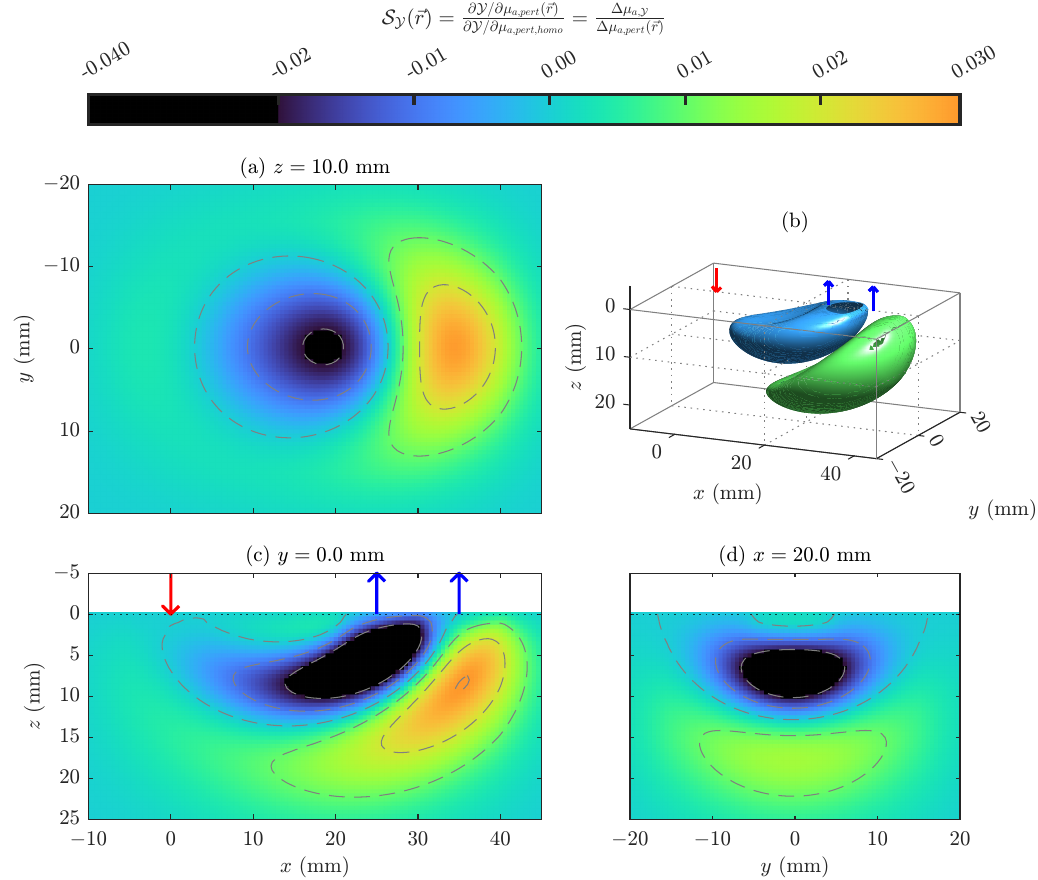}
	\end{center}
	\caption{Third angle projection of the \acrfull{sen} to a $\SI{10.0}{\milli\meter}\times\SI{10.0}{\milli\meter}\times\SI{2.0}{\milli\meter}$ perturbation scanned \SI{0.5}{\milli\meter} measured by \acrfull{TD} \acrfull{SS} \acrfull{var}. (a) $x$-$y$ plane sliced at $z=\SI{10.0}{\milli\meter}$. (b) Iso-surface sliced at $\as{S}=0.010$ and $\as{S}=-0.005$. (c) $x$-$z$ plane sliced at $y=\SI{0.0}{\milli\meter}$. (d) $y$-$z$ plane sliced at $x=\SI{20.0}{\milli\meter}$. Generated using \acrfull{DT}.\\ 
	\Acrfullpl{rho}: [25.0, 35.0]~\si{\milli\meter}\\ 
	\Acrfull{n} inside: \num{1.333};\quad	\Acrfull{n} outside: \num{1.000}\\ 
	\Acrfull{musp}: \SI{1.10}{\per\milli\meter};\quad	\Acrfull{mua}: \SI{0.011}{\per\milli\meter}\\ 
	}\label{fig:TD_SS_V_3rd}
\end{figure*}

%% file: TD_DS_V_3rd.tex
\begin{figure*}
	\begin{center}
		\includegraphics{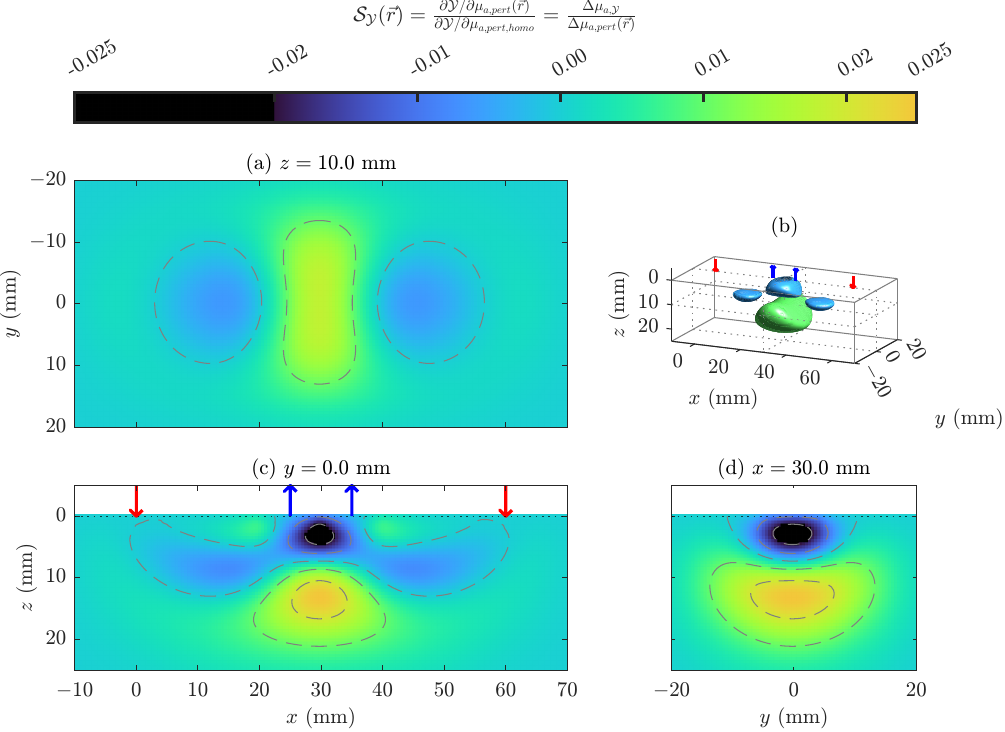}
	\end{center}
	\caption{Third angle projection of the \acrfull{sen} to a $\SI{10.0}{\milli\meter}\times\SI{10.0}{\milli\meter}\times\SI{2.0}{\milli\meter}$ perturbation scanned \SI{0.5}{\milli\meter} measured by \acrfull{TD} \acrfull{DS} \acrfull{var}.(a) $x$-$y$ plane sliced at $z=\SI{10.0}{\milli\meter}$. (b) Iso-surface sliced at $\as{S}=0.010$ and $\as{S}=-0.005$. (c) $x$-$z$ plane sliced at $y=\SI{0.0}{\milli\meter}$. (d) $y$-$z$ plane sliced at $x=\SI{30.0}{\milli\meter}$. Generated using \acrfull{DT}.\\ 
	\Acrfullpl{rho}: [25.0, 35.0, 35.0, 25.0]~\si{\milli\meter}\\ 
	\Acrfull{n} inside: \num{1.333}\\ 
	\Acrfull{n} outside: \num{1.000}\\ 
	\Acrfull{musp}: \SI{1.10}{\per\milli\meter}\\ 
	\Acrfull{mua}: \SI{0.011}{\per\milli\meter}\\ 
	}\label{fig:TD_DS_V_3rd}
\end{figure*}

%% file: TD_SS_V_mrho.tex
\begin{figure*}
	\begin{center}
		\includegraphics{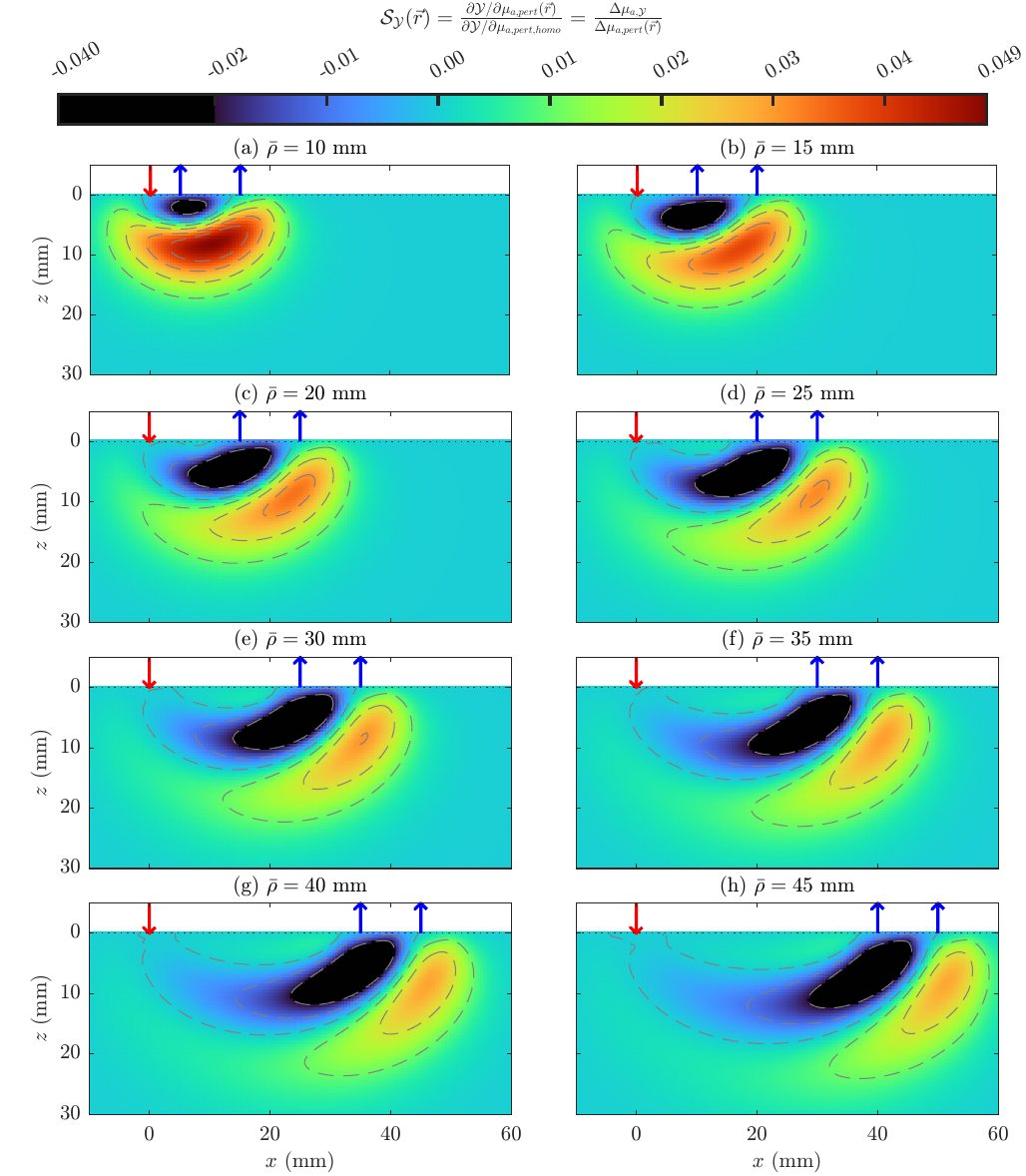}
	\end{center}
	\caption{$x$-$z$ plane of the \acrfull{sen} to a $\SI{10.0}{\milli\meter}\times\SI{10.0}{\milli\meter}\times\SI{2.0}{\milli\meter}$ perturbation scanned \SI{0.5}{\milli\meter} measured by \acrfull{TD} \acrfull{SS} \acrfull{var}. (a)-(h) Different values of \acrfull{mrho}. Generated using \acrfull{DT}.\\ 
	\Acrfull{drho}: \SI{10.0}{\milli\meter}\\ 
	\Acrfull{n} inside: \num{1.333};\quad	\Acrfull{n} outside: \num{1.000}\\ 
	\Acrfull{musp}: \SI{1.10}{\per\milli\meter};\quad	\Acrfull{mua}: \SI{0.011}{\per\milli\meter}\\ 
	}\label{fig:TD_SS_V_mrho}
\end{figure*}

%% file: TD_DS_V_mrho.tex
\begin{figure*}
	\begin{center}
		\includegraphics{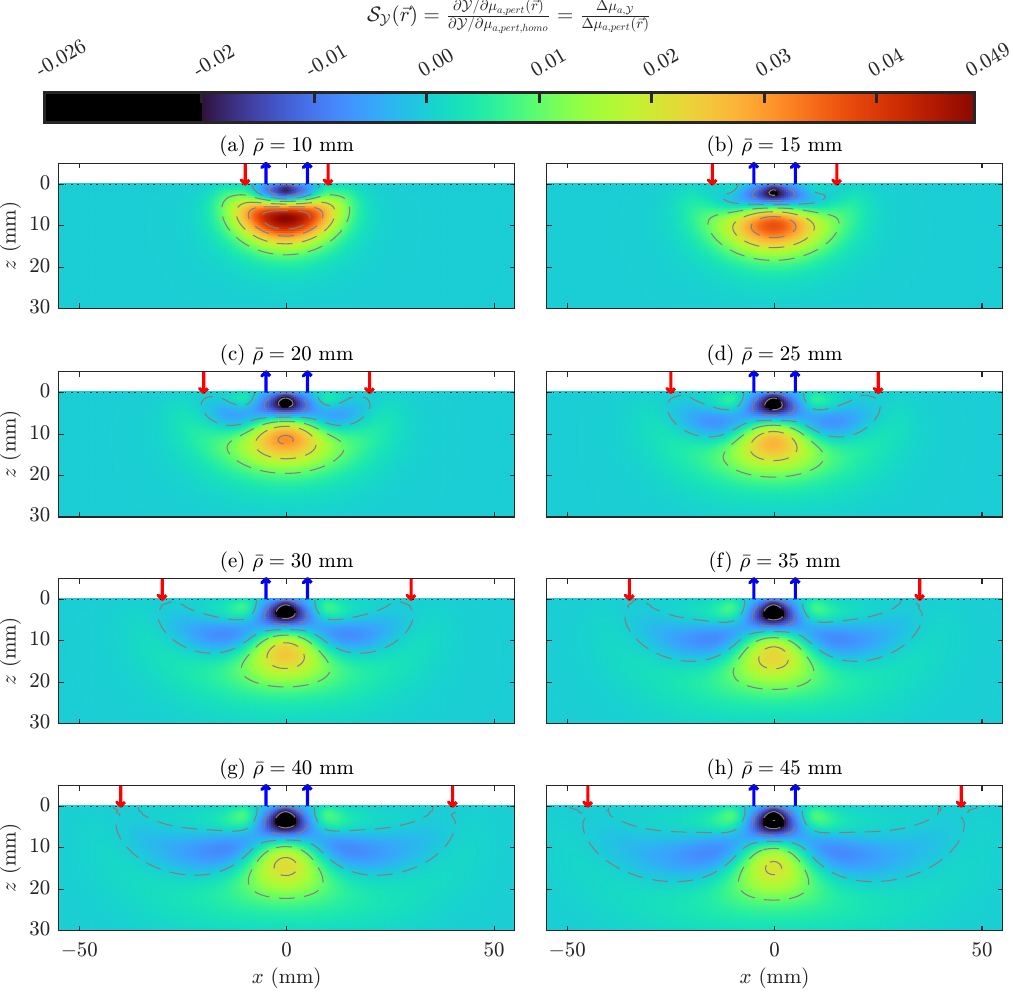}
	\end{center}
	\caption{$x$-$z$ plane of the \acrfull{sen} to a $\SI{10.0}{\milli\meter}\times\SI{10.0}{\milli\meter}\times\SI{2.0}{\milli\meter}$ perturbation scanned \SI{0.5}{\milli\meter} measured by \acrfull{TD} \acrfull{DS} \acrfull{var}. (a)-(h) Different values of \acrfull{mrho}. Generated using \acrfull{DT}.\\ 
	\Acrfull{drho}: \SI{10.0}{\milli\meter}\\ 
	\Acrfull{n} inside: \num{1.333};\quad	\Acrfull{n} outside: \num{1.000}\\ 
	\Acrfull{musp}: \SI{1.10}{\per\milli\meter};\quad	\Acrfull{mua}: \SI{0.011}{\per\milli\meter}\\ 
	}\label{fig:TD_DS_V_mrho}
\end{figure*}

%% file: TD_SS_V_drho.tex
\begin{figure*}
	\begin{center}
		\includegraphics{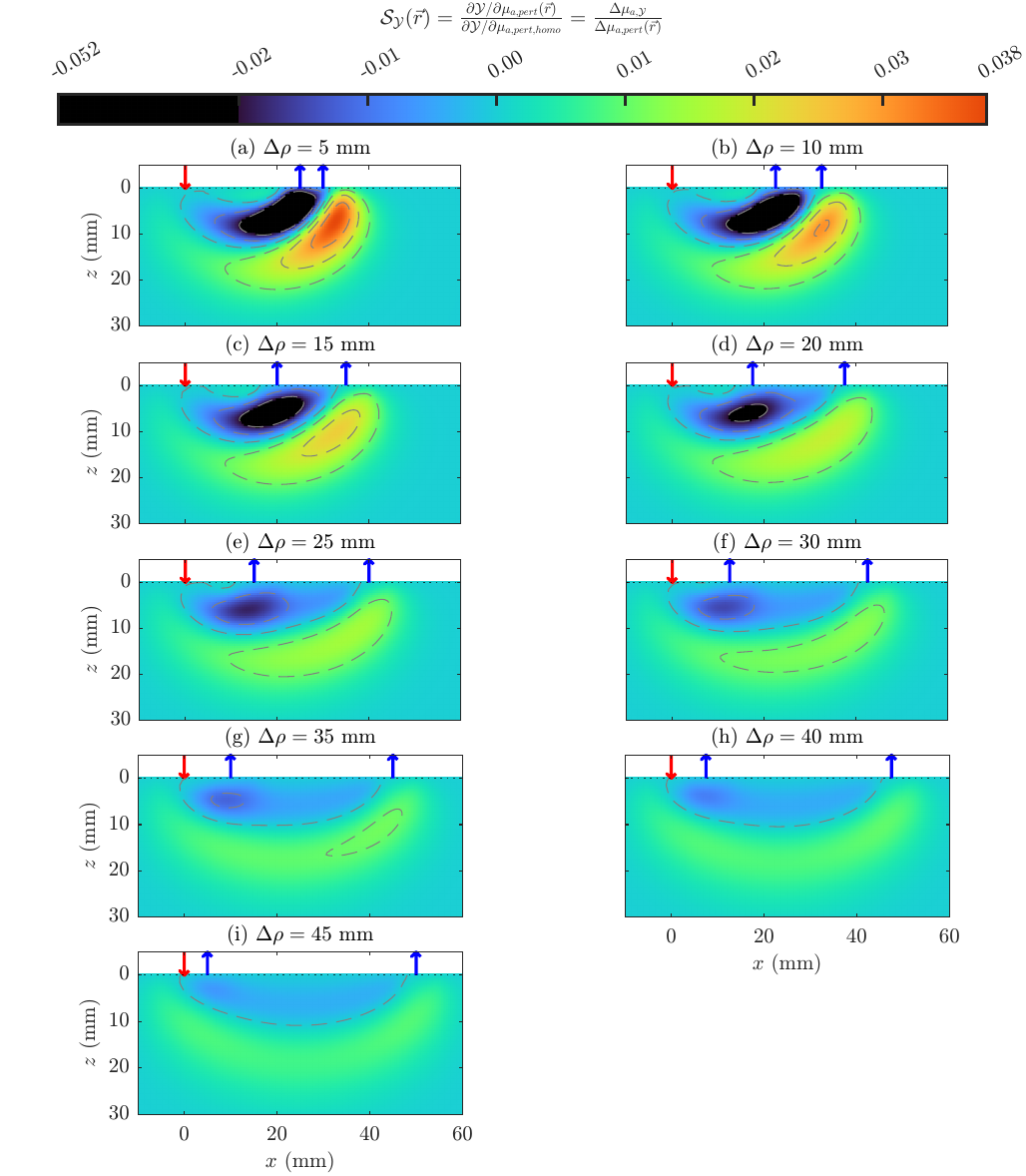}
	\end{center}
	\caption{$x$-$z$ plane of the \acrfull{sen} to a $\SI{10.0}{\milli\meter}\times\SI{10.0}{\milli\meter}\times\SI{2.0}{\milli\meter}$ perturbation scanned \SI{0.5}{\milli\meter} measured by \acrfull{TD} \acrfull{SS} \acrfull{var}. (a)-(i) Different values of \acrfull{drho}. Generated using \acrfull{DT}.\\ 
	\Acrfull{mrho}: \SI{27.5}{\milli\meter}\\ 
	\Acrfull{n} inside: \num{1.333};\quad	\Acrfull{n} outside: \num{1.000}\\ 
	\Acrfull{musp}: \SI{1.10}{\per\milli\meter};\quad	\Acrfull{mua}: \SI{0.011}{\per\milli\meter}\\ 
	}\label{fig:TD_SS_V_drho}
\end{figure*}

%% file: TD_DS_V_drho.tex
\begin{figure*}
	\begin{center}
		\includegraphics{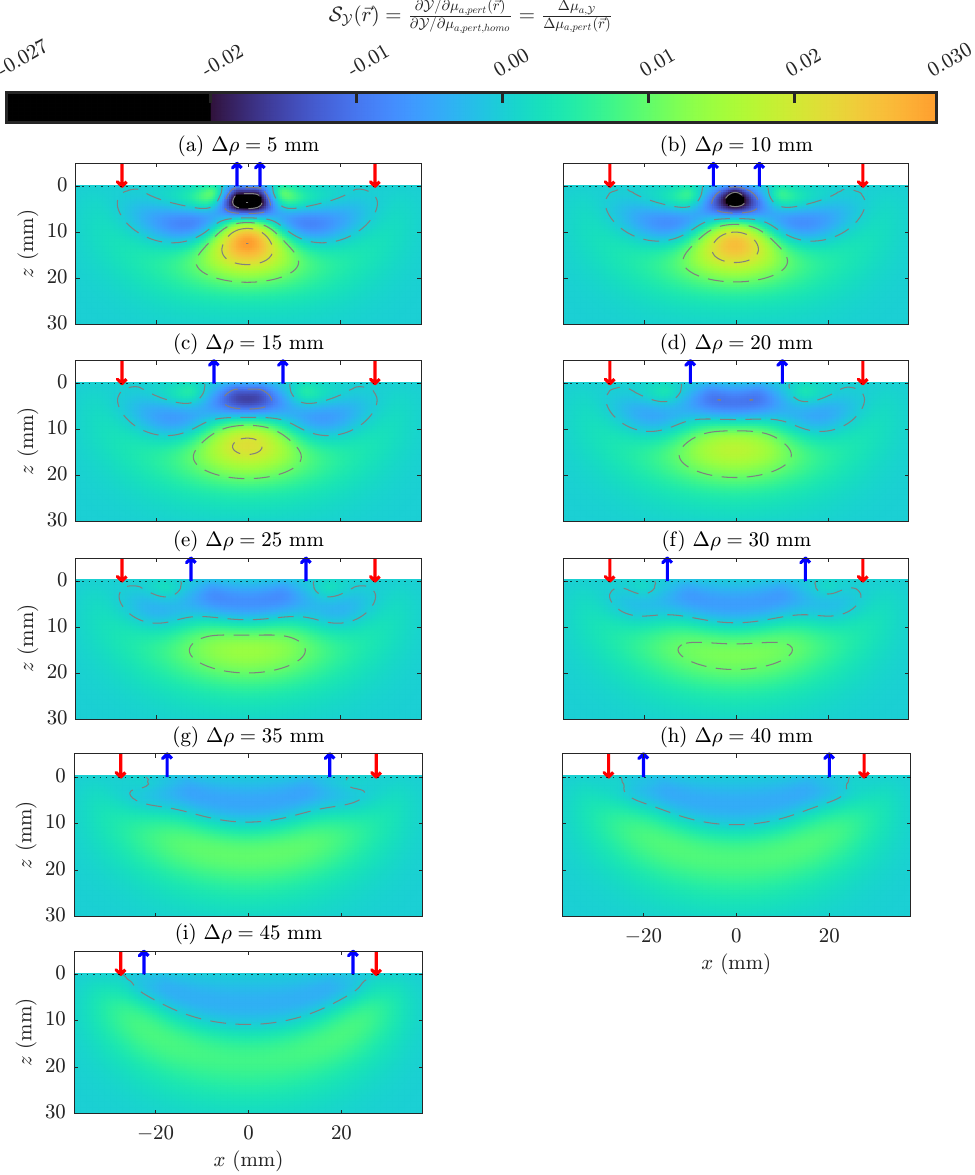}
	\end{center}
	\caption{$x$-$z$ plane of the \acrfull{sen} to a $\SI{10.0}{\milli\meter}\times\SI{10.0}{\milli\meter}\times\SI{2.0}{\milli\meter}$ perturbation scanned \SI{0.5}{\milli\meter} measured by \acrfull{TD} \acrfull{DS} \acrfull{var}. (a)-(i) Different values of \acrfull{drho}. Generated using \acrfull{DT}.\\ 
	\Acrfull{mrho}: \SI{27.5}{\milli\meter}\\ 
	\Acrfull{n} inside: \num{1.333};\quad	\Acrfull{n} outside: \num{1.000}\\ 
	\Acrfull{musp}: \SI{1.10}{\per\milli\meter};\quad	\Acrfull{mua}: \SI{0.011}{\per\milli\meter}\\ 
	}\label{fig:TD_DS_V_drho}
\end{figure*}